%% file: main.tex
\pdfoutput=1
\documentclass[12pt,a4paper]{article}
\usepackage{tikz}
\usetikzlibrary{backgrounds,fit,positioning}
\usetikzlibrary{shapes.geometric, arrows}

\DeclareGraphicsExtensions{.jpg,.pdf,.png}

\tikzstyle{startstop} = [rectangle, rounded corners, minimum width=3cm, minimum height=1cm, text centered, text width=3cm, draw=black, fill=black!5!white, font=\footnotesize]
\tikzstyle{process} = [rectangle, minimum width=3cm, minimum height=1cm, text centered, text width=3cm, draw=black, fill=black!5!white, font=\footnotesize]
\tikzstyle{decision} = [diamond, aspect=3, minimum width=0.8cm, minimum height=0.3cm, text centered, text width=3cm, draw=black, fill=white!90!yellow, font=\footnotesize, yshift=-0.2cm]

\tikzstyle{optional_decision} = [diamond, dashed, aspect=3, minimum width=0.8cm, minimum height=0.3cm, text centered, text width=2.5cm, draw=black, fill=white!90!yellow, font=\footnotesize, yshift=-0.2cm]
\tikzstyle{process_description} = [rectangle, minimum width=3cm, minimum height=1cm, text centered, text width=3cm, font=\footnotesize]
\tikzstyle{process_cpu} = [rectangle, minimum width=3cm, minimum height=1cm, text centered, text width=3cm, draw=black, fill=white!85!blue, font=\footnotesize]

\tikzstyle{arrow} = [thick,->,>=stealth]
\tikzstyle{dashed_arrow} = [thick,->,>=stealth,dashed]

\usepackage{ifthen} % for conditional statements
\newboolean{pdflatex}
\setboolean{pdflatex}{true} % False for eps figures 

\newboolean{articletitles}
\setboolean{articletitles}{true} % False removes titles in references

\newboolean{uprightparticles}
\setboolean{uprightparticles}{false} %True for upright particle symbols

\newboolean{inbibliography}
\setboolean{inbibliography}{false} %True once you enter the bibliography

\def\paperauthors{LHCb collaboration} % Leave as is for PAPER, CONF and FIGURE
\def\paperasciititle{The LHCb Upgrade} % Set ASCII title here !! MAKE sure it's only ASCII characters !! 
\def\papertitle{The LHCb VELO Upgrade Module Construction} % Latex formatted title
\def\paperkeywords{{High Energy Physics}, {LHCb}} % Comma separated list
\def\papercopyright{\the\year\ CERN for the benefit of the LHCb collaboration} % new since 9/Apr/2018
\def\paperlicence{CC BY 4.0 licence}
\def\paperlicenceurl{https://creativecommons.org/licenses/by/4.0/}

\newcommand{\lhcborcid}[1]{\href{https://orcid.org/#1}{\hspace*{0.1em}\raisebox{-0.45ex}{\includegraphics[width=1em]{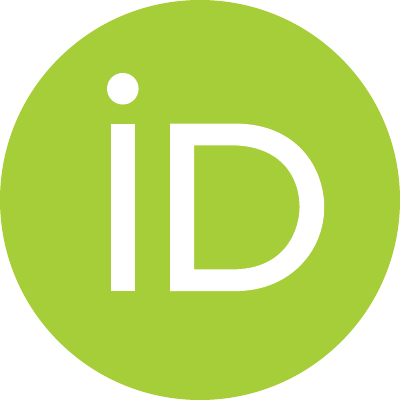}}}}

\input{preamble}
\usepackage{longtable} % only for template; not usually to be used in PAPERs
\usepackage{tablefootnote}
\usepackage{caption}
\usepackage{subcaption}
\usepackage{wrapfig}
\usepackage{gensymb}
\usepackage[capitalise, noabbrev]{cleveref}
\begin{document}

%%%%%%%%%%%%%%%%%%%%%%%%%
%%%%% Title     %%%%%%%%%
%%%%%%%%%%%%%%%%%%%%%%%%%
\renewcommand{\thefootnote}{\fnsymbol{footnote}}
\setcounter{footnote}{1}

% %%%%%%% CHOOSE TITLE PAGE--------
%\onecolumn
%\input{title-LHCb-INT}
%\input{title-LHCb-ANA}
%\input{title-LHCb-CONF}
%\input{title-LHCb-FIGURE}
\input{title-LHCb-PAPER}

%\twocolumn
% %%%%%%%%%%%%% ---------

\renewcommand{\thefootnote}{\arabic{footnote}}
\setcounter{footnote}{0}

%%%%%%%%%%%%%%%%%%%%%%%%%%%%%%%%
%%%%%  Table of Content   %%%%%%
%%%%%%%%%%%%%%%%%%%%%%%%%%%%%%%%
%%%% Uncomment if desired
\tableofcontents
\cleardoublepage

%%%%%%%%%%%%%%%%%%%%%%%%%
%%%%% Main text %%%%%%%%%
%%%%%%%%%%%%%%%%%%%%%%%%%

\pagestyle{plain} % restore page numbers for the main text
\setcounter{page}{1}
\pagenumbering{arabic}

%% Uncomment during review phase. 
%% Comment before a final submission.
%\linenumbers

%% This is the main body
%% It is useful to have a single file so comments are not missed in overleaf.
\input{body}

% Do not include this in any draft (just for information in the template)
%\input{acknowledgements_intro}
% Comment this in for paper drafts; do not include this in analysis note, conference and figure reports
%\input{acknowledgements}

%\input{supplementary}

%\input{appendix}

% This should be taken out in the final paper
%\input{supplementary-app}

\addcontentsline{toc}{section}{References}
\setboolean{inbibliography}{true}
\bibliographystyle{LHCb}
%\bibliography{main,standard,LHCb-PAPER,LHCb-CONF,LHCb-DP,LHCb-TDR,VELO/Velo}
%\bibliography{VELO/Velo}
\bibliography{main}

\end{document}

%% file: preamble.tex
\usepackage[top=1in, bottom=1.25in, left=1in, right=1in]{geometry}

\usepackage{microtype}
\usepackage{lineno}  % for line numbering during review
\usepackage{xspace} % To avoid problems with missing or double spaces after
\usepackage{caption} %these three command get the figure and table captions automatically small

\usepackage{graphicx}  % to include figures (can also use other packages)
\usepackage{color}
\usepackage{colortbl}
\graphicspath{{./figs/}} % Make Latex search fig subdir for figures
\usepackage{amsmath} % Adds a large collection of math symbols
\usepackage{amssymb}
\usepackage{amsfonts}
\usepackage{upgreek} % Adds in support for greek letters in roman typeset

\newcommand*\patchAmsMathEnvironmentForLineno[1]{%
\expandafter\let\csname old#1\expandafter\endcsname\csname #1\endcsname
\expandafter\let\csname oldend#1\expandafter\endcsname\csname
end#1\endcsname
 \renewenvironment{#1}%
   {\linenomath\csname old#1\endcsname}%
   {\csname oldend#1\endcsname\endlinenomath}%
}
\newcommand*\patchBothAmsMathEnvironmentsForLineno[1]{%
  \patchAmsMathEnvironmentForLineno{#1}%
  \patchAmsMathEnvironmentForLineno{#1*}%
}
\AtBeginDocument{%
\patchBothAmsMathEnvironmentsForLineno{equation}%
\patchBothAmsMathEnvironmentsForLineno{align}%
\patchBothAmsMathEnvironmentsForLineno{flalign}%
\patchBothAmsMathEnvironmentsForLineno{alignat}%
\patchBothAmsMathEnvironmentsForLineno{gather}%
\patchBothAmsMathEnvironmentsForLineno{multline}%
\patchBothAmsMathEnvironmentsForLineno{eqnarray}%
}

\usepackage{hyperxmp}

\usepackage[pdftex,
            pdfauthor={\paperauthors},
            pdftitle={\paperasciititle},
            pdfkeywords={\paperkeywords},
            pdfcopyright={Copyright (C) \papercopyright},
            pdflicenseurl={\paperlicenceurl}]{hyperref}
\usepackage[colorinlistoftodos,textsize=scriptsize]{todonotes}

\usepackage[bottom,flushmargin,hang,multiple]{footmisc}

\usepackage[all]{hypcap} % Internal hyperlinks to floats.

\input{lhcb-symbols-def} % Add in the predefined LHCb symbols

\usepackage{cite} % Allows for ranges in citations
\usepackage{mciteplus}
\usepackage{lipsum} 

%% file: lhcb-symbols-def.tex
\usepackage{xspace} 
\usepackage{upgreek}

\def\velo   {VELO\xspace}

\def\MagUp {\mbox{\em Mag\kern -0.05em Up}\xspace}

\ifthenelse{\boolean{uprightparticles}}%
{

 \def\PDelta      {\ensuremath{\Delta}\xspace}                 
 \def\PXi         {\ensuremath{\Xi}\xspace}                 
 \def\PLambda     {\ensuremath{\Lambda}\xspace}                 
 \def\PSigma      {\ensuremath{\Sigma}\xspace}                 
 \def\POmega      {\ensuremath{\Omega}\xspace}                 
 \def\PUpsilon    {\ensuremath{\Upsilon}\xspace}

 \def\PB      {\ensuremath{\mathrm{B}}\xspace}                 
                  
 \def\PD      {\ensuremath{\mathrm{D}}\xspace}

 \def\PK      {\ensuremath{\mathrm{K}}\xspace}

 \def\Pi      {\ensuremath{\mathrm{i}}\xspace}

 \def\Ps      {\ensuremath{\mathrm{s}}\xspace}

 \def\thebaroffset{0.0em}
}
{

 \mathchardef\PDelta="7101
 \mathchardef\PXi="7104
 \mathchardef\PLambda="7103
 \mathchardef\PSigma="7106
 \mathchardef\POmega="710A
 \mathchardef\PUpsilon="7107
                  
 \def\PB      {\ensuremath{B}\xspace}                 
                  
 \def\PD      {\ensuremath{D}\xspace}

 \def\PK      {\ensuremath{K}\xspace}

 \def\Pi      {\ensuremath{i}\xspace}

 \def\Ps      {\ensuremath{s}\xspace}

 \def\thebaroffset{0.18em}
}
\newcommand{\offsetoverline}[2][\thebaroffset]{\kern #1\overline{\kern -#1 #2}}%

\makeatletter
\ifcase \@ptsize \relax% 10pt
  \newcommand{\miniscule}{\@setfontsize\miniscule{4}{5}}% \tiny: 5/6
\or% 11pt
  \newcommand{\miniscule}{\@setfontsize\miniscule{5}{6}}% \tiny: 6/7
\or% 12pt
  \newcommand{\miniscule}{\@setfontsize\miniscule{5}{6}}% \tiny: 6/7
\fi
\makeatother

\DeclareRobustCommand{\optbar}[1]{\shortstack{{\miniscule (\rule[.5ex]{1.25em}{.18mm})}
  \\ [-.7ex] $#1$}}

\def\squark    {{\ensuremath{\Ps}}\xspace}

\def\KorKbar {\kern \thebaroffset\optbar{\kern -\thebaroffset \PK}{}\xspace}

\def\D       {{\ensuremath{\PD}}\xspace}

\def\DorDbar {\kern \thebaroffset\optbar{\kern -\thebaroffset \PD}\xspace}

\def\Dp      {{\ensuremath{\D^+}}\xspace}
\def\Dm      {{\ensuremath{\D^-}}\xspace}

\def\DpDm    {\ensuremath{\Dp {\kern -0.16em \Dm}}\xspace}

\def\B       {{\ensuremath{\PB}}\xspace}

\def\BorBbar {\kern \thebaroffset\optbar{\kern -\thebaroffset \PB}\xspace}

\def\Bd      {{\ensuremath{\B^0}}\xspace}

\def\BdorBdbar {\kern \thebaroffset\optbar{\kern -\thebaroffset \Bd}\xspace}

\def\Bs      {{\ensuremath{\B^0_\squark}}\xspace}

\def\BsorBsbar {\kern \thebaroffset\optbar{\kern -\thebaroffset \Bs}\xspace}

\def\Y#1S{\ensuremath{\PUpsilon{(#1S)}}\xspace}

\def\LorLbar     {\kern \thebaroffset\optbar{\kern -\thebaroffset \PLambda}\xspace}

\def\CP                {{\ensuremath{C\!P}}\xspace}

\def\AT#1     {\ensuremath{A_{\mathrm{T}}^{#1}}\xspace}           % 2

\def\C#1      {\ensuremath{\mathcal{C}_{#1}}\xspace}                       % 9
\def\Cp#1     {\ensuremath{\mathcal{C}_{#1}^{'}}\xspace}                    % 7
\def\Ceff#1   {\ensuremath{\mathcal{C}_{#1}^{\mathrm{(eff)}}}\xspace}        % 9  
\def\Cpeff#1  {\ensuremath{\mathcal{C}_{#1}^{'\mathrm{(eff)}}}\xspace}       % 7
\def\Ope#1    {\ensuremath{\mathcal{O}_{#1}}\xspace}                       % 2
\def\Opep#1   {\ensuremath{\mathcal{O}_{#1}^{'}}\xspace}                    % 7

\newcommand{\nospaceunit}[1]{\ensuremath{\text{#1}}}       
\newcommand{\aunit}[1]{\ensuremath{\text{\,#1}}}       
\newcommand{\tev}{\aunit{Te\kern -0.1em V}\xspace}
\newcommand{\gev}{\aunit{Ge\kern -0.1em V}\xspace}
\newcommand{\mev}{\aunit{Me\kern -0.1em V}\xspace}
\newcommand{\kev}{\aunit{ke\kern -0.1em V}\xspace}
\newcommand{\ev}{\aunit{e\kern -0.1em V}\xspace}
 
\newcommand{\mevc}{\ensuremath{\aunit{Me\kern -0.1em V\!/}c}\xspace}
\newcommand{\gevc}{\ensuremath{\aunit{Ge\kern -0.1em V\!/}c}\xspace}
\newcommand{\mevcc}{\ensuremath{\aunit{Me\kern -0.1em V\!/}c^2}\xspace}
\newcommand{\gevcc}{\ensuremath{\aunit{Ge\kern -0.1em V\!/}c^2}\xspace}
\def\m    {\aunit{m}\xspace}

\def\cm   {\aunit{cm}\xspace}

\def\mm   {\aunit{mm}\xspace}

\def\mum  {\ensuremath{\,\upmu\nospaceunit{m}}\xspace}

\def\nm   {\aunit{nm}\xspace}

\def\fb   {\ensuremath{\aunit{fb}}\xspace}
\def\invfb   {\ensuremath{\fb^{-1}}\xspace}

\def\ns   {\ensuremath{\aunit{ns}}\xspace}

\def\mhz  {\ensuremath{\aunit{MHz}}\xspace}

\def\gsim{{~\raise.15em\hbox{$>$}\kern-.85em
          \lower.35em\hbox{$\sim$}~}\xspace}
\def\lsim{{~\raise.15em\hbox{$<$}\kern-.85em
          \lower.35em\hbox{$\sim$}~}\xspace}

\def\murad{\ensuremath{\,\upmu\nospaceunit{rad}}\xspace}

\def\tell1  {TELL1\xspace}
\def\ukl1   {UKL1\xspace}

%% file: title-LHCb-PAPER.tex
% ===============================================================================
% Purpose: LHCb-PAPER journal paper title page template
% Author: 
% Created on: 2010-09-25
% ===============================================================================

%%%%%%%%%%%%%%%%%%%%%%%%%
%%%%%  TITLE PAGE  %%%%%%
%%%%%%%%%%%%%%%%%%%%%%%%%
\begin{titlepage}
\pagenumbering{roman}

% Header ---------------------------------------------------
\vspace*{-1.5cm}
\centerline{\large EUROPEAN ORGANIZATION FOR NUCLEAR RESEARCH (CERN)}
\vspace*{1.5cm}
\noindent
\begin{tabular*}{\linewidth}{lc@{\extracolsep{\fill}}r@{\extracolsep{0pt}}}
\ifthenelse{\boolean{pdflatex}}% Logo format choice
{\vspace*{-1.5cm}\mbox{\!\!\!\includegraphics[width=.14\textwidth]{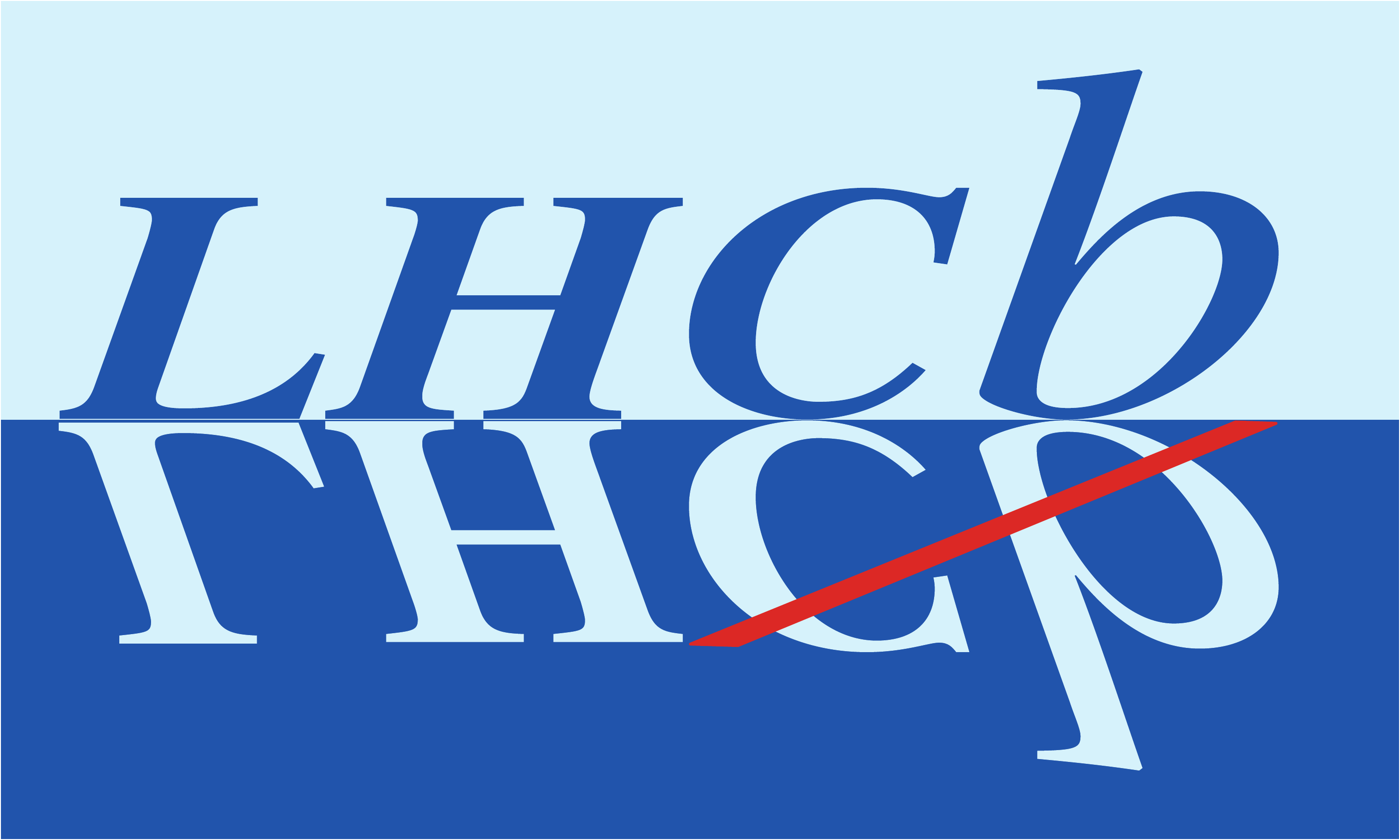}} & &}%
{\vspace*{-1.2cm}\mbox{\!\!\!\includegraphics[width=.12\textwidth]{VELO/figs/lhcb-logo.pdf}} & &}%
\\
% & & CERN-EP-20XX-ZZZ \\  % ID 
 & & LHCb-DP-2024-001 \\  % ID 
% & & \today \\ % Date - Can also hardwire e.g.: 23 March 2010
 & & April 19, 2024 \\ % Date - Can also hardwire e.g.: 23 March 2010
 & & \\
% not in paper \hline
\end{tabular*}

\vspace*{4.0cm}

% Title --------------------------------------------------
{\normalfont\bfseries\boldmath\huge
\begin{center}
% DO NOT EDIT HERE. Instead edit macro in main.tex to keep metadata correct
  \papertitle 
\end{center}
}

%\vspace*{1.0cm}

% Authors -------------------------------------------------
\begin{center}
%In the footnote, replace 'paper' by 'Letter' in case of submission to PRL or PLB 
% Edit macro in main.tex to keep metadata correct
%\paperauthors\footnote{Authors are listed at the end of this paper.}
%LHCb VELO Group, author list to be determined
\input{VELO/Authorship_ModuleConstructionPaper}

\end{center}

\vspace{\fill}

% Abstract -----------------------------------------------
\begin{abstract}
  \noindent

 The LHCb detector has undergone a major upgrade for LHC Run~3. This Upgrade I detector facilitates operation at higher luminosity and utilises full-detector information at the LHC collision rate, critically including the use of vertex information. A new vertex locator system, the VELO Upgrade, has been constructed. The core element of the new VELO are the double-sided pixelated hybrid silicon detector modules which operate in vacuum close to the LHC beam in a high radiation environment. The construction and quality assurance tests of these modules are described in this paper. The modules incorporate 200\mum thick, n-on-p silicon sensors bump-bonded to 130\nm technology ASICs. These are attached with high precision to a silicon microchannel substrate that uses evaporative CO$_2$ cooling. The ASICs are controlled and read out with flexible printed circuits that are glued to the substrate and wire-bonded to the chips. The mechanical support of the module is given by a carbon fibre plate, two carbon fibre rods and an aluminium plate. The sensor attachment was achieved with an average precision of 21$\,\mathrm{\mu m}$, more than 99.5\% of all pixels are fully functional, and a thermal figure of merit of  3$\,\mathrm{Kcm^{2}W^{-1}}$ was achieved. The production of the modules was successfully completed in 2021, with
the final assembly and installation completed in time for data taking in
2022.
  
\end{abstract}

\vspace*{2.0cm}

\begin{center}
  Submitted to JINST
\end{center}

\vspace{\fill}

{\footnotesize 
% Edit macro in main.tex to keep metadata correct
\centerline{\copyright~\papercopyright. \href{\paperlicenceurl}{\paperlicence}.}}
\vspace*{2mm}

\end{titlepage}

%%%%%%%%%%%%%%%%%%%%%%%%%%%%%%%%
%%%%%  EOD OF TITLE PAGE  %%%%%%
%%%%%%%%%%%%%%%%%%%%%%%%%%%%%%%%

%  empty page follows the title page ----
\newpage
\setcounter{page}{2}
\mbox{~}
%\newpage
%
%% Author List ----------------------------
%%  You need to get a new author list!
%\input{LHCb_authorlist.tex}
%
%The author list for journal publications is provided by the Membership Committee shortly after 'approval to go to paper' has been given.
%%It will be made available on the page
%%\verb!http://www.physik.uzh.ch/~strauman/forMemCo/LHCb-PAPER-XXXX-XXX/! .
%It will be sent to you by email shortly after a paper number has beens assigned.
%The author list should be included already at first circulation, 
%to allow new members of the collaboration to verify whether they have been included correctly.
%Occasionally a misspelled name is corrected or associated institutions become full members.
%In that case, a new author list will be sent to you.
%In case line numbering doesn't work well after including the authorlist, try moving the \verb!\bigskip! after the last author to a separate line.
%
%
%The authorship for Conference Reports should be ``The LHCb
%  collaboration'', with a footnote giving the name(s) of the contact
%  author(s), but without the full list of collaboration names.

%% file: VELO/Authorship_ModuleConstructionPaper.tex
% VELO Module Construction author list
%\centerline{\large\bf The LHCb VELO group}
\begin
{flushleft}
\small
K.~Akiba$^{2}$\lhcborcid{0000-0002-6736-471X},
M.~Alexander$^{10}$\lhcborcid{0000-0002-8148-2392},
C.~Bertella$^{12}$\lhcborcid{0000-0002-3160-147X},
A.~Biolchini$^{2}$\lhcborcid{0000-0001-6064-9993},
A.~Bitadze$^{12}$\lhcborcid{0000-0001-7979-1092},
G.~Bogdanova$^{5}$\lhcborcid{0009-0000-9524-4670},
S.~Borghi$^{12}$\lhcborcid{0000-0001-5135-1511},
T.J.V.~Bowcock$^{11}$\lhcborcid{0000-0002-3505-6915},
K.~Bridges$^{11}$,
M.~Brock$^{13}$,
A.T.~Burke$^{12}$\lhcborcid{0000-0003-0243-0517},
J.~Buytaert$^{7}$\lhcborcid{0000-0002-7958-6790},
W.~Byczynski$^{7}$\lhcborcid{0009-0008-0187-3395},
J.~Carroll$^{11}$,
V.~Coco$^{7}$\lhcborcid{0000-0002-5310-6808},
P.~Collins$^{7}$\lhcborcid{0000-0003-1437-4022},
A.~Davis$^{12}$\lhcborcid{0000-0001-9458-5115},
O.~De~Aguiar~Francisco$^{12}$\lhcborcid{0000-0003-2735-678X},
K.~De~Bruyn$^{14}$\lhcborcid{0000-0002-0615-4399},
S.~De~Capua$^{12}$\lhcborcid{0000-0002-6285-9596},
K.~De~Roo$^{2}$,
F.~Doherty$^{10}$\lhcborcid{0000-0001-6470-4881},
L.~Douglas$^{10}$,
L.~Dufour$^{7}$\lhcborcid{0000-0002-3924-2774},
R.~Dumps$^{7}$,
D.~Dutta$^{12}$\lhcborcid{0000-0002-1191-3978},
L.~Eklund$^{15}$\lhcborcid{0000-0002-2014-3864},
A.~Elvin$^{12}$\lhcborcid{0009-0006-0882-3900},
S.~Farry$^{11}$\lhcborcid{0000-0001-5119-9740},
A.~Fernandez~Prieto$^{6}$\lhcborcid{0000-0003-1984-6367},
V.~Franco~Lima$^{11}$\lhcborcid{0000-0002-3761-209X},
J.~Freestone$^{12}$,
C.~Fuzipeg$^{12}$\lhcborcid{0009-0009-5347-9354},
M.D.~Galati$^{2}$\lhcborcid{0000-0002-8716-4440},
A.~Gallas~Torreira$^{6}$\lhcborcid{0000-0002-2745-7954},
R.E.~Geertsema$^{2}$\lhcborcid{0000-0001-6829-7777},
E.~Gersabeck$^{12}$\lhcborcid{0000-0002-2860-6528},
M.~Gersabeck$^{12}$\lhcborcid{0000-0002-0075-8669},
F.~Grant$^{10}$\lhcborcid{0009-0009-0474-0317},
T.~Halewood-leagas$^{11}$\lhcborcid{0000-0001-9629-7029},
K.~Hennessy$^{11}$\lhcborcid{0000-0002-1529-8087},
W.~Hulsbergen$^{2}$\lhcborcid{0000-0003-3018-5707},
D.~Hutchcroft$^{11}$\lhcborcid{0000-0002-4174-6509},
D.~Hynds$^{2}$\lhcborcid{0009-0009-0976-2312},
E.~Jans$^{2}$\lhcborcid{0000-0002-5438-9176},
D.~John$^{2}$,
M.~John$^{13}$\lhcborcid{0000-0002-8579-844X},
N.~Jurik$^{7}$\lhcborcid{0000-0002-6066-7232},
T.~Ketel$^{2}$\lhcborcid{0000-0002-9652-1964},
S.~Klaver$^{3}$\lhcborcid{0000-0001-7909-1272},
P.~Kopciewicz$^{4}$\lhcborcid{0000-0001-9092-3527},
I.~Kostiuk$^{2}$\lhcborcid{0000-0002-8767-7289},,
M.~Kraan$^{2}$,
M.~Langstaff$^{12}$\lhcborcid{0009-0006-8036-6716},
T.~Latham$^{9}$\lhcborcid{0000-0002-7195-8537},
A.~Leflat$^{5}$\lhcborcid{0000-0001-9619-6666},
E.~Lemos~Cid$^{6}$\lhcborcid{0000-0003-3001-6268},
V.~Lukashenko$^{2}$\lhcborcid{0000-0002-0630-5185},
M.~Merk$^{2}$\lhcborcid{0000-0003-0818-4695},
M.~Milovanovic$^{7}$\lhcborcid{0000-0003-1580-0898},
M.~Monk$^{9}$\lhcborcid{0000-0003-0484-0157},
D.~Murray$^{12}$\lhcborcid{0000-0002-5729-8675},
I.~Nasteva$^{1}$\lhcborcid{0000-0001-7115-7214},
A.~Oblakowska-Mucha$^{4}$\lhcborcid{0000-0003-1328-0534},
T.~Pajero$^{13}$\lhcborcid{0000-0001-9630-2000},
C.~Parkes$^{12}$\lhcborcid{0000-0003-4174-1334},
A.~Pazos~Alvarez$^{6}$\lhcborcid{0000-0003-3153-8084},
E.~Perez~Trigo$^{6}$\lhcborcid{0000-0001-7650-6639},
M.~Perry$^{12}$\lhcborcid{0009-0003-2617-0964},
F.~Reiss$^{12}$\lhcborcid{0000-0002-8395-7654},
K.~Rinnert$^{11}$\lhcborcid{0000-0001-9802-1122},
E.~Rodriguez~Rodriguez$^{6}$\lhcborcid{0000-0002-7973-8061},
J.~Rovekamp$^{2}$,
F.~Sanders$^{2}$,
L.G.~Scantlebury~Smead$^{13}$\lhcborcid{0000-0001-8702-7991},
M.~Schiller$^{10}$\lhcborcid{0000-0001-8750-863X},
T.~Shears$^{11}$\lhcborcid{0000-0002-2653-1366},
N.A.~Smith$^{11}$\lhcborcid{0000-0002-3638-809X},
A.~Snoch$^{2}$\lhcborcid{0000-0001-6431-6360},
P.~\v Svihra$^{12}$\lhcborcid{0000-0002-7811-2147},
T.~Szumlak$^{4}$\lhcborcid{0000-0002-2562-7163},
M.~van~Beuzekom$^{2}$\lhcborcid{0000-0002-0500-1286},
M.~van~Overbeek$^{2}$,
P.~Vazquez~Regueiro$^{6}$\lhcborcid{0000-0002-0767-9736},
V.~Volkov$^{5}$\lhcborcid{0009-0005-3500-5121},
M. ~Wormald$^{11}$,
G.~Zunica$^{12}$\lhcborcid{0000-0002-5972-6290}.\bigskip

{\footnotesize \it
$^{1}$Universidade Federal do Rio de Janeiro (UFRJ), Rio de Janeiro, Brazil\\
$^{2}$Nikhef National Institute for Subatomic Physics, Amsterdam, Netherlands\\
$^{3}$Nikhef National Institute for Subatomic Physics and VU University Amsterdam, Amsterdam, Netherlands\\
$^{4}$AGH - University of Science and Technology, Faculty of Physics and Applied Computer Science, Krak{\'o}w, Poland\\
$^{5}$Affiliated with an institute covered by a cooperation agreement with CERN\\
$^{6}$Instituto Galego de F{\'\i}sica de Altas Enerx{\'\i}as (IGFAE), Universidade de Santiago de Compostela, Santiago de Compostela, Spain\\
$^{7}$European Organization for Nuclear Research (CERN), Geneva, Switzerland\\
$^{8}$H.H. Wills Physics Laboratory, University of Bristol, Bristol, United Kingdom\\
$^{9}$Department of Physics, University of Warwick, Coventry, United Kingdom\\
$^{10}$School of Physics and Astronomy, University of Glasgow, Glasgow, United Kingdom\\
$^{11}$Oliver Lodge Laboratory, University of Liverpool, Liverpool, United Kingdom\\
$^{12}$Department of Physics and Astronomy, University of Manchester, Manchester, United Kingdom\\
$^{13}$Department of Physics, University of Oxford, Oxford, United Kingdom\\
$^{14}$Van Swinderen Institute, University of Groningen, Groningen, Netherlands, associated to $^{2}$\\
$^{15}$Department of Physics and Astronomy, Uppsala University, Uppsala, Sweden, associated to $^{10}$\\
\medskip
$ ^{*}$Contact author: sdecapua@cern.ch
}
\end{flushleft}

%% file: body.tex
%\setcounter{section}{2}
%\section{Vertex Locator}
\label{sec:velo}
\input{VELO/Velo}

%% file: VELO/Velo.tex
\newboolean{showvelo}
\setboolean{showvelo}{false}

\newcommand\velobudgetandauthor[2]{
\ifthenelse{\boolean{showvelo}}{\vspace{-3.5mm}{\footnotesize\color{blue}{#1 pages, authors: #2\\ \noindent}}\vspace{-1mm}}{}}
\newcommand\velotitle[1]{
\ifthenelse{\boolean{showvelo}}{\pagebreak}{}
\subsection{#1}
}
\def\um{\xspace\ensuremath{\mathrm{\mu m}}\xspace}

%% Introduction %%
\section{Introduction}
\label{sec:introduction}
\input{VELO/introduction}
\\\clearpage

%% Requirements %%
\section{Requirements}
\label{sec:requirements}
\input{VELO/requirements}
\clearpage

%% Design %%
\section{Detector design}
\label{sec:design}
\input{VELO/design}
\clearpage

%% Assembly %%
\section{Module assembly procedures}
\label{sec:assembly}
\input{VELO/assembly}
\clearpage

%% Mechanical %%
\section{Module assembly measurements}
\label{sec:metrology}
\input{VELO/metrology}
\clearpage

%% Module validation %%
\section{Module validation}
\label{sec:validation}
\input{VELO/validation}
\clearpage

%% Summary of production %%
\section{Overview of module production}
\label{sec:summary}
\input{VELO/summary}
\clearpage

%% conclusions %%
\section{Conclusions}
\label{sec:conclusions}
\input{VELO/conclusions}
\clearpage

\section{Acknowledgements}
\label{sec:acknowledgements}
\input{VELO/acknowledgements}
\clearpage

%% file: VELO/introduction.tex
The LHCb detector is a forward spectrometer located at the Large Hadron Collider (LHC) optimized for the reconstruction of b- and c- hadrons decays. The physics programme includes searches for effects beyond the Standard Model, mostly through precision studies of \CP Violation and rare decays, searches for new hadronic states, electroweak precision measurements and a broad QCD phenomenology in the forward region. A major change of the experiment has been undertaken for the LHCb Upgrade I~\cite{LHCb-DP-2022-002} which will operate for the next decade (Run 3 and Run 4). This detector incorporates a new vertex locator (VELO) system\cite{LHCb-TDR-013}, which provides crucial input to the trigger with  precise primary and secondary vertices reconstruction. The production and quality assurance of the modules of this system are described in this paper.

The LHCb Upgrade will allow the experiment to run at an instantaneous luminosity of  $2\times$10$^{33}$cm$^{-2}$s$^{-1}$, a factor five higher than in the previous runs. The experiment is expected to collect an integrated luminosity of 50 fb$^{-1}$ by the end of Run~4. While the previous experiment relied on a purely hardware based first level trigger, with a maximum readout rate of 1 MHz~\cite{LHCb-TDR-005}, a purely software trigger is now utilised. This includes the identification of heavy-flavour decays with vertex reconstruction, to expand the physics programme. 
The new trigger~\cite{LHCb-TDR-016} requires all detector systems to be upgraded for 40\,MHz readout and all front- and back-end electronics to be replaced in order to cope with the increased data rates. The Upgrade was installed during the LHC long shutdown LS2 (2019-2021), with the first data taken in 2022.
 
\begin{figure}[ht]
 \centering
     \includegraphics[width=0.99\textwidth]{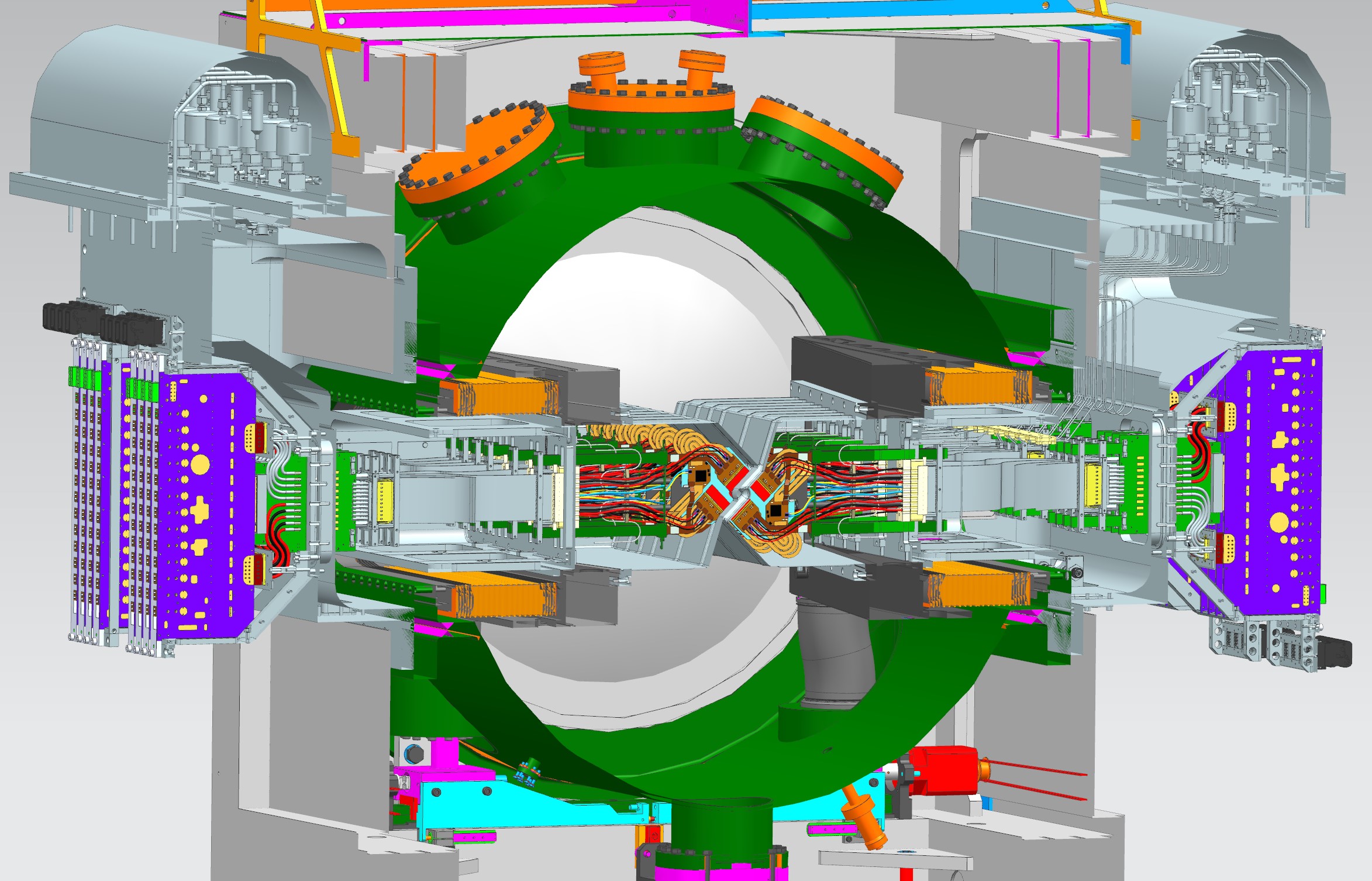}
    \caption[]{A CAD model of the layout of the VELO Upgrade detector, showing the RF foil in the centre (grey), the detector modules,  Vacuum Feedthrough Board (light green) and the Opto-and-Power Board (purple).}
  \label{fig:velo}
\end{figure}

Using the same mechanics as its predecessor, the upgraded VELO is separated into two retractable halves which surround the interaction point, as shown in Fig.~\ref{fig:velo}. Each half is enclosed in a thin corrugated RF aluminium box, that keeps the detector vacuum separated from the main LHC vacuum, as well as shielded from beam-induced currents. During LHC beam injection, the two VELO halves are kept in a retracted position at a safe distance of 30~mm from the beams and only moved inward to enclose the beams once the conditions have been declared stable. In the closed position, the sensitive elements of the detectors are as close as 5.1~mm to the beams. Each VELO half consists of 26 VELO modules, mounted on an aluminium base which allows the required movements in both horizontal and vertical directions. The electrical signals and service voltages are routed to each module through an Opto-and-Power board (OPB) and a vacuum feedthrough board (VFB). The data and signals between the OPB and LHCb data acquisition system are carried by optical fibres.

%% file: VELO/requirements.tex
The VELO plays an essential role in the LHCb tracking system, by precisely reconstructing  the location of $pp$ collision vertices and secondary displaced vertices. This information is fed to the LHCb reconstruction algorithm and provides crucial information for the event selection. 
In order to fulfill its crucial role, a complete redesign and replacement of the VELO silicon sensors and electronics was necessary. The LHCb collaboration has opted for a solution based on hybrid pixel sensors. A new radiation hard ASIC (VeloPix ~\cite{VANBEUZEKOM201392}), based on the design of the Timepix/Medipix family of chips~\cite{Gromov:2011zz}, was developed to cope with the higher data rates. The main changes with respect to the original detector are summarised in \cref{tab:mainChanges}.
\begin{table}[h]
    \centering
     \caption{Specifications of the upgraded VELO compared to those of the original version.}
    \begin{tabular}{l|c|c}
       & $2009-2018$ & $2022-2031$\\
    \hline
       RF-foil inner radius (minimum thickness) & $5.5\,\mm$ ($300\,\um$) & $3.5\,\mm$ ($150\,\um$) \\
       Inner distance of active silicon detector & $8.2\,\mm$  & $5.1\,\mm$ \\
       Total fluence (silicon tip) [$ n_{\rm eq}/\cm^{2}$]\tablefootnote{1\,MeV neutron equivalent per cm$^2$.} & $4\times 10^{14}$ &  $\sim8\times 10^{15}$ \\
       Sensor segmentation & $r-\phi$ strips & square pixels\\
       Pitch &  $37 - 97\,\mum$ & $55\,\mum$\\
       Technology & n-on-n & n-on-p \\
       Areas of active silicon detectors & $0.22\,\m ^2$ & $0.12\,\m ^2$ \\
       Number of modules & 42 & 52 \\
       Total number of channels & 172~k  & 41~M \\
       Readout rate [\mhz] & 1, analogue & 40, binary \\
       & & (zero suppressed)\\
       Whole-\velo data rate & 150~Gbit/s & $\sim2$~Tbit/s\\
       Total power consumption & $800\,$W & $\sim2$~kW
    \end{tabular}
    \label{tab:mainChanges}
\end{table}

The requirements of the VELO Upgrade system are described in~\cite{LHCb-DP-2022-002} and some key elements are summarised here. The distinctive signature of b- and c- hadrons decaying in LHCb is given by tracks that originate from secondary vertices with a large impact parameter to all primary vertices. 
The vertex resolution is improved by reducing the pixel size, the amount of material traversed by the track, the distance between the active area and the interaction point, and by increasing the distance between the measured points. The spatial resolution requirements led to the $55\,\mum$ pitch pixels, with a minimum distance of only $5.1\,\mm$ from the beams, operated in vacuum with 52 modules spread just over 1\,m, each with an individual slim support structures, and arranged to close around the beams. The choice of a $55\,\mum$ pitch led to the need of a $<30\,\mum$ (i.e. half pixel size) precision assembly and metrology, as well as stability in the operation under the thermal and mechanical environment.  

It is expected that the sensor with highest occupancy will receive on average 8.5 charged particles per bunch crossing. The bunch crossing rate in LHCb is expected to be 27~MHz on average, with a 40~MHz peak (25\ns spacing) due to variations in the LHC filling scheme. In the high occupancy regions of the detector, this leads to an expected peak pixel hit rate of $\sim900\,{\rm MHz}/{\rm ASIC}$, with the whole VELO reaching a total data rate of 2.85~Tbit/s. After 50~\invfb of integrated luminosity exposure, it is expected that some chips will have accumulated an integrated flux of $8 \times 10^{15}$~$n_{\mathrm{eq}}/\mathrm{cm}^{2}$. With such a dose, leakage currents of about $200\,\upmu{\rm A}/\cm^2$, or $\sim7$\,nA per 55$\times 55\mum^2$ square pixel, are expected at $-25\,^\circ{\rm C}$, with a bias voltage of 1000~V. For these reasons, both the VeloPix and the silicon sensors are designed to cope with high data rates and large non-uniform irradiation.

Such demanding front-end processing requires a significant amount of power and highly efficient on-detector cooling to protect the silicon from thermal runaway. The new VELO modules have a large heat generation in the front-end electronics (up to 30~W per module).
The adopted cooling solution must ensure that all modules are kept below $-20\,^{\circ}$C. Since the operating temperature of the CO$_2$ in the cooling plant is approximately $-30\,^{\circ}$C, the temperature drop between the sensor and the coolant is required to be kept below 8\,$^{\circ}$C. The solution must also ensure a minimal amount of material and avoid leaks of the coolant in the detector vacuum. The module cooling design is fully integrated within the module. The coolant, evaporative CO$_2$, is circulated within microchannels directly etched into thin silicon substrates that provide the mechanical support to the silicon sensors in the module. Silicon features an excellent heat transfer coefficient, has a high radiation length and matches the coefficient of thermal expansion of the readout electronics. The module design and its components are discussed in the following chapter. 

%% file: VELO/design.tex
\subsection{General layout}
\label{sec:velo_layout}
\begin{figure}[htbp]
\centering
\includegraphics[width=0.99\textwidth]{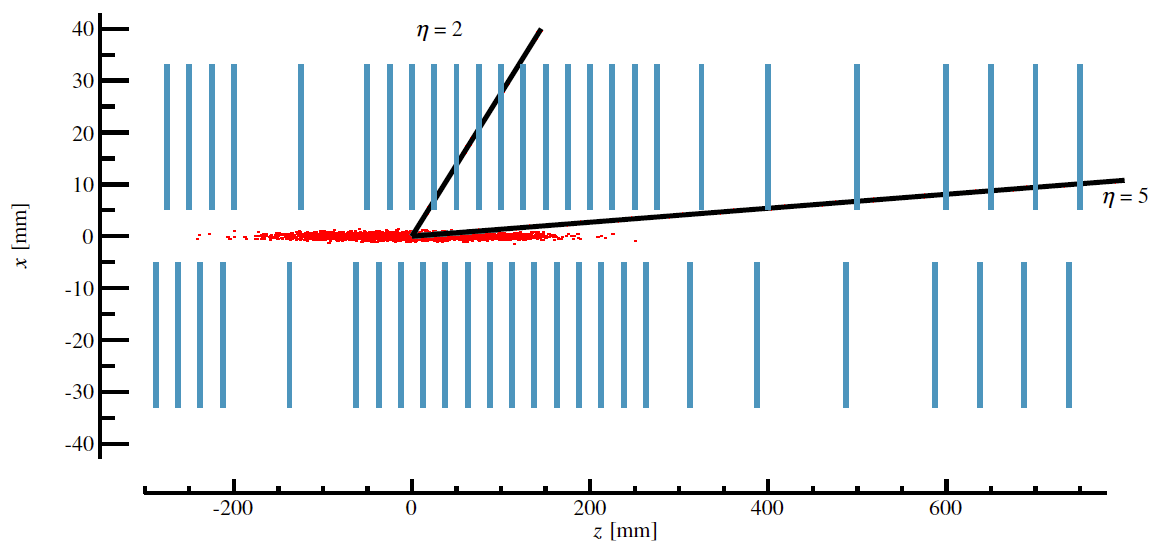}
\caption{Schematic cross-section at $y = 0$ with an illustration of the $z$-extent of the luminous region and the nominal LHCb acceptance. }
\label{fig:layout_from_PUB-2019-008}
\end{figure}

The new detector consists of 52 modules. Each module is identical in order to simplify the assembly process and quality control.
They are grouped into two retractable halves of 26 modules, as illustrated in Fig.~\ref{fig:layout_from_PUB-2019-008}. Modules are distributed perpendicular to the beamline and arranged to fully cover the LHCb pseudorapidity acceptance ($2<\eta<5$) and ensure that at least four pixel layers are crossed by any track from the interaction region, for all azimuthal directions~\cite{Bird:2256124}. A number of modules are placed upstream of the interaction point in order to improve the reconstruction of primary vertices. The minimum spacing between modules is 25~mm for both halves, although in one half they are displaced by +12.5~mm in $z$ to allow for overlap when the detector is in the closed configuration and thus provide a full azimuthal coverage. 

\begin{figure}[htbp]
    \centering
    \includegraphics[width=0.4\textwidth]{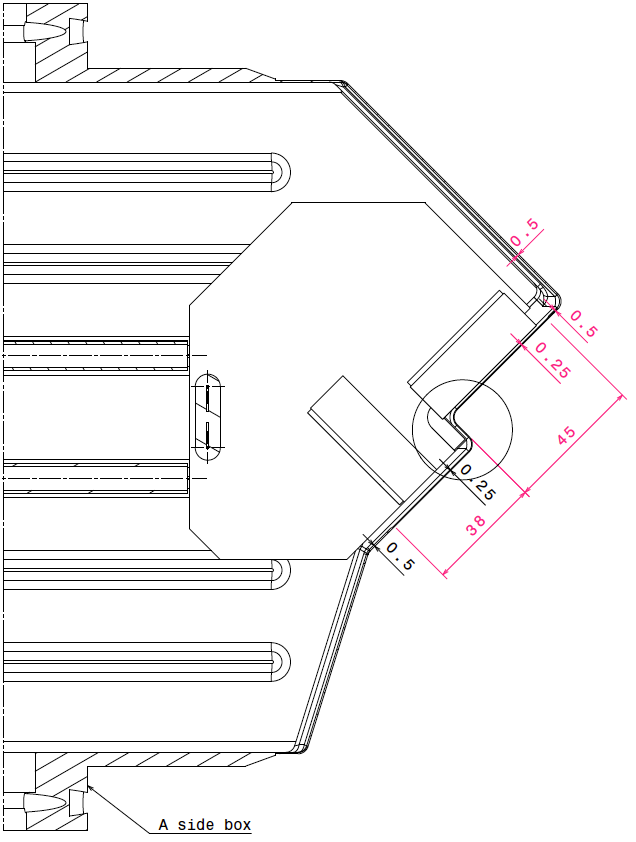}~~~~~~
    \includegraphics[width=0.5\textwidth]{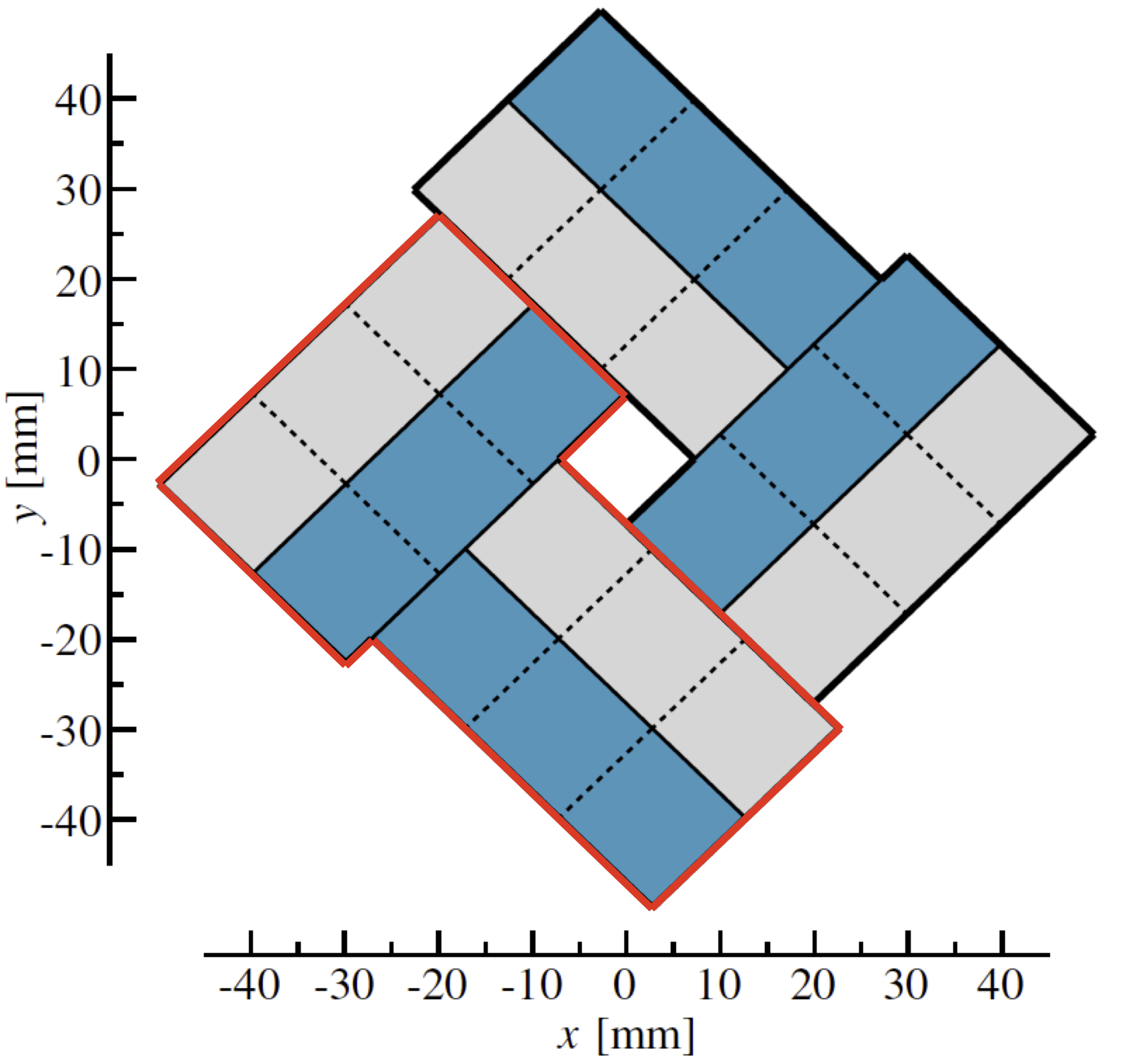}
    \caption{Left: engineering drawing showing the clearance, in mm, between the RF foil and the modules. Right: schematic layout of the chips in the $x-y$ plane with the VELO in the closed configuration. Half the chips are placed on one side of the module (grey) and half on the opposite side (blue). The countour in red highlights the VeloPix ASICs belonging to the same module.}
    \label{fig:module_layout} 
\end{figure}

The detector is operated in vacuum, separated from the LHC vacuum by a thin aluminium alloy RF (Radio Frequency) foil. The RF foil provides protection to the front-end electronics from beam-induced currents as well as providing beam wakefield suppression.  The RF foil surface is corrugated to follow the L-shaped active area of the modules, which serves the purpose of creating a rectangular active area in the $x-y$ plane split across both modules, as shown in Fig.~\ref{fig:module_layout}. The silicon detectors are rotated $45^\circ$ around the $z$~axis, this minimises any risk of the detectors grazing the RF foil during installation. 

\begin{figure}[htbp]
   \centering
  \includegraphics[width=0.9\textwidth]{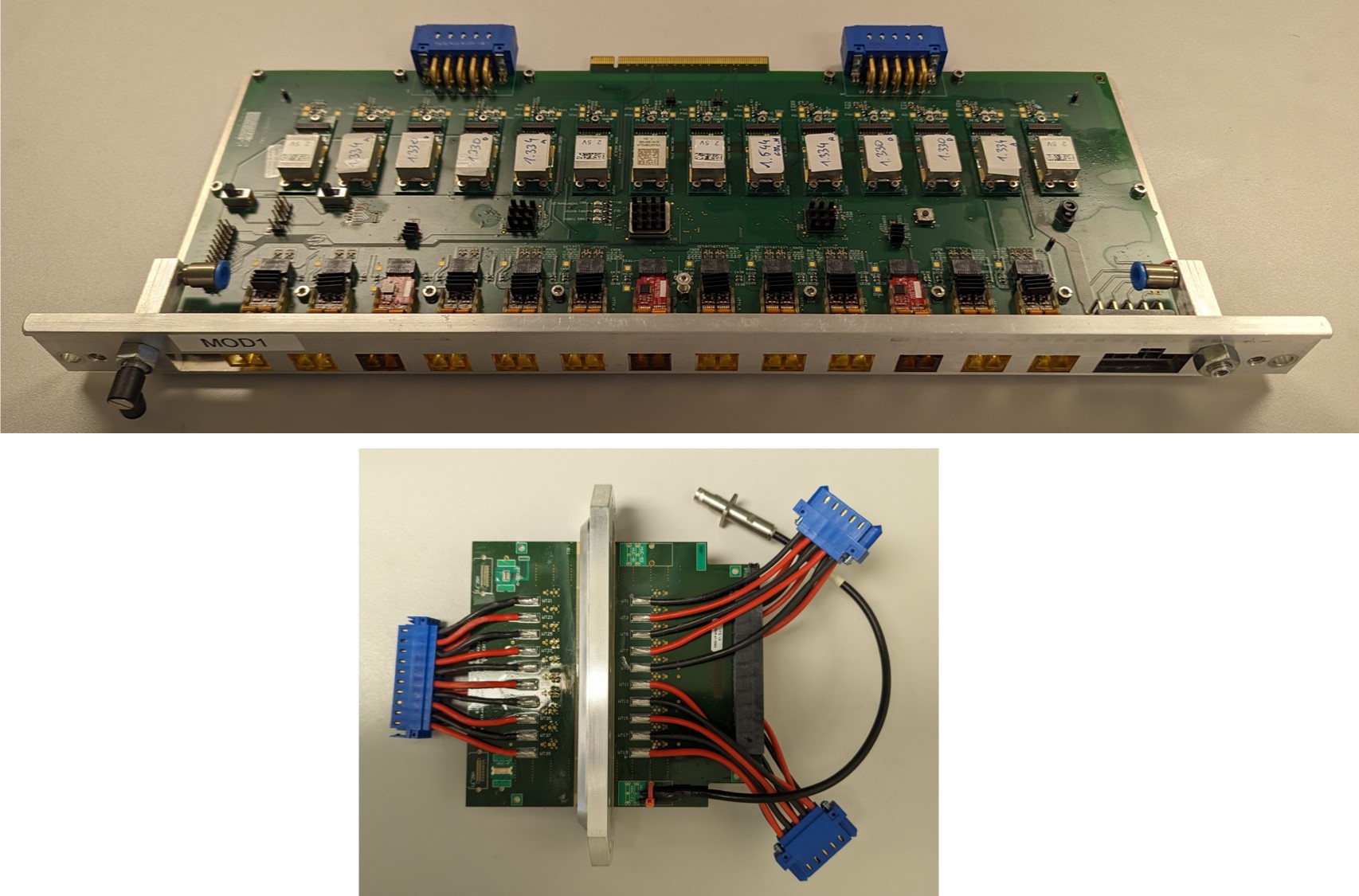}
   \caption{A photograph of an opto-and-power board (top) and of a vacuum feedthrough board (bottom) glued to its aluminium flange (vacuum side on the left).} 
   \label{fig:vfb}
\end{figure}

A custom-made vacuum feedthrough board (VFB) is used to route the high speed data signals, the control signals, the LV power supply ($\sim 2.5$~A) and the bias voltage ($<1000$~V) inside the vacuum. Due to the high rate data requirement, a careful design of the custom transmission lines was needed~\cite{high_speed_links}. The VFB is a 2~mm thick 12-layer printed circuit, which is glued to an aluminium flange with Araldite 2011 on the air side and Araldite 2020 on the vacuum side. A photograph of the vacuum feedthrough board, glued to an aluminium flange, is shown in Fig.~\ref{fig:vfb}. This board was also used in the module quality assurance vacuum tank test setup.

The feed-through board has connectors soldered on both sides of the vacuum barrier, allowing for a fast connection to the module inside the vacuum, and to the OPB~\cite{velo:opbnote} in the air side. The OPB is the interface between the detector modules and the off-detector electronics. Its main purpose is to perform the electrical to optical conversion of the data links and of the control and monitoring signals, as well as to perform the DC/DC conversion of the supply voltages and distribute them to the front-end hybrid.

\subsection{Module layout}
\label{sec:module_layout}
\begin{figure}[htbp]
   \centering
    \includegraphics[width=0.99\linewidth]{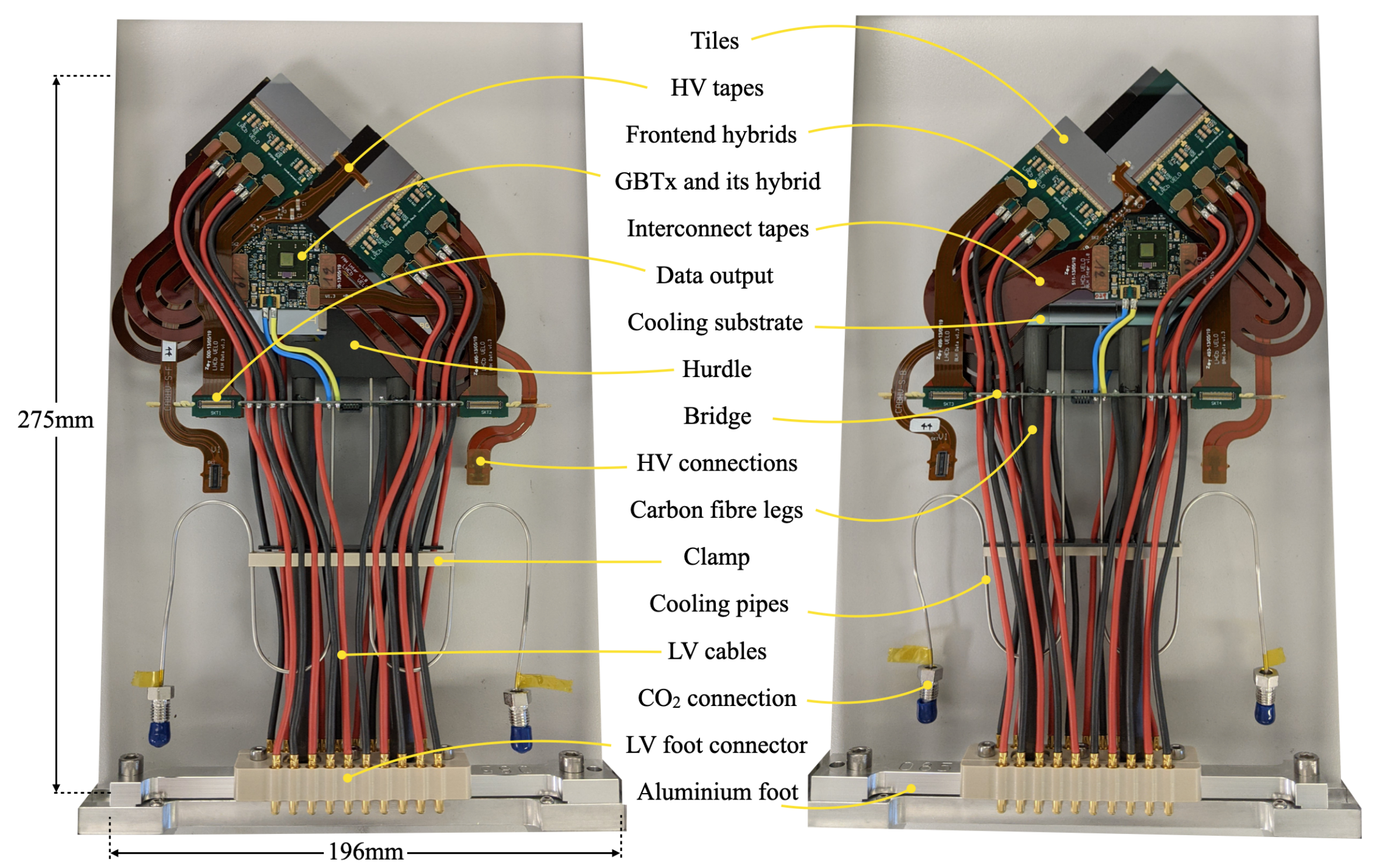}
    \caption{Left: connector side and right: non-connector side of a fully-assembled VELO module.}
    \label{fig:velo:module_sides}
\end{figure}

A VELO pixel module is a double-sided detector structure with a microchannel plate at its core and two tiles arranged in an L shape attached on either side of the microchannel plate. Each tile consists of three VeloPix ASICs bump-bonded to a silicon sensor. A custom-made invar cooling connector, with CO$_2$ supply and return capillaries, is soldered to the microchannel substrate. The invar connector is glued to a \textit{hurdle}, which consists of a carbon fibre plate, two carbon fibre poles and an aluminium foot. On both module sides, two VeloPix chips and one GBTx control chip are hosted each on a front-end hybrid. These carry services to the chips and route their data out. Interconnect cables connect the VeloPix hybrids to the GBTx hybrid. 
The GBTx chip~\cite{Moreira:GBTx1,Moreira:GBTx2} ensures the configuration and synchronisation of the VeloPix ASICs by decoding and distributing the LHCb clock and other control signals. VeloPix data are routed through four thin PCB cables each with up to seven differential links, for a total of 20 links per double-sided module. Power is delivered via an assembly of 20 silicone-coated copper cables and a PCB transition bridge that reduces the thickness of the cables and thus the material budget close to the beamline. The transition bridge also provides mechanical support for the various cables. The bias voltage is delivered via thin PCB tapes which are wire-bonded to the top surface of the sensors. All parts of a VELO module are shown in Fig.~\ref{fig:velo:module_sides} and described in more detail in the next sections.

\subsubsection{Microchannel plate}
\label{sec:microchannels}
\begin{figure}[ht]
 \centering
     \includegraphics[width=0.6\textwidth]{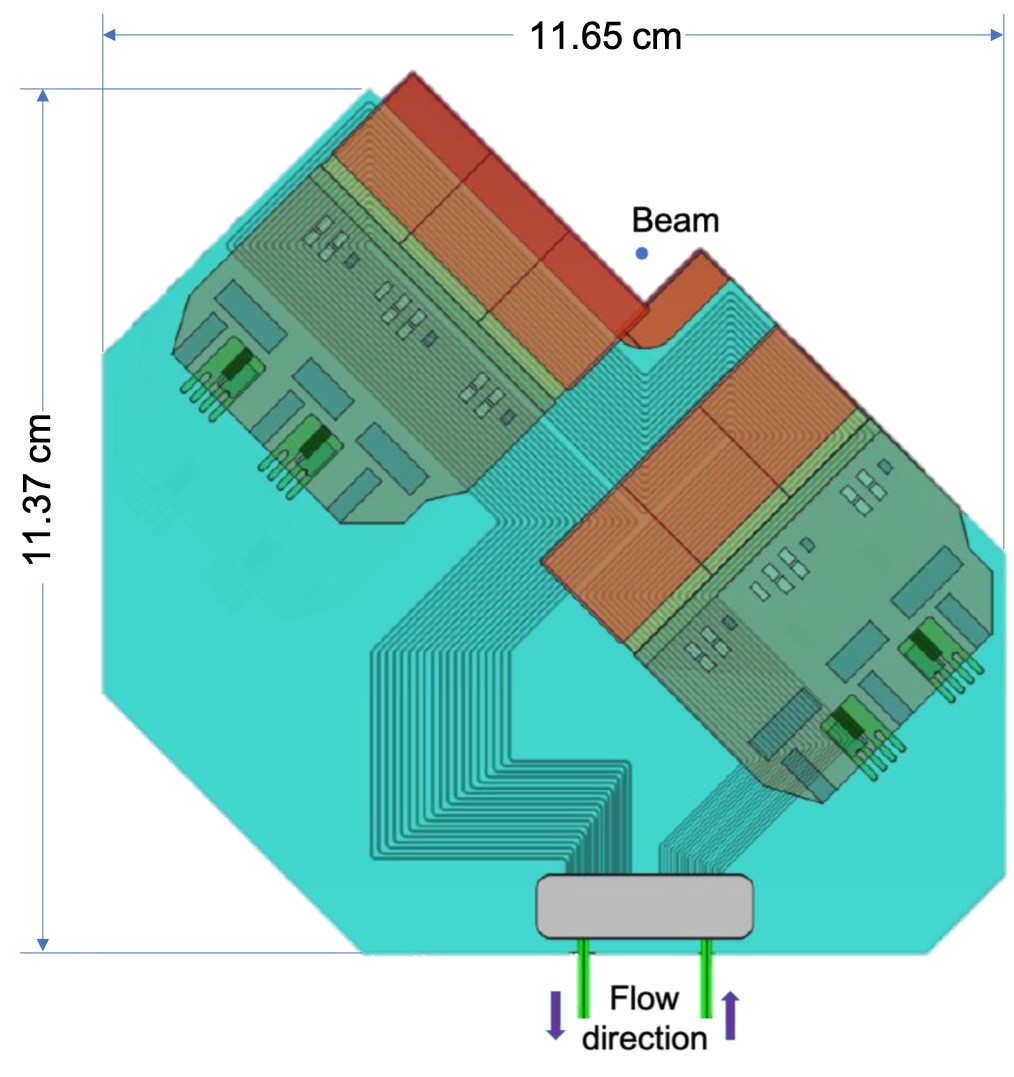}
    \caption[]{Layout of the final microchannel design, overlaid with the hybrid pixel tiles and front-end hybrids. The tiles on the front and back of the cooler are shown in brown, and the position of the front-end hybrids on the front side of the module in green.}
  \label{fig:mc_lhcb_design}
\end{figure}

Since the module is operated in vacuum, power dissipated in the electronics must be efficiently removed. The microchannel cooling technology was chosen since it provides an excellent thermal performance, minimum material budget and no mismatch of thermal expansion with respect to the tiles. Fig.~\ref{fig:mc_lhcb_design} shows a drawing of the microchannels.
The power dissipation of the electronics components on the module is removed by the evaporative $\text{CO}_2$ coolant running through the track-like cooling channels. This technology builds on the experience of operating the first evaporative $\text{CO}_2$ cooling system in the original VELO~\cite{LHCb-DP-2014-001}.
In total, 19 channels are embedded into a silicon plate of 500\mum thickness. Each channel has its own input and output and the coolant distribution is done under the fluidic connector. The connector is made of invar and provides interface between the microchannels and the cooling line. An extensive R\&D programme on the soldering of the connector to the microchannel plate was needed, in order to establish a reliable connection and adequate quality control procedure \cite{Francisco:2021tda}. The initial section of the channels (roughly 4\,cm long) is narrower ($60\mum\times60\mum$) to ensure an even distribution of flow and prevent instabilities. The main channels are $120\mum\times200\mum$ and around 26~cm long. Overall, the cooling substrate outer dimensions are around 11.6~cm$\times$11.4~cm. 

\subsubsection{Hurdle}
\label{sec:hurdle}
The hurdle provides the microchannel plate with a mechanical support that is light and thermally insulates the microchannel from the mounting point. It consists of a carbon fibre plate and  two carbon fibre rods fixed to an aluminium foot.

The module foot is machined out of aluminium 5083. It contains a precision-drilled hole for the mounting bolts and the alignment dowel, providing the module with a solid and accurate mounting point on the detector base. Two additional holes are used to insert and glue the carbon fibre rods. These are cylindrical tubes made of T700 carbon fibres, aligned along the length of the cylinder and glued together with a high performance epoxy resin. The choice of the alignment direction of the fibres ensures minimal deformation due to thermal stress. At the other end of the rods, a flattened section cut is used to glue them to a carbon fibre plate (midplate). The midplate is a thin sheet of carbon fibres aligned in the 0$^{\circ}$/90$^{\circ}$ and -45$^{\circ}$/+45$^{\circ}$ 
directions. 
The midplate is directly glued to the invar cooling connector with Araldite 2011 and it is the only point of contact between the microchannel assembly and its mechanical support. A 3D-printed clamp, made of polyether ether ketone (PEEK) and attached to the carbon fibre rods, secures the LV cables and cooling pipes and relieves them from mechanical and thermal stress.

\subsubsection{Tiles}
\label{sec:tiles}
\begin{figure}[ht]
 \centering
     \includegraphics[width=0.99\textwidth]{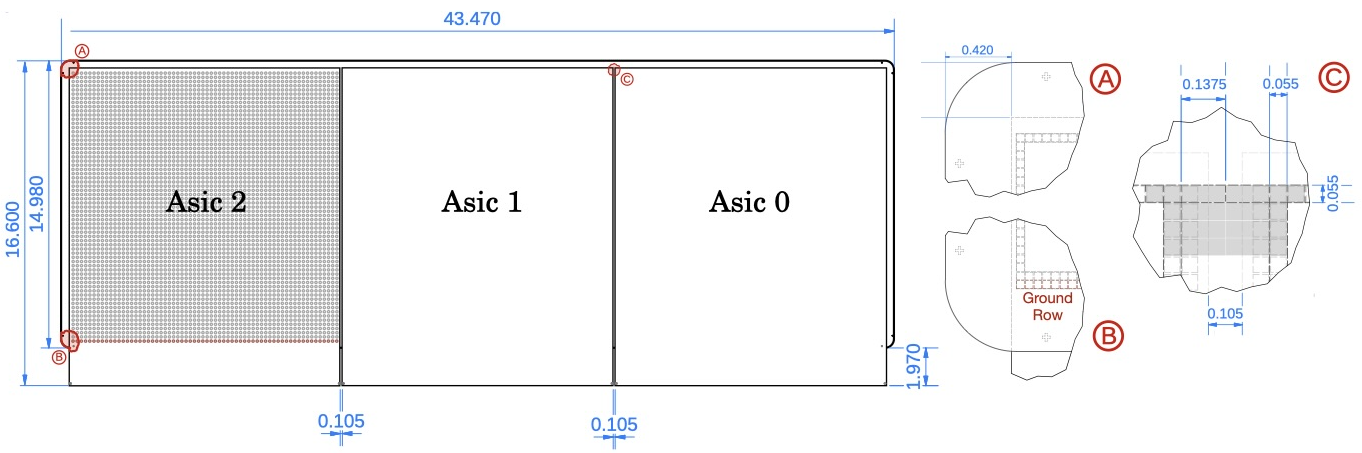}
    \caption[]{Schematic of the sensor tile, showing the overall dimensions of the sensor and its pixel layout. An additional row of pixels, indicated in dark red, provides a connection between the ASIC ground and the innermost guard ring of the sensor.}
  \label{fig:asic_schematic}
\end{figure}

A VELO tile consists of a pixelated planar silicon sensor bump-bonded to three pixelated VeloPix chips (ASICs). The VeloPix~\cite{VANBEUZEKOM201392} contains an active matrix of 256$\times$256 (55$\times$55\,$\mum^2$) pixels, for a total sensitive area of 1.98\,cm$^2$. The chip is fabricated in 130\,nm technology\footnote{by Taiwan Semiconductor Manufacturing Company, Hsinchu 300-096, Taiwan, R.O.C.}, which has proven radiation hardness above 400\,MRad~\cite{plackett2009}. The sensor is a 200$\mum$ thick, n-on-p silicon sensor\footnote{made by Hamamatsu Photonics K.K., Hamamatsu, Shizuoka 435-8558, Japan.}, with an overall dimension of 43.470$\times$14.980$\mm^2$ (including a 450$\mum$ wide inactive region for the guard rings). This builds on the experience of operating the first n-on-p silicon sensor in the original VELO~\cite{LHCb-DP-2014-001}.
The sensor is implanted with square pixels with a pitch of 55~$\mu$m to match the pixel array of the read-out electronics. The flip chip bonding\footnote{carried out at ADVACAM Oy, Tietotie 3, FI-02150 Espoo, Finland.} consists of almost two hundred thousand solder bumps per VeloPix triplet. Since the sensors are attached to three ASICs in line, the pixels at the inter-chip regions are elongated to the size of $55 \times 137.5\mum$ to minimize the dead area. The schematic of a sensor tile is illustrated in Fig.~\ref{fig:asic_schematic}.

\begin{figure}[ht]
 \centering
    \includegraphics[width=0.7\textwidth]{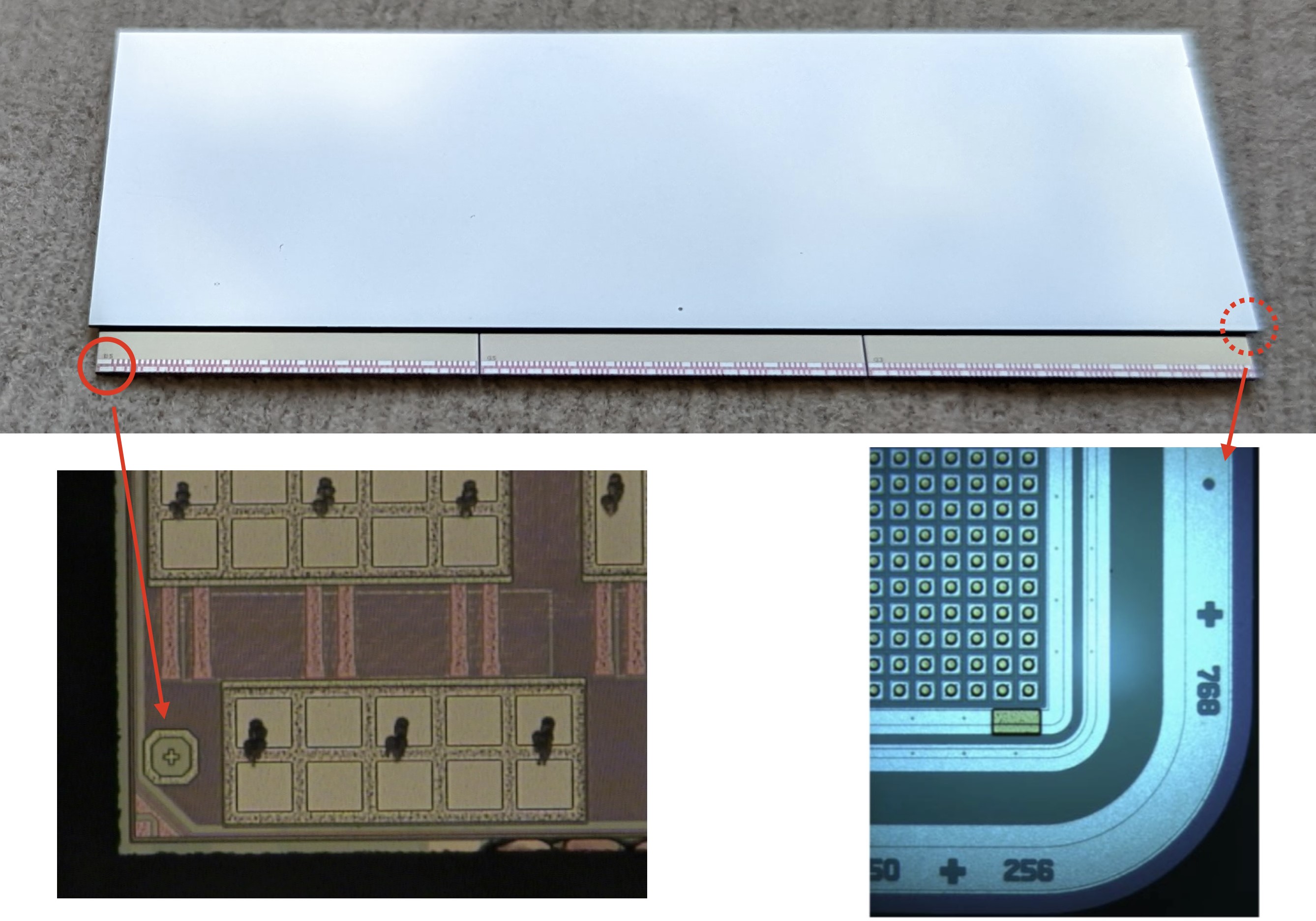}
    \caption[]{Top: a photograph of a VELO tile. Bottom left: microscopic image of the ASIC corner and its alignment marker. Bottom right: microscopic image of the sensor round corner with its cross- and disk-like shaped alignment markers.}
  \label{fig:tile_photos}
\end{figure}

Both the ASICs and the sensors come with etched alignment markers, which are used during the positioning and attachment of the tile to the microchannel plate, as well as during metrology measurements after attachment. Fig.~\ref{fig:tile_photos} shows a photograph of a VELO tile and two microscopic images of the corner of the ASIC and of the sensor.  

\begin{figure}[htbp]
  \centering
  \includegraphics[width=0.7\linewidth]{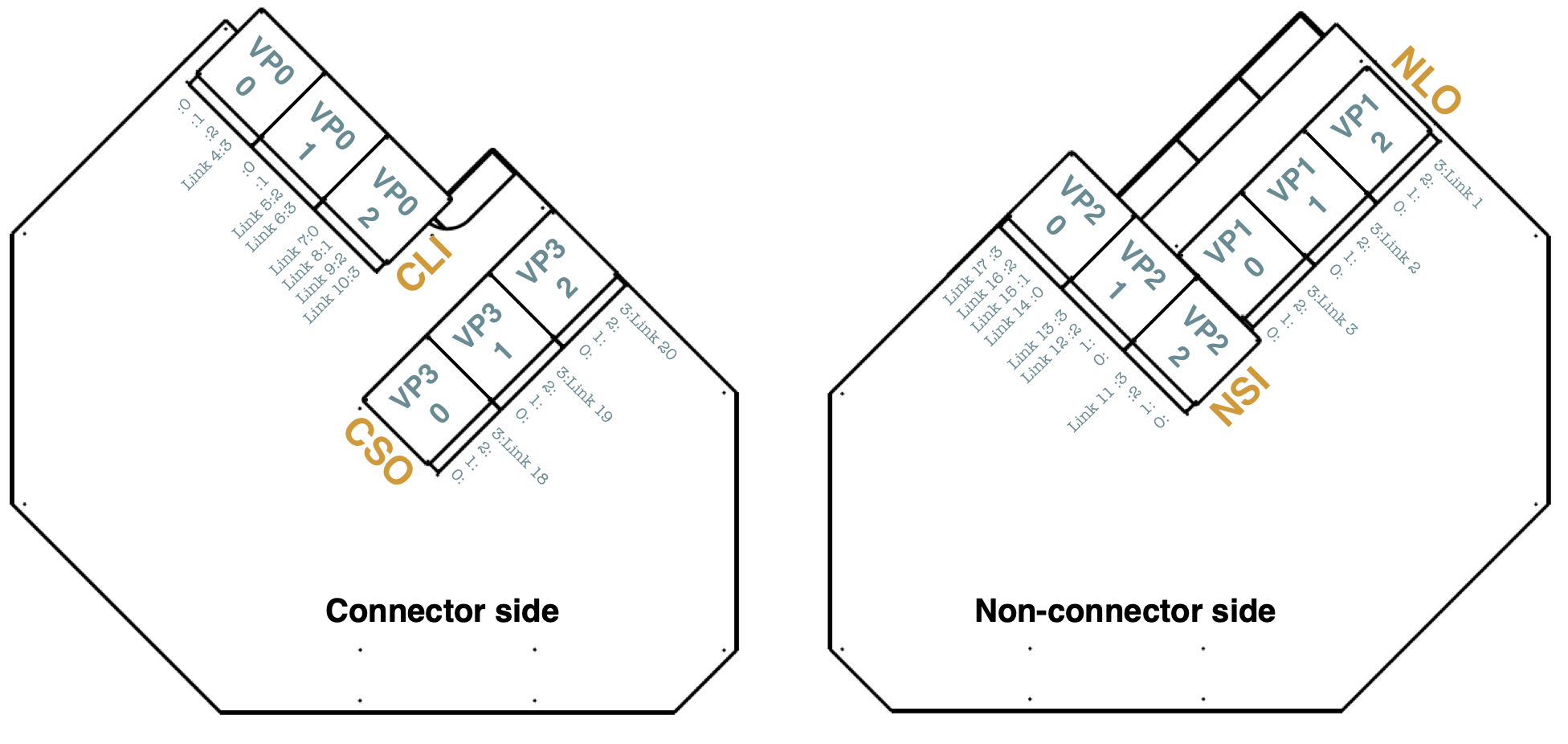} \caption{Mapping of the tiles and ASICs. The schematics also show how the 20 data links are distributed among the 12 VeloPix chips.}
  \label{fig:tiles_nomenclature}
\end{figure}

The nomenclature~\cite{velo_nomenclature} of the tiles and the ASICs is shown in Fig.~\ref{fig:tiles_nomenclature}. On the fluidic connector side, the tiles are named CLI (VP0) and CSO (VP3), while on the non-connector side they are named NLO (VP1) and NSI (VP2). Within each tile, the ASICs are labelled with a number that ranges from 0 to 2. While CSO and NLO are fully enclosed in the microchannel plate, tile CLI overhangs the plate by 5\,mm along its full length, and tile NSI by the same amount along its ASIC 0. 
Each VeloPix has four
output links, however the number of links that are enabled is configurable to just one link, or two links or four links. The VeloPix chips in the hottest region of the detector (VP0-2 and VP2-0) have all four links enabled, while those located in the lower occupancy regions operate with either two or one link. In a VELO module, there are a total of 20 output links.

\subsubsection{Hybrids and cables}
\label{sec:hybrids}
\begin{figure}[htbp]
  \centering
  \includegraphics[width=0.7\linewidth]{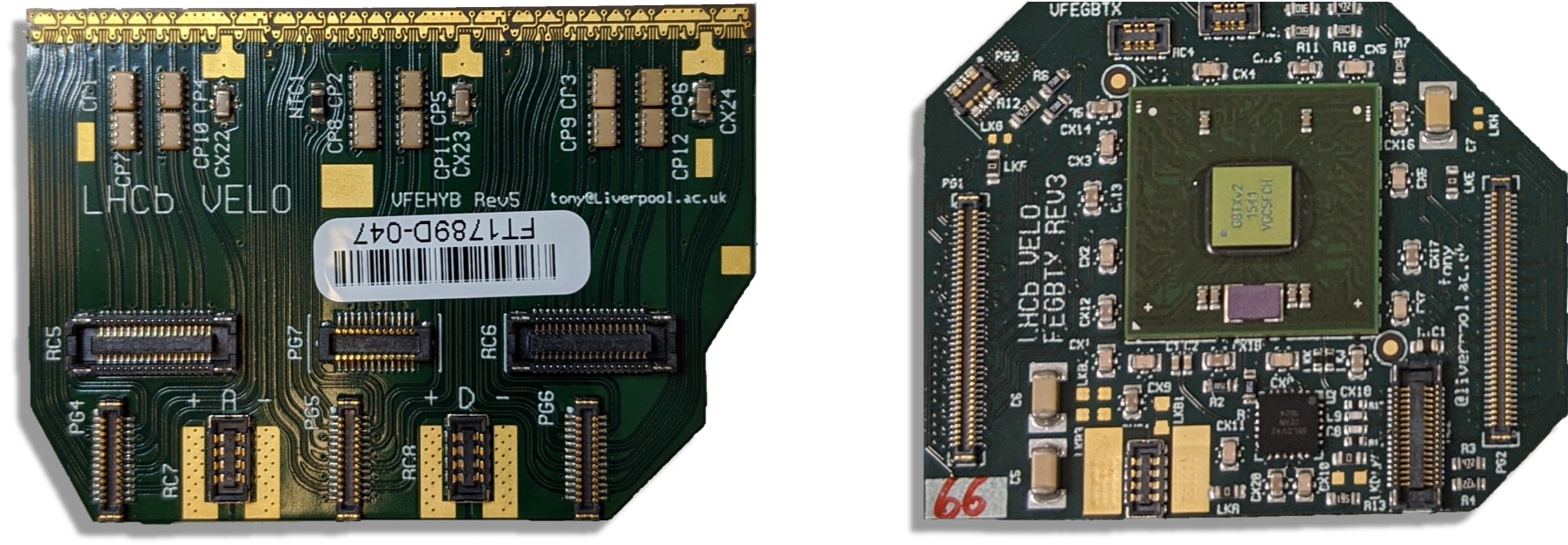}
  \caption{A photograph of VeloPix and GBTx hybrids.}
\label{fig:hybrids}
\end{figure}

On a VELO module, the routing of the the outward-bound data and the distribution of control signals and services is provided by a set of hybrids. These are 390$\mum$ thick, four-layer, flexible printed circuits and come in two types: the front-end hybrids and the GBTx hybrids, as shown in Fig.~\ref{fig:hybrids}. The front-end hybrids provide the electronic interface to the VeloPix chips through wire-bonds. The GBTx hybrids carries the gigabit transceiver chip~\cite{Moreira:GBTx2} and is responsible for the distribution of timing and fast controls to the ASICs, as well as providing slow controls and monitoring of the bias voltage. Each hybrid also carries a temperature probe (NTC).

\begin{figure}[htbp]
  \centering
  \includegraphics[width=0.99\linewidth]{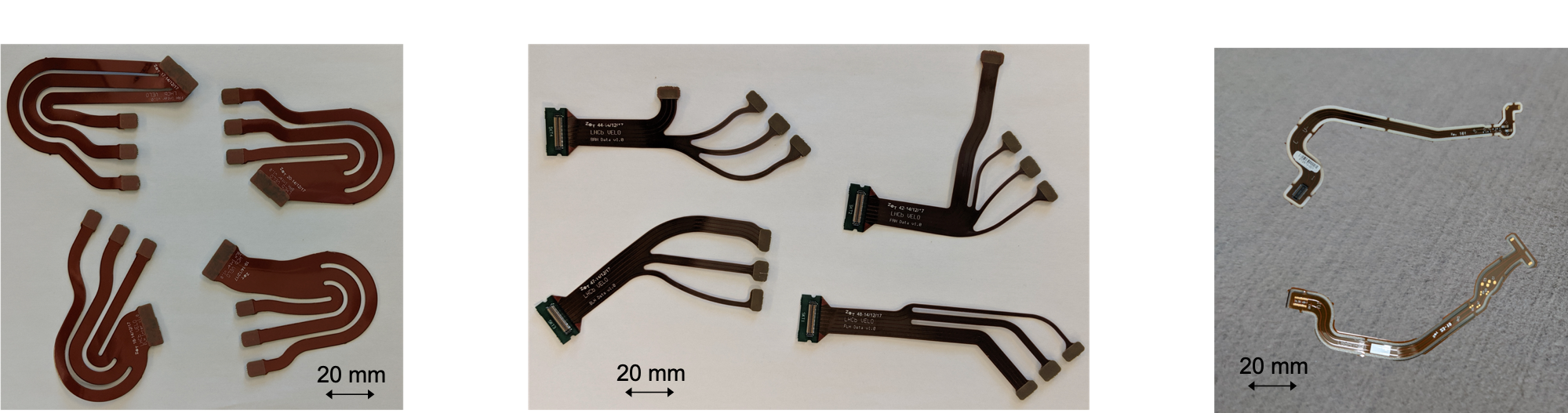}
  \caption{A photograph of a set of interconnect cables (left), data cables (middle) and HV tapes (right).}
\label{fig:cables}
\end{figure}

Each hybrid connects to two different cables: the interconnect cables, which connect the front-end hybrids to the GBTx hybrid, and the data cables, which route the outputs of each tile to the off-module electronics. Each VELO module is equipped with a set of four front-end hybrids, two GBTx hybrids, four interconnect cables and four data cables, evenly split onto the two sides of the module. The high voltage (HV) is carried to the silicon sensors by the HV tapes, which are connected to the microchannel plate with a fast-curing glue (Araldite 2012) and afterwards wire-bonded to the surface of the sensor. Data, interconnect and HV cables are shown in Fig.~\ref{fig:cables}.

\subsubsection{LV harness}
\label{sec:LVharness}
\begin{figure}[htbp]
  \centering
  \includegraphics[width=0.6\linewidth]{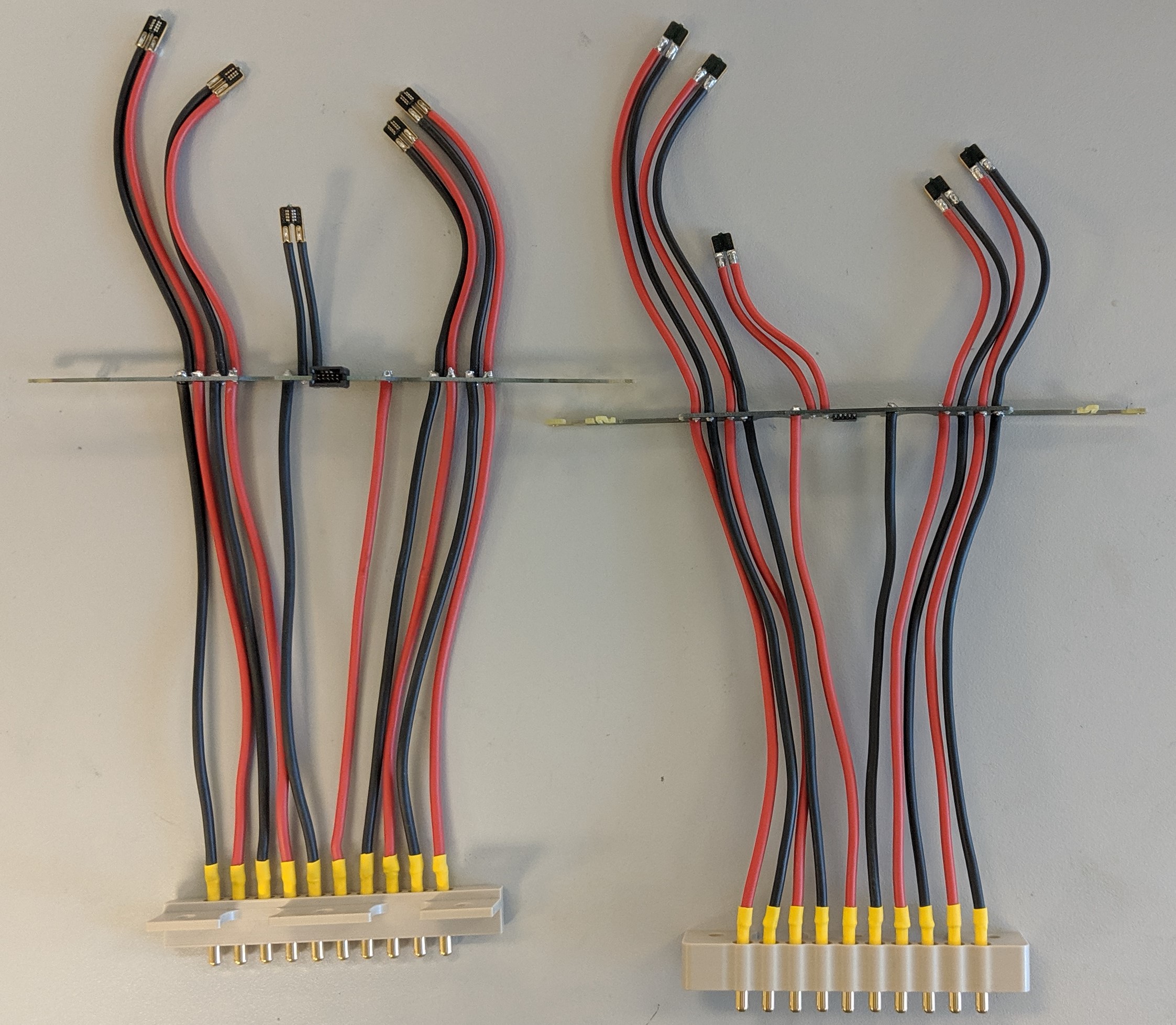}
  \caption{A photograph of a LV harness, showing the cables, the custom-made PCB connectors, the LV bridge pieces, the LV pins and the LV foot connector.}
\label{fig:LVharness}
\end{figure}

Power is delivered to the chips via an assembly of 20 silicone-coated copper cables and a PCB transition bridge. The transition bridge is glued to the carbon fibre legs where it establishes a mechanical link while changing from thinner gauge cables close to the readout, to thicker ones going back to the module foot. The LV cables are connected to the front-end and control hybrids via custom-made PCB connectors. At the other end of the LV harness, pins are inserted in the LV foot connector. A photograph of a LV harness is shown in Fig.~\ref{fig:LVharness}. 

\subsection{Alternative designs and changes}
During the VELO Upgrade R\&D programme, a range of solutions were investigated for many elements of the module. Here we provide brief comments on some of the design choices that were made and changes made during the process.

\subsubsection{Cooling substrate}
\label{sec:cooling_choice}
\begin{figure}[htbp]
  \centering
  \includegraphics[width=0.95\linewidth]{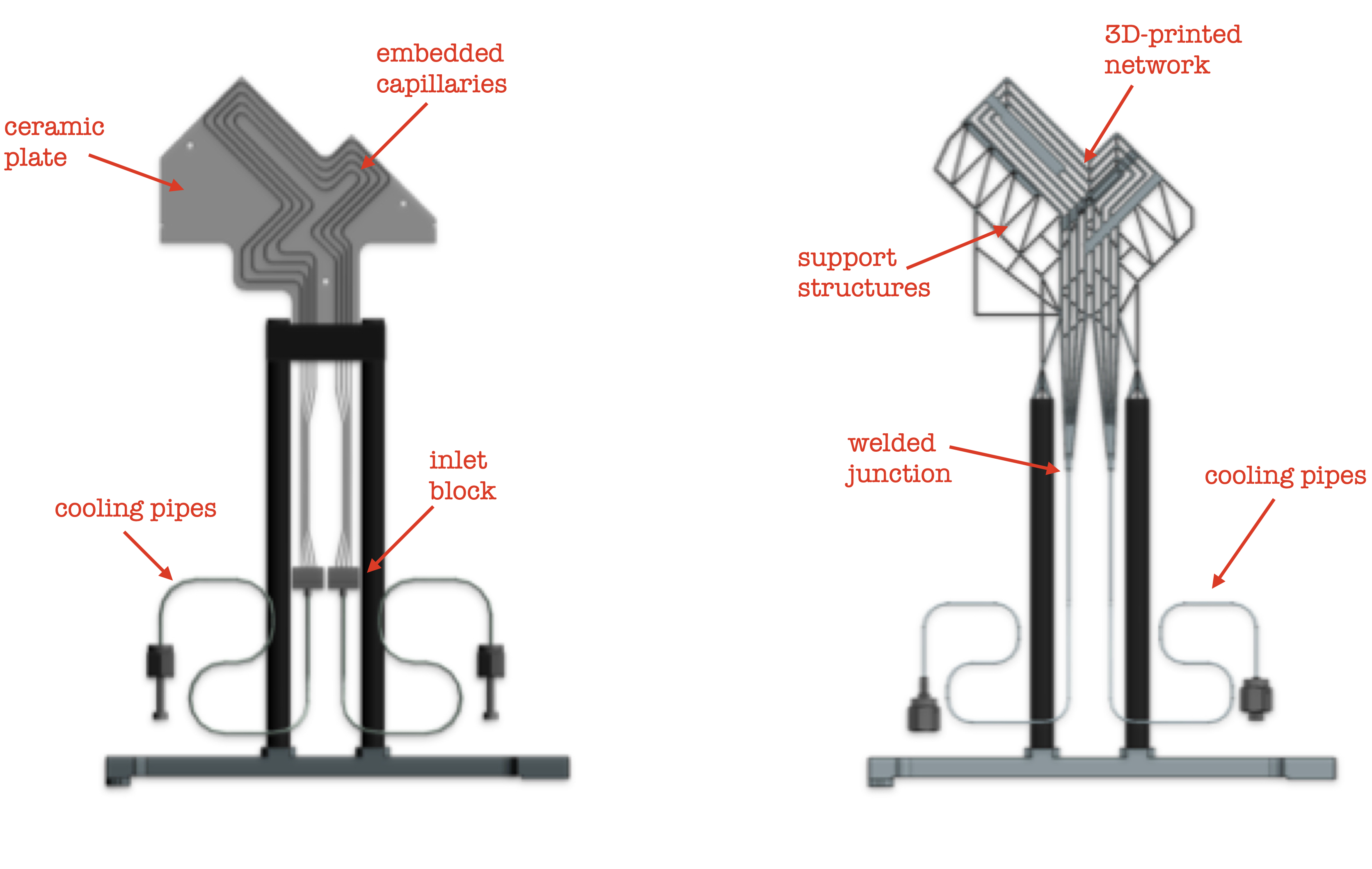}
\caption{CAD models of the 3D printed titanium substrate (left) and the aluminium nitride plate with embedded cooling pipes (right).}
  \label{fig:titanium_and_ceramic}
\end{figure}

The R\&D focussed on the microchannel solutions as it is well suited to the needs of this project. Due to the complexity of the silicon microchannel technology, described above, two other microchannel implementations were considered: 3D printed titanium substrates and ceramic plates with embedded stainless-steel capillaries (see Fig.~\ref{fig:titanium_and_ceramic}). \\
The 3D printed titanium option offers a cheap solution, flexible layout and fast turnaround for prototyping. Cooling pipes can be brazed or welded directly to it and the restrictions needed at the inlet to trigger the CO$_2$ evaporation can be implemented easily in the design of the substrate. A wall thickness of 100\,$\mu$m was already achievable at the time of our R\&D programme. \\
The ceramic solution was investigated by manufacturing a plate in aluminium nitride, with precisely machined grooves to house stainless steel cooling pipes of 0.64(0.4)\,mm outer(inner) diameter. The inlet block is equipped with orifices of 140\mum diameter for the expansion of the coolant, which is then routed to the four cooling pipes. The ceramic offers an excellent heat transfer (about 100\,W/mK at 25\,$^{\circ}$C) and a robust support for the tiles. \\
Although both options had the potential to be viable solutions for the VELO Upgrade, they were rejected in favour of the microchannel technology. The use of a thick ceramic plate, with embedded steel capillaries, increased the material budget in the LHCb acceptance; deviations from planarity were observed on the surface, although smaller than the expected thickness of the glue layer used to attach the ASICs; the manufacturing procedure consisted of a several-step manual process. Microscope inspections of the 3D-printed titanium substrates showed that the surface was covered with partially melted powder particles, sometimes reaching a height of tens of microns; the possible formation of these structures on the internal walls, posed a risk of perturbation to the flow of the coolant. Taking into account these considerations, the additional R\&D required to achieve an optimal cooling performance from either of these alternative solutions was incompatible with the project timeline with the effort available.

\subsubsection{Mechanical support}
Alternative designs of the module mechanical support were considered. The original VELO used a carbon fibre structure with an open triangular cross-section~\cite{LHCb-TDR-005}. A support made from a single 1\,mm thick carbon fibre sheet was considered, with the microchannel wafer glued to the carbon fibre support. This provided good stability in X-Y and by design had flexibility along Z. This flexibility led to individual module vibrational stability issues, with a carbon fibre constraint system along the VELO length to position and secure the modules. In the hurdle design an alternative mid-plate was considered, made of a silica glass mid-plate instead of carbon fibre, which was rejected after cracking occurred during production of prototypes. Alternatives were studied under stability in vibration tests, thermal studies, in vacuum operation, and in the construction of prototypes. The final design was chosen based on its low-material, ease of construction and advantages of the open geometry for routing of the cables and cooling pipes.

\subsubsection{Front-end and control hybrids layout}
A single hybrid for each side of the module was originally envisaged. 
The choice of separating the front-end electronics into three parts, rather than one large single hybrid, was implemented to reduce the stress on the module. The thickness of the hybrids, driven by the copper cross section required to bring sufficient power to the chips, leaves them significantly stiff. Consequently, a single stiff hybrid with a coefficient of thermal expansion different to that of silicon, would exert an unacceptable force on the microchannel plate over the 50\,$^\circ$C thermal change between room temperature and the operational temperature of the detector. 

\subsubsection{Glue choices}
\label{sec:glue_choices}
Significant studies were carried out to select and test suitable adhesives for each of the different gluing steps in the module assembly and optimise the way in which they were used. The gluing of the sensor tiles to the module, the hybrid to the module and the high voltage connection required particular attention. Thermal tests, including thermal cycling, test after irradiation and mechanical shear tests were performed. The deposition patterns were optimised as further discussed in Sect.~\ref{sec:glue_parameters}. 

During prototyping it was observed that glue between the ASICs increases the capacitance of the system increasing the electronic noise, so any glue excess filling the space between the ASICs must be avoided. The glue selected for the tiles' attachment was a thermally conductive epoxy encapsulant\footnote{Stycast 2850FT with catalyst 9} \cite{glue_note}. Detailed tests showed that the mechanical connection between the tiles and the microchannel substrate was not ideal, leading to adhesion failures on test samples as well as on an early module. The cause was traced to the hygroscopic nature of the chosen catalyst, causing the accumulation of a layer of humidity on the deposited patterns that hinders the bonding properties of the glue. The moisture layer becomes visible only after thermal cycling and worsens over time, progressively weakening the adhesive strength and shortening the lifetime of the modules.
For this reason, an extensive study was carried out, to test alternative epoxies for their mechanical strength, radiation hardness and thermal performance. At the end of this campaign, the glue chosen for the module production was the same as originally proposed, but in combination with a different catalyst\footnote{Stycast 2850FT with catalyst 23LV}, and the additional use of a heat gun treatment (see Sect.~\ref{sec:glue_parameters}).

The main requirements on the glue for the attachment of the hybrids to the substrate were elasticity and good radiation hardness. The elasticity of the adhesive is needed due to the different coefficient of thermal expansion (CTE) of the silicon substrate and the hybrid PCB (due to its copper layers). There was some evidence in prototypes of microchannel rupture when testing adhesives without flexibility. Since the cooling of these components is not as critical as for the tiles (less produced heat), an alkoxy silicone adhesive\footnote{Loctite SI 5145.} was chosen, which gave suitable results.

The use of conductive adhesives for the attachment of the high voltage tapes to the back side of the silicon sensors was briefly investigated. No reliable candidate suitable for a heavy radiation environment was identified. Hence, the HV tapes were glued\footnote{using Huntsman Araldite 2012.} and the electrical connection made by wire-bonding.

%% file: VELO/assembly.tex
Module production took place in two assembly sites, the University of Manchester (Manchester, UK) and Nikhef (Amsterdam, NL). The assembly process requires high precision and consistency between the two sites, thus a detailed construction procedure was developed to ensure excellent quality across the whole production (see also~\cite{Murray:2808484,Svihra:2806219,Zunica:2869815}).
In order to achieve this goal, a number of checks are conducted after each of the assembly steps. They include visual inspections, metrology, electrical tests, thermal and mechanical performance measurements in air and in vacuum. The mounting of the modules onto the VELO bases, as well as the testing of the full DAQ chain and cooling system, took place at the University of Liverpool (Liverpool, UK). From there, the two fully assembled VELO halves were transported to CERN (Geneva, CH) for final commissioning of the whole detector.

\begin{figure}[htbp]
  \centering
  \includegraphics[width=0.99\linewidth]{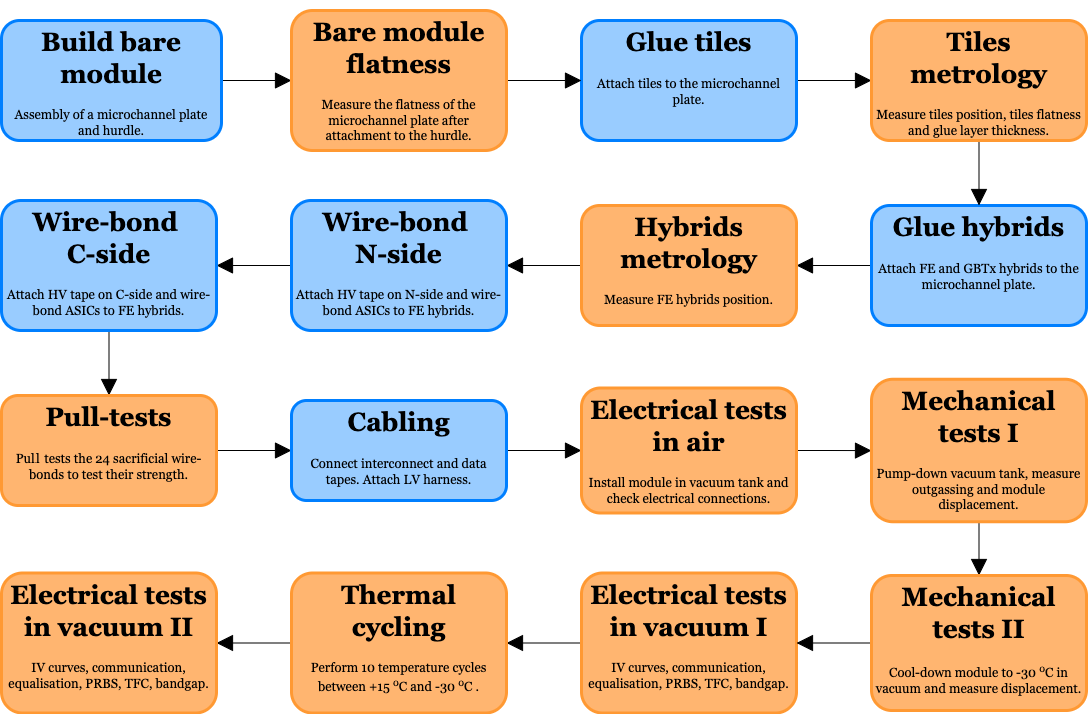}
  \caption{A flow chart of the assembly steps. Processes are shown in blue and measurements in orange.}
  \label{fig:assembly_flow}
\end{figure}

\begin{figure}[htbp]
  \centering
  \includegraphics[width=0.19\linewidth]{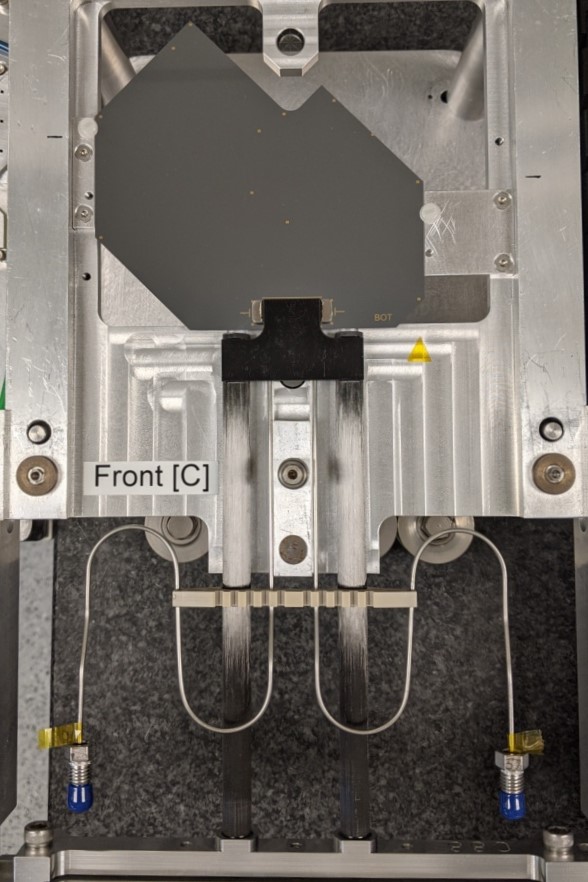}
  \includegraphics[width=0.19\linewidth]{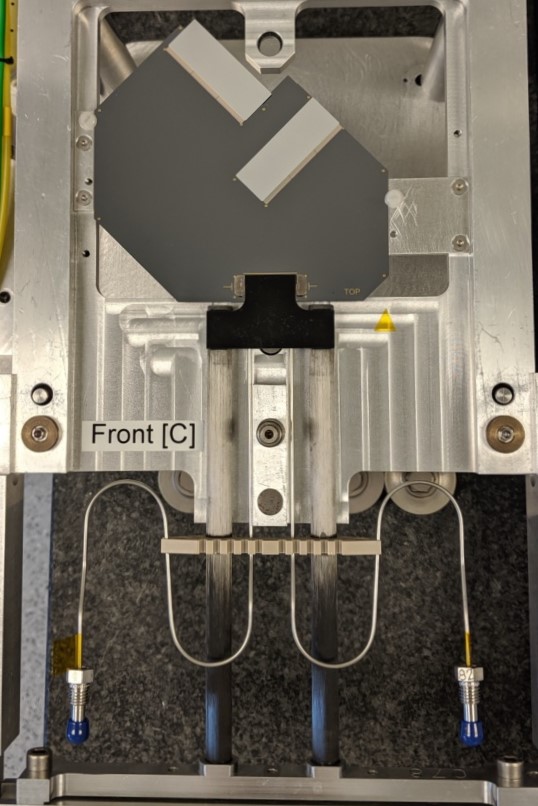}
  \includegraphics[width=0.19\linewidth]{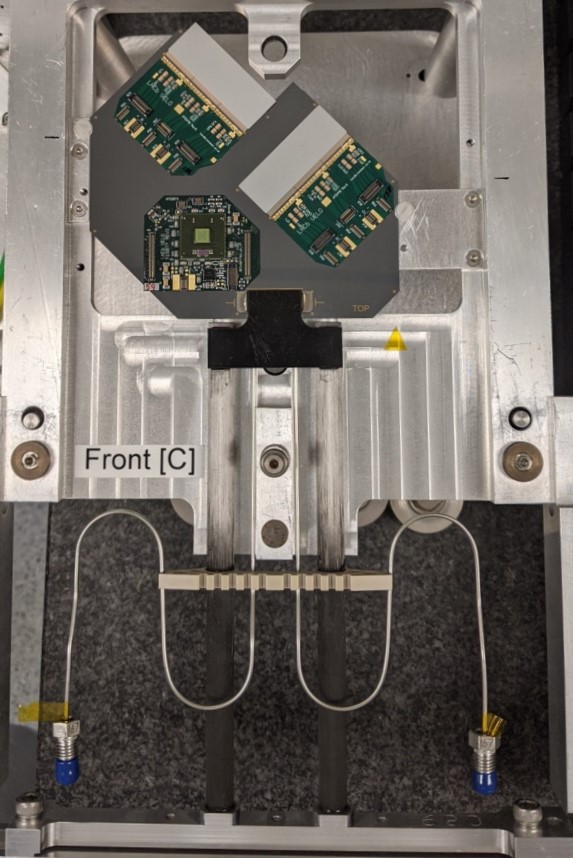}
  \includegraphics[width=0.19\linewidth]{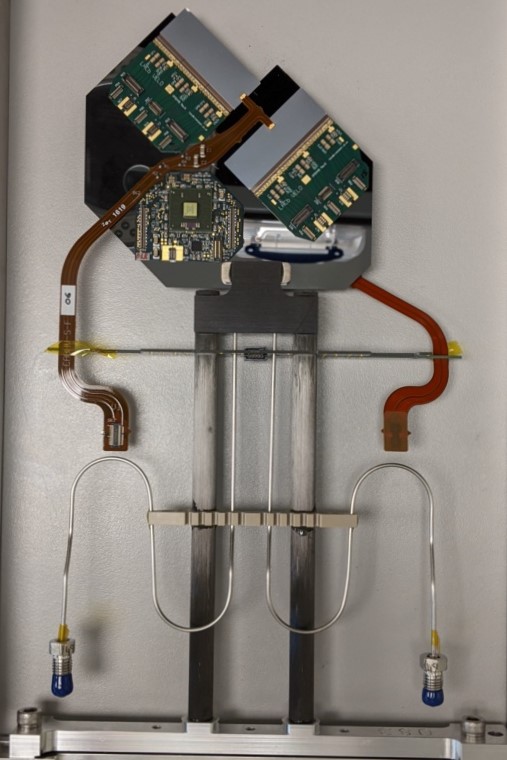}
  \includegraphics[width=0.19\linewidth]{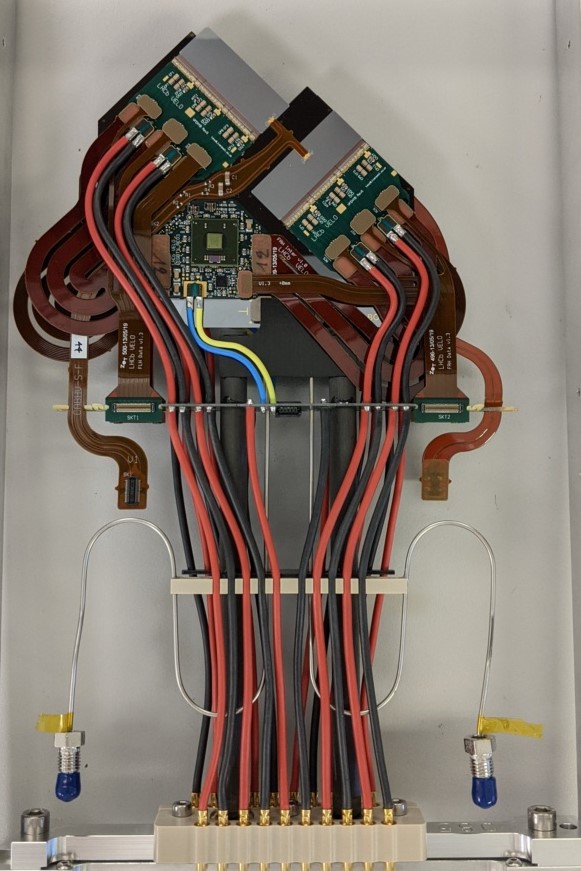}
  \caption{Pictures of a VELO upgrade module after each of the five assembly steps. From left to right: bare module assembly, tiles attachment, hybrids attachment, wire-bonding and cabling.}
  \label{fig:module_assembly}
\end{figure}

The module assembly process consists of five major steps, shown as blue blocks in the flow diagram in Fig.~\ref{fig:assembly_flow}: construction of the bare module, attachment of tiles, attachment of hybrids, wire-bonding and finally cabling. After each of these steps, a number of tests is carried out and they are shown as orange blocks in the flow diagram. The information from both the assembly and testing steps is uploaded to an online database and automatically analysed, thus providing an instantaneous feedback about the quality of the performed task. The database also guarantees an off-site record of all the actions performed on a module, which will serve as future reference for the long term, and is discussed below. The progress of the module construction through the five module assembly steps is shown in Fig.~\ref{fig:module_assembly} and these steps are described in detail in the following sections.

\subsection{Database} 
\label{sec:database}
The database provides tracking of the module assembly as it progresses, allowing to store all relevant data. It contains information about all the components that are used to build a module, along with every process and test carried out on individual parts as well as modules.

\begin{figure}[htbp]
    \centering
    \includegraphics[width=0.99\linewidth]{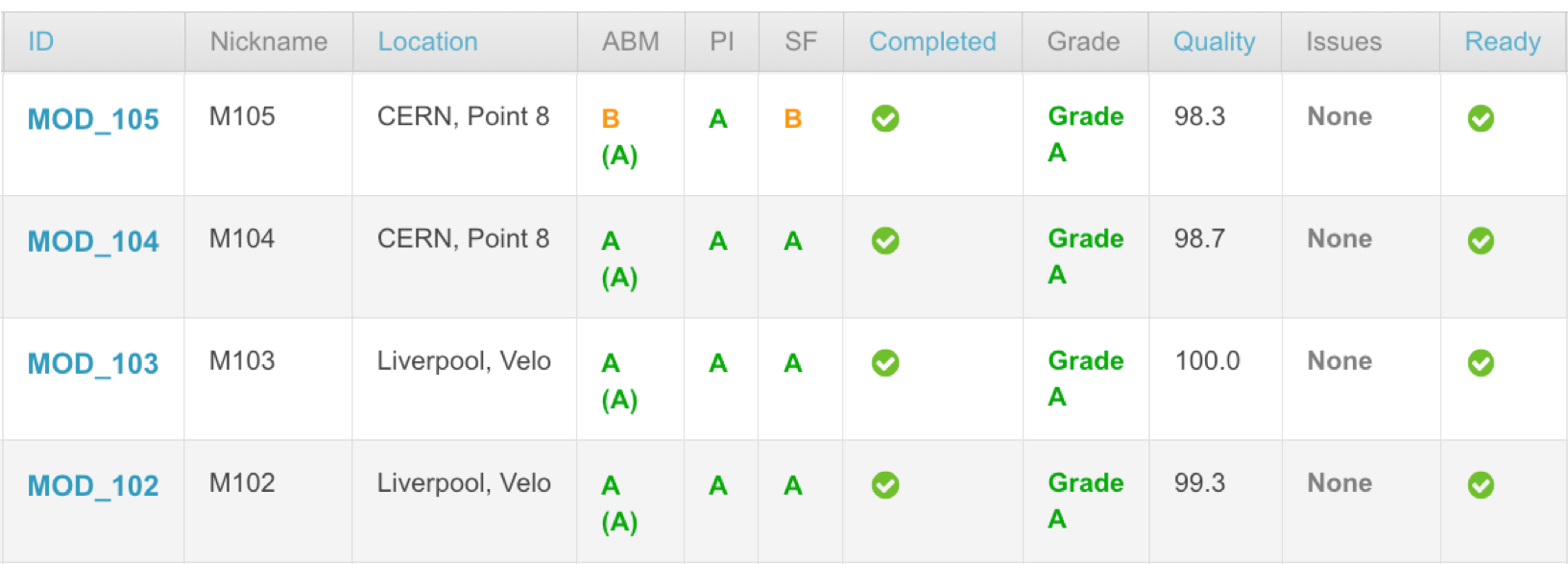}
    \caption{Extract from an overview page of the construction database showing a few assembled modules. Each row corresponds to a module. A few assembly and testing steps are shown as columns: the bare module assembly (ABM), a photo inspection of the bare module (PI) and the substrate flatness measurement (SF). The table also contains a grade and a quality number.}
    \label{fig:database}
\end{figure}

The database uses PostgreSQL~\cite{postgresql}, accessed via Django framework~\cite{django} based on Python~\cite{python}. The Python libraries Bokeh and Holoviews~\cite{holoviews} are used to enable the interactive control of the generated plots, such that a simplified data analysis can be performed immediately after uploading the results of a test. This provides a quick visual representation of the measurements and allows for an immediate feedback on the quality of the assembly procedure at any stage. An example of the modules page in the database is shown in Fig.~\ref{fig:database}, where the rows represent individual modules and the columns correspond to the assembly/testing steps, arranged in the same order as the tests are performed during construction.

Each assembly or test step is graded automatically by an algorithm that assigns a letter based on the performance, defined as follows: [A] excellent; [B] very good, [C] good, [D] one missing functionality, [F] multiple missing functionalities. Grades A, B and C are assigned to fully functional modules, which meet the production quality requirements and are recommended to be suitable for installation in the VELO detector. The automatic gradings for the validation tests are based on the quality criteria described in Sect.~\ref{sec:validation}.

\begin{figure}[htbp]
    \centering
    \includegraphics[width=0.99\linewidth]{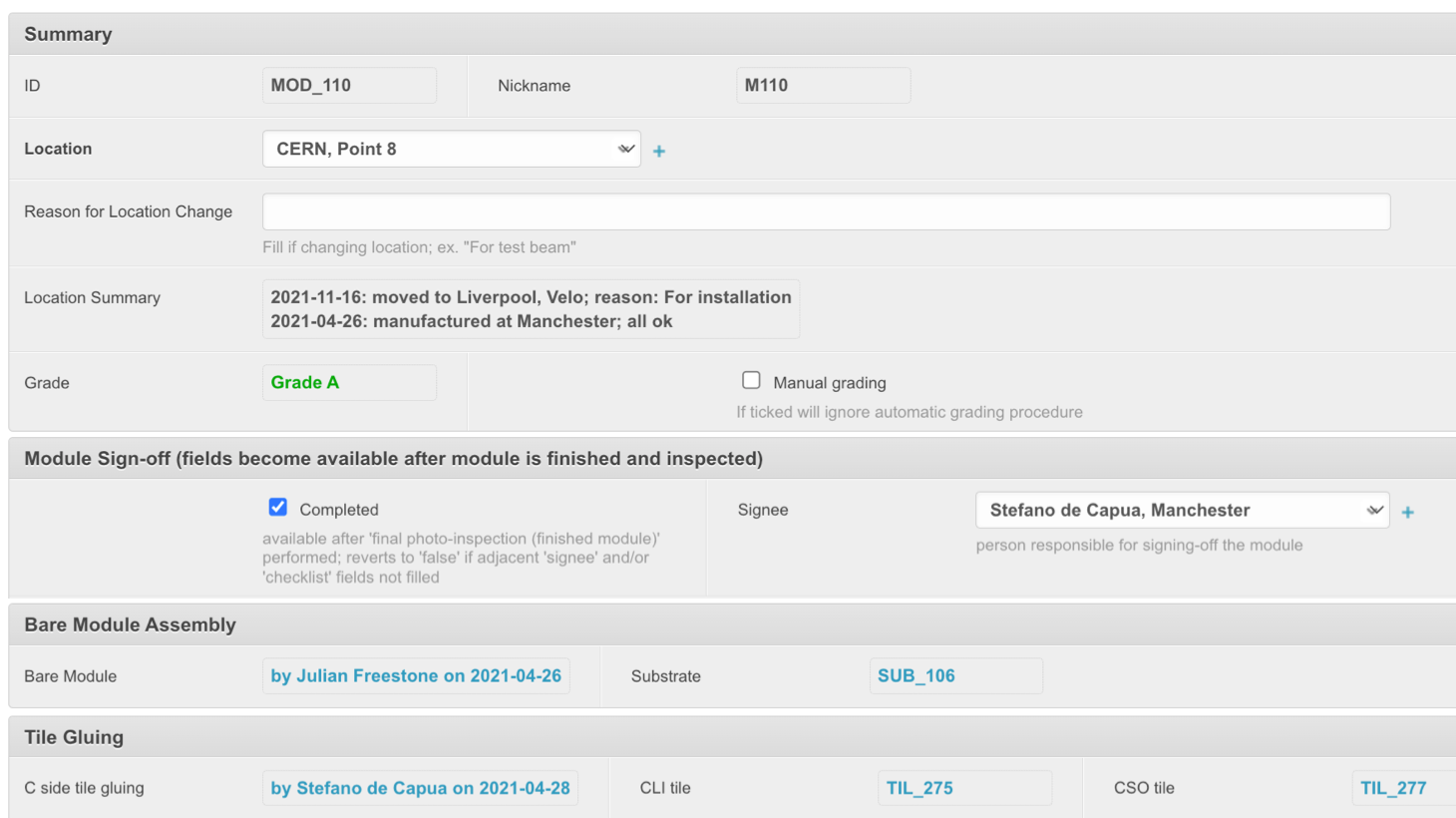}
    \caption{Extract from a database module page showing the summary and two of the assembly and measurements steps performed on a module.}
    \label{fig:module_page}
\end{figure}

The table shown in Fig.~\ref{fig:database} also provides quick access to any process or test performed, by simply clicking on the corresponding cell and landing on a page with all the information concerning that particular step. In addition, the first column gives access to a more detailed module page, containing a comprehensive list of every action taken during the assembly. A small section of a module page is shown, as an example, in Fig.~\ref{fig:module_page}.

\subsection{Bare module assembly}
\label{sec:bare_module_assembly}
\begin{figure}[htbp]
  \centering
  \includegraphics[width=0.8\linewidth]{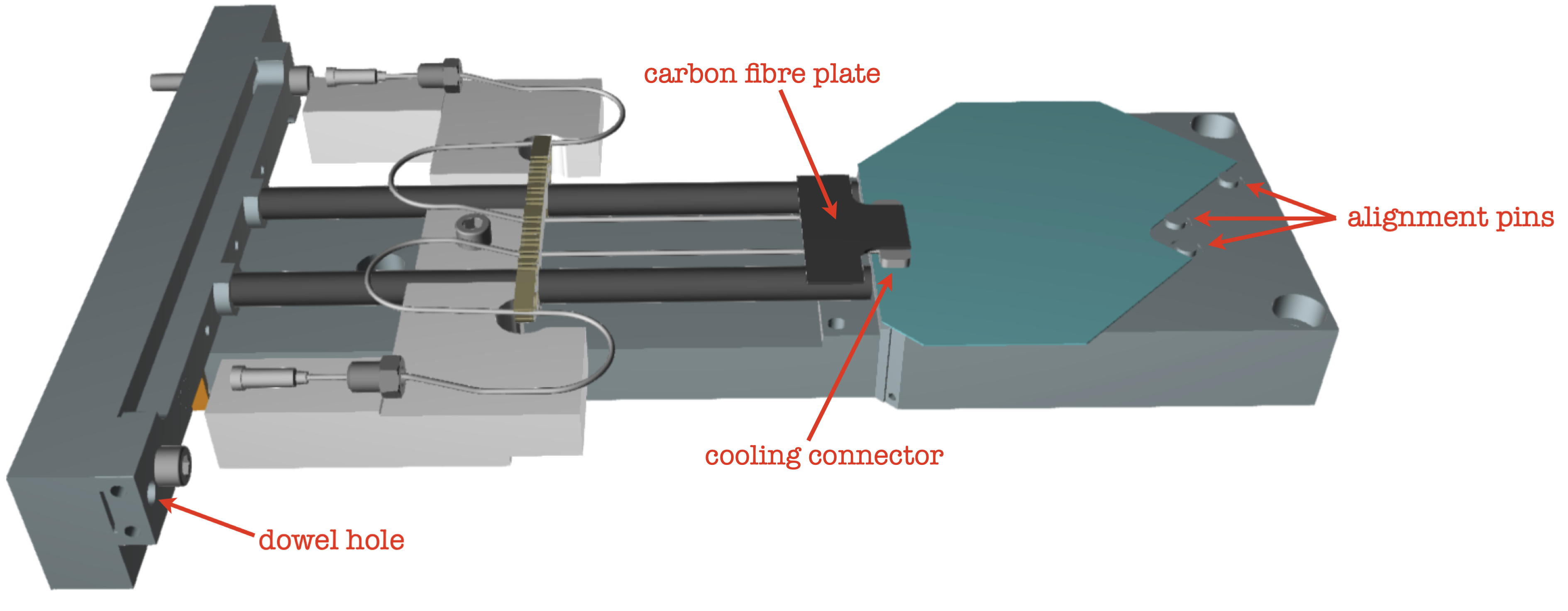}
  \caption{Drawing of the bare module jig. The jig is used to align the microchannel plate to the mechanical support (hurdle) and hold it in place during curing of the glue.}
  \label{fig:bare_module_jig}
\end{figure}

The first step in the construction of a VELO module consists of assembling together the mechanical and cooling structure, forming a bare module onto which the other components will then be attached. The base of the module is an aluminium foot with a steel dowel pin on one side, used for alignment purposes. Two carbon fibre rods are glued into the foot and a carbon fibre plate is glued at the other end of the rods. This structure is then inserted in the bare module jig, shown in Fig.~\ref{fig:bare_module_jig}, together with the microchannel plate and a cooling clamp which relieves the microchannel plate from thermal and mechanical stress from the rest of the system. 

The bare module jig ensures that the microchannel plate is placed within tolerance with respect to the dowel hole in the foot. Three alignment pins are used for this purpose (which the L-shaped edge of the microchannel plate fits against). After correctly positioning it in the jig, the microchannel plate is pushed flush against these three pins. A small amount of glue\footnote{Huntsman Araldite 2011.} is deposited on the top surface of the cooling connector and the carbon fibre plate is pressed onto it and clamped. 

\begin{figure}[htbp]
  \centering
  \includegraphics[width=0.3\linewidth]{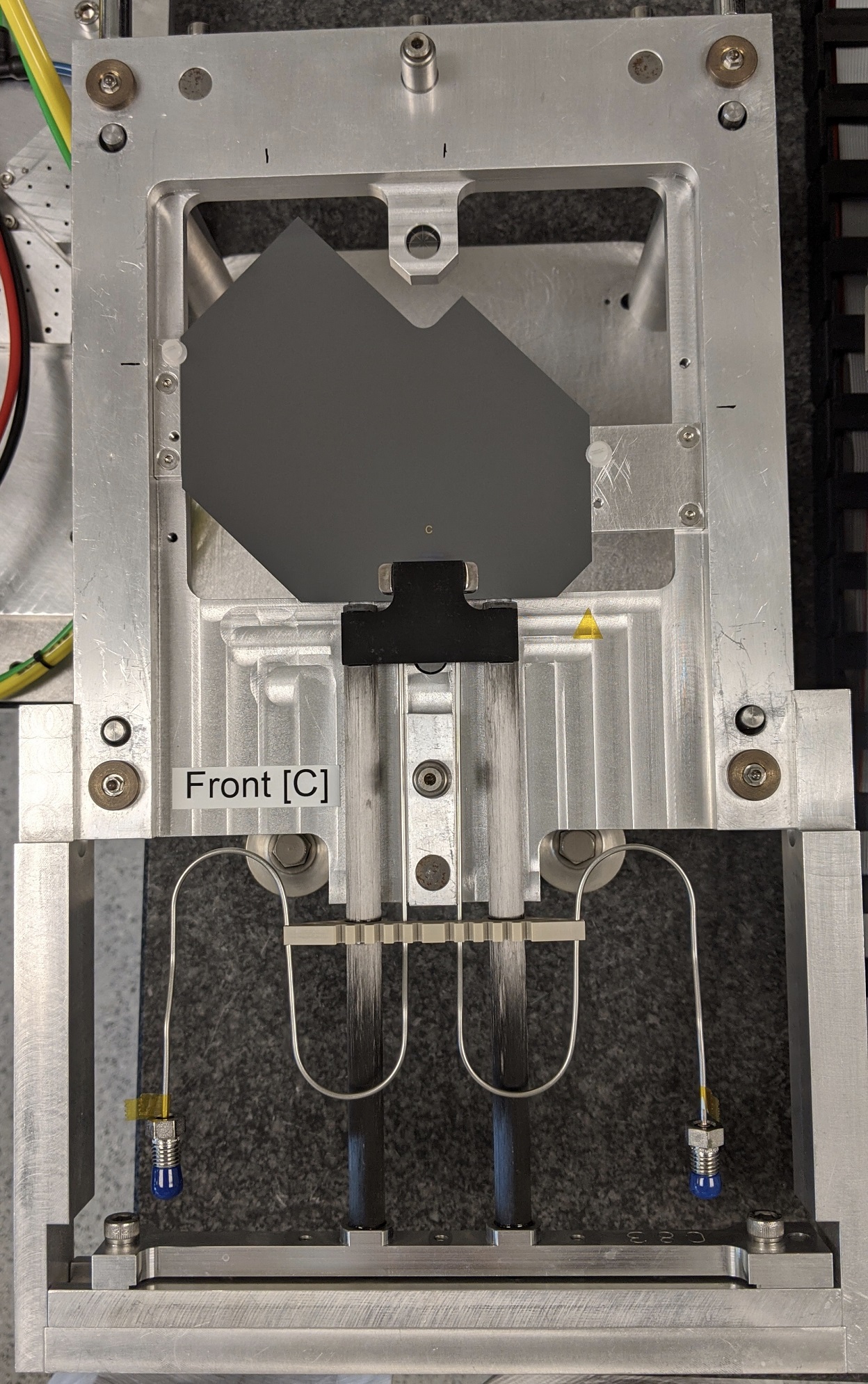}
  \includegraphics[width=0.3\linewidth]{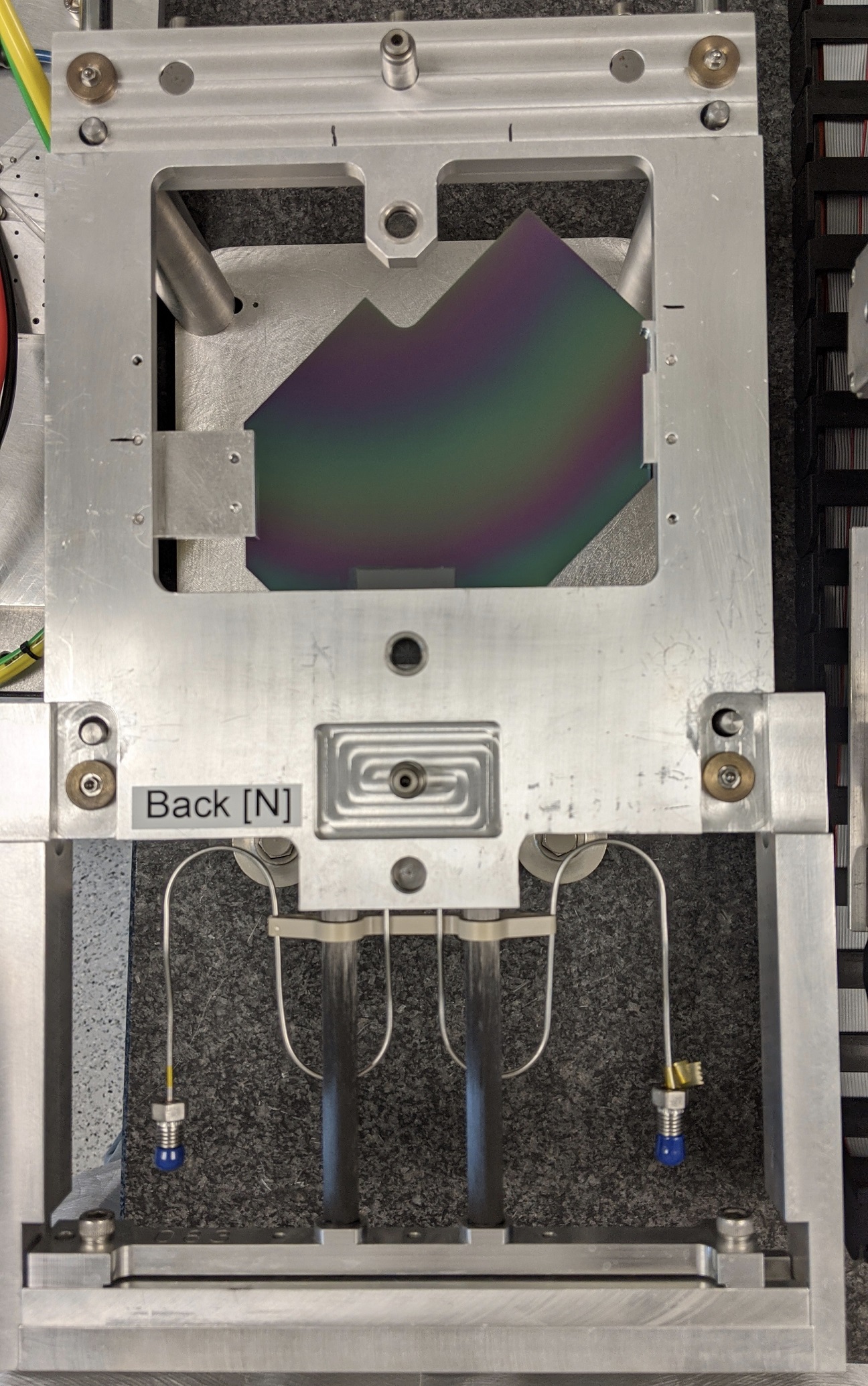}
  \caption{A bare module installed in the turn-plate. Both the connector and non-connector side of a module are shown.}
  \label{fig:bare_module}
\end{figure}

After curing, the module is moved into a different jig, called the turn-plate, and held in place by the frame at two fixation points at its sides. The turn-plate is used to carry out the attachment of tiles and hybrids, as shown in Fig.~\ref{fig:bare_module}. A local reference system is defined using a glass marker located at the top of the turn-plate, and visible from either side, and the dowel hole at the bottom of the turn-plate. A transformation is then performed to translate all coordinates into the LHCb reference frame, in which the origin coincides with the interaction point, $x$ is oriented along the long side of the module, $y$ is oriented along the short side of the module, and $z$ along the beam line.  

\subsection{Tiles placement}
\label{sec:tiles_placement}
Placement of the tiles on a module is one of the assembly steps that requires the highest precision, aiming at achieving a precision better than $\pm$30$\mum$ relative to the dowel hole on the module foot. In their nominal position, the tiles attached on opposing sides of the module should have an overlap of 200$\mum$. In addition, the absolute position of the tiles with respect to the module foot in the horizontal plane is also critical, given that the smallest nominal clearance between sensors and RF foil is 890\mum.

\begin{figure}[htbp]
  \centering
  \includegraphics[width=0.4\linewidth]{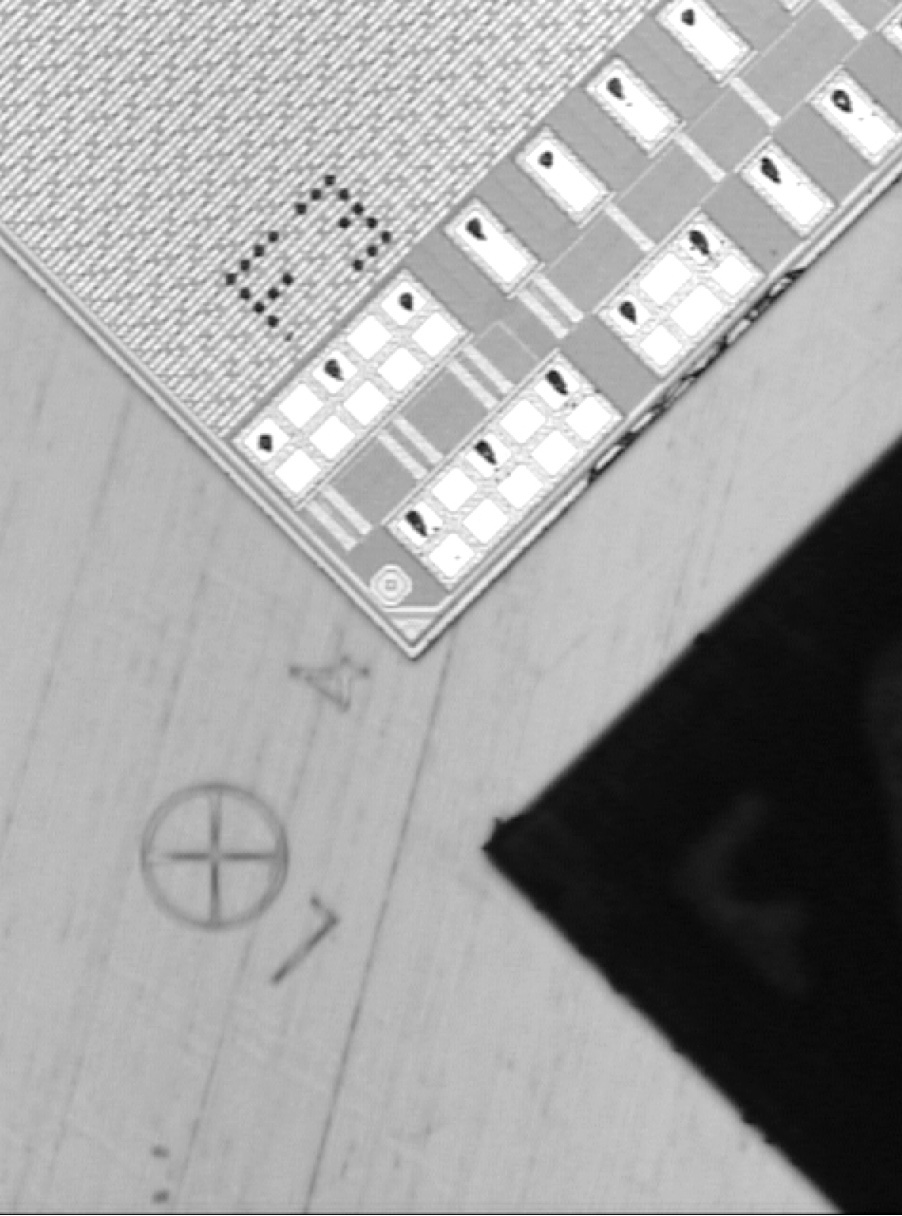}
  \includegraphics[width=0.4\linewidth]{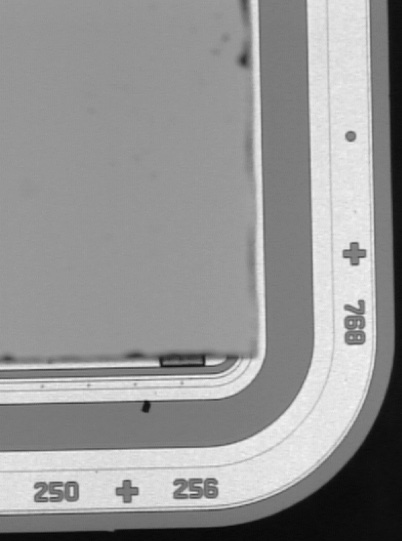}
  \caption{Markers used by the alignment system camera for pattern recognition. Left: Marker in the corner of an ASIC. Right: markers on the back of a sensor.}
  \label{fig:markers}
\end{figure}

The tiles are first placed in a dedicated alignment system, held in place by vacuum chucks (see Sect.~\ref{sec:tiles_attachment_manchester} and \ref{sec:tiles_attachment_nikhef}). The alignment system comprises a camera for pattern recognition and motorized motion stages that allow for micron-precision positioning. The alignment process locates the markers etched on the corners of each VeloPix ASIC (see Fig.~\ref{fig:markers}) and uses this information to automatically move the alignment stages to the target position. Afterwards, the two tiles are transferred to a dedicated frame system for adhesive deposition and attachment to the microchannel substrate. Fig.~\ref{fig:markers} also shows the markers present on the back of the sensor that, after attachment, are visible in the two overhanging tiles (one per side) and can be used for front-back cross-reference.

The attachment procedure followed by the two module production sites, although based on the same approach, slightly differs in terms of equipment employed and assembly sequence. Both procedures are described in the following sections.

\subsubsection{Glue parameters and pattern} 
\label{sec:glue_parameters}
The glue layer connecting the ASICs with the microchannel substrate is the only path for heat dissipation in the final detector: maximising the glue coverage, while minimising its thickness, is critical to ensure sufficient cooling performance at the overhanging sensor tip. Although a thin, uniform layer of glue underneath the tile ensures the best cooling performance, the glue layer also needs to be thick enough to absorb any mechanical stress induced by the temperature changes. In addition, the maximum size of the filler particles is about 60$\mum$, which needs to be taken into account to prevent damages to the silicon substrate when pressing the tiles down. Taking into account these considerations, the targeted glue layer thickness underneath each tile was set to 80$\mum$. \\
Furthermore, the glue layer should ideally cover as much area of the tile as possible, to maximise the heat transfer, but at the same time spillage in the gaps between adjacent ASICs should be avoided. In addition, since the detector is operated in vacuum, air pockets trapped in the glue layer may lead to damage and an appropriate glue pattern needs to be chosen in order to minimise the chances of trapping air. For these reasons, a glue dispenser robot was chosen to achieve the accuracy and control required to meet all the aforementioned requirements. The glue deposition was performed using a programmable robot\footnote{Fisnar F5200N.} coupled to a syringe and a needle. The fluid dispenser runs on compressed air.\\

\begin{figure}[htbp]
  \centering
  \begin{tabular}{m{6cm} m{6cm}}
    \includegraphics[width=1.0\linewidth]{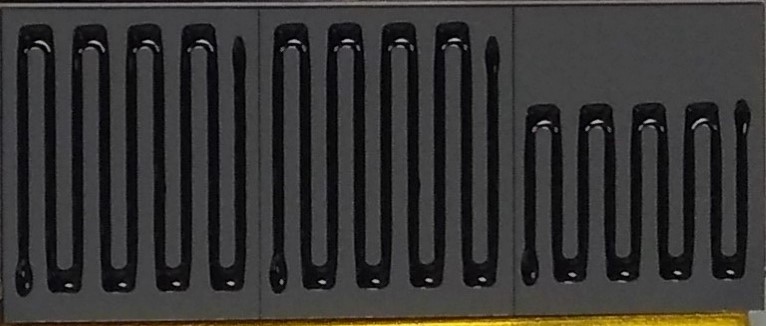} &
    \includegraphics[width=1.0\linewidth]{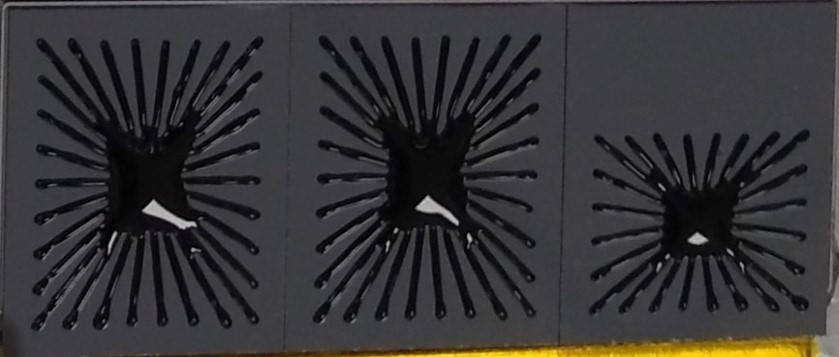} \\
    \includegraphics[width=1.0\linewidth]{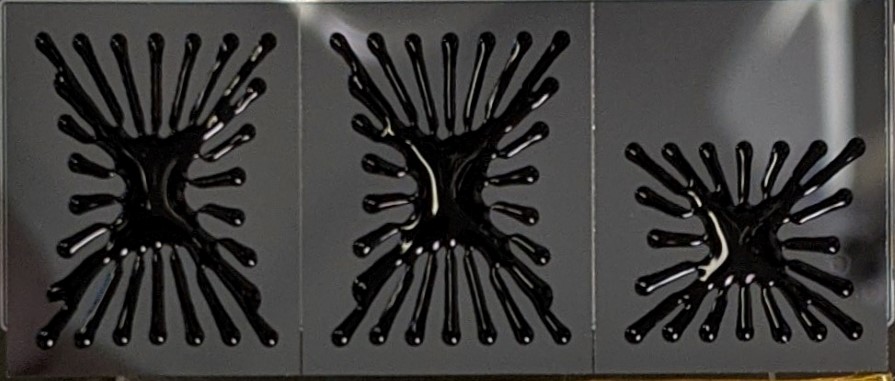} &
  \includegraphics[width=0.75\linewidth]{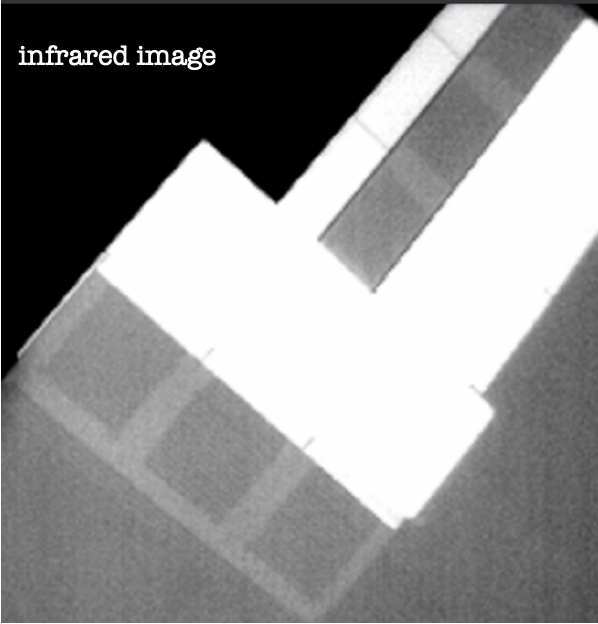} \\
  \end{tabular}
  \caption{Comparison of glue patterns deposited on tiles. Top left: snake pattern. Top right: star pattern. Bottom left: star pattern with concave inner sides. Bottom right: infrared image taken after tiles attachment, showing the resulting squared shape patterns.}
  \label{fig:glue_patterns}
\end{figure}

An extensive R$\&$D campaign was carried out to investigate different adhesives and deposition patterns~\cite{glue_note}. Initially, a snake pattern (see Fig~\ref{fig:glue_patterns}) was proposed, which allows for the air to escape through the openings and ensures a good coverage. However, in case of accidental connections between the points near the openings, an air pocket could be created. Hence, a star shaped pattern (see Fig.~\ref{fig:glue_patterns}) was evaluated and proved to be inherently better at preventing air trapping in the pattern. The star pattern has also a lower chance to cause glue spillage, as the endpoints of the star can be kept at a safe distance from the ASICs, while still providing an excellent coverage. The chances of spillage were further reduced by concavely shaping the core of the star in a hourglass shape (see Fig.~\ref{fig:glue_patterns}), achieving a square shaped pattern after attachment. 

The glue selected for the tiles attachment was a thermally conductive epoxy~\cite{glue_note} (see Sect.~\ref{sec:glue_choices}). The initial step of the glueing procedure consists of heating the epoxy to 50\,$^{\circ}$C for one hour and then leaving it to cool down to room temperature. This procedure helps to standardise the properties of the glue before the attachment is performed, and it also ensures a proper mixing between the main resin and the aluminium oxide filler. 
The epoxy is then mixed to the catalyst in a planetary mixer, poured into a syringe and its viscosity is checked to ensure it lies within an acceptable range. The viscosity is determined before every attachment by measuring the amount of glue dispensed per unit time at a given force. This allows to set the line deposition speed according to the actual viscosity and hence ensures the same amount of glue is deposited in every module. The deposition is then started, bringing the tip of the needle at 200$\mum$ from the surface of the tiles, which are held in place by vacuum chucks. After the glue pattern has been deposited, a heat gun pointing towards the tile is used to heat up the top layer of the glue to 60\,$^{\circ}$C for 1 minute, such that any moisture that might have accumulated on the surface of the glue during deposition, and that can lead to adhesion failures, is removed.
\begin{figure}[htbp]
  \centering
  \includegraphics[width=0.45\linewidth]{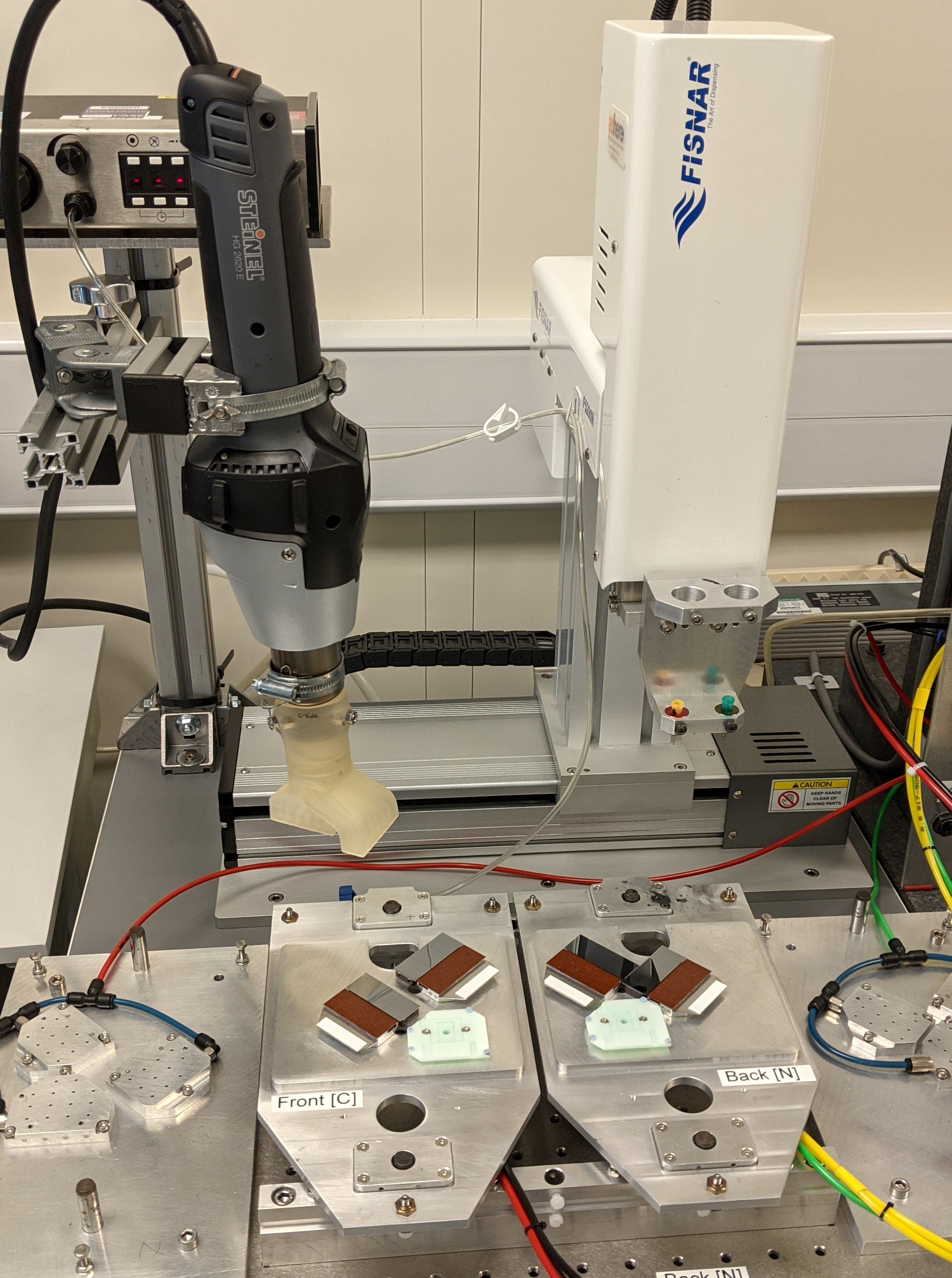}
  \caption{Photograph showing the glue dispenser, the heat gun and four silicon tiles.}
  \label{fig:glue_robot}
\end{figure}

\subsubsection{Tile attachment procedure at Manchester}
\label{sec:tiles_attachment_manchester}
\begin{figure}[htbp]
  \centering
  \includegraphics[width=0.95\linewidth]{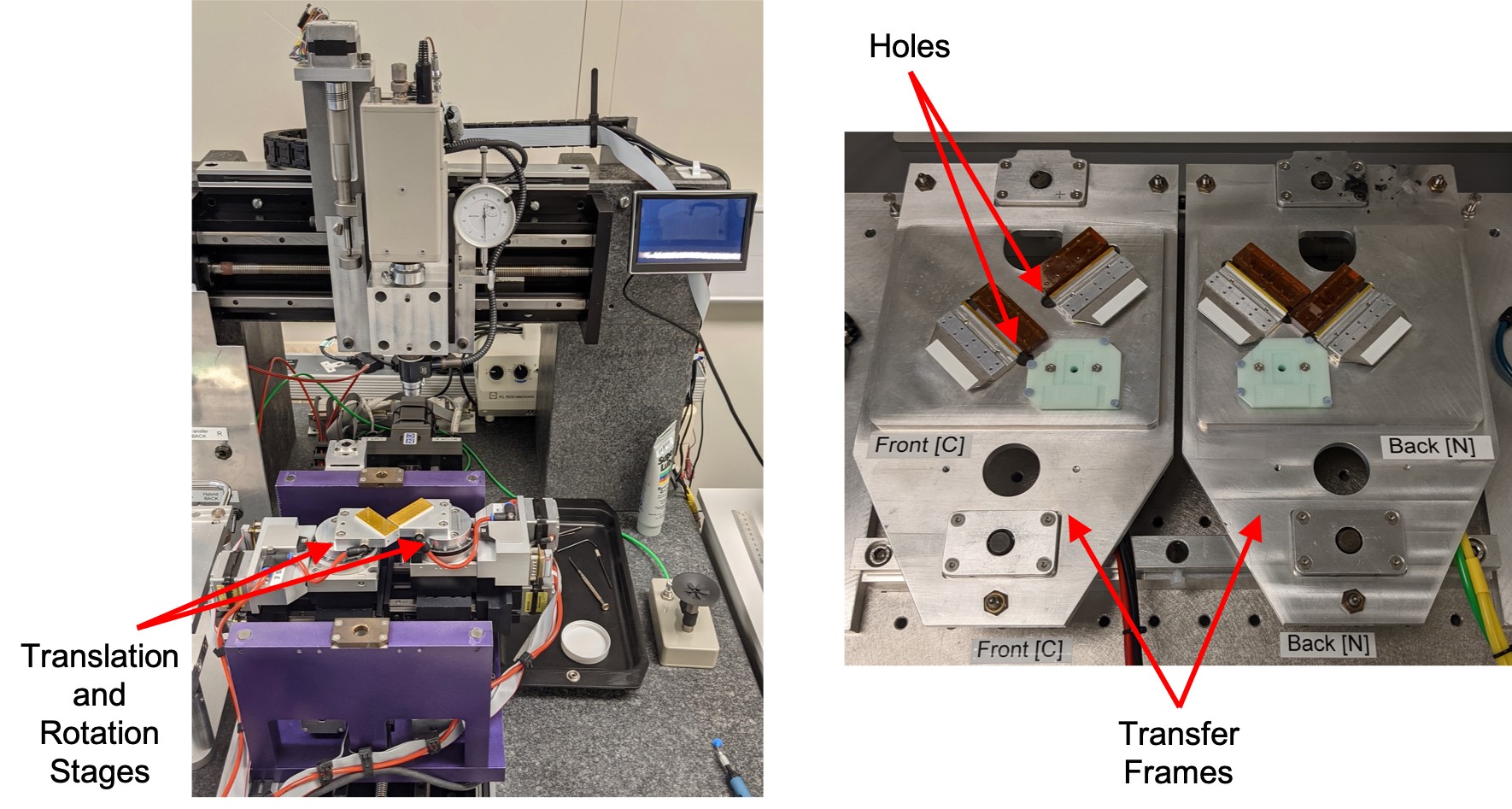}
  \caption{Photographs of the custom-made set of jigs and alignment system used for tiles attachment in Manchester. Left: the alignment stages and camera. Right: the transfer plates, installed under the glue dispenser.}
  \label{fig:Stages}
\end{figure}

The alignment and attachment of the tiles in Manchester is performed using a custom-made set of jigs and an alignment system (Fig.~\ref{fig:Stages}). The system consists of two translation and one rotation stage and uses a camera for pattern recognition. A pair of tiles are placed on the stages and held by vacuum. The alignment process locates the markers on the ASICs of each tile and adjusts the $x$, $y$ and $\phi$\footnote{defined as the angle in the $xy$ plane.} coordinates of each individual stage with respect to a turn-plate, which will hold the bare module during tiles attachment. The turn-plate has the same mechanical fixing to the bare module foot as in the final VELO base. 

\begin{figure}[htbp]
  \centering
  \includegraphics[width=0.8\linewidth]{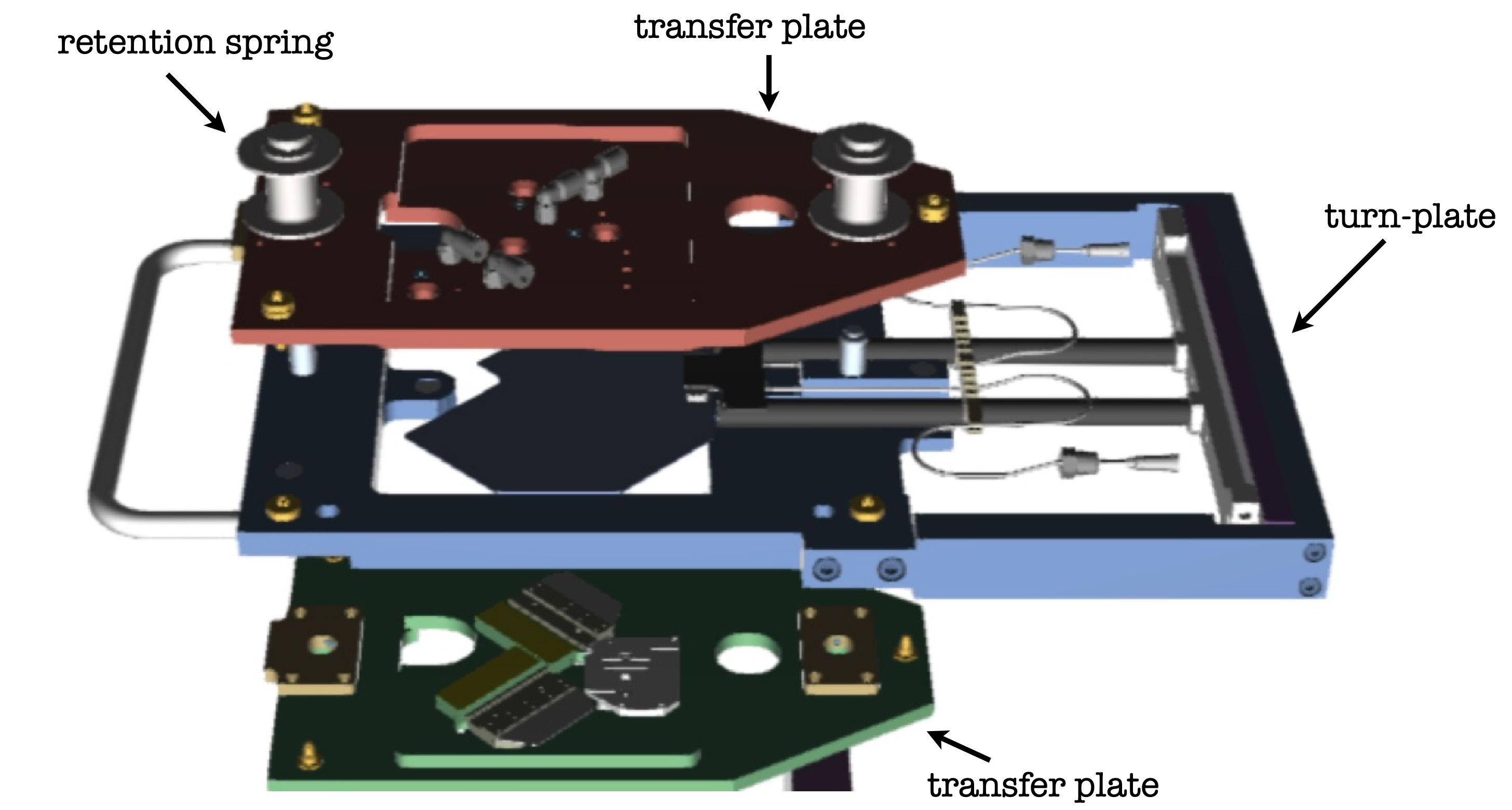}  \caption{A CAD model showing how the transfer plates (in green and red) are brought together and attached to the turn-plate (in blue) which holds the bare module.}
  \label{fig:TransferPlates}
\end{figure}

After alignment, the two tiles are transferred to a dedicated transfer frame (see Fig.~\ref{fig:TransferPlates}) and the transfer is validated visually by looking through holes in the transfer plate which are located in front of the ASIC markers. The same procedure is then repeated for the second pair of tiles.
The two transfer-plates, carrying the tiles for both the connector and non-connector sides, are places under the glue dispenser and the glue patterns are deposited on all four tiles, on the VeloPix side. The bare module is then installed in the turn-plate and the two transfer plates are placed either side of the turn-plate and held in place with retention springs. The three jigs are left in this configuration for 24 hours while the glue cures.

\begin{figure}[htbp]
  \centering
  \includegraphics[width=0.6\linewidth]{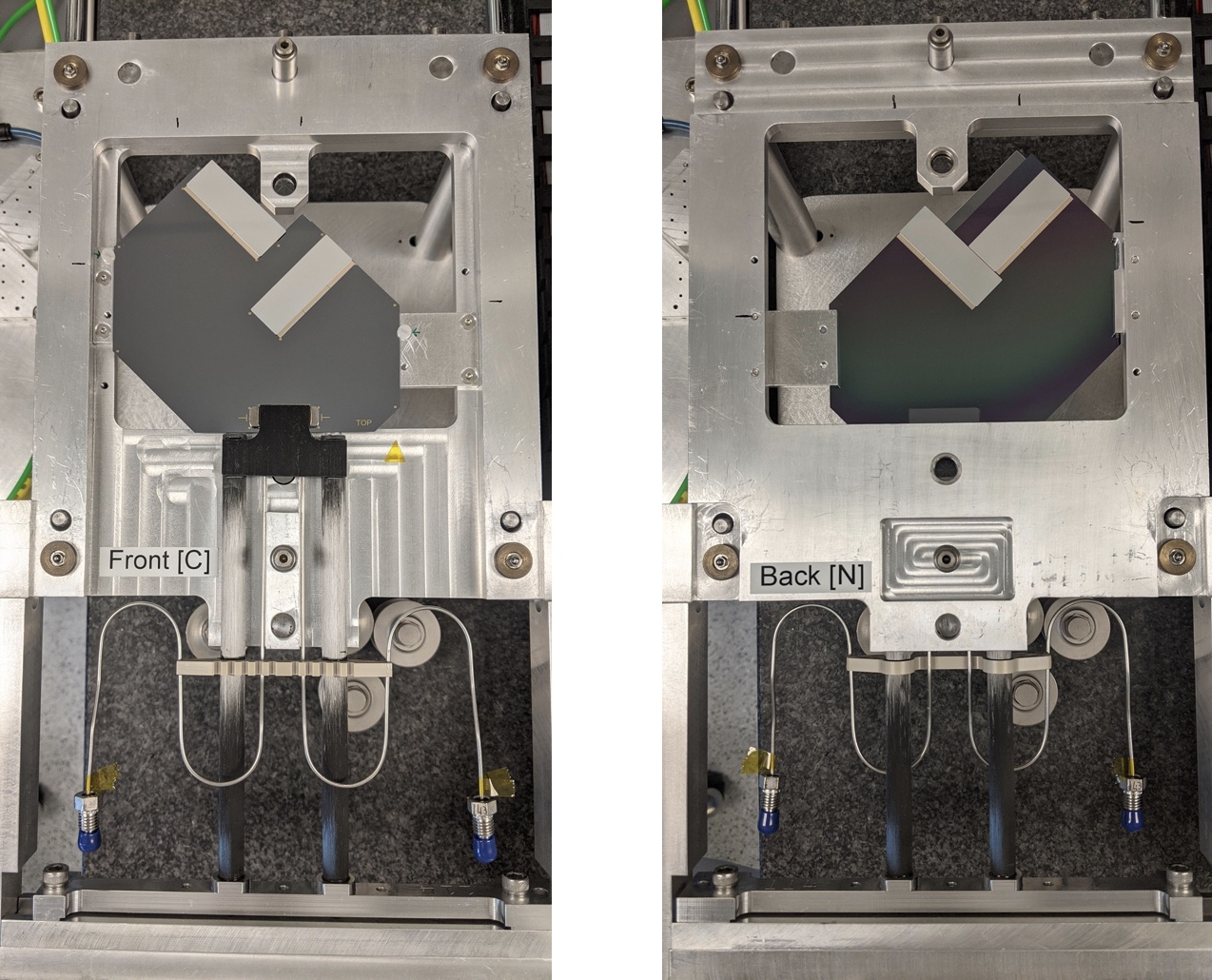}  \caption{Photographs of the two sides of a module installed in the turn-plate, after tile attachment.}
  \label{fig:tiles_module}
\end{figure}

Photographs of the two sides of a module, after tiles attachment, are shown in Fig.~\ref{fig:tiles_module}.

\subsubsection{Tile attachment procedure at Nikhef} 
\label{sec:tiles_attachment_nikhef}
\begin{figure}[htbp]
  \centering
  \includegraphics[width=0.5\linewidth]{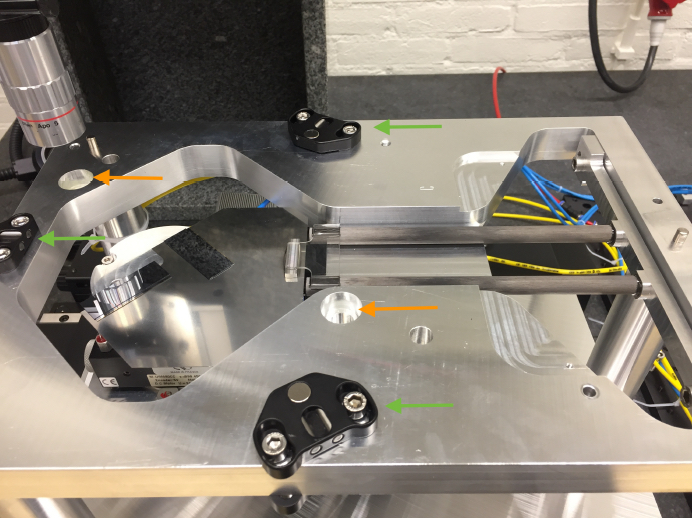}
  \caption{The turn-plate, shown inside the pick and place machine. The orange arrows indicate the position of the glass markers used to determine the position of the jig relative to the camera. The green arrows point to the kinematic mounts that align the transfer plate relative to the turn-plate.}
  \label{fig:Nikhef_turnjig}
\end{figure}

As in Manchester, the placement of the tiles on the bare module is performed using custom-built mechanical jigs, motorized motion stages and a camera. 
Fig.~\ref{fig:Nikhef_turnjig} shows the turn-plate, the mechanical frame used to hold the module during tiles attachment. 

\begin{figure}[htbp]
  \centering
  \includegraphics[width=0.4\linewidth]{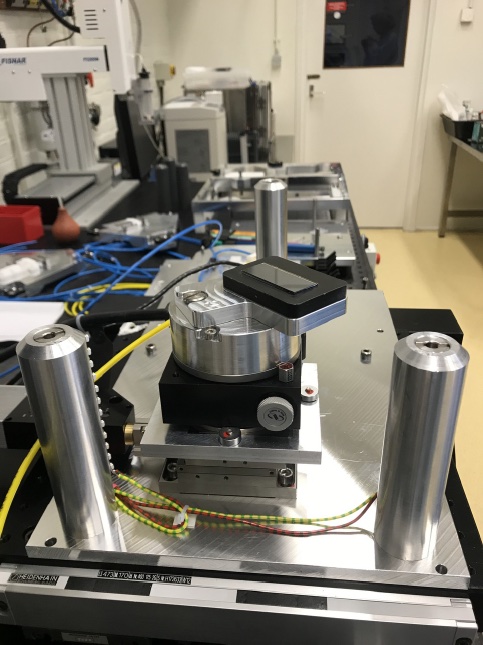}
  \caption{Tile vacuum positioning jig on the $x$-$y$-$\phi$ stack, with a dummy tile in the centre and the three pillars that hold the turn-plate in the pick and place machine.}
  \label{fig:Nikhef_positioningjig}
\end{figure}

Two main stages allow to move the turn-plate in $x$ and $y$, relative to the camera, while three smaller stages move the tile relative to the turn-plate in $x$, $y$ and $\phi$~(Fig.~\ref{fig:Nikhef_positioningjig}). Finally, one stage is responsible for controlling the camera focus.  

\begin{figure}[htbp]
  \centering
  \includegraphics[width=0.4\linewidth]{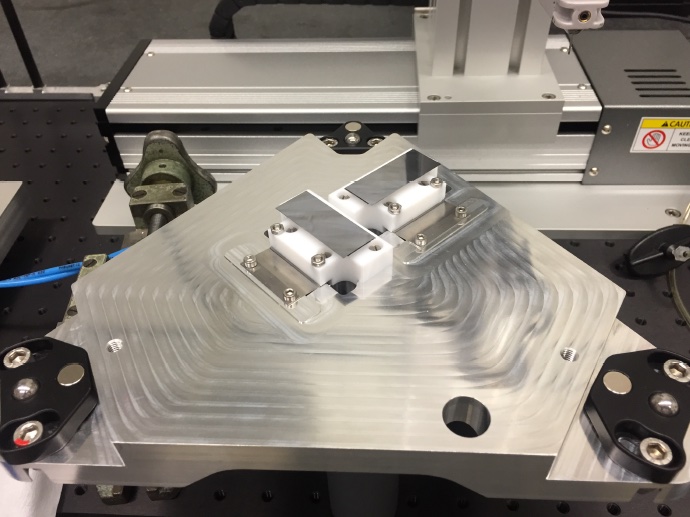}
  \caption{The transfer plate, shown with two dummy sensor tiles attached to the vacuum chucks.}
  \label{fig:Nikhef_transferjig}
\end{figure}

Fig.~\ref{fig:Nikhef_transferjig} shows the transfer plate used for holding the tiles during attachment to the bare module. There are two types of transfer plates, one for the attachment to the connector side of the bare module and one for the non-connector side. The relative alignment between the turn-plate and the transfer plates is determined by a set of kinematic mounts that allow to reposition the two jigs with an accuracy of a few micrometers.

Initially, the turn-plate is installed in the pick and place machine with the VeloPix side facing upwards. Then, the positioning jig is placed in the turn-plate and moved to the target position according to the tile to be aligned. The tile is placed on the tile positioning jig and the jig itself is installed on the turn-plate. The camera is used to properly align the two jigs. Finally, the tile is aligned to its target position by using the two outer ASIC markers and then transferred to the transfer plate. The procedure is repeated for the second tile, such that the transfer plate carries both tiles that need to be attached to one side of the module. The procedure is repeated for the other side of the module. The attachment, including the glue curing time (24 hours), is performed one side at the time.

\subsection{Hybrid placement}
\label{sec:hybrids_placement}
\begin{figure}[htbp]
  \centering
  \includegraphics[width=0.4\linewidth]{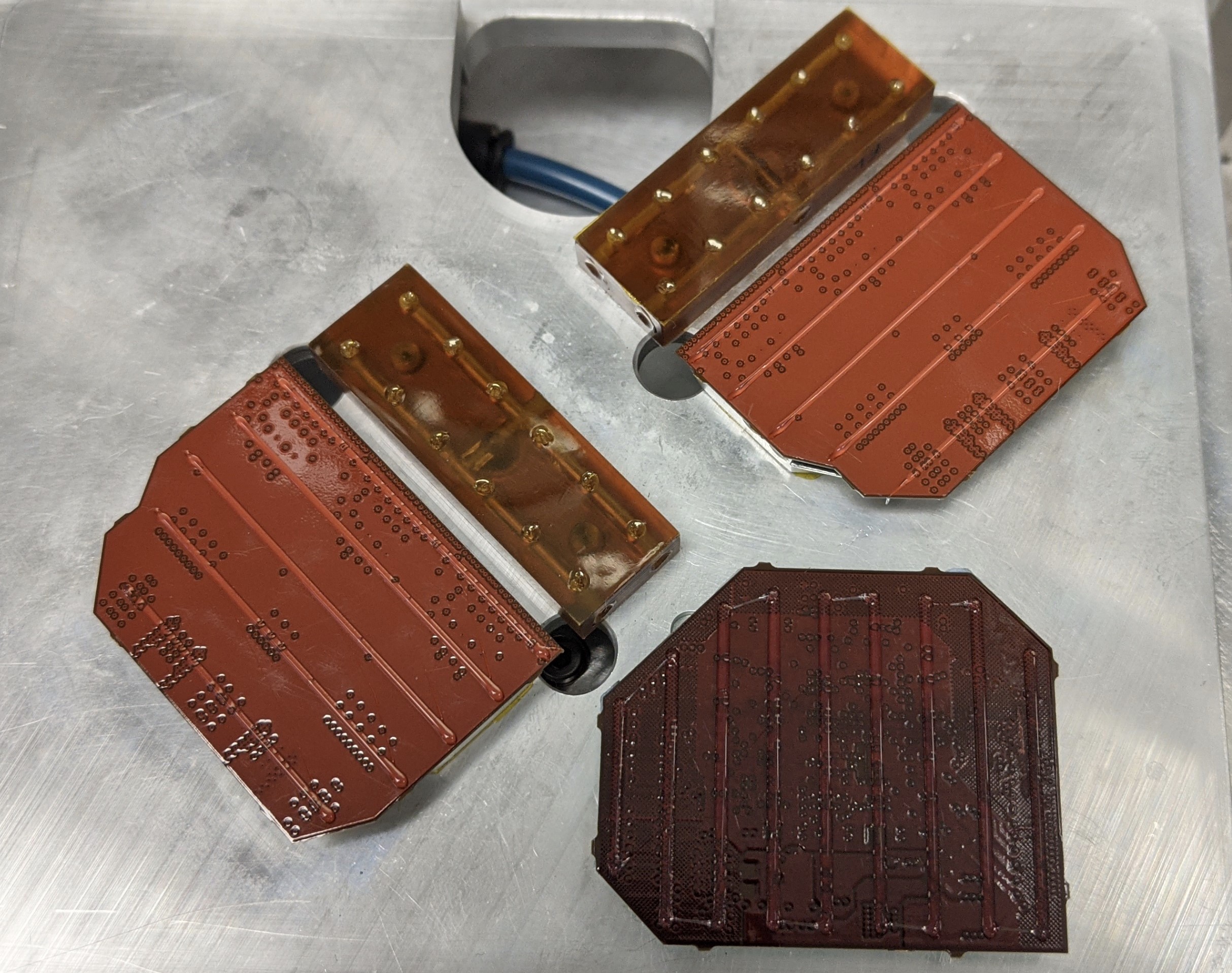}
  \includegraphics[width=0.52\linewidth]{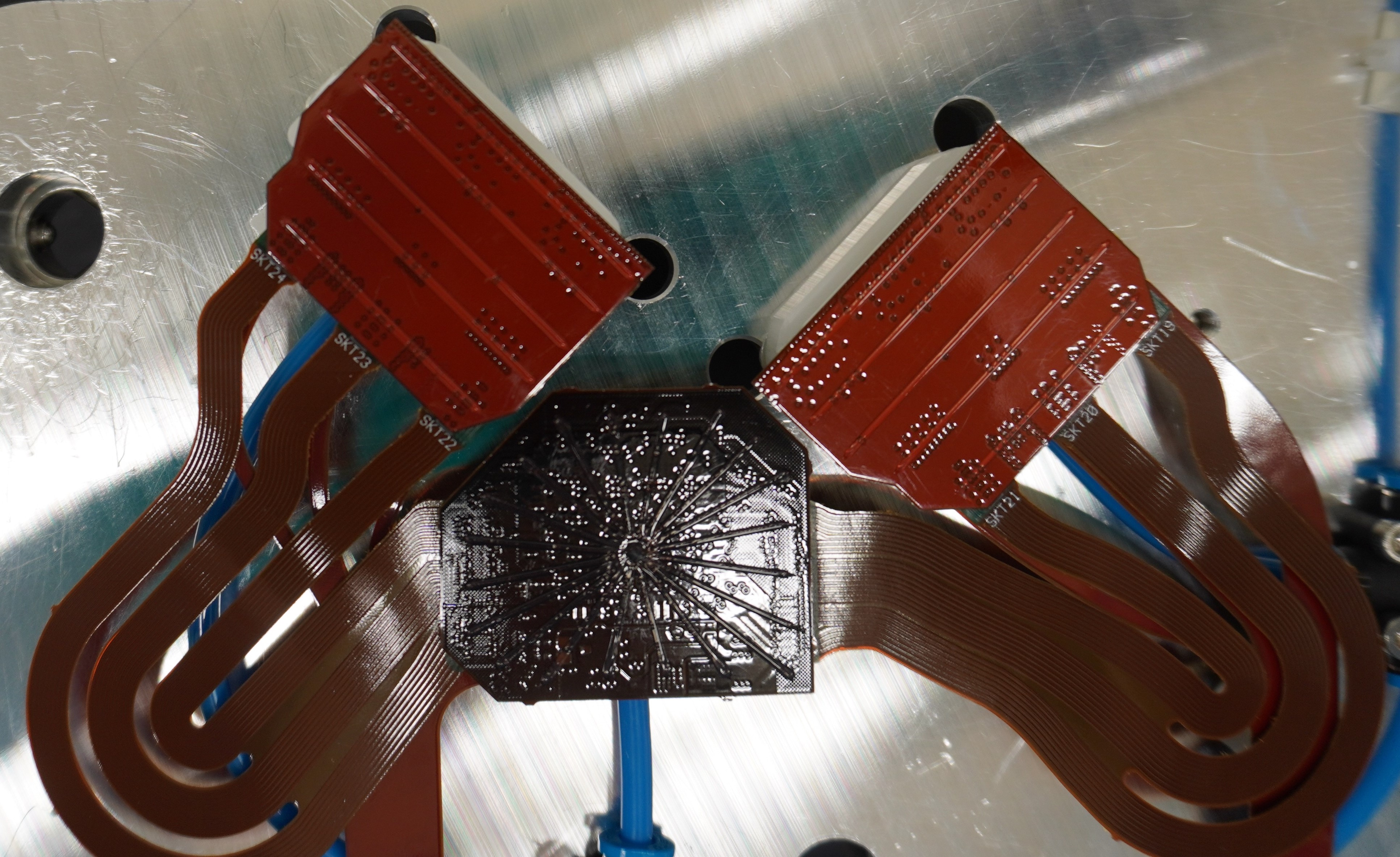} 
  \caption{The glue pattern of the FE and GBTx hybrids as implemented in Manchester (left) and Nikhef (right).}
  \label{fig:hybrid_glue_patterns}
\end{figure}

The procedure for attaching the front-end and control hybrids to a module is performed similarly to that of the tiles. However, the hybrid placement does not require the same level of precision, with an alignment of 100$\mum$ with respect to the tile being enough to ensure a successful wire-bonding.
The choice of the glue is mainly dictated by the need for good radiation hardness and flexibility. 
The cooling requirements are not as critical as for the tiles, as discussed in Sect.~\ref{sec:glue_choices}. 
The glue is deposited with the same robot, but using a different syringe and needle. Since the cooling requirements are less stringent, the glue pattern does not need to maximise the area covered, but rather to provide mechanical stability during the wire-bonding process (for the FE hybrids) and a reliable adhesion to the microchannel substrate. For these reasons, the chosen pattern consists of a small number of lines: five evenly spaced lines for the front-end hybrids, with one of them directly underneath the bond pads, and nine lines for the GBTx hybrids. At Nikhef, a star pattern was implemented for the GBTx hybrid, in order to potentially provide a better heat transfer to the control chip. Since the thermal performance measurements did not show a significant difference between the star and the line patterns, the star pattern was not adopted by Manchester (see Fig.~\ref{fig:hybrid_glue_patterns}). The front-end hybrids are also plasma cleaned before attachment, in order to ensure good cleanliness of the bonding pads and of the pins in their connectors. 

\subsubsection{Procedure at Manchester}
\begin{figure}[htbp]
  \centering
  \includegraphics[width=0.3\linewidth]{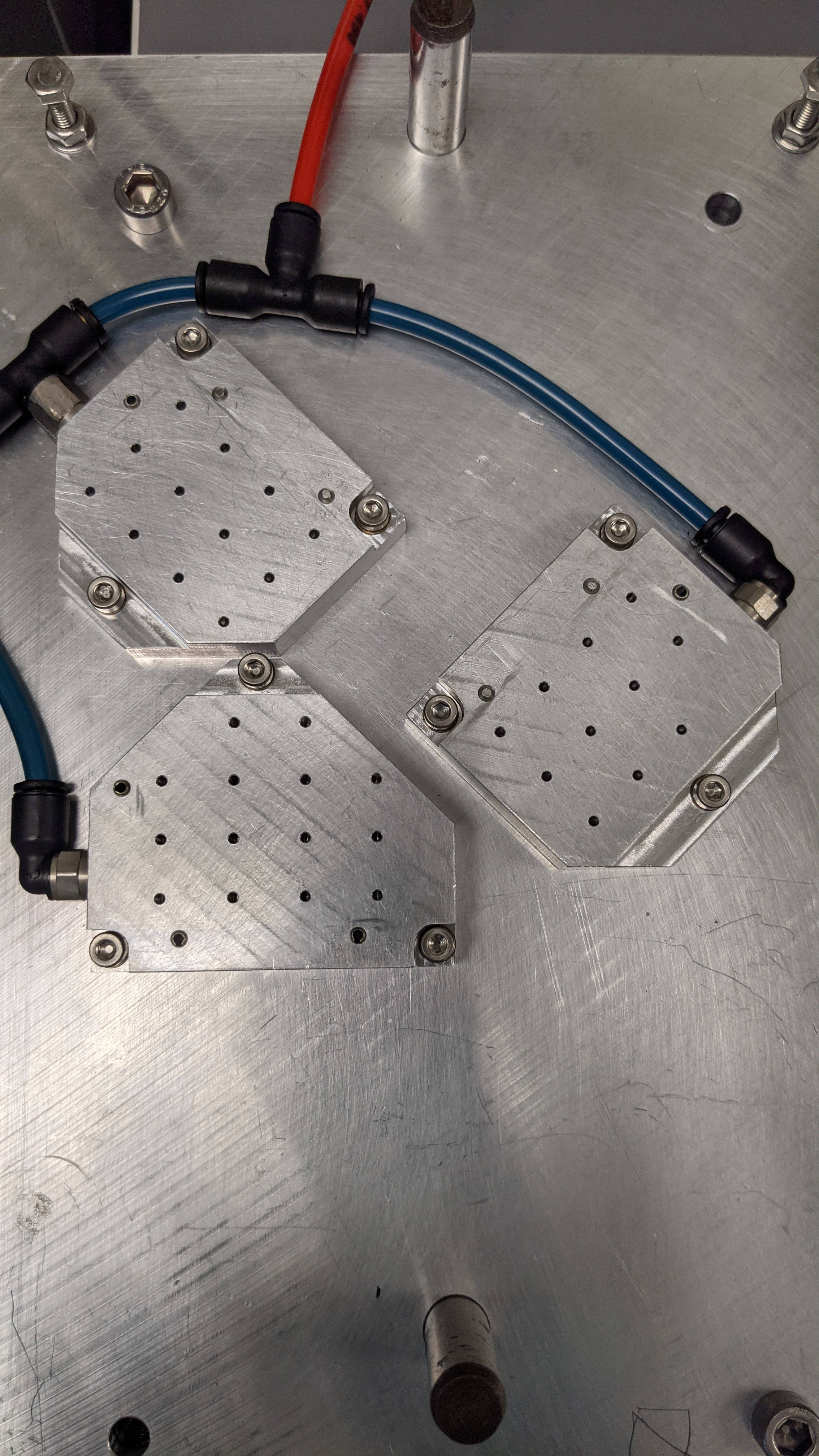}
  \includegraphics[width=0.3\linewidth]{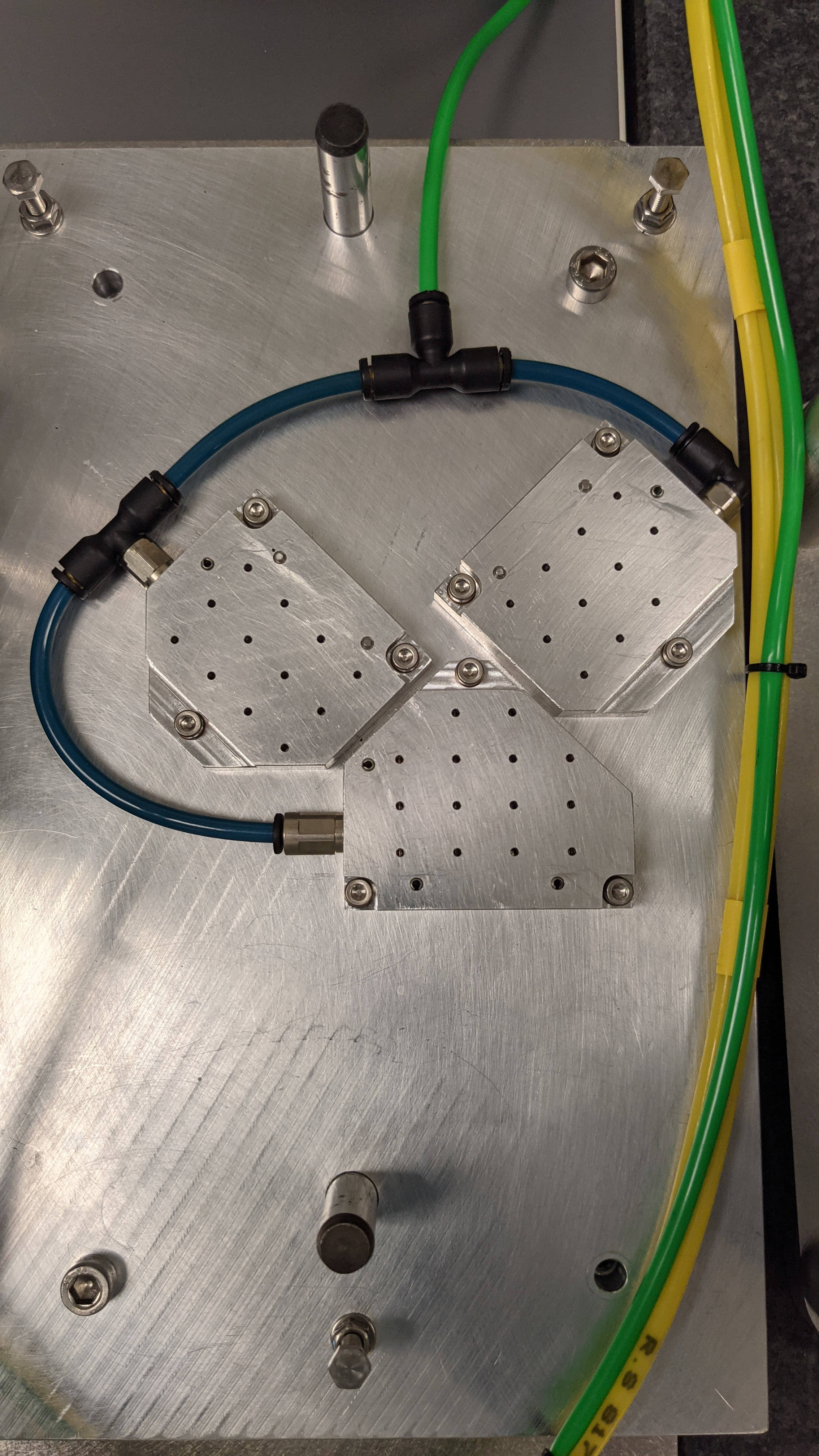}
  \caption{Hybrids alignment jigs, for the connector side (left) and non-connector side (right).}
  \label{fig:hybrids_align_jigs}
\end{figure}

The front-end and GBTx hybrids are positioned in dedicated alignment frames, shown in Fig.~\ref{fig:hybrids_align_jigs}, where they are mechanically aligned against a set of pins, which ensure the precision required of about 100$\mum$. Once the alignment is completed, vacuum is activated to hold the hybrids in position. 

The attachment of the hybrids to the module is performed  following the same strategy used for the tiles. The hybrids for the connector side of the module are transferred to the corresponding transfer plate, shown in Fig.~\ref{fig:Stages}, and installed under the glue dispenser where the glue is deposited. This procedure is then repeated for the non-connector side hybrids and finally the six hybrids are attached to the module on both sides simultaneously. In Manchester, the interconnect and data cables are attached at a later stage.

\begin{figure}[htbp]
  \centering
  \includegraphics[width=0.5\linewidth]{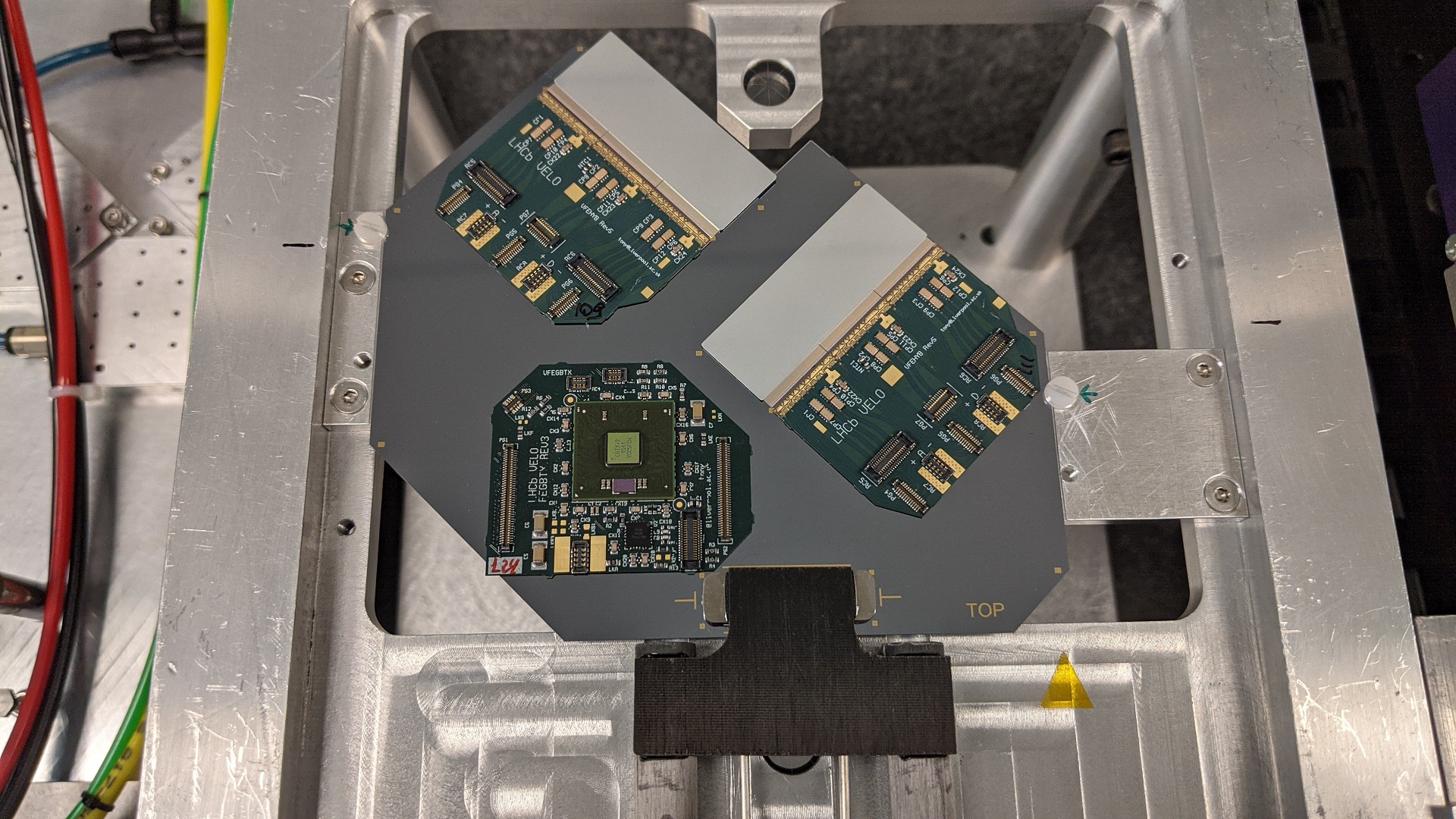}
  \includegraphics[width=0.42\linewidth]{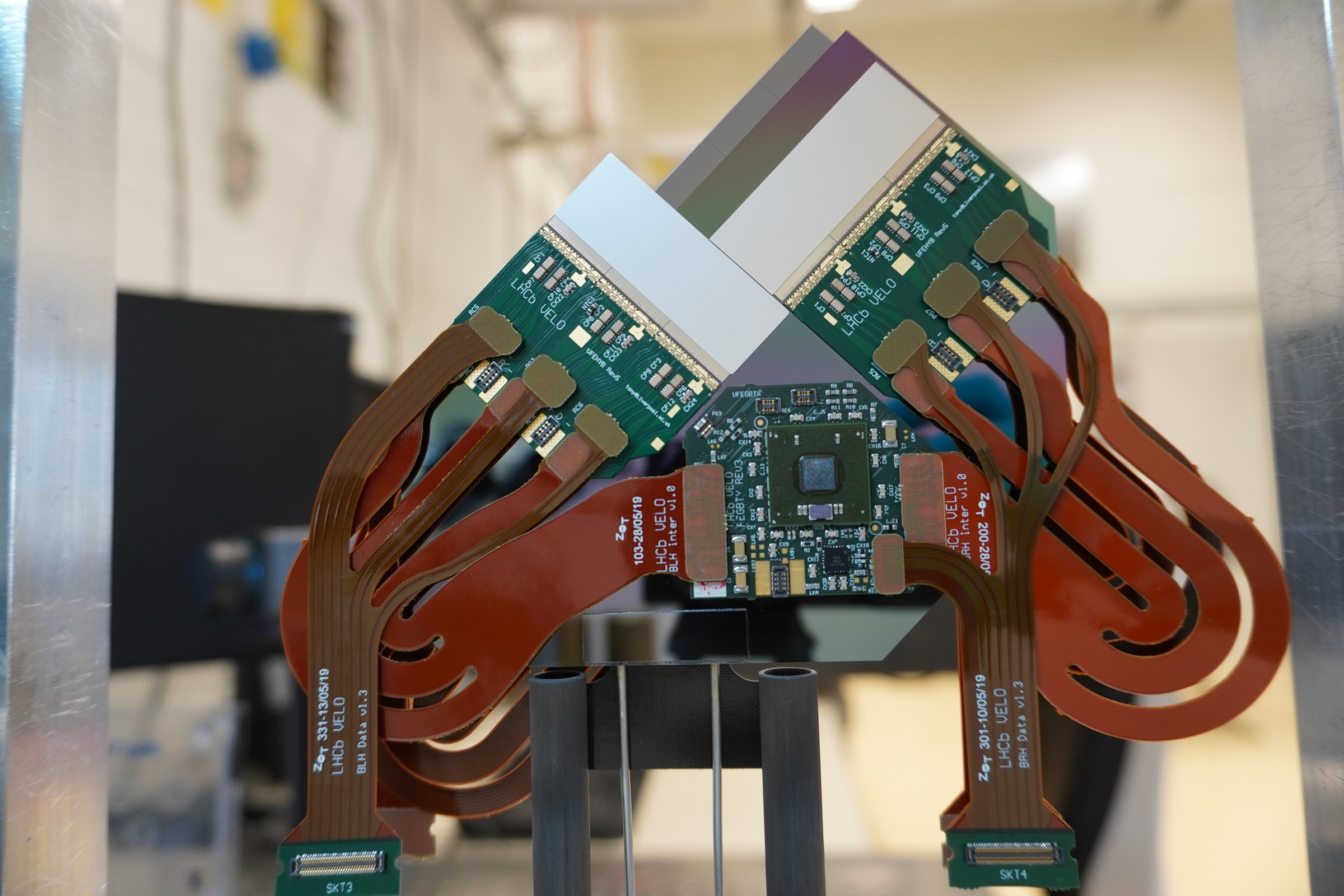}
  \caption{Photographs of both sides of a module with hybrids attached. Left: connector side of a module after hybrids' attachment in Manchester. Right: non-connector side of a module after hybrids' attachment at Nikhef.}
  \label{fig:hybrids_module}
\end{figure}

Photographs of both sides of a module with hybrids attached are shown in Fig.~\ref{fig:hybrids_module}. On the left hand side, the connector side of a module after hybrids' attachment in Manchester. On the right hand side, the non-connector side of a module after hybrids' and cables' attachment at Nikhef.

\subsubsection{Procedure at Nikhef}
\begin{figure}[htbp]
  \centering
  \includegraphics[width=0.8\linewidth]{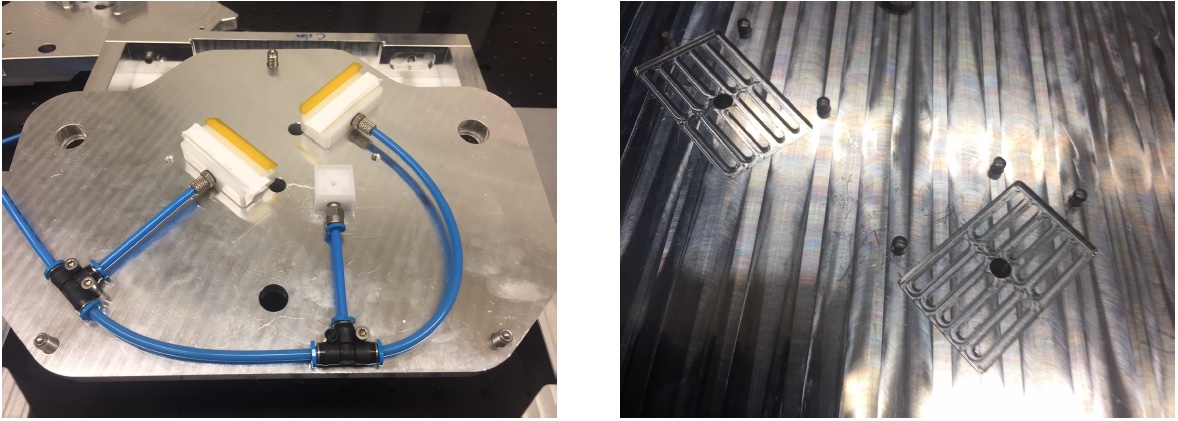} 
  \caption{Left: the hybrid transfer jig, showing the vacuum chucks for the two FE hybrids and the GBTx hybrid. Right: the alignment jig for the connected hybrids before being lifted by the transfer jig.}
  \label{fig:vacuum_pickup_tool}
\end{figure}

The main difference in the procedure at Nikhef is that all the flat cables are connected to the hybrids before gluing. This was motivated by concerns that the relative rigidity of the flat cables could cause difficulties in their attachment, though in practice this did not prove to be an issue. Nevertheless, at Nikhef, the placement of the hybrids and cables is performed in a single step.
A vacuum pickup tool is used to hold the hybrids, and is shown in Fig.~\ref{fig:vacuum_pickup_tool} left. As the alignment of the hybrid
placement is not as critical as for the placement of the tiles, a simplified alignment scheme is used without the need for an optical camera. A mechanical jig with alignment pins against which the FE hybrids can be positioned is used (Fig.~\ref{fig:vacuum_pickup_tool} right). 

The GBTx hybrid is not strongly constrained, in order to prevent additional stresses in the interconnect cables. Once the hybrids are in place, they are held by vacuum and the transfer jig placed on top. The hybrids and flat cables assembly is picked up and placed under the same automated glue dispenser described previously. Glue is then deposited in a line (star) pattern on the backside of the front-end (GBTx) hybrids. To prevent large forces on the microchannel substrate, distance screws are adjusted when the hybrids are picked up. Corresponding spacers on the jig where the module is held during gluing prevent the hybrids from pressing on the substrate.

\subsection{Wire-bonding and HV cables attachment}
\begin{figure}[htbp]
  \centering
  \begin{tabular}{m{7cm} m{7cm}}
  \includegraphics[width=1.0\linewidth]{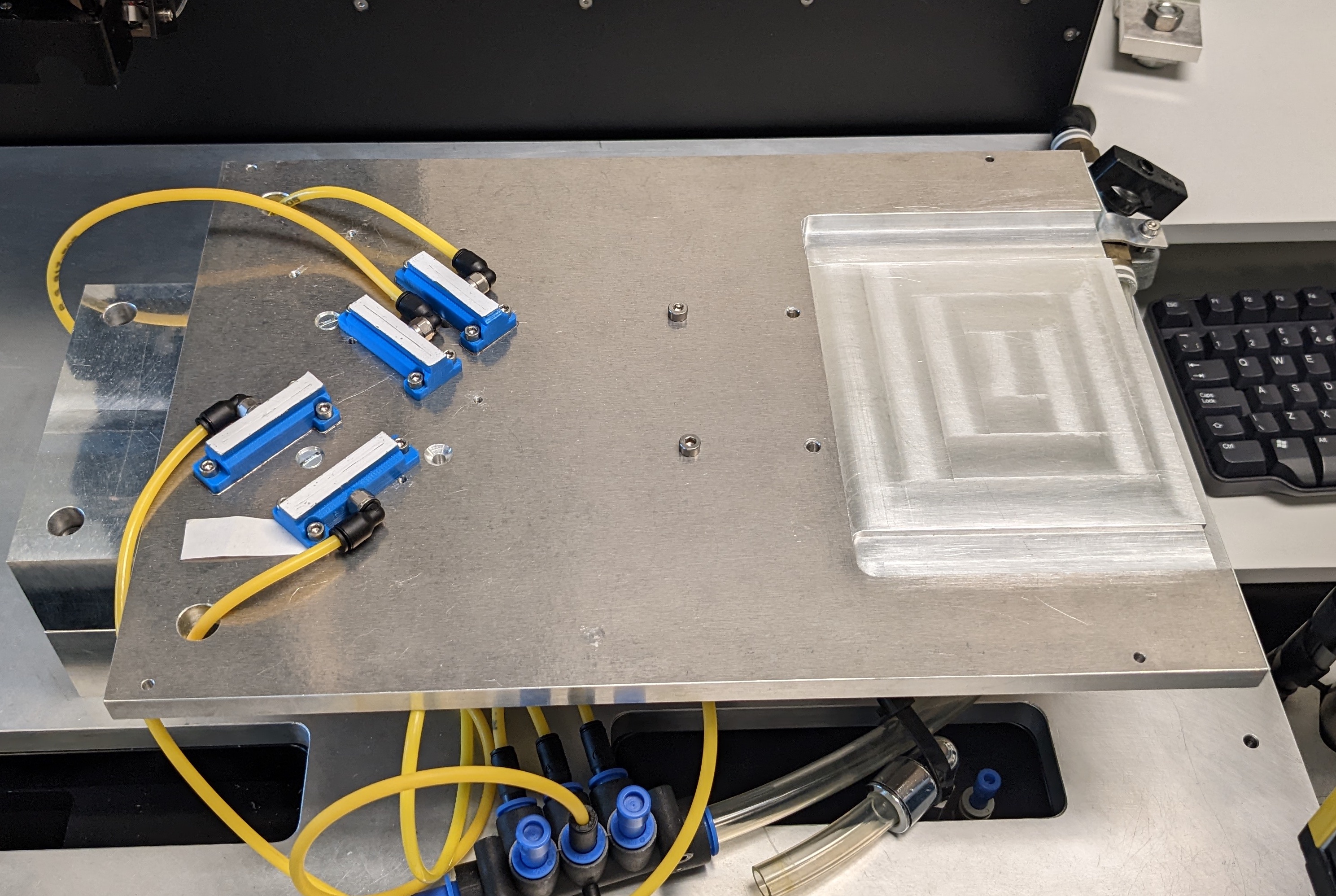} &
  \includegraphics[width=1.0\linewidth]{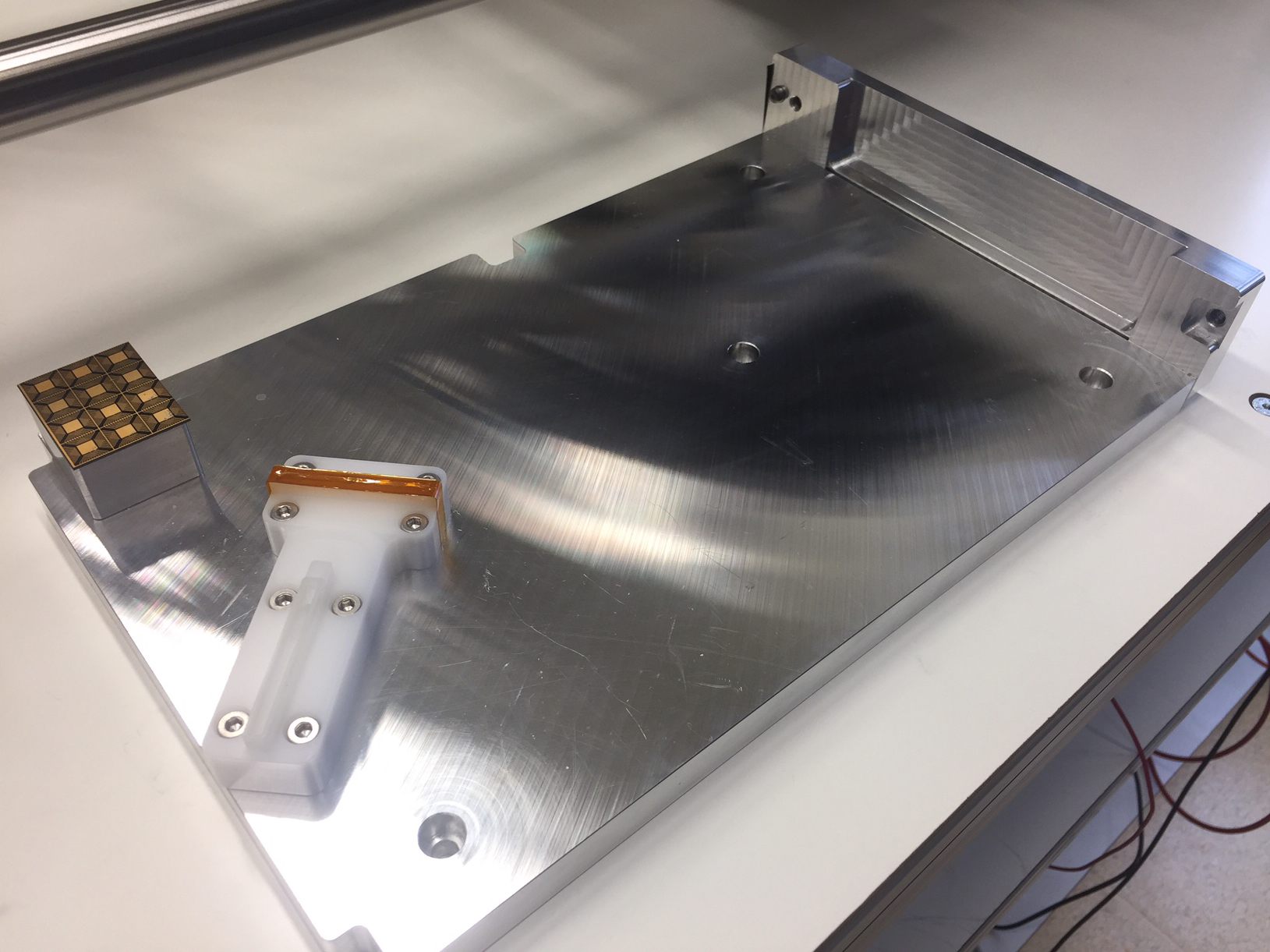}
  \end{tabular}
  \caption{Wire-bonding support jig used in Manchester (left) and Nikhef (right).}
  \label{fig:wire_bonding_jigs}
\end{figure}

After they have been attached to a module, the VeloPix front-end ASICs are wire-bonded to their readout hybrids. In order to support the microchannel substrate during this process, a jig has been manufactured in each assembly site to hold the module from the opposite side, as shown in Fig.~\ref{fig:wire_bonding_jigs}. The vacuum chucks, used to hold the module in place, are positioned such as to avoid contact with the sensors on the supporting side.

The first step of the procedure consists of attaching the HV cable to one side of the module, as shown in Fig.~\ref{fig:cabling_stand} left. The glue is left to cure and then the module is installed in the wire-bonding jig and fixed in place under the wire-bonding machine. The machine is aligned to the module by using image recognition and the alignment validated. The automatic bonding program is executed and all 840 wire-bonds on that side of the module. Finally, the HV tape is manually wire-bonded to the sensor surface. The same procedure is then repeated for the other side of the module.

\subsection{Cables attachment}
The final step in the assembly consists of attaching all remaining cables to the module. In Manchester, this includes the interconnect cables, data cables and LV cables. At Nikhef, since the flat cables are attached together with the FE and GBTx hybrids, this step only concerns the attachment of the LV cables.

\begin{figure}[htbp]
  \centering
  \includegraphics[width=0.3\linewidth]{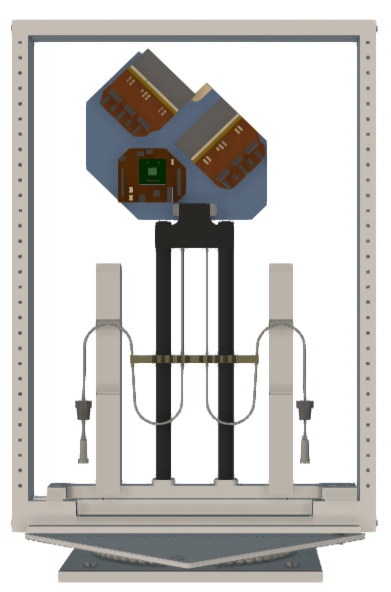}
  \includegraphics[width=0.3\linewidth]{VELO/figs/m4.jpg}
  \caption{ Left: drawing of a module mounted on a free-rotating stand for cables attachment. Right: a photograph of a module after HV cables attachment.}
  \label{fig:cabling_stand}
\end{figure}

The flat cables are thin tape-like copper-polyimide cables, equipped with high density connectors. As for the front-end hybrids, these cables are also plasma cleaned to ensure optimal connectivity of the pins in the connectors. During attachment, the module is installed in a frame and held vertically on a stand, free to rotate around the vertical axis such that the optimal angle can be reached for each individual connection, as shown in Fig.~\ref{fig:cabling_stand}.

\begin{figure}[htbp]
  \centering
  \includegraphics[height=0.5\linewidth]{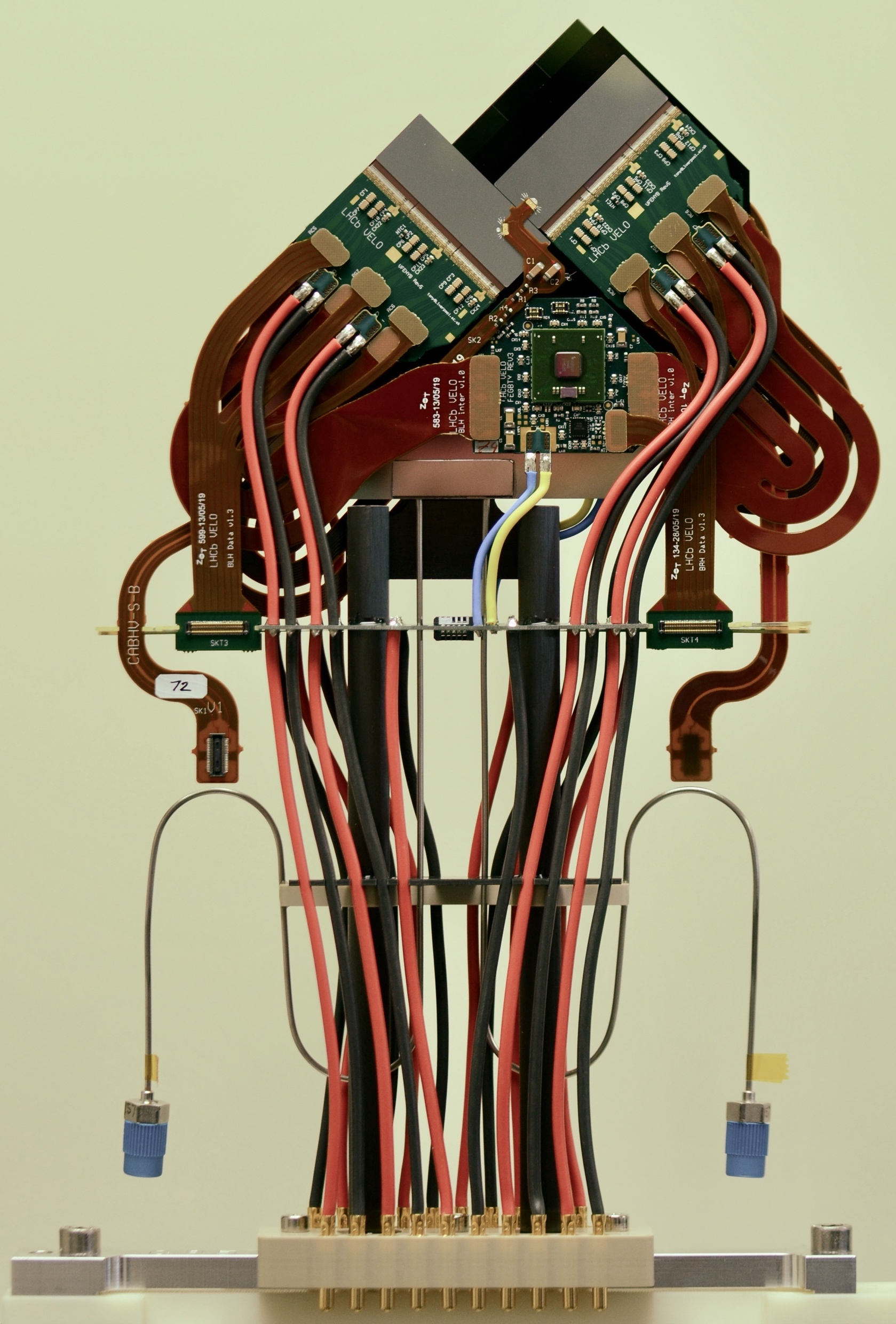}
  \includegraphics[height=0.5\linewidth]{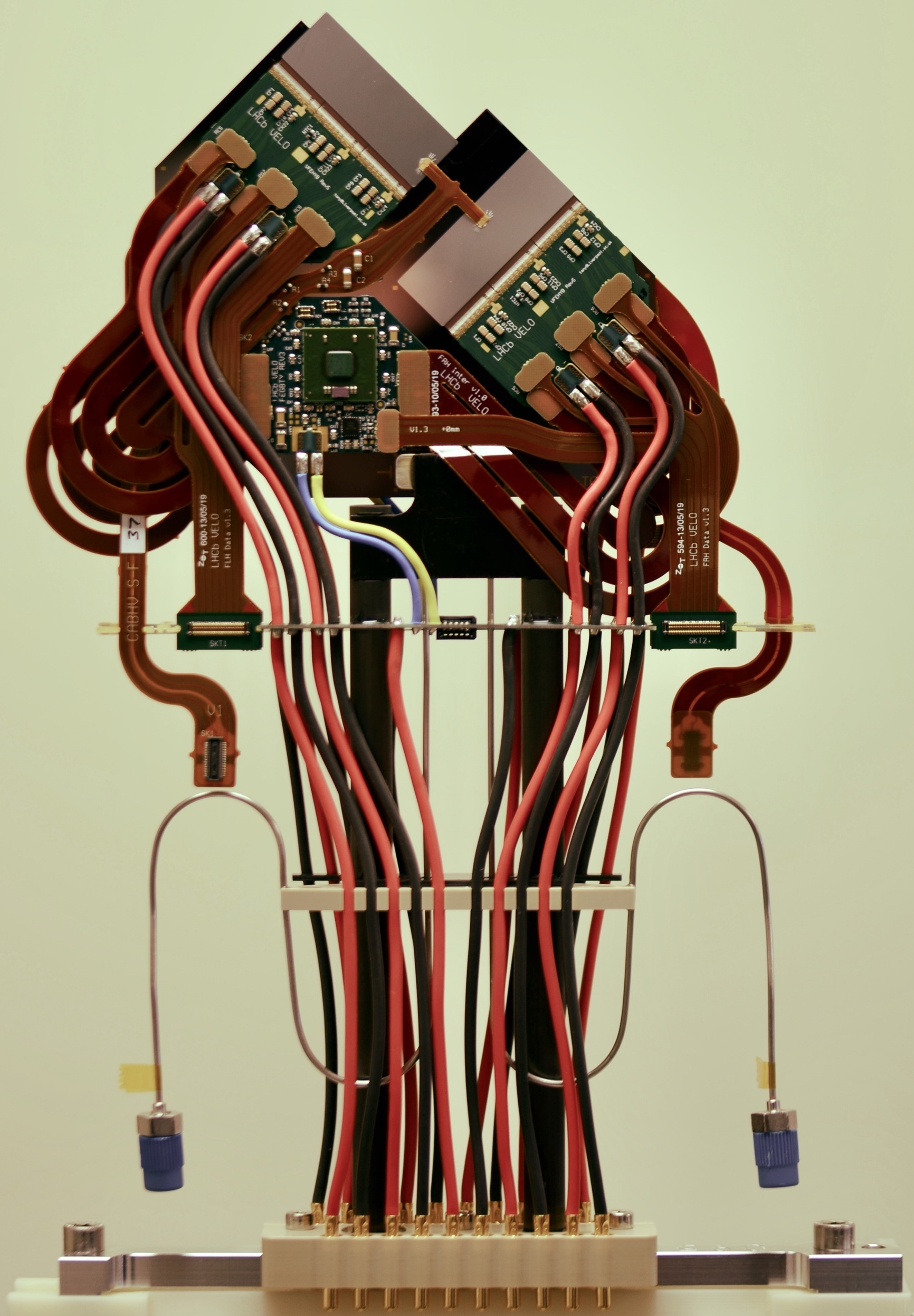}
  \caption{Photographs of the front and rear side of a fully assembled module.}
  \label{fig:module_assembly_final}
\end{figure}

The same frame and stand are used to hold the module during attachment of the LV harness. The transition bridges, fully equipped with LV cables, are placed on top of the two plastic pillars shown in Fig.~\ref{fig:cabling_stand}, such that the right distance from the module foot is set. The two parts of the LV foot connector are screwed to the aluminium foot. Then, the transition bridges are  glued to the carbon fibre legs and left to cure. Finally, the LV cable connectors are mated to the FE and GBTx hybrids and the pins are inserted into the LV foot connector. The finished module is shown in Fig.~\ref{fig:module_assembly_final}.

%% file: VELO/metrology.tex
Each individual assembly step is followed by a corresponding measurement to assess the quality achieved and assign a grade to the step. These measurements are performed at both assembly sites with similar techniques. The results are analysed with the same software and graded upon the same set of criteria, and stored in the database (see Sect.~\ref{sec:database}). While the two assembly sites employed different equipment, the test procedures (and their outcome) are fairly comparable. We choose to describe here the procedures used in Manchester, and point out the differences, whenever relevant.

\subsection{Bare module flatness}
\label{sec:bare_module_flatness}
Before shipment to the assembly sites, a flatness measurement of the microchannel substrate is carried out at CERN, in order to confirm that no deformations were introduced during the fluidic connector soldering procedure. The survey is performed with a 3D optical profilometer\footnote{Keyence model VR-3200.}, 
a 3D optical system that uses light reflectometry to reach an accuracy of $\pm$3$\mum$~\cite{Francisco:2021tda}.

The flatness of the cooling substrate after being assembled into a bare module is measured again at the assembly sites to ensure that no damage has occurred during transport, storage and assembly of the substrate. The measurement is important also to verify the presence of any mechanical stress introduced in the silicon substrate by the assembly procedure, which might cause failures in the attachment of the tiles.

Here, the flatness is measured using a 3D optical system\footnote{OGP SmartScope Flash 500.} that includes a camera and a laser. This device can record data with micrometer precision in the $xy$ plane and along the $z$ axis. It is also capable of automated feature-finding methods and can locate shapes according to their different contrasts. During the survey, the module is held by the turn-plate (as shown in Fig.~\ref{fig:bare_module}) and placed under the optical system. The turn-plate carries markers on a glass surface, such that they are visible from either side of the plate. These markers are used to define a reference system with respect to the precision hole in the module foot. The survey is performed using using the laser of the optical system, which scans the surface of the microchannel plate along $x$ and $y$ while recording the $z$ values. In Manchester, only the connector side substrate flatness is measured, while both sides are surveyed at Nikhef.

\begin{figure}[htbp]
  \centering
  \includegraphics[width=0.45\linewidth]{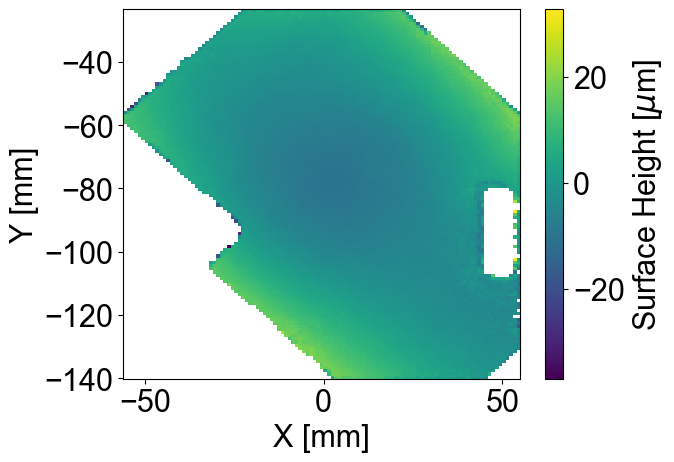}
  \includegraphics[width=0.45\linewidth]{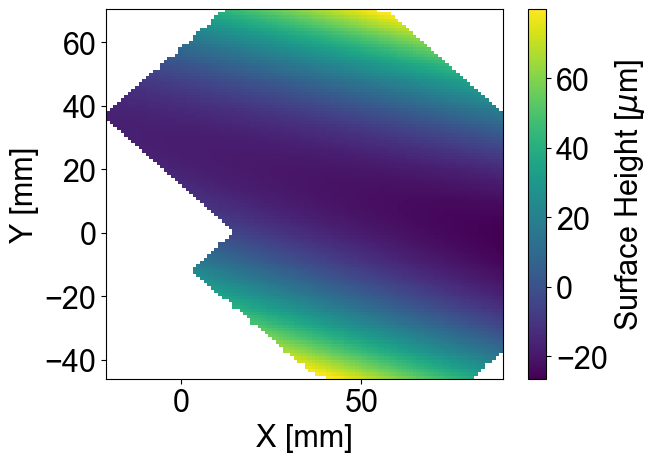}
  \caption{Comparison of a flatness measurement performed on an unconstrained cooling substrate (left) and as a bare module (right).}
  \label{fig:bare_module_survey}
\end{figure}

The results obtained from these surveys are compared with those from CERN. They are also used to generate a substrate flatness grade. An example of a comparison between a survey performed at CERN (left) and the one carried out at an assembly site (right) is shown in Fig.~\ref{fig:bare_module_survey}. As illustrated by this example, the measurements performed at the assembly site after bare module assembly and those carried out at CERN on the microchannel substrate, show a significant difference at the sides of the substrate. This disagreement is due to the fact that, in the assembly sites, the substrate on the bare module is held in place by the holding frames at two fixation points at its sides, whereas at CERN the substrate is freely placed on a flat surface. However, no issue due to this constraint has been observed during module assembly.

The two-dimensional scan of $z$ measurements across the surface of the plate is taken and the data points acquired are used to perform a fit with a least-square plane. Finally, the distribution of variations between the measured points and the fitted plane is used to calculate the four quartiles. The quality criteria assigned are based on the value of the first and third quartiles, as indicated in \cref{tab:metrology_substrate}. In general, grades are assigned as being A,B,C,D or F, although some grades are not used in specific cases. 
\begin{table}[h]
    \centering
    \caption{Quality criteria and assigned grading for the microchannel substrate flatness. In this test, the grade F is not used.}
    \label{tab:metrology_substrate}
    \begin{tabular}{c | c}
            grade & absolute value of quartiles $Q_1$ and $Q_3$\\
            \hline
            \textbf{A} & max($|Q_1|, |Q_3|$) $< 50~\mum$\\
            \textbf{B} & $50~\mum <$ max($|Q_1|, |Q_3|$) $< 75~\mum$\\
            \textbf{C} & $75~\mum <$ max($|Q_1|, |Q_3|$) $< 100~\mum$\\
            \textbf{D} & max($|Q_1|, |Q_3|$) $> 100~\mum$\\
            \textbf{F} & -- \\
        \end{tabular}
\end{table}

\subsection{Tile metrology}
The tiles represent the active part of a VELO module. Their precise positioning is required to avoid mechanical interference, while a precise prior knowledge of their relative positions is a crucial starting point for the alignment of the whole detector in LHCb. The glue layer underneath each tile is also vital for a good alignment in $z$ and for ensuring an optimal flow of the heat produced by the ASICs. For these reasons, an accurate metrology of the tiles after attachment is performed.

\subsubsection{Tile position}
\label{sec:metrology_tiles_position}
The positions of the tiles are measured with the 3D optical system. As in the case of the bare module flatness measurement, the glass markers on the turn-plate are used to define a reference system relative to the dowel hole in the module foot. Using pattern recognition features, the left-most and right-most fiducials of the two tiles facing the camera are found and their ($x$,$y$) position values are recorded. The same technique is used to measure the fiducials on the back of the sensor attached to the opposite side of the module (visible because of its overhang). The procedure is then repeated for the other side of the module.

\begin{figure}[htbp]
    \centering
    \includegraphics[width=0.49\linewidth]{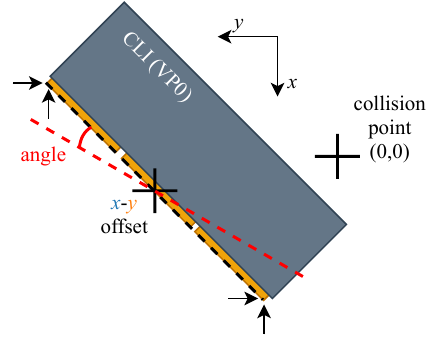}
    \includegraphics[width=0.49\linewidth]{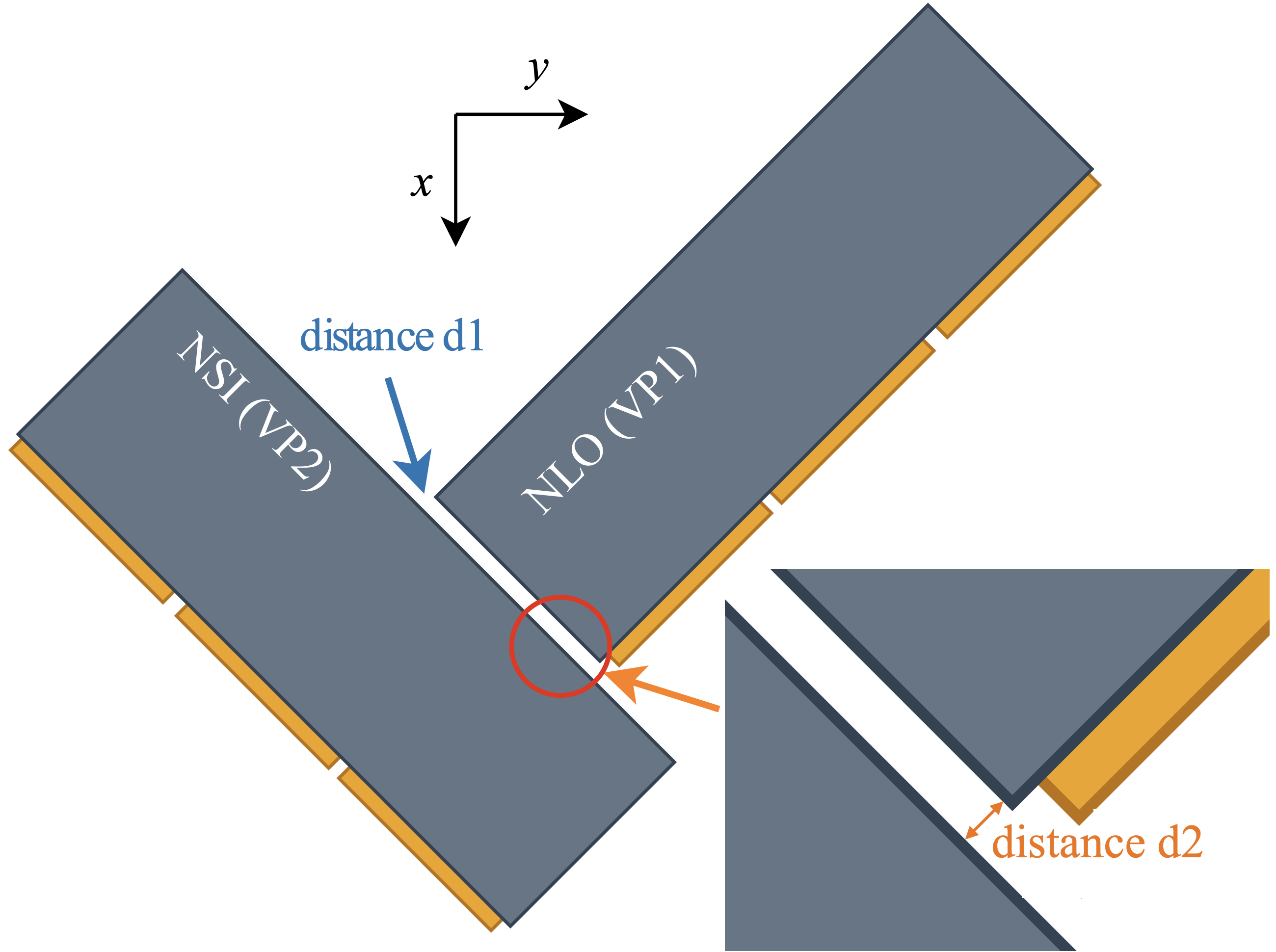}
    \caption{Left: diagram showing the middle point $x$-$y$ offset and angle which are measured for every tile. Right: diagram showing the closest distance between the two tiles on the non-connector side of the module.}
    \label{fig:tiles_position}
\end{figure}

The data are used to calculate the middle point along the tile and its rotation with respect to its nominal orientation, as illustrated in Fig.~\ref{fig:tiles_position}. On the non-connector side of the module, where the two tiles are placed at only 145~\um from each other, their relative distance is also measured. The quality criteria applied to assign a grade are listed in Tab.~\ref{tab:metrology_tiles_position}.

\begin{table}[h]
    \centering
    \caption{Quality criteria and assigned grading for the tile positions. The grade is determined by the worst result of the four parameters.}
    \label{tab:metrology_tiles_position}
    \begin{tabular}{c | c | c | c}
            grade & middle point position & angle & distance 1 and 2\\
            \hline
            \textbf{A} & $< 55~\mum$  & $< 1000~\murad$ & $> 130~\mum$ \\
            \textbf{B} & $< 110~\mum$  & $< 2000~\murad$ & $> 100~\mum$ \\
            \textbf{C} & $> 110~\mum$  & $> 2000~\murad$ & $> 50~\mum$ \\
            \textbf{D} & --  & -- & $< 50~\mum$ \\
            \textbf{F} & --  & -- & $< 10~\mum$ \\
        \end{tabular}
\end{table}

\subsection{Glue layer thickness and tile flatness}
\label{sec:metrology_glue}
This measurement is performed using the laser of the 3D optical system, one tile at a time, and is made immediately after the tile position measurement. The glue layer thickness will determine both the adhesion quality and the thermal performance, while the tile flatness checks for any irregularities.

\begin{figure}[htbp]
  \centering
  \includegraphics[width=0.45\linewidth]{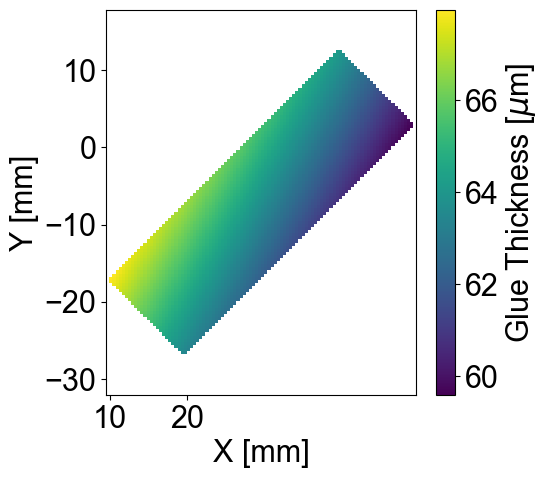}
  \includegraphics[width=0.45\linewidth]{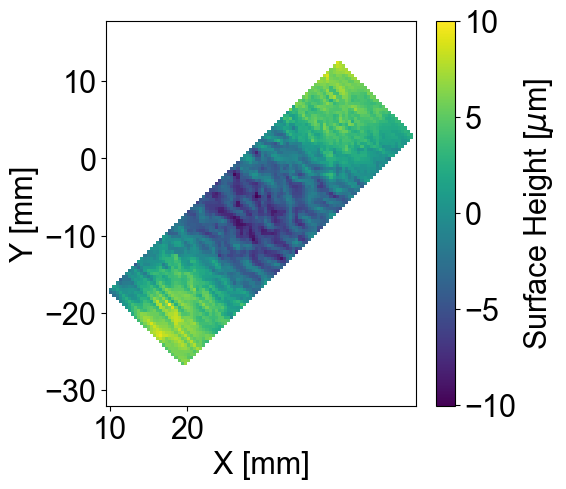}
  \caption{Example measurements of the glue layer thickness (left) and tile flatness (right). These heat maps show the variation of the $z$ measurements relative to the cooling substrate.}
  \label{fig:glue_thickness}
\end{figure}

The module is in its turn-plate and the glass markers on the turn-plate are used to define the reference system. The laser scans an area covering the tile as well as the regions on the cooling substrate that surround the tile. In this way, the glue thickness and its variation can be calculated by comparing the measurements obtained on the tile and on the substrate. The scan over the tile surface also shows the variations in the height of the tile itself, and it is used as a measurement of the tile flatness. The difference of the two-dimensional scan of $z$ measurements across the four tiles is determined with respect to a least-square fit with a plane. An example of the glue layer thickness and tile flatness measurements are shown in Fig.~\ref{fig:glue_thickness}. The quality criteria used to assign a module grading to both measurements are listed in \cref{tab:metrology_tiles}.

\begin{table}[h]
    \centering
    \caption{Quality criteria and assigned grading for the tile flatness and glue layer thickness. The grade is determined by the worst result of the two parameters}
    \label{tab:metrology_tiles}
    \begin{tabular}{c | c | c}
            grade & tile flatness: $|$mean$|$+$2\cdot$stdev & glue layer thickness: mean$\pm2\cdot$stdev\\
            \hline
            \textbf{A} & $< 30~\mum$ & $ 40-120~\mum$\\
            \textbf{B} & $< 60~\mum$ & $ 20-160~\mum$\\
            \textbf{C} & $< 120~\mum$ & $ 0-200~\mum$\\
            \textbf{D} & $> 120~\mum$ & $> 200~\mum$\\
            \textbf{F} & -- & -- \\
        \end{tabular}
\end{table}

\subsection{Hybrid metrology}
\label{sec:metrology_hybrids}
\begin{figure}[htbp]
  \centering
  \includegraphics[width=0.4\linewidth]{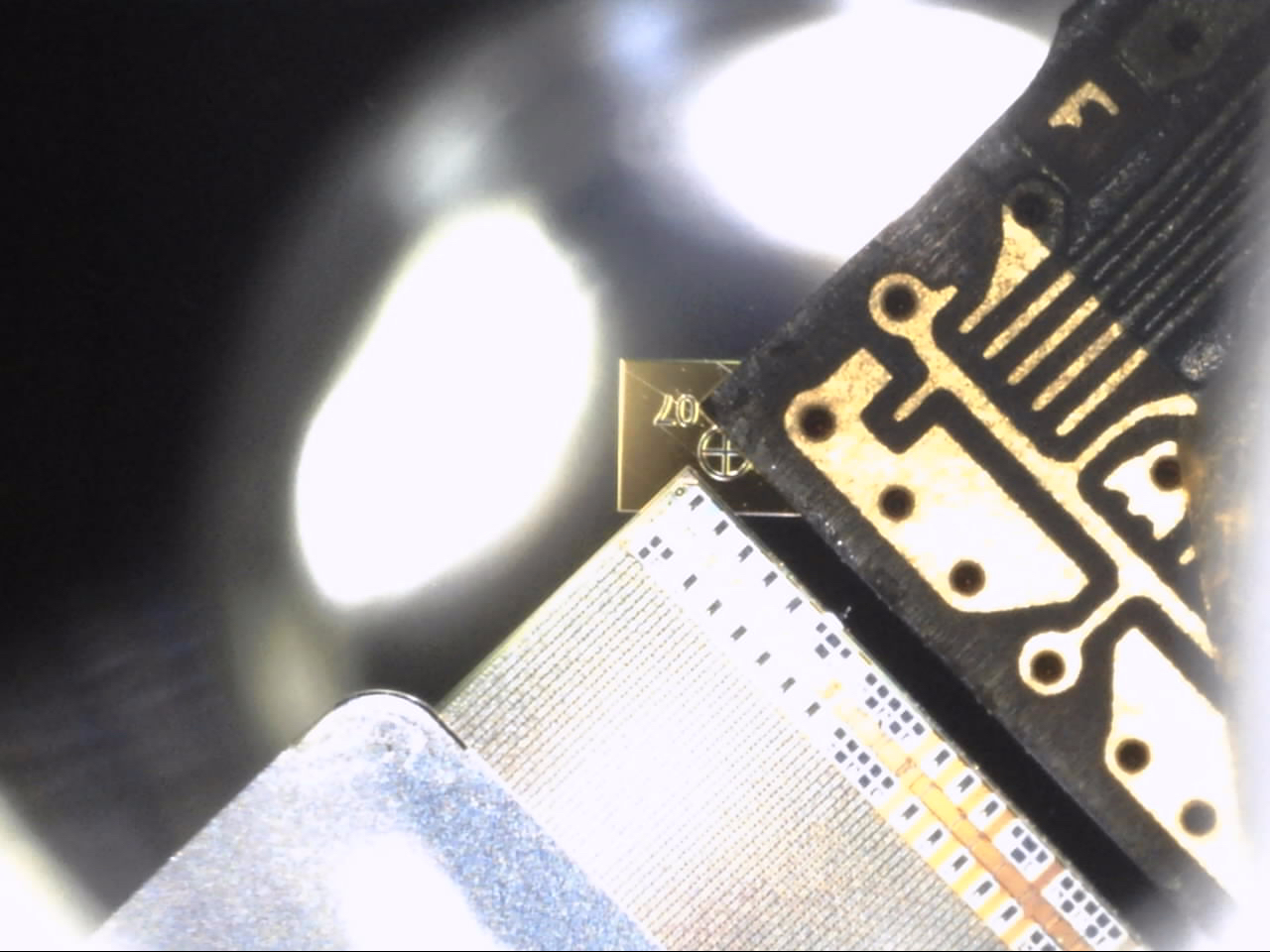}
  \includegraphics[width=0.4\linewidth]{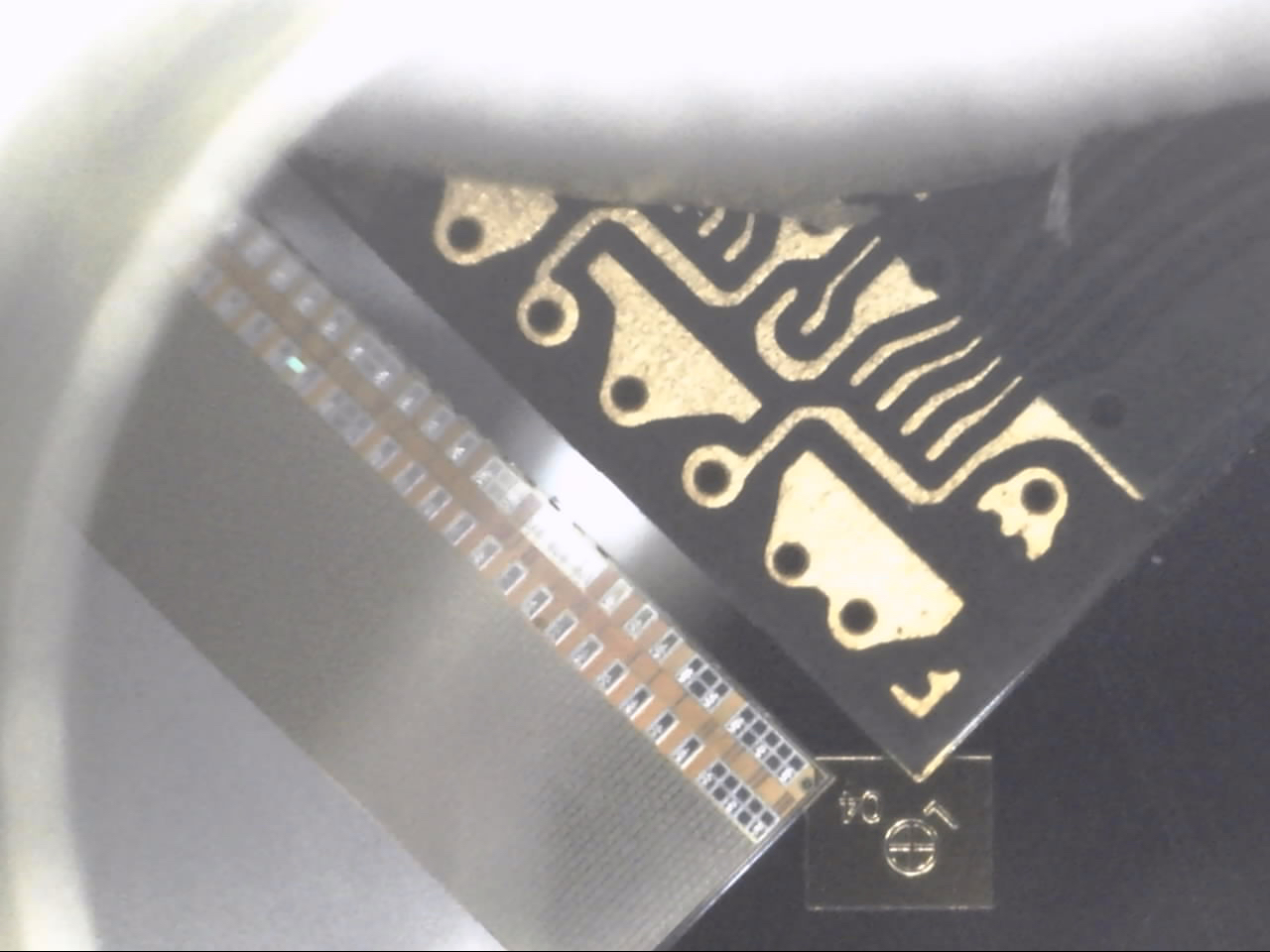}
  \caption{High magnification photographs showing the alignment at both corners of a tile relative to its front-end hybrid.}
  \label{fig:hybrids_alignment}
\end{figure}

As mentioned in Sect.~\ref{sec:hybrids_placement}, the accuracy required in the positioning of the hybrids is mainly dictated by the wire-bonding process, which requires the bond pads on the hybrids to be well aligned to those on the ASICs, and to be at a consistent distance from each other over the full length of the tile. For this reason, only the front-end hybrids are checked. After their attachment, this alignment is verified by taking high magnification photographs of the region where hybrids and tiles are facing each other. These pictures also allow for spotting any glue spillage that might affect the quality of the wire-bonding. An example of these photographs is shown in Fig.~\ref{fig:hybrids_alignment}. 

\subsection{Wire-bond pull test}
\label{sec:metrology_wirebonds}
\begin{figure}[htbp]
  \centering
  \includegraphics[width=0.35\linewidth]{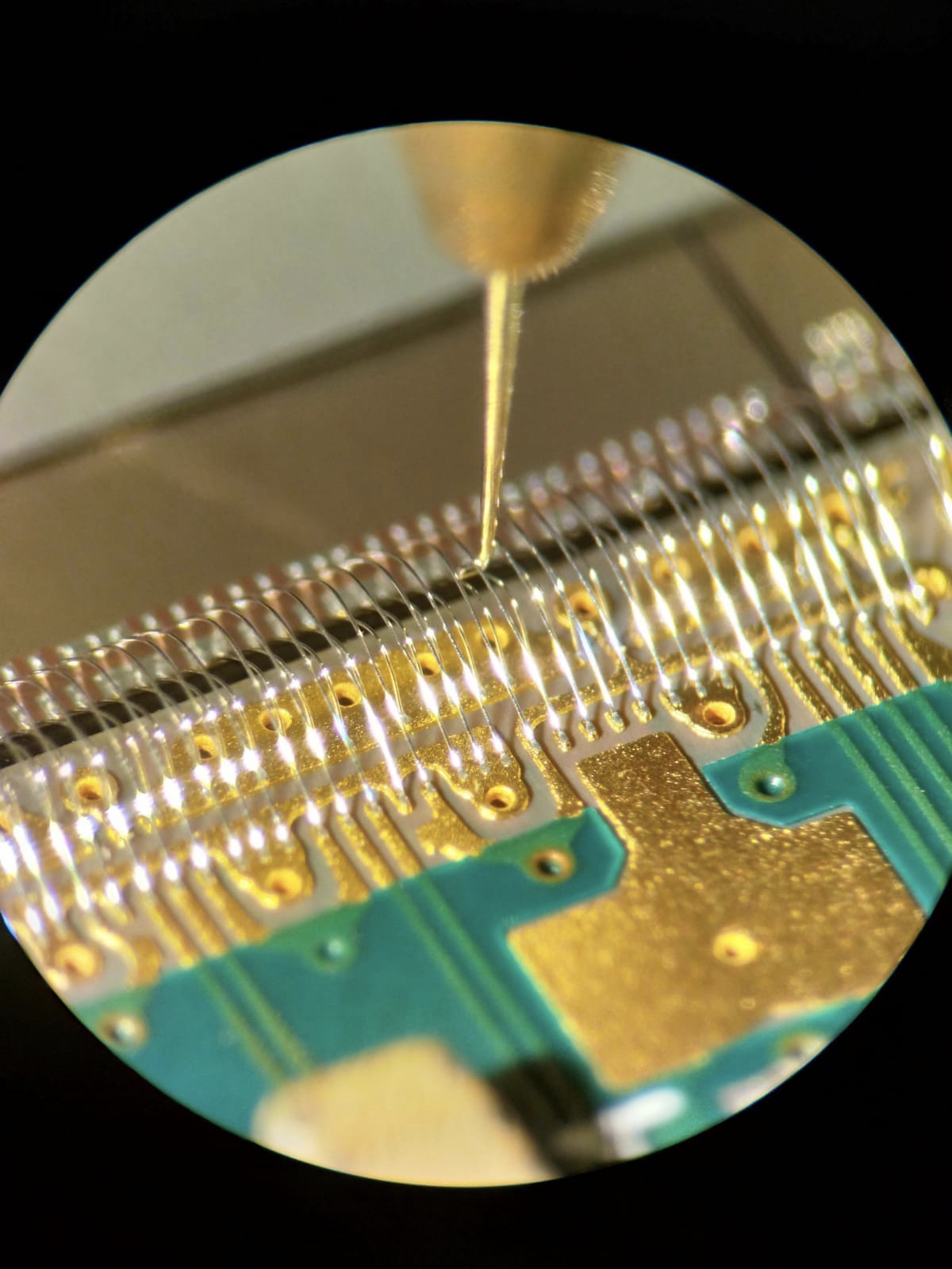}
  \caption{Photograph showing the hook of the pull-testing device catching the sacrificial wire-bond to be pulled on.}
  \label{fig:pull_test}
\end{figure}

After the wire-bonding process, the quality of the wire-bonds is assessed by measuring the strength of their attachment. This is a destructive test which is performed on wire-bonds that serve no electrical purpose and are bonded solely to serve for this indication of the strength of the wire-bonding process in that particular module. Each ASIC contains four sacrificial wire-bonds, which are pulled by a dedicated device that records the breaking force. The device uses a small hook that is lowered next to the sacrificial wire-bond to catch the wire and pull it upwards until it breaks, as shown in Fig.~\ref{fig:pull_test}. The breaking force is then recorded and the hook is moved to the next sacrificial wire-bond. After all sacrificial bonds have been pulled, the broken wires are removed. The quality criteria used to assign a grade to this step are summarised in \cref{tab:metrology_wirebond}.

\begin{table}[h]
    \centering
    \caption{Quality criteria and assigned grading for the wire-bond pull force.}
    \label{tab:metrology_wirebond}
    \begin{tabular}{c | c }
            grade & pull force F\\
            \hline
            \textbf{A} & at most two bonds with F $<$ 5 g, but none with F $<$ 4 g \\
            \textbf{B} & at least one bond with F $< 4$~g, but none with F $<$ 2 g \\
            \textbf{C} & -- \\
            \textbf{D} & at least one bond with F $< 2$~g\\
            \textbf{F} & -- \\
    \end{tabular}
\end{table}

%% file: VELO/validation.tex
Once the assembly process is completed, each module is validated through a sequence of mechanical, thermal and electrical tests, such that its overall quality is assessed. The test setup and the tests performed are described in the following sections, with the results obtained discussed in Sect.~\ref{sec:summary} . While differing in some details, the setup at each of the two assembly sites provided information for all the essential tests. For brevity we choose to describe the Manchester setup here. 

\subsection{Test setup}
\label{sec:test_setup}
\begin{figure}[htbp]
  \centering
  \includegraphics[width=0.8\linewidth]{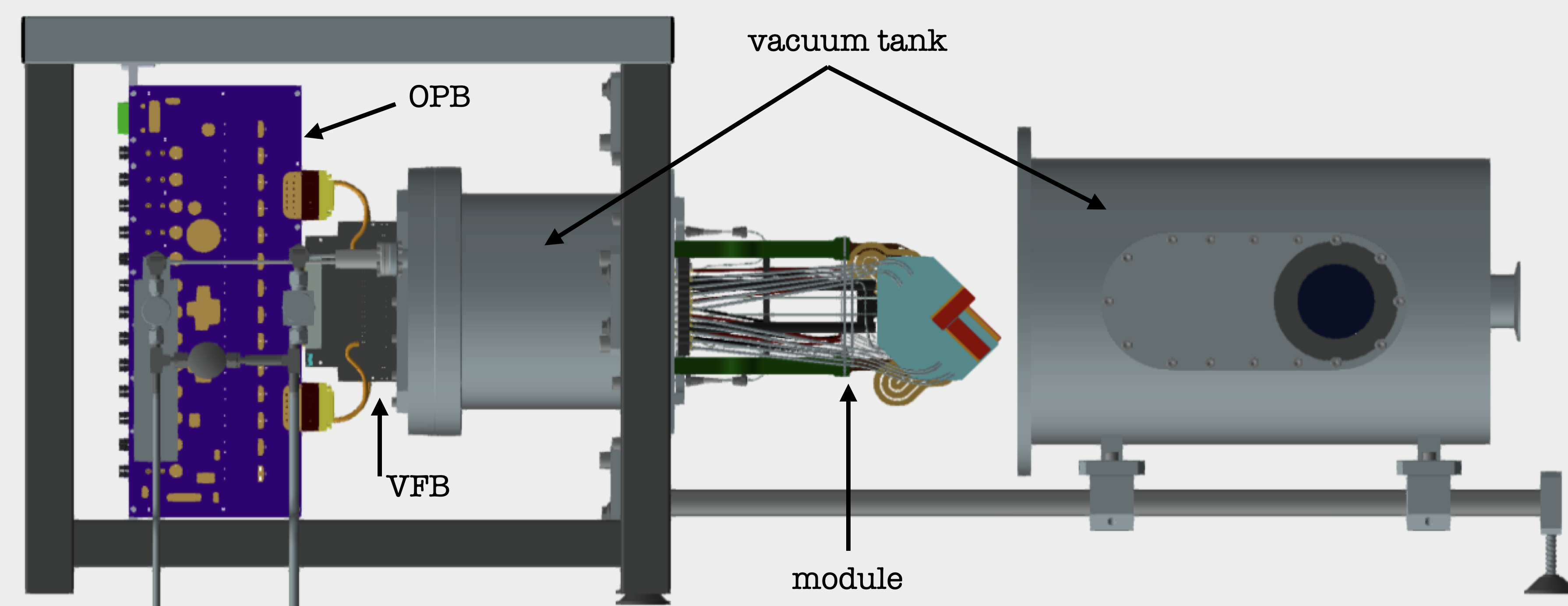} \caption{A schematic drawing of the test setup used at the assembly site for module validation.}
  \label{fig:test_setup}
\end{figure}

 Each assembly site is equipped with a test setup that replicates the running conditions of the detector in the final experiment. The setup consists of a vacuum tank that can accommodate a single module. It is connected to a CO$_2$ cooling system and to the full electronics chain. A schematic drawing of the setup is shown in Fig.~\ref{fig:test_setup}.

The vacuum tank consists of two parts, that can be brought together, sealed with a rubber O-ring. One part is an empty cylinder that can move on two rails and it is attached to a vacuum pump\footnote{Pfeiffer Vacuum D-35614 Asslar Turbo Pump.} through a bellow. The cylinder is also equipped with two germanium windows to facilitate photography with an infra-red camera. The other part of the vacuum tank houses a number of dedicated vacuum feed-throughs, in order to feed LV and HV power inside the vacuum, to enable data transfer from the module to the data acquisition system, to allow for the coolant to be circulated inside the module and to control a number of thermal and mechanical probes. This part of the tank also houses a frame for the installation of the module. The pressure inside the tank is kept below 10$^{-3}$~mbar. 

In the experiment, the VELO detectors are operated in vacuum and kept cold by circulating CO$_2$ at a flow of 0.4 g/s and evaporated at a pressure of about 12 bar. The same conditions are required at the assembly sites, so that the modules can be evaluated in realistic running conditions. The assembly sites also use evaporative CO$_2$ systems to cool down the modules.

In the module validation setups, the electrical chain consists of the same electronics system as in the final experiment, wherever possible. In particular, the vacuum feed-through board (VFB) and the opto- and power board (OPB) are the same components which are employed in the experiment, as described in Sect.~\ref{sec:velo_layout}. The module data cables are connected, via high speed tapes, to the vacuum side of the VFB, which then carries these signals out of the vacuum to the OPB. The OPB is then connected to the data acquisition system (MiniDAQ) via fibres. The HV is provided to the module directly via the VFB, while the LV is instead provided through the OPB, which also has a connection to read out the temperature probes (NTC) in the front-end and GBTx hybrids and on the OPB itself. These NTC sensors are continuously monitored, logged and connected to a hardware interlock system. The MiniDAQ~\cite{minidaq} is a Linux server running CentOS~7, containing a special PCIe40 card specifically adapted for the assembly sites. This card differs from the readout PCIe40s installed in the experiment as it combines the functionalities otherwise performed by three separate cards for the slow control, the timing and fast control and the readout. In the experiment, these functions are carried out by the Experiment Control System (ECS), Time \& Fast Controls (TFC) and TELL40 respectively~\cite{LHCb-TDR-016}. The MiniDAQ also carries the drivers, the analysis software and the SCADA~\cite{wincc:web} projects required to run the experiment. 

In Manchester, the validation setup is powered using two programmable power supplies\footnote{Rohde \& Schwarz HMP2020.}, each of which is a dual-channel system which can supply up to 188~W. 
Three channels are required to power the OPB and the VeloPix digital and analogue supplies. Each channel carries a voltage of 6V to power an array of DC-DC converters on the OPB, which then step down the voltage level to that required by the front-end components.

A VELO module requires four HV lines, one for each tile. In Manchester, this is provided by a 3kV/1mA multichannel high voltage module\footnote{CAEN AG536N board, installed in a CAEN SY4527LC crate.}, which delivers the required bias voltage to the module sensors during testing. However, these boards have a current resolution of about 50~nA, which is not accurate enough when measuring the I-V characteristics of a silicon sensor and for this reason, when performing I-V measurements, the HV cables are switched to a power supply\footnote{Keithley model 2410.} which can deliver up to $\pm 1100$~V and has a current resolution of 1~nA. 

\begin{figure}[htbp]
  \centering
  \includegraphics[width=0.8\linewidth]{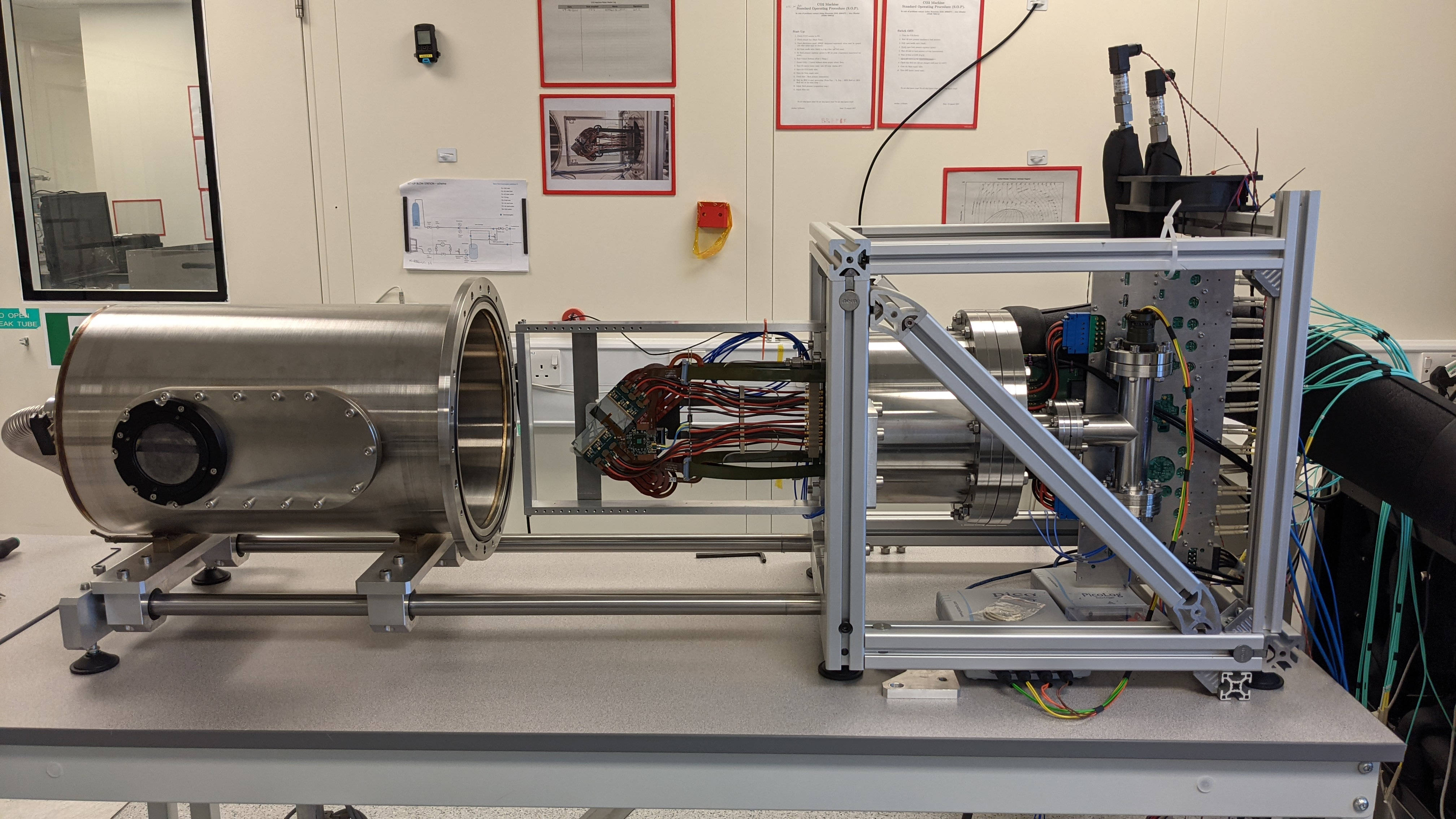} \caption{Photograph of a module installed in the validation setup, with the vacuum chamber cap retracted on the left.}
  \label{fig:module_tank}
\end{figure}

After assembly, the module is still held in the frame used during cable attachment; this frame can also be attached to a mounting point inside the tank. The module is first positioned against the tank flange by aligning the LV foot connector to its counterpart inside the tank and then screwed to the mounting point. The two LV connectors are then mated. An aluminium bar housing two displacement sensors is attached to the module frame, such that the displacement sensors face the non-connector side of the module. Finally, the two HV cables, the four data tapes and the two cooling capillaries are mated to the corresponding long HV, long data cables and cooling pipes installed inside the tank. Fig.~\ref{fig:module_tank} shows a module after installation in the vacuum tank.

\subsection{Mechanical tests}

\subsubsection{Module displacement}
\label{validation_displacement}
\begin{figure}[htbp]
  \centering
  \includegraphics[width=0.5\linewidth]{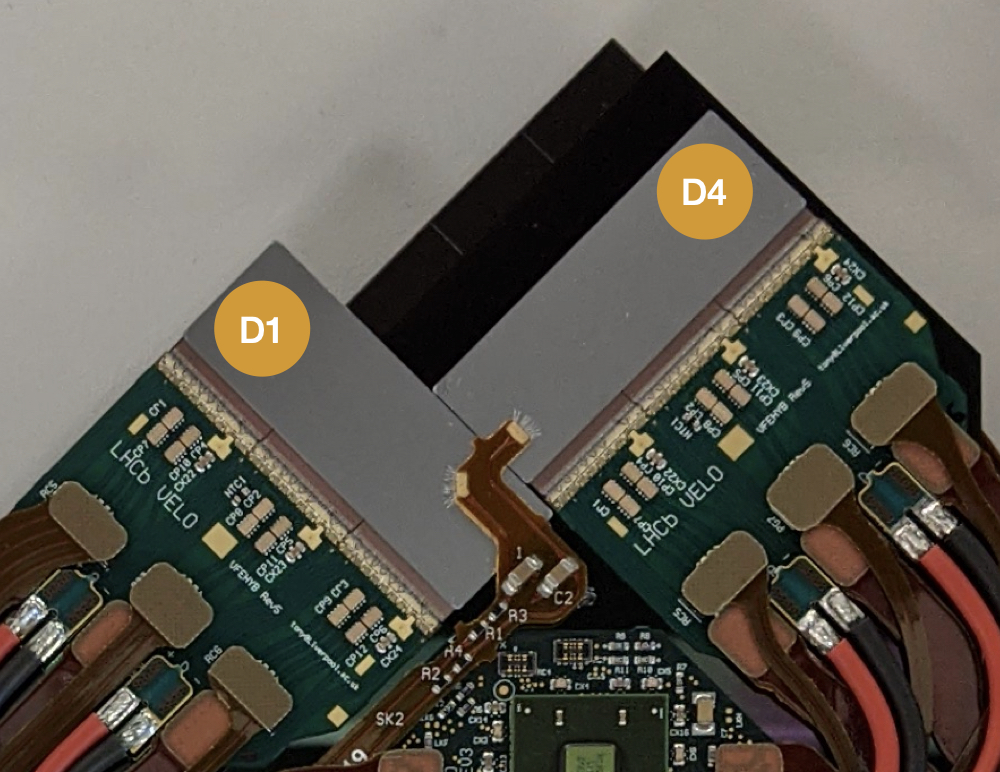} 
  \caption{The displacement sensors are positioned in front of the non-connector side of the module substrate, in the two locations marked as D1 and D4.}
  \label{fig:displacement_sensors}
\end{figure}

A VELO module is an assembly of various parts made of different materials (silicon, copper, kapton, adhesives,...) with different CTE. It is therefore expected that the module will distort when its temperature is changed from room to operating values. Thermal distortions can put a strain on the substrate, and represent a risk for the integrity of the module. Moreover, it is vital to ensure that these deformations do not significantly reduce the clearance between the modules and the RF foil. Finally, knowing the final position of each module at its operating temperature of -30$^{\circ}$C is critical for the alignment of the detector during operation. In addition to thermal distortions, there are other factors that can lead to module displacements: the mating of the HV tapes and data cables during installation, the connection of the cooling capillaries to the tank cooling pipes and the evacuation of the tank. 
Displacements are measured at two different locations on the non-connector side of the substrate, as shown in Fig.~\ref{fig:displacement_sensors}. 

\begin{figure}[htbp]
  \centering
  \includegraphics[width=0.7\linewidth]{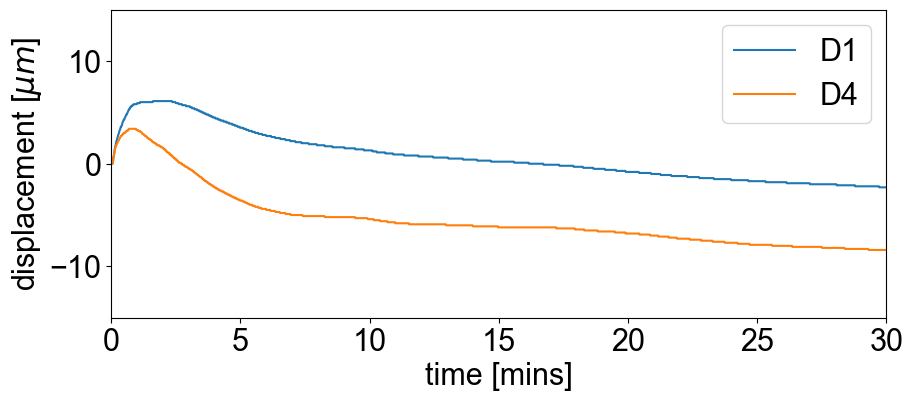}\\
  \includegraphics[width=0.7\linewidth]{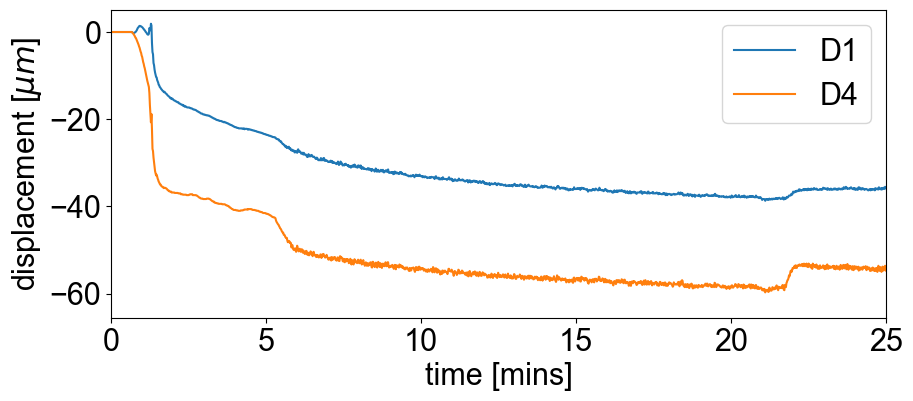} \\
  \includegraphics[width=0.7\linewidth]{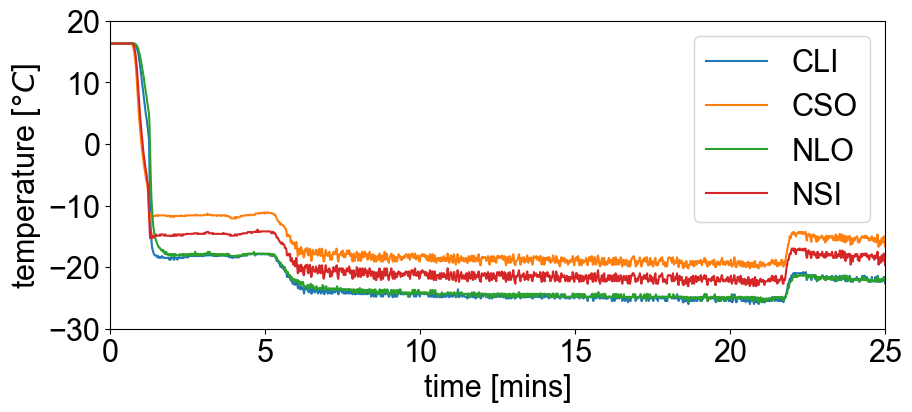}
  \caption{Displacements of the module measured with the two displacement sensors (D1 and D4). Top: displacement versus time during evacuation of the vacuum tank from atmospheric pressure to below $10^{-3}$~mbar. Middle: displacement versus time when cooling down the modules. Lower: temperature versus time when cooling down the modules, for the same procedure and with the same abscissa as the middle plot.}
  \label{fig:displacement_results}
\end{figure}

This displacement measurement is performed by using two capacitive sensors\footnote{Micro-Epsilon CSH1-CAm1,4.}, housed in a dedicated aluminium holder that is directly attached to the frame holding the module. These capacitive sensors have a range of about 1.4~mm and therefore are placed at about 500$\mum$ from the silicon sensor surface. The displacement measurements are continuously recorded during the tightening of the cooling connectors, the evacuation of the vacuum tank, and the cooling down from room temperature to -30 $^{\circ}$C. An example of results from these measurements for a single module are shown in Fig.~\ref{fig:displacement_results}. The displacements from cooling pipe connection and from evacuation are comparatively small, while the  displacement under cool down is larger and clearly correlated with the temperature of the four sensors (CLI,CSO,NLO,NSI). The displacement of each module is measured and a grade assigned according to the criteria in Tab.~\ref{tab:metrology_displacement}.

\begin{table}[h]
    \centering
    \caption{Quality criteria and assigned grading for the module displacement tests.}
    \label{tab:metrology_displacement}
    \begin{tabular}{c | c | c | c}
            grade & cooling pipes connection & pump-down & cool-down\\
            \hline
            \textbf{A} & $< 50~\mum$ & $< 50~\mum$ & $< 100~\mum$\\
            \textbf{B} & $< 80~\mum$ & $< 80~\mum$ & $< 150~\mum$\\
            \textbf{C} & $< 100~\mum$ & $< 100~\mum$ & $< 200~\mum$\\
            \textbf{D} & $> 100~\mum$ & $> 100~\mum$ & $> 200~\mum$\\
            \textbf{F} & -- & -- & -- \\
    \end{tabular}
\end{table}

\subsubsection{Thermal cycles}
\begin{figure}[htbp]
  \centering
  \includegraphics[width=0.7\linewidth]{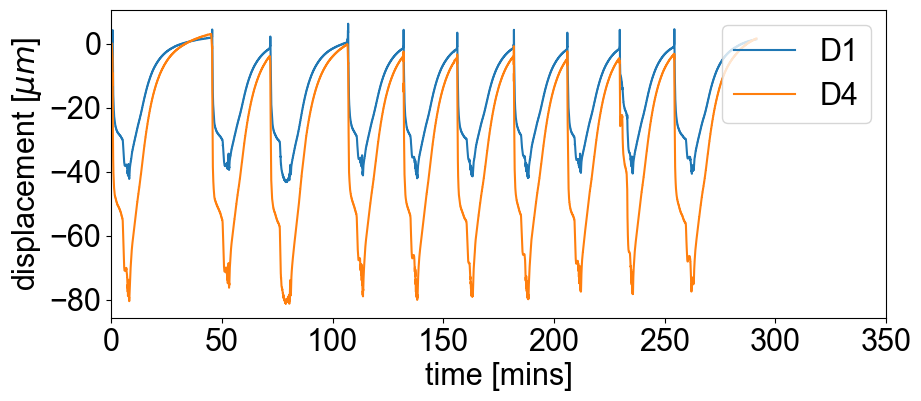}
  \includegraphics[width=0.7\linewidth]{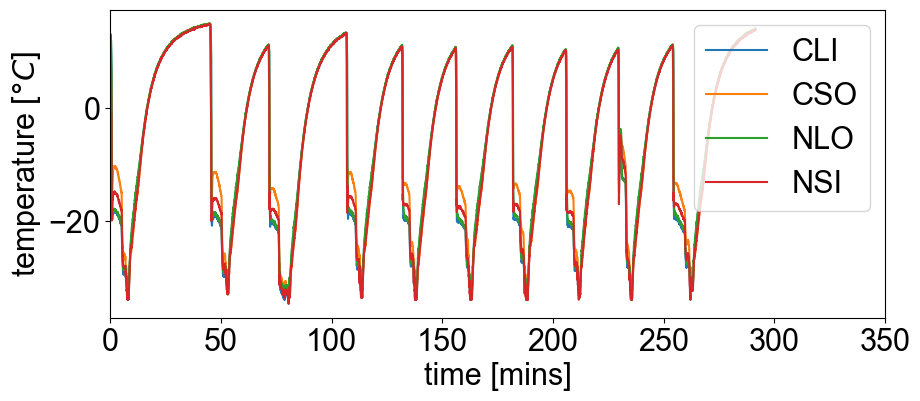}
  \caption{Displacement (top) and temperature (bottom) of the four tiles versus time during the thermal cycling step.}
  \label{fig:thermal_cycles}
\end{figure}

Although during normal operation of the LHCb experiment the modules are kept cold to reduce radiation damage, it is expected that they will undergo several tens of thermal cycles between -30 and +20 $^{\circ}$C during LHC Run~3 and 4. All modules have therefore to be able to withstand the effects of a large number of $\Delta$T=50 $^{\circ}$C cycles, without any significant impact on their performance. For this purpose, a set of ten thermal cycles is performed for every assembled module. All electrical tests are carried out twice, before and after thermal cycling, such that any degradation of the module performance is detected. The cycling process consists of cooling down the module to -30 $^{\circ}$C, then switching the cooling off and allowing the system to warm up back to room temperature. Each cycle takes about 30 minutes. An example of a set of ten thermal cycles, together with the displacements observed in the module, is shown in Fig.~\ref{fig:thermal_cycles} where the displacements observed are seen to be highly reproduceable between cycles.

\subsection{Electrical tests}
\label{sec:electrical_tests}
Immediately after installation of a module in the vacuum tank, a quick connection test is performed before closing the tank and starting the evacuation. This quick test checks the basic functionalities of the ASICs and ensures that all on-module and off-module cables have been connected properly. Afterwards, the vacuum tank is evacuated to a pressure below 10$^{-3}$~mbar and the CO$_2$ coolant circulated in the module to bring its temperature down to -30$^{\circ}$~C. The full chain of electrical tests is then carried out, as detailed in the following sections. The main purpose of the electrical tests is to assess the quality of the chips (VeloPix chips and GBTx's) and of the silicon sensors on the final module, but they are also useful to detect any problems with the interconnecting and data cables or with the wire-bonds between the tiles and the front-end hybrids. At the end of the first run of electrical tests, the thermal cycles are performed and then a second run of electrical tests is carried out. The results of the two runs of tests are then compared. The electrical tests are controlled using the SCADA~\cite{wincc:web}. 

\subsubsection{IV curves}
\label{sec:iv_curves}
\begin{figure}[htbp]
  \centering
  \includegraphics[width=0.8\linewidth]{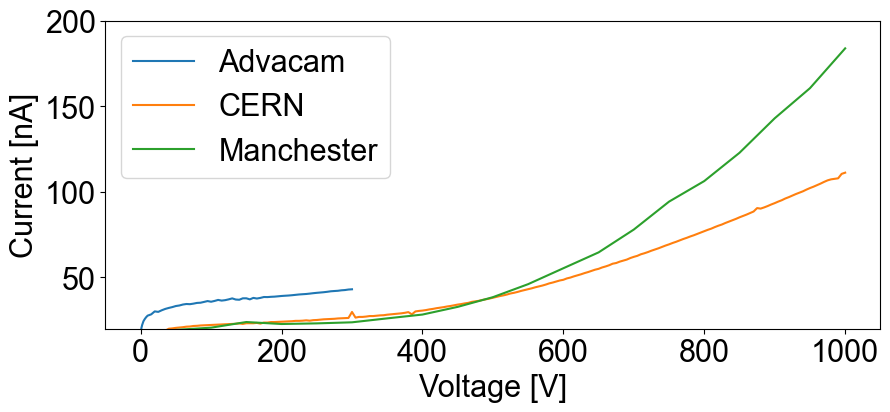}
  \caption{IV curves of an example VELO tile measured after bump-bonding at Advacam, for tile quality assurance at CERN, and after module assembly at Manchester. All were measured at around room temperature.}
  \label{fig:iv_curves}
\end{figure}

The I$-$V characteristics of each sensor tile, at room temperature, is measured by the tile manufacturer and again at CERN on delivery in order to assess the quality of the tiles before shipment to the assembly sites. The same characterization is then repeated for each module after construction, in order to ensure that no degradation has occurred during assembly. 
The voltage scan is performed with automated software, customized independently by the two assembly sites, in the range 0-1000~V, in steps of 5~V and with a ramp-up speed of 1~V/s. The maximum operating voltage of the modules is 1000~V. This voltage may be needed to be applied for some tiles at the end of life in the experiment because of radiation damage. At each step of the scan, a waiting time of 10~s is set before the software acquires five consecutive current measurements and outputs their average. An example of the I-V characteristics of a tile, as measured by the tile manufacturer, tile quality assurance site and module assembly site, is shown in Fig.~\ref{fig:iv_curves}. As the test setups and environmental conditions are not the same in the different tests a precise agreement in results is not expected. 

In addition to this test, a ramp-up test at the nominal cooling conditions (-30$^{\circ}$~C) is performed, in order to ensure that every non-irradiated tile can reach the minimum bias voltage of 140~V, while drawing a current lower than 1~$\mu$A.

The grading criteria for both the IV and HV tests are listed in Tab.~\ref{tab:iv_hv_test}. The criteria for the IV measurements are based on the measurements performed in a probe station after reception from the manufacturer, as explained in Sect.~\ref{sec:summary_iv}.

\begin{table}[h]
    \centering
    \caption{Quality criteria and assigned grading for the IV test and HV ramp-up test.}
    \label{tab:iv_hv_test}
    \begin{tabular}{c | c | c }
            grade & IV curve & lowest voltage before 1~$\mu$A\\
            \hline
            \textbf{A} & $<$ high quality  & $>$ 700~V \\
            \textbf{B} & $<$ good quality  & $<$ 700~V \\
            \textbf{C} & $<$ good quality below 140~V & -- \\
            \textbf{D} & $>$ good quality below 140~V & --  \\
            \textbf{F} & -- & $<$ 250~V \\
    \end{tabular}
\end{table}

\subsubsection{VeloPix communication test}
\label{sec:communication_test}
\begin{figure}[htbp]
  \centering
  \includegraphics[width=0.7\linewidth]{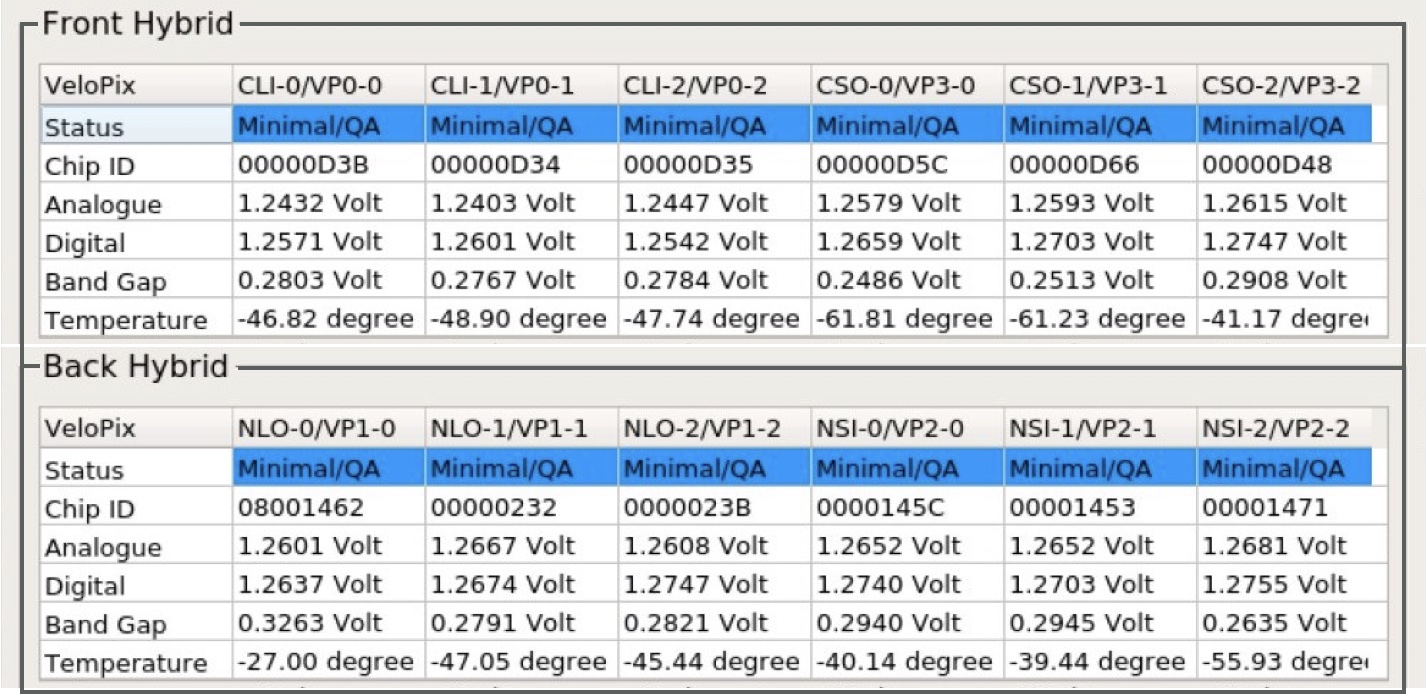}
  \caption{Typical values of the VeloPix chips internal analogue and digital voltages and uncalibrated temperatures.}
  \label{fig:comm_test}
\end{figure}

The VeloPix chips receive their LV power through dedicated power line wire-bonds. The power is supplied by 1.3~V DC-DC converters~\cite{faccio:DCDC}, but due to losses over the OPB traces, VFB connector and LV cables, the voltage drops to a typical value of about 1.2~V at the VeloPix chips. Therefore, the purpose of this test is to ensure that communication can be established with all the chips of a module, and that all chips are supplied with the required minimum power (1.18~V). The VeloPix chips are equipped with an internal monitoring of their analogue and digital voltages, and temperature. They can also report their Chip ID, which is burnt to an eFuse on the chip itself. The test is carried out by reading back these values from each ASIC, thus verifying the integrity of the electrical chain from the VeloPix through the wire-bonds, the front-end hybrids, the interconnecting cables, the GBTx chip and the long data tapes. Typical values of the supplied voltages and internal temperatures of the VeloPix chips are shown in Fig.~\ref{fig:comm_test} for one module.

\subsubsection{Time and fast control test}
\label{sec:tfc}
\begin{figure}[htbp]
  \centering
  \includegraphics[width=0.95\linewidth]{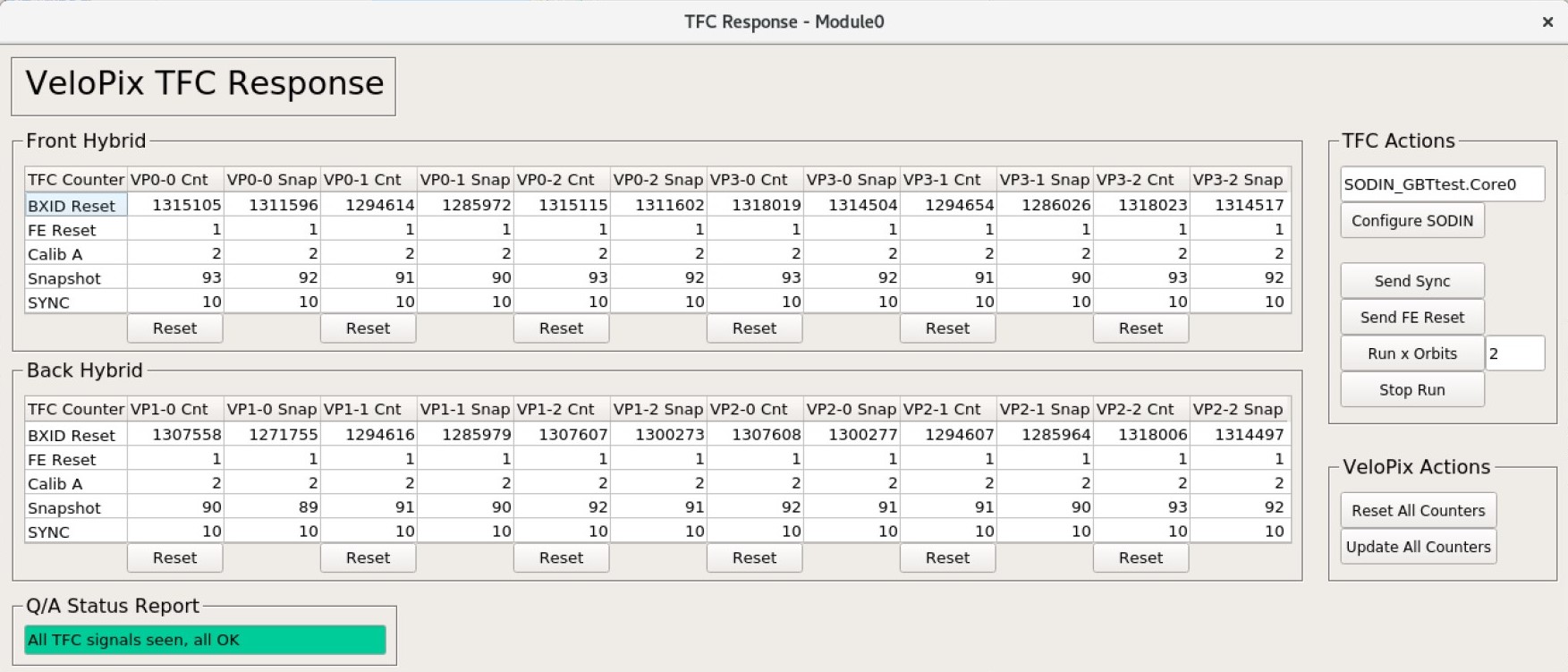}
  \caption{The WinCC panel used to carry out TFC tests. Signals are sent to the VeloPix chips and specific registers are then read back to verify that the increase in the corresponding counters matches the expectation.}
  \label{fig:tfc_panel}
\end{figure}

The TFC system~\cite{TFC} is responsible for the proper synchronization of the VELO modules with the rest of the experiment. The TFC signals reach the VeloPix chip through a single differential pair of wire-bonds. Any issue with this pair would result in the module being unusable by the experiment, as it would be operating asynchronously to the rest of the detector. The TFC signals are sent out from the PCIe40 card~\cite{pciegen3} to the GBTx chips on the module, and from there they are distributed to all VeloPix chips, where counters are incremented according to the signals received. A dedicated WinCC panel (see Fig.~\ref{fig:tfc_panel}) reads these counters from a set of specific registers and displays the results for each individual VeloPix, as well as returning a green (red) flag for successful (unsuccessful) tests. 

The gradings based on the internal digital and analog voltages and the TFC test are assigned according to the criteria listed in Tab.~\ref{tab:communication}.

\begin{table}[h]
    \centering
    \caption{Quality criteria and assigned grading for electrical communication tests.}
    \label{tab:communication}
    \begin{tabular}{c | c | c}
            grade & analog and digital voltage & TFC result\\
            \hline
            \textbf{A} & $> 1.20$~V & all signals seen\\
            \textbf{B} & $> 1.18$~V & --\\
            \textbf{C} & $< 1.18$~V & --\\
            \textbf{D} & one missing & one missing\\
            \textbf{F} & multiple missing & multiple missing\\
    \end{tabular}
\end{table}

\subsubsection{Pseudo random bit stream test}
\label{sec:prbs}
\begin{figure}[htbp]
  \centering
  \includegraphics[width=0.5\linewidth]{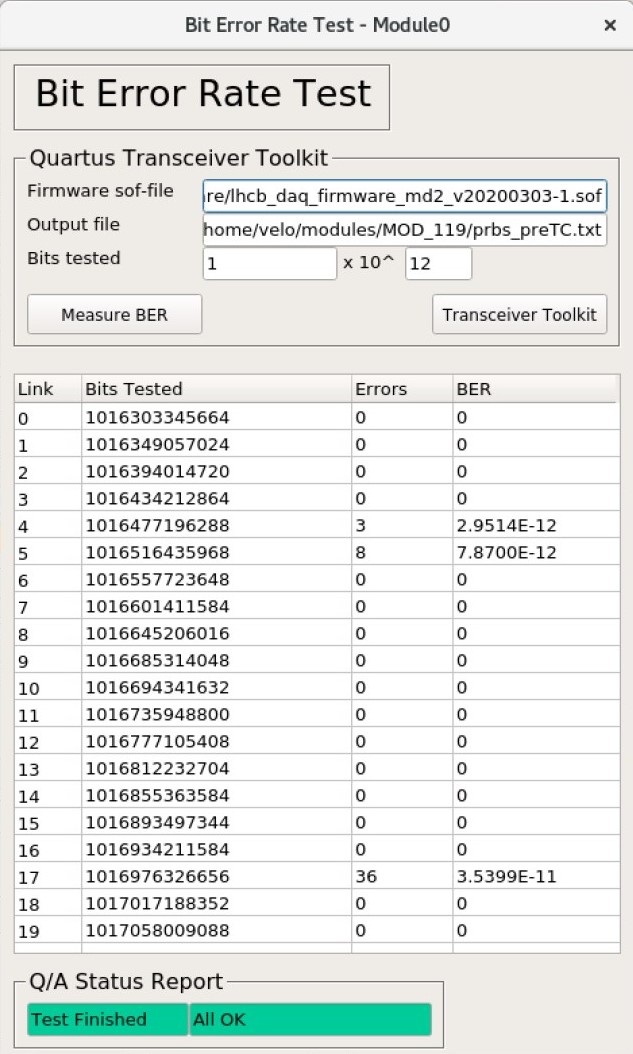}
  \caption{Typical result of a PRBS test.}
  \label{fig:prbs_test}
\end{figure}

The data stream coming from the VeloPix chips is routed to the DAQ directly via dedicated links in the hybrid data cables, without passing through the GBTx chip. Any error in the stream can lead to corrupted data packets and affect the performance of the final experiment. In order to verify the integrity of this path, a bit error rate test is carried out by using an industrial standard, the Pseudo-Random Bit Stream (PRBS). A stream is generated by the VeloPix using logic within the chip, and then propagated down the data links to the MiniDAQ, where it is collected and decoded. The result is then compared to the known pattern of the type of PRBS chosen, and the number of bits not matching the expected pattern recorded. This allows to assess the overall quality of the path carrying data signals from the point of PRBS generation on chip to the long data cables. However, this test is not sensitive to issues in the bump-bonds or with the silicon sensor itself, and cannot detect any damage in the pixel matrix. A typical result of a PRBS test is shown in Fig.~\ref{fig:prbs_test}.

The PRBS test returns the number of bits tested and the corresponding number of errors for each of the twenty links in a module. The database uses this data to calculate the bit error rate (BER), which is then used to assign a grade according to the criteria summarised in \cref{tab:prbs}.
\begin{table}[h]
    \centering
    \caption{Quality criteria and assigned grading for PRBS test.}
    \label{tab:prbs}
    \begin{tabular}{c | c | c}
            grade & BER of all links & BER of more than half of the links\\
            \hline
            \textbf{A} & $< 1 \times 10^{-11}$ & $< 1 \times 10^{-12}$\\
            \textbf{B} & $< 1 \times 10^{-10}$ & $< 1 \times 10^{-11}$\\
            \textbf{C} & $> 1 \times 10^{-10}$ & $> 1 \times 10^{-11}$\\
            \textbf{D} & one link down & --\\
            \textbf{F} & multiple links down & --\\
    \end{tabular}
\end{table}

\subsubsection{Equalisation and Noise Scan}
\label{sec:equalisation}
\begin{figure}[htbp]
  \centering
  \includegraphics[width=0.6\linewidth]{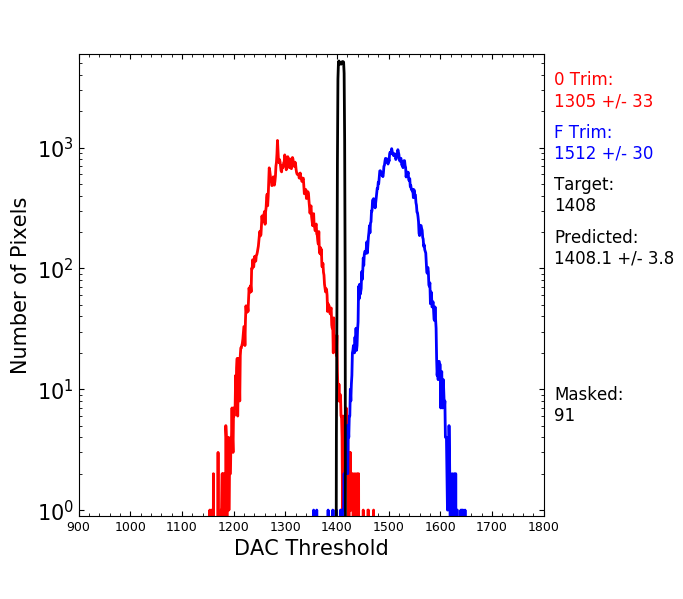}
  \caption{An equalisation scan showing the result for trim-0 (red), trim-F (blue) and the global equalisation (black).}
  \label{fig:equalisation}
\end{figure}

The noise level in the ASICs is a critical parameter that affects the performance of the detector and thus needs to be characterised. For reference, a MIP in a 200\,$\mum$ thick silicon sensor generates about 16,000 electron-hole pairs, which in the VeloPix correspond to about 1070 DAC counts ($\sim$15 e-h pairs per DAC count~\cite{poikela2017}). Moreover, because of imperfections in the silicon of the sensor and of the ASIC, each pixel may respond slightly differently to a given deposited energy and threshold. In order to correct for these differences, each pixel in the VeloPix is equipped with two configurable threshold values: the threshold value and the trim value. The threshold value is a 14-bit global threshold that is applied to all pixels on the same chip, while the trim is a 4-bit offset that is used to tune the differences between pixels in the same chip. Therefore, to equalise the response of the whole matrix, a threshold scan is performed and used to set a specific threshold trim value for each pixel. These scans are carried out automatically by a dedicated WinCC panel, which records the response of each pixel in the matrix to the minimal (trim=0) and maximal (trim=F) configurations. A typical result from these scans is shown in Fig.~\ref{fig:equalisation}.

\begin{figure}[htbp]
  \centering
  \includegraphics[width=0.45\linewidth]{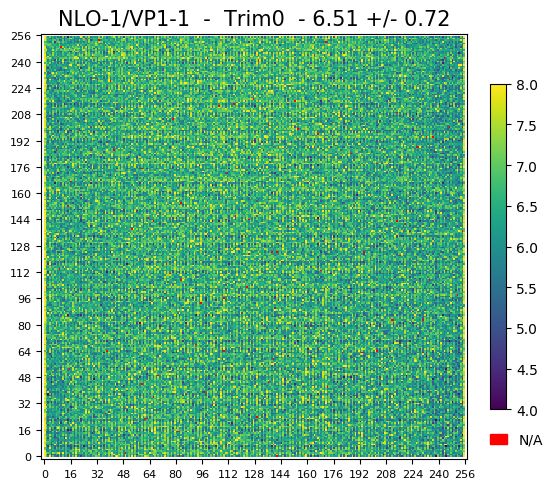}
  \includegraphics[width=0.45\linewidth]{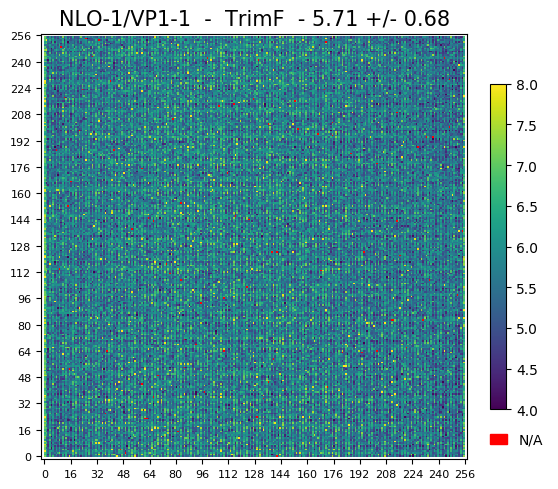}
  \caption{2D noise maps, in DAC counts, evaluated at trim-0 and trim-F; the average noise and its uncertainty is also shown in the tile of the plot.}
  \label{fig:noise_maps}
\end{figure}

The 0-trim scan is performed by setting the pixel trim value to the lowest of its 16 settings and then varying the global threshold value in a predefined range. At each step, the number of pixels that go from not firing to firing is counted, resulting in the red distribution shown in Fig.~\ref{fig:equalisation}. This procedure is then repeated with the trim value set to its highest value of 15 (trim-F), leading to the blue curve shown in Fig.~\ref{fig:equalisation}. Finally, the midpoint of the two trims is used to build the global equalisation threshold that is shown in black in Fig.~\ref{fig:equalisation}. During these scans, the noise of each pixel is also measured and a noise distribution obtained for each pixel. The sigma of the distribution is then used to build a 2D noise map of each VeloPix. These maps are important not only to assess the noise level of the chip, but also to highlight any issue that might have been introduced during tiles attachment to the substrate. An example of a 2D noise map for trim-0 and trim-F is shown in Fig.~Fig.~\ref{fig:noise_maps}. Finally, the average of all the values in the trim-F noise map, and their standard deviation, is used to assigned an average noise value and uncertainty to each VeloPix.

The noise and equalisation scans produce a large amount of data (about 15\,MB/chip), specific for each VeloPix chip. The grading criteria are based on the number of masked pixels per ASIC and per tile, and on the average noise and its standard deviation, as summarised in \cref{tab:eq_noise}. 
\begin{table}[h]
    \centering
    \caption{Quality criteria and assigned grading for the equalisation and noise scan.}
    \label{tab:eq_noise}
    \begin{tabular}{c | c | c | c | c}
            grade & tile masked & ASIC masked & noise & noise stdev\\
            \hline
            \textbf{A} & $< 1 \%$ & $< 5 \%$ & -- & --\\
            \textbf{B} & $< 5 \%$ & -- & -- & --\\
            \textbf{C} & $> 5 \%$ & -- & -- & --\\
            \textbf{D} & -- & -- & $> 8$ & $> 2$\\
            \textbf{F} & -- & -- & -- & --\\
    \end{tabular}
\end{table}

\subsection{Thermal performance}
\label{sec:thermal_performance}
Assessing the cooling performance of a module is important to ensure that the heat transfer capabilities from the tiles meet the requirements discussed in Sect.~\ref{sec:requirements}. The cooling performance is measured using the VeloPix band-gap circuit. This circuit generates a reference voltage of around 300~mV on the periphery of the analogue read-out, independently of temperature changes. Therefore, a proportional-to-absolute-temperature (PTAT) circuit is used to monitor the temperature changes on the chip and to feed them back into the band-gap circuit, allowing the generation of the temperature-independent reference voltage. These two voltages are written into two registers of the VeloPix, labelled as band-gap and temperature, and can be used to monitor the variation of the internal temperature. In parallel, the NTC temperature sensor values are also logged and used as an independent measurement of the temperature observed near the tiles. 

\begin{figure}[htbp]
  \centering
  \includegraphics[width=0.8\linewidth]{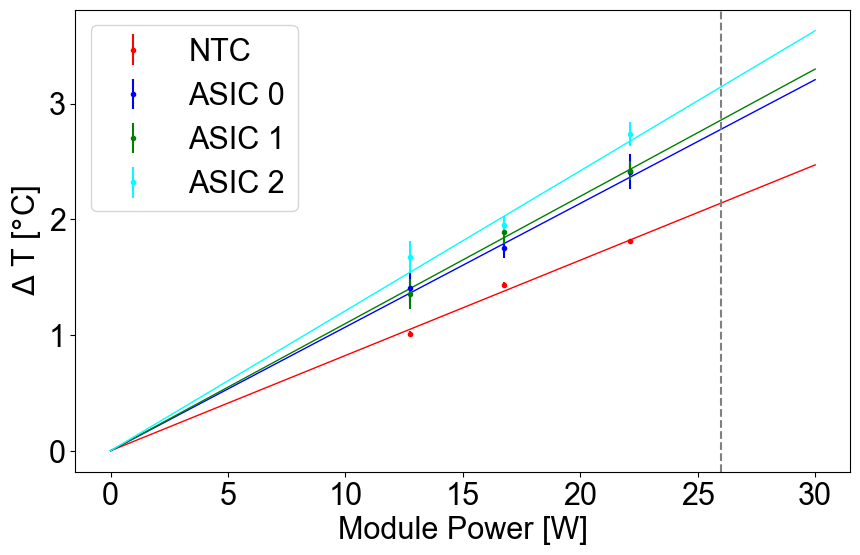}
  \caption{Temperature variation in a tile as function of the power consumption in its three ASICs. The dashed line at 26~W indicates the estimated end-of-life power consumption.}
  \label{fig:cooling_performance}
\end{figure}

The cooling performance of a tile is thus determined by varying the power consumption in the three ASICs, measuring the corresponding temperature variations, and extrapolating the tile $\Delta$T to the end-of-life power consumption (26~W). The variation of the power consumption is achieved by enabling different functionalities in the chip (power modes). An example of the results obtained from the cooling performance test is shown in Fig.~\ref{fig:cooling_performance}.  The NTC temperature is typically lower than that on the ASICs due to the thermal gradient between the ASICs and hybrid.

The measurements are repeated before and after ten thermal cycles to identify any potential thermal degradation. The quality criteria applied and grading assigned are listed in Tab.~\ref{tab:bandgap}.
\begin{table}[h]
    \centering
    \caption{Quality criteria and assigned grading for the band-gap test and thermal performance change before and after thermal cycles are performed. The mean and maximum variations are computed over all the 12 ASICs of a module.}
    \label{tab:bandgap}
    \begin{tabular}{c | c | c | c}
            grade & $\Delta$T at 26~W & mean variation & max variation\\
            \hline
            \textbf{A} & $< 8~^{\circ}$C & $< 1~^{\circ}$C & -- \\
            \textbf{B} & $< 10~^{\circ}$C & $> 1~^{\circ}$C & -- \\
            \textbf{C} & $> 10~^{\circ}$C & -- & -- \\
            \textbf{D} & missing ASIC & -- & $> 5~^{\circ}$C \\
            \textbf{F} & -- & $> 5~^{\circ}$C & --\\
    \end{tabular}
\end{table}

%% file: VELO/summary.tex
\subsection{Production rate}
\begin{figure}[htbp]
    \centering
    \includegraphics[width=0.99\linewidth]{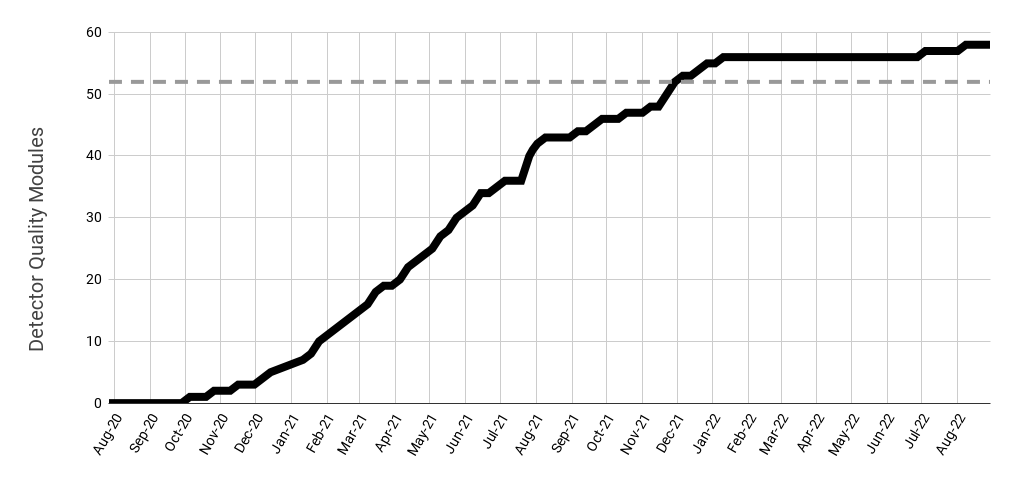}
    \caption{Module production rate (only detector quality modules are shown here). The dotted grey line indicates the target of 52 detector quality modules needed for the VELO Upgrade.}
    \label{fig:module_production}
\end{figure}

The module production started in July 2020 in the two assembly sites, the University of Manchester and Nikhef. The production thus occurred during the period of the COVID-19 pandemic which led to significant additional organisational and logistical challenges. Despite these difficulties, the production was completed in time to allow installation for the start of LHC Run~3. 
The rate at which modules have been assembled and qualified is shown in Fig.~\ref{fig:module_production}. Only modules that have passed all quality criteria and were declared of production quality are shown in this plot. After an initial learning curve of about six months, the production rate increased to about 1.5 modules per week. However, in August 2021 the rate slowed down again, mainly due to delays in the production of a new batch of front-end and control hybrids. The target number of required modules (52) was achieved in December 2021. 

\begin{figure}[htbp]
    \centering
    \includegraphics[width=0.99\linewidth]{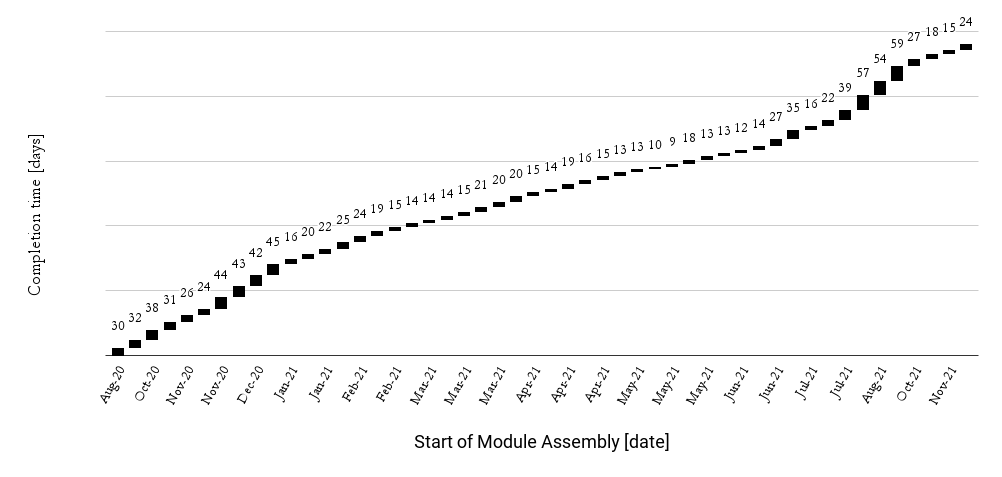}
    \caption{The time taken to fully assemble and perform the qualification tests on each module at one of the production sites. The days elapsed from the start to the completion are given for each module are shown on the y-axis. The x-axis indicates the start date of the module assembly.}
    \label{fig:module_production_days}
\end{figure}

In Fig.~\ref{fig:module_production_days}, the number of days needed to fully assemble and qualify each module is shown. This plot also illustrates how the amount of time spent on each module decreased over time, and reached a minimum of about two weeks per module during the central phase of production, when all procedures were well established and production was running smoothly. This completion time then increased again towards the end of the production, for the reason explained above. However, a number of steps in the module assembly and qualification could be carried in parallel, thus reducing the effective time per module: the production of the modules shown in the plot took approximately 16 months (493 days), which means that, in average, a new module was ready for the experiment every 9.5 days. 

\subsection{Summary plots}
During module production, the results of each individual test performed on a module were uploaded to the database, where the data were automatically processed and a grade assigned to the test. The analysis also allowed us to monitor any drift in the measured quantities such that any required action could be taken immediately to rectify the issue. A collection of summary plots for all the measured quantities is presented in the following sections, where these are shown for consistency for all modules produced at a single production site and ordered by date of completion. In addition some numbers characterising the overall module assembly accuracy and performance are given, where these are calculated for the set of 52 modules installed into the experiment.

\subsubsection{Summary of bare module flatness measurements}
\begin{figure}[htbp]
    \centering
    \includegraphics[width=0.99\linewidth]{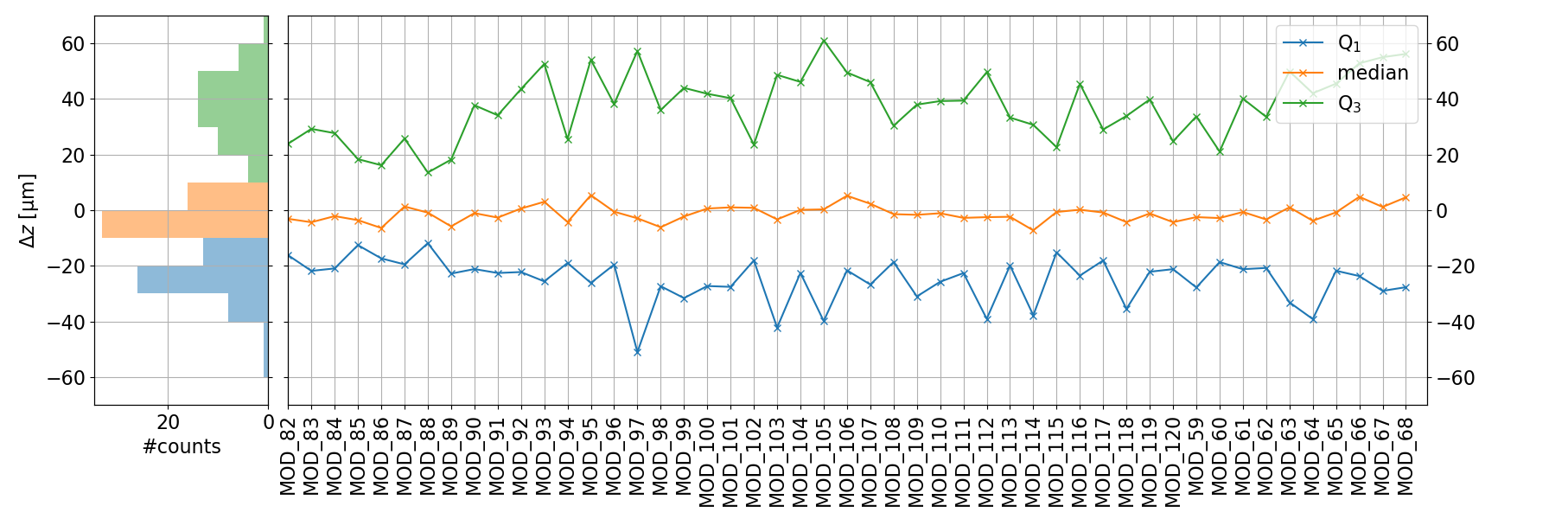}
    \caption{The bare module flatness results as a function of module number across the production, showing the quartiles of the height distribution after the subtraction of a fit of a plane ($\Delta$z~=~z$_{\rm meas}$~-~z$_{\rm fit}$).}
    \label{fig:summary_bare_module_flatness}
\end{figure}

The flatness of a bare module is measured  as described in Sect.~\ref{sec:bare_module_flatness}. The results are shown in Fig.~\ref{fig:bare_module_survey}. Given the large area that needs to be covered during the measurements, a number of outlying data points are present in each data set due to failures in the focusing of the optical devices. Thus, for the comparison between different modules, the $\Delta$z~=~z$_{\rm meas}$~-~z$_{\rm fit}$ residual distributions are split into four ranges (quartiles), each containing 25\% of the data points. The first three quartiles are then used to monitor the trend of the bare module flatness measurements, as shown in Fig.~\ref{fig:summary_bare_module_flatness}, while the fourth quartile is more affected by the mismeasured points. As expected, the median of the flatness distribution ($Q_2$) lies around $\Delta$z~=~0, while quartiles $Q_1$ and $Q_3$ are fairly consistent along the whole module production.
This test demonstrates that none of the substrates were damaged during transportation and that there were no irregular surface features that would prevent the attachment of the tiles.

\subsubsection{Summary of tile position measurements}
\begin{figure}[htbp]
    \centering
        \centering
        \includegraphics[width=0.99\linewidth]{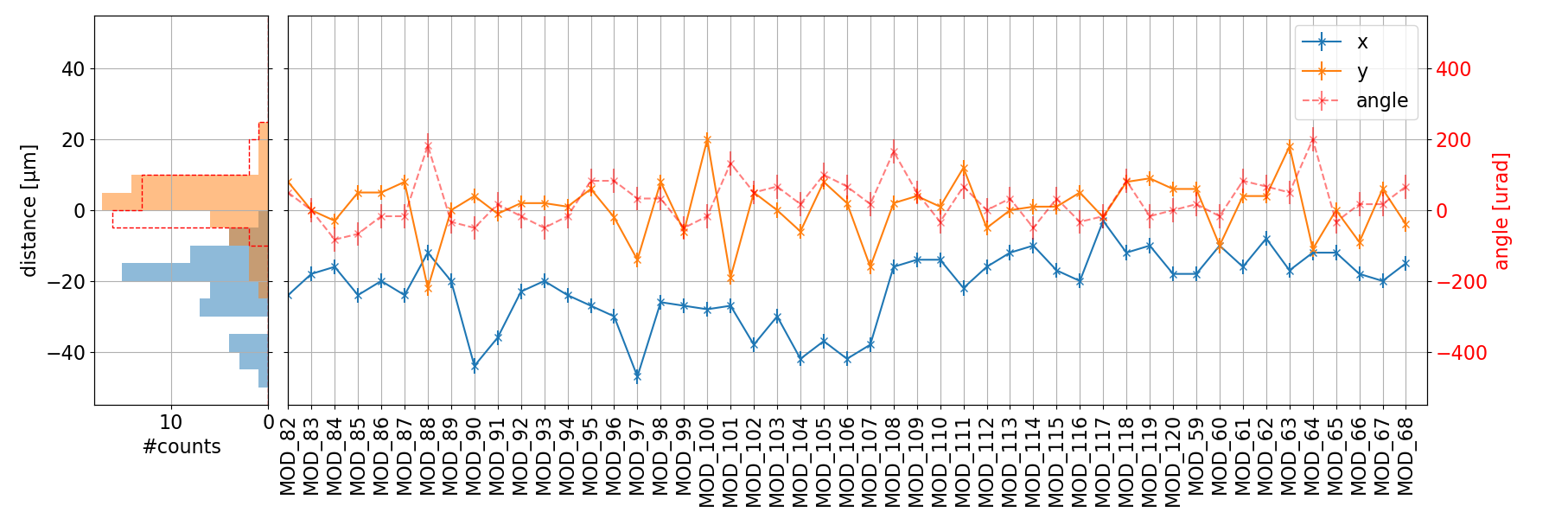} \\
        \includegraphics[width=0.99\linewidth]{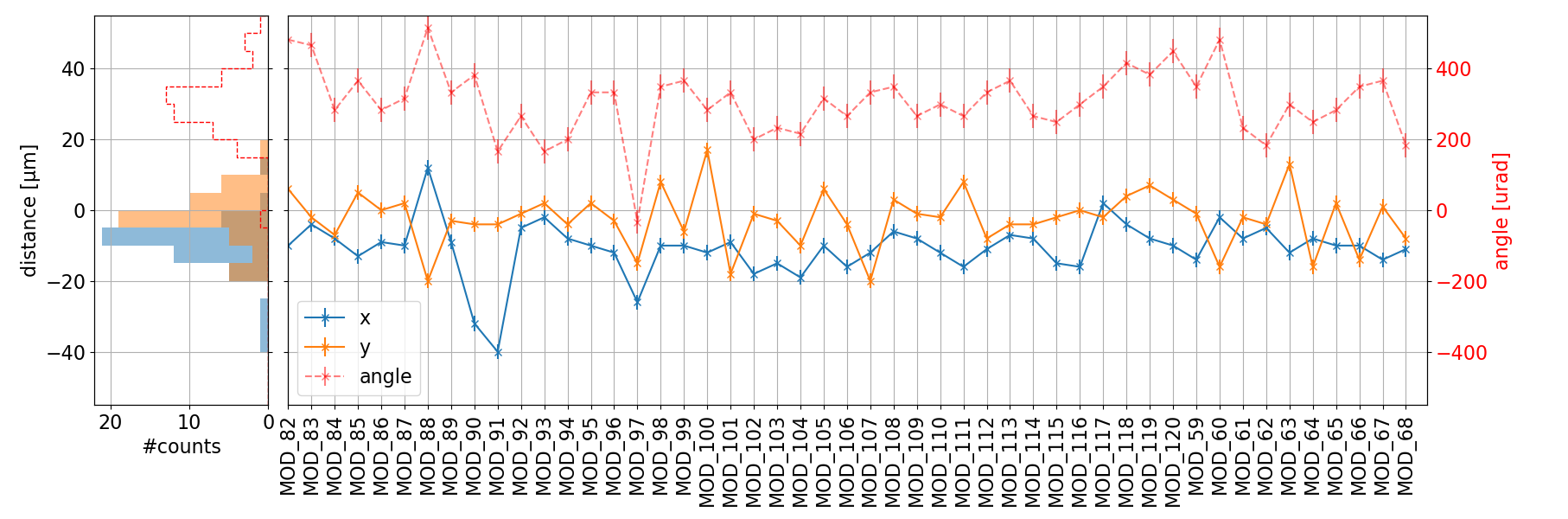} \\
        \includegraphics[width=0.99\linewidth]{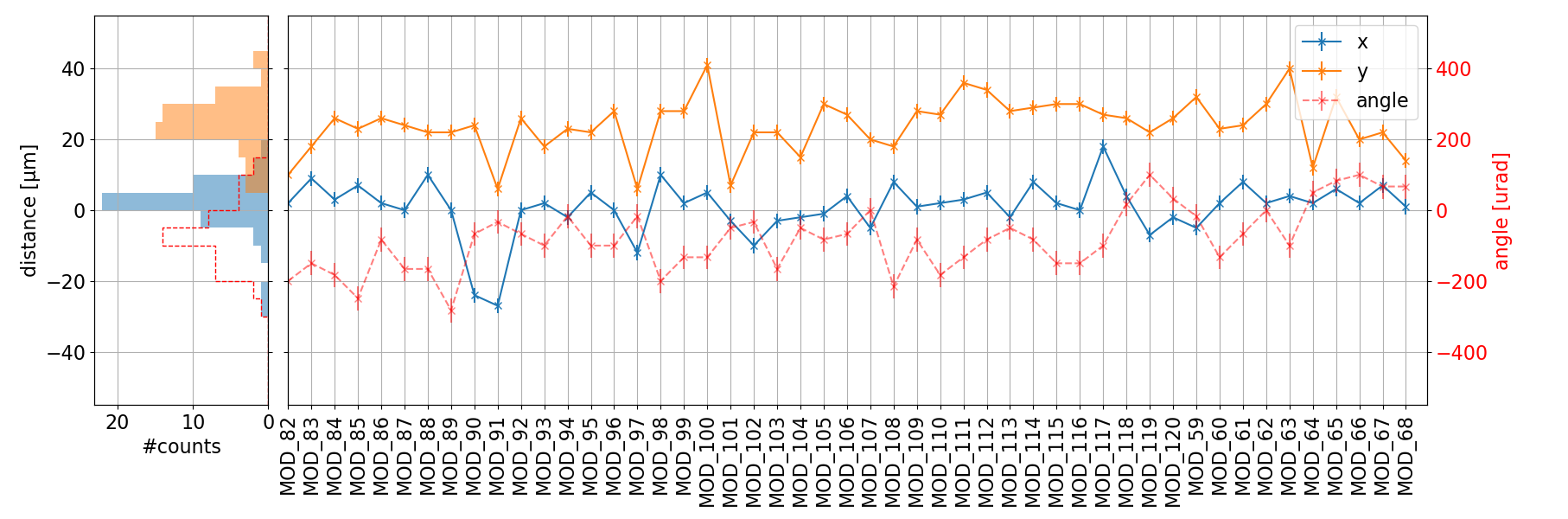} \\
        \includegraphics[width=0.99\linewidth]{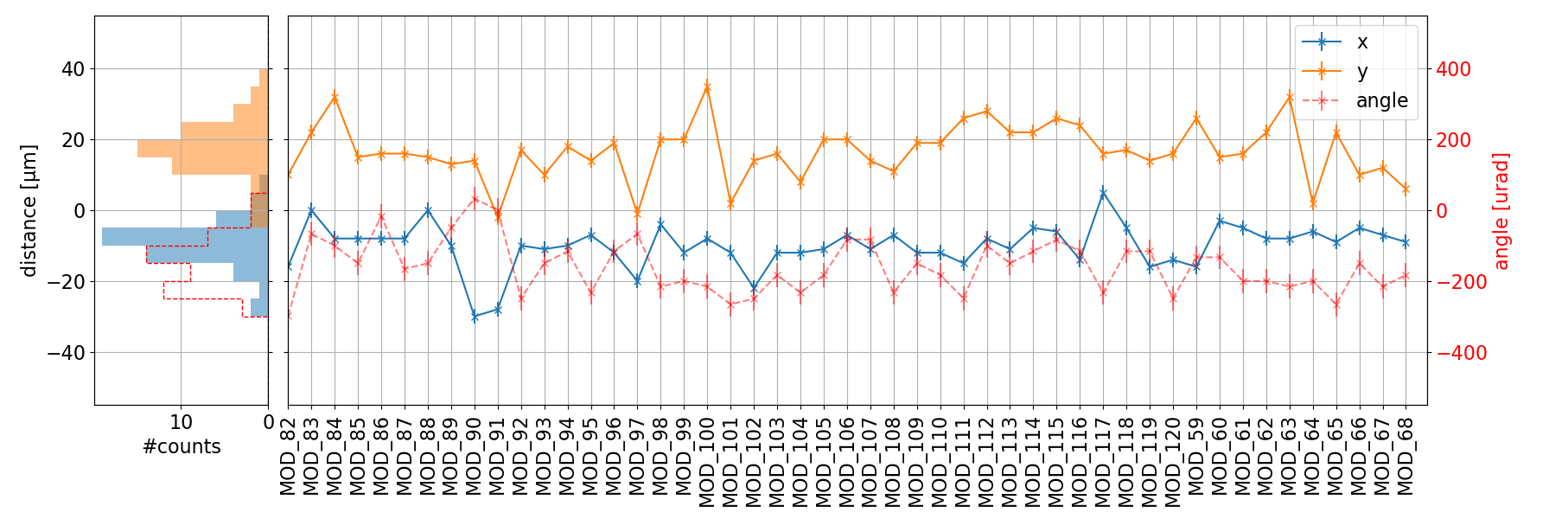} \\
        \caption{The achieved tile positions relative to their ideal location as a function of module number across the production. The plots are for the four tiles on each module, arranged from top to bottom as CLI, CSO, NLO, NSI.  In each plot the offset of the middle point of the tile in $x$ and $y$ and the rotation of the tile is shown.}
    \label{fig:summary_tile_position1}
\end{figure}

As described in Sect.~\ref{sec:metrology_tiles_position}, the position of a tile after attachment is evaluated by measuring the $x$ and $y$ coordinates of the left fiducial in the leftmost ASIC and of the right fiducial of the rightmost ASIC. In addition, also the fiducials visible on the back of the over-hanging tiles are measured. These values can then be compared to the target position and the absolute differences used to assess the quality of the tile placement. This approach treats the measurements as independent while the tile is a rigid object. In order to better estimate the placement, the $x$ and $y$ coordinates of the middle point between the measured fiducials, together with the angle of rotation of the tile, is considered. The trends of the $x$ and $y$ offsets and the angle are shown in Fig.~\ref{fig:summary_tile_position1} for the four tiles. A 20\um drift from module 93 until 107 can be observed in the trend of the $x$ coordinate for CLI, moving from around 20\um to 40\um. Having this careful monitoring procedure in place allowed this to be identified during production. This drift was corrected back to below 20\um by adjusting the alignment constants used for the positioning of the tiles from module 108 onwards. 

The mean offset of the tiles from their ideal position is computed by taking, for all tiles on all installed modules,  the average of the absolute distance for the centre of the tiles from its ideal position. This yields 21\um with a standard deviation of 10\um. The mean angular deviation of the tiles is 4~$\mathrm{\mu rad}$ with a standard deviation of 243~$\mathrm{\mu rad}$. 
The tile positioning is thus far better than is needed from mechanical constraints and  provides an excellent starting point for the software alignment of the detector with tracks.

\begin{figure}[htbp]
    \centering   
        \centering
        \includegraphics[width=0.99\linewidth]{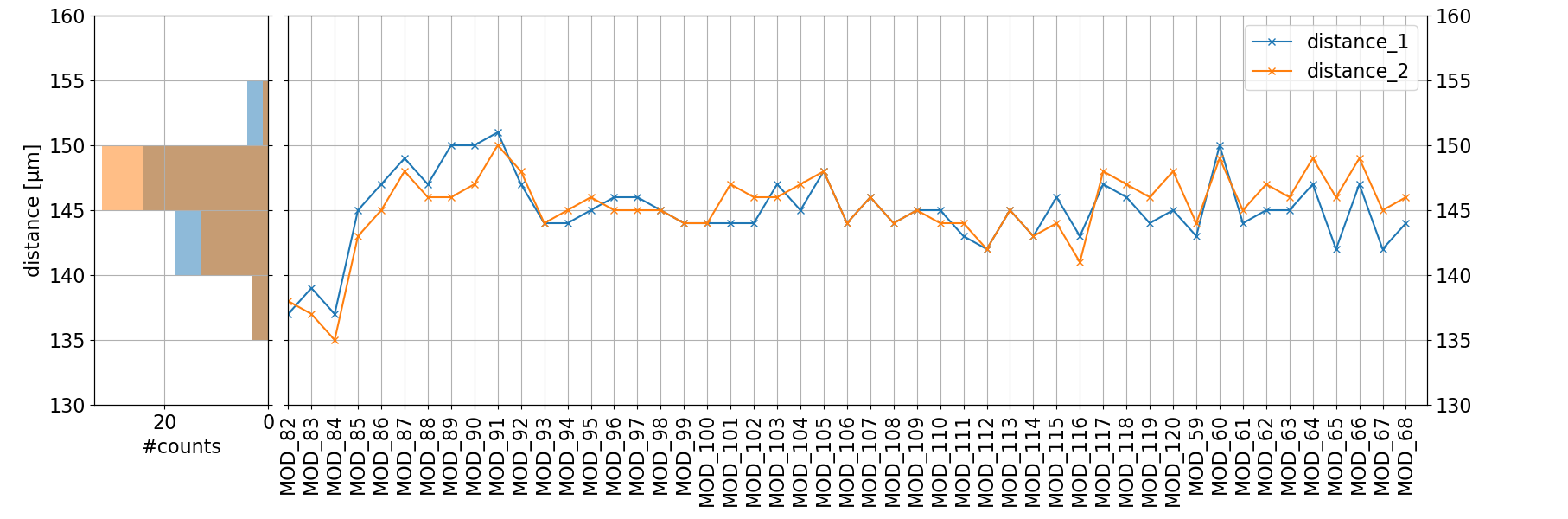}
        \caption{The distances d1 and d2 between the two tiles on the non-connector side as a function of module number across the production.}
    \label{fig:summary_tile_position2}
\end{figure}

On the non-connector side of a module, the distances d1 and d2 between the two tiles are also measured. The trend for those quantity is shown in Fig.~\ref{fig:summary_tile_position2}. The distance is intended to be 145\um and is a key parameter in the assembly. As shown in the plot values close to this were achieved throughout the production.

\subsubsection{Summary of glue layer thickness and tile flatness measurements}
\label{sec:summary_glue}
\begin{figure}[htbp]
    \centering
    \includegraphics[width=0.99\linewidth]{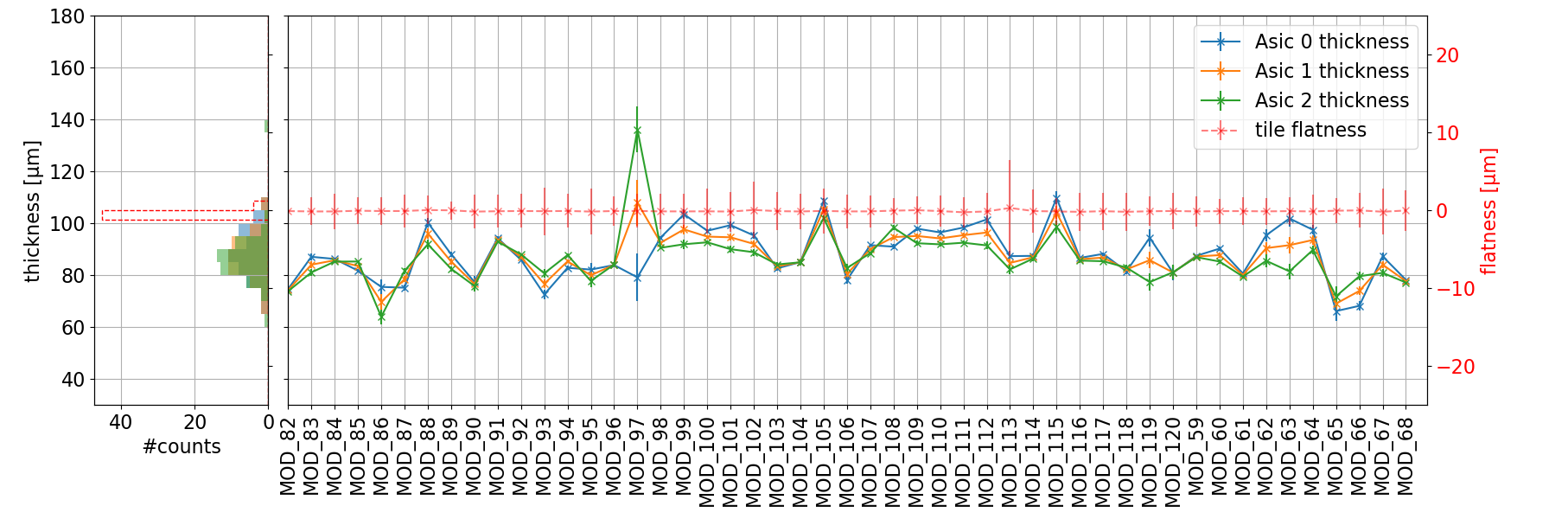}
    \includegraphics[width=0.99\linewidth]{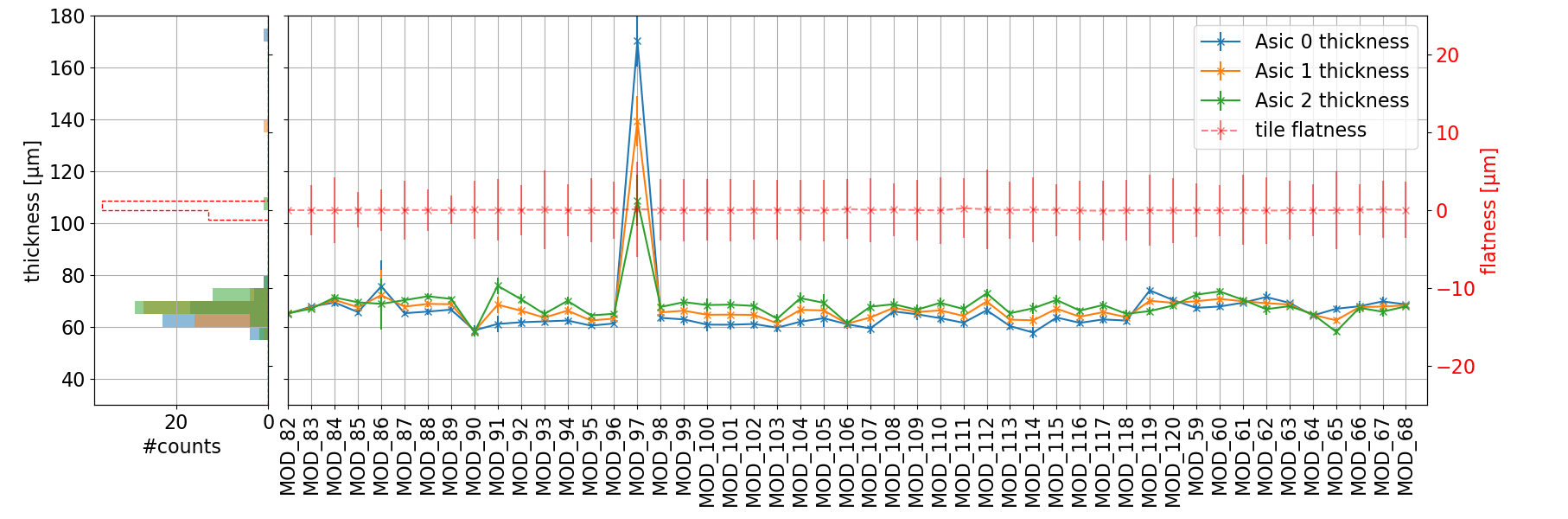}
    \includegraphics[width=0.99\linewidth]{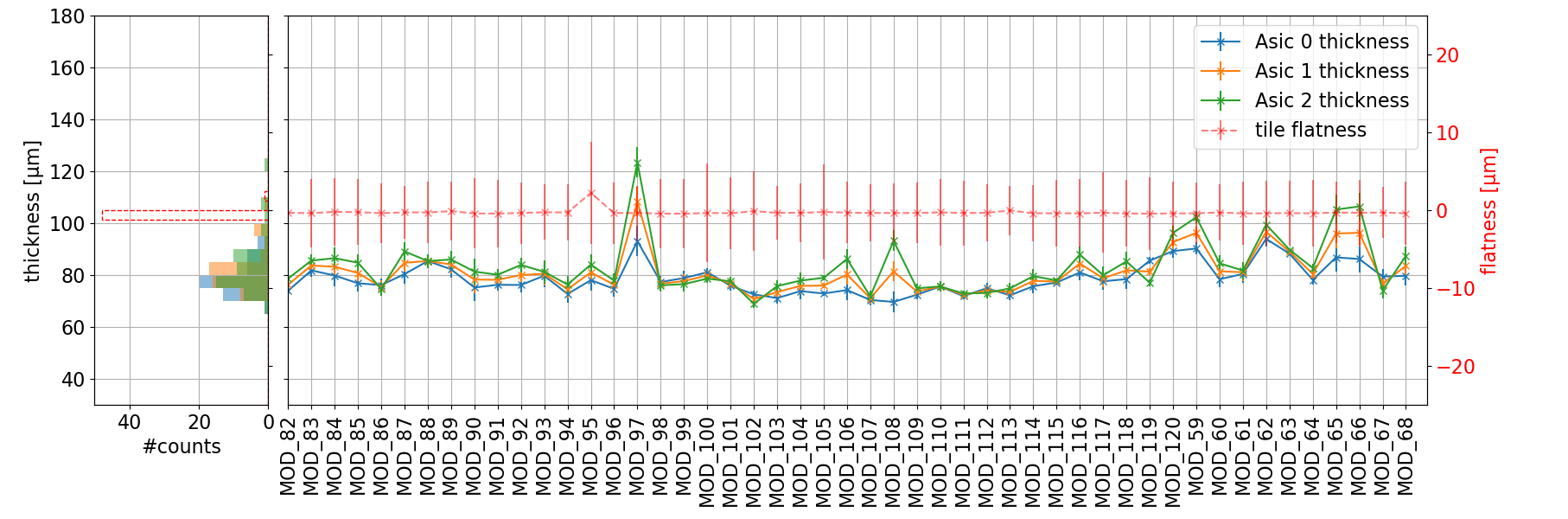}
    \includegraphics[width=0.99\linewidth]{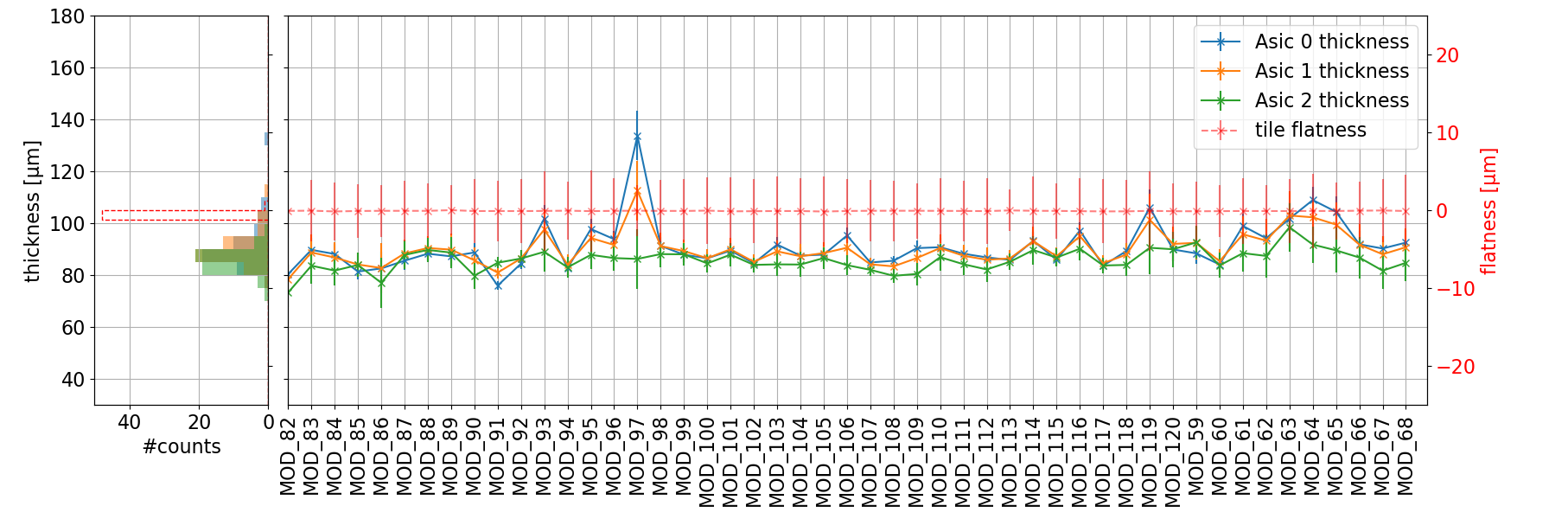}
    \caption{The glue thickness and tile flatness measurements are shown as a function of module number across the production. The points show the mean value while the error bars indicate the standard deviation. The plots are for the four tiles on each module, arranged from top to bottom as CLI, CSO, NLO, NSI.  In each plot the mean glue thickness for each of the four ASICs on the module is shown (left axis) and the mean tile flatness (right axis).}
    \label{fig:summary_glue_thickness}
\end{figure}

The glue layer thickness and the tile flatness are measured using deviations from a plane fitted to measured data points. The same data set is used to both evaluate the glue thickness and the tile flatness, as discussed in Sect.~\ref{sec:metrology_glue}. The mean and standard deviation of the distributions obtained for each tile, together with those from the flatness measurements, are used to compare the quality of the attachment of the tiles during module production. The trends of these are shown in Fig.~\ref{fig:summary_glue_thickness} for each of the four tiles as a function of the module number. Module 97 shows a larger than typical glue thickness. The hypothesis is that a dust particle was deposited along with the glue in the CSO tile which caused the microchannel to deform in the jig and led to variations in the glue thickness under all tiles. This leads to a somewhat worse thermal performance in the CSO tile (see Fig~\ref{fig:summary_bandgap_corrected}) but still within the acceptable range.

The mean deviation of the tile surfaces from a fitted plane for all points measured across all tiles on all installed modules is 0\um with a standard deviation of 4\um, thus showing that no significant bowing was introduced during the tile assembly or attachment. The glue layer thickness is an important parameter as it determines both the mechanical integrity and thermal performance of the module. Given the complex glue patterns, required treatment and evolving viscosity, the achieved average glue thickness of 78\,$\mu$m, with 5\,$\mu$m standard deviation, can be considered to be in excellent agreement with the target of 80\,$\mu$m.

\subsubsection{Summary of wire-bond pull tests}
\begin{figure}[htbp]
    \centering
    \includegraphics[width=0.99\linewidth]{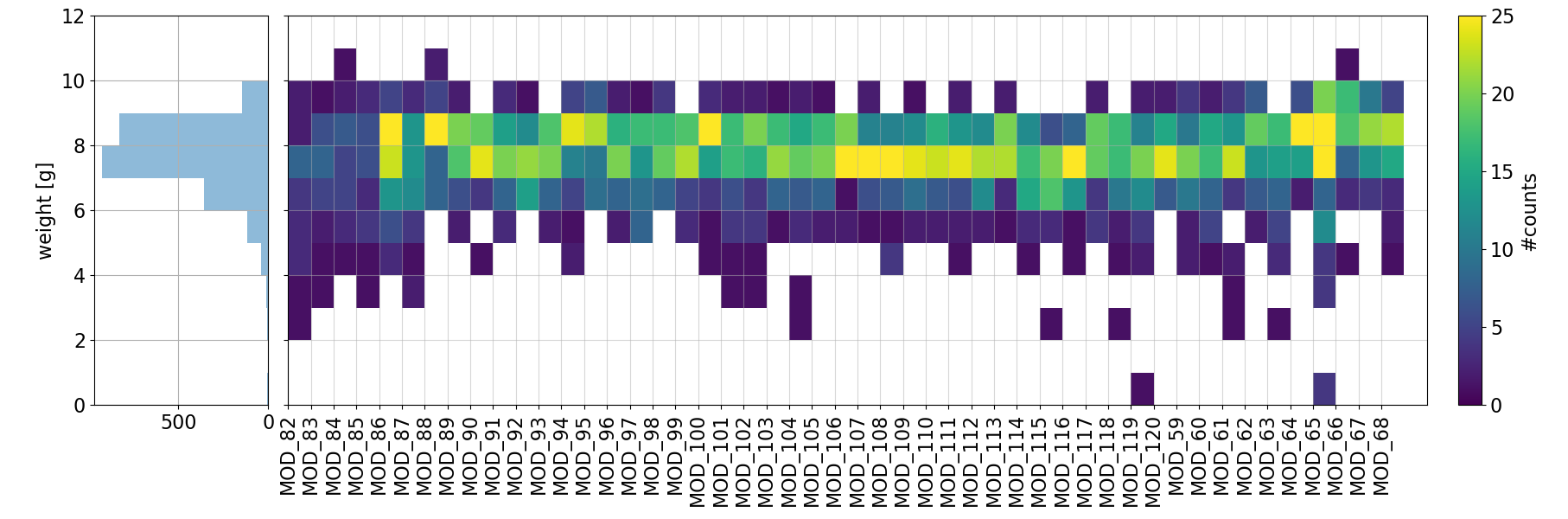}
    \caption{The measured wire-bond strength, determined from pull tests on bonds on the production modules, as a function of module number across the production. Less measurements were made directly after the wire bonding on the first production modules. }
    \label{fig:summary_pull_test}
\end{figure}

As described in Sect.~\ref{sec:metrology_wirebonds}, after a module has been fully wire-bonded, its four sacrificial bonds are pulled and the breaking force recorded. The results for all modules are shown in Fig.~\ref{fig:summary_pull_test}. As the  VELO modules are double-sided, obtaining sufficient support for the bonding was a concern during the design and R\&D. However, the bonding quality met the specifications throughout the production without any trends in bonding performance observed.

\subsubsection{Summary of displacement measurements}
\begin{figure}[htbp]
    \centering
    \includegraphics[width=0.99\linewidth]{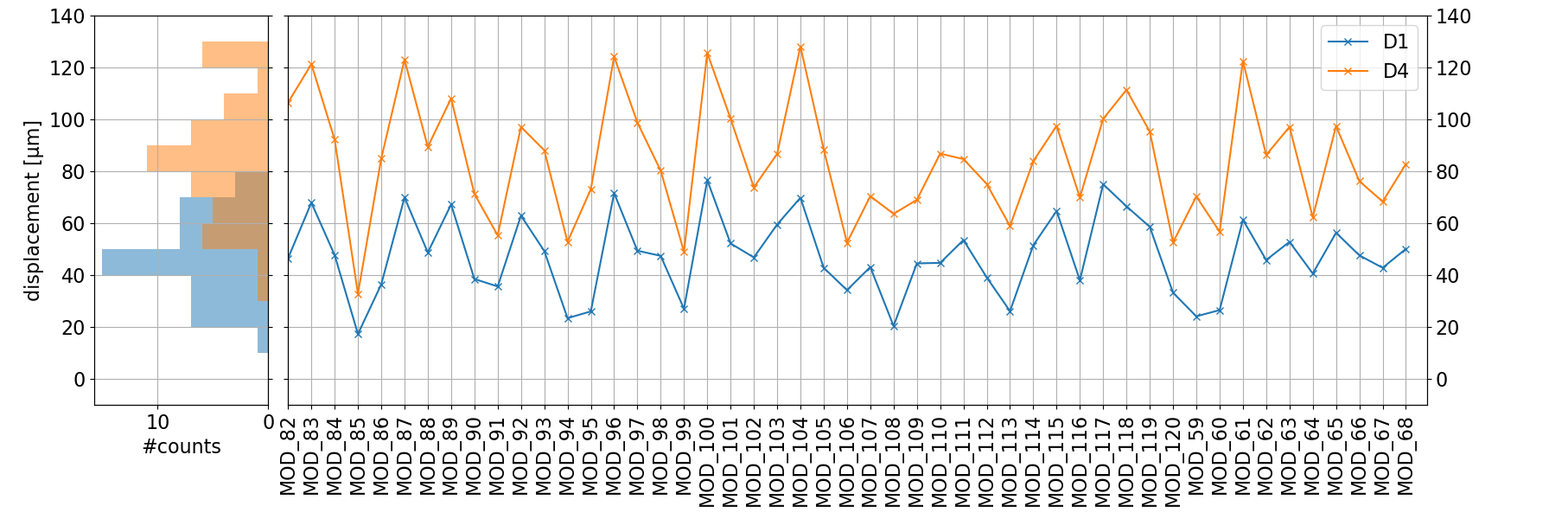}
    \includegraphics[width=0.99\linewidth]{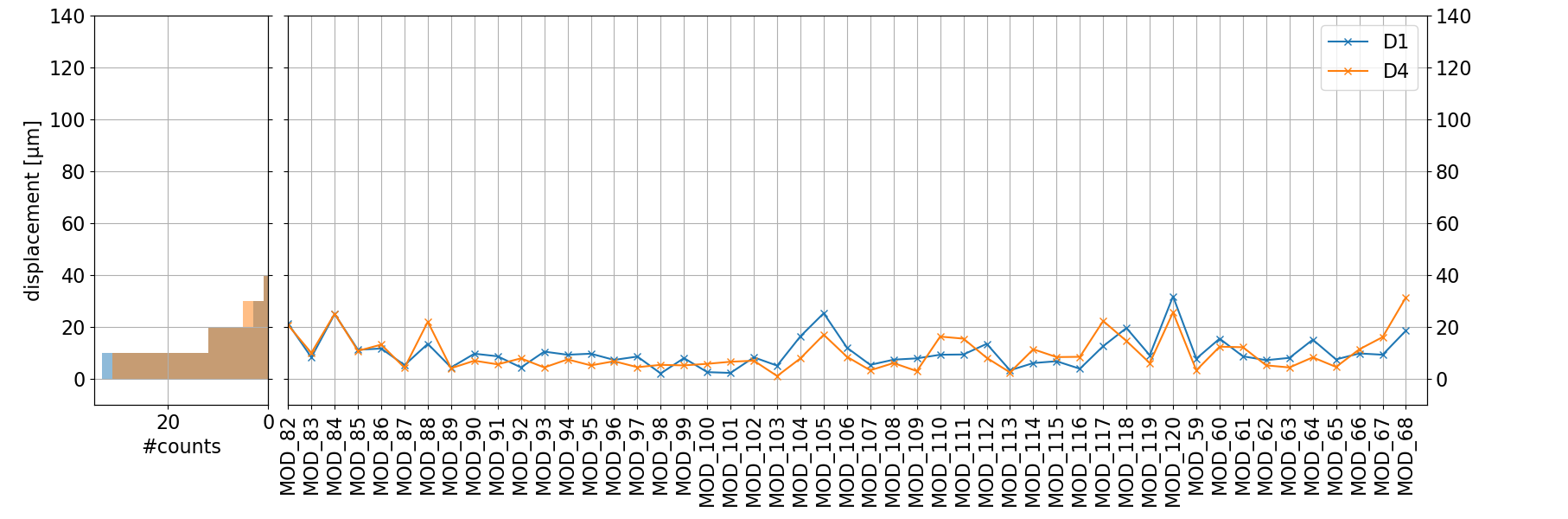}
    \caption{Displacements measured during cooling down the module (top) and evacuating the test tank (bottom), as a function of module number across the production. The results for both displacement sensors (see text) are shown.}
    \label{fig:summary_displacements}
\end{figure}

As discussed in Sect.~\ref{validation_displacement}, the displacement of each module is evaluated during tightening of the cooling connectors, the evacuation of the test tank and the cooling down from room temperature to -30$^{\circ}$~C. The trends are shown in Fig.~\ref{fig:summary_displacements}. While the first two displacements are usually small and fairly consistent across the whole lot of modules, the cooling down process induces larger deformations that can also vary significantly from one module to the other. Nonetheless, the displacement during cooling down is always in the same direction and very consistent when repeated on the same module. These results are a useful input to the software alignment that is performed using track reconstruction, and can provide a starting point, especially for the $z$ position of the installed modules.

 \subsubsection{Summary of communication tests}
 \label{sec:summary_communication_test}
\begin{figure}[htbp]
    \centering
    \includegraphics[width=0.95\linewidth]{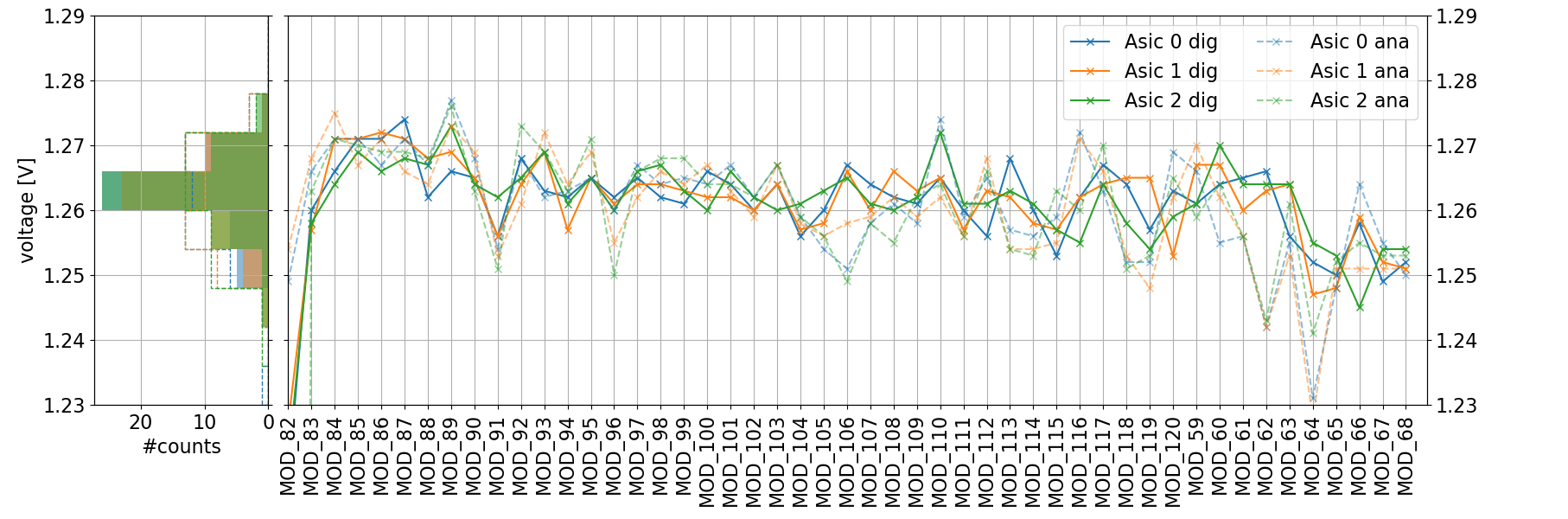}
    \includegraphics[width=0.95\linewidth]{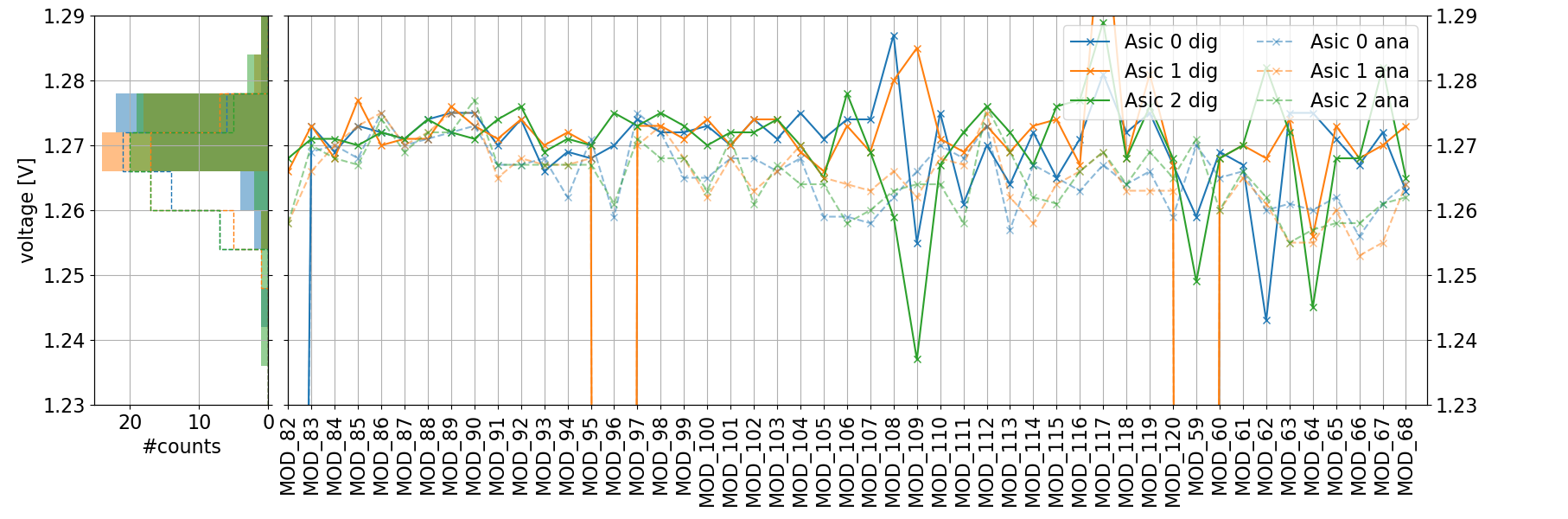}
    \includegraphics[width=0.95\linewidth]{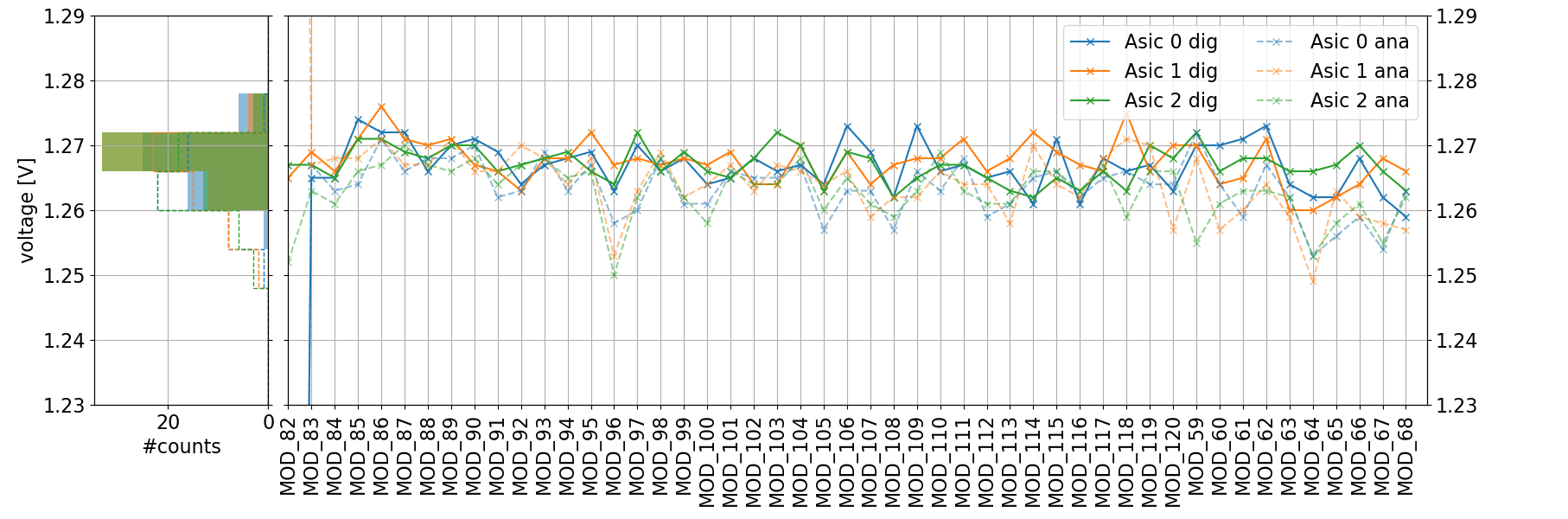}
    \includegraphics[width=0.95\linewidth]{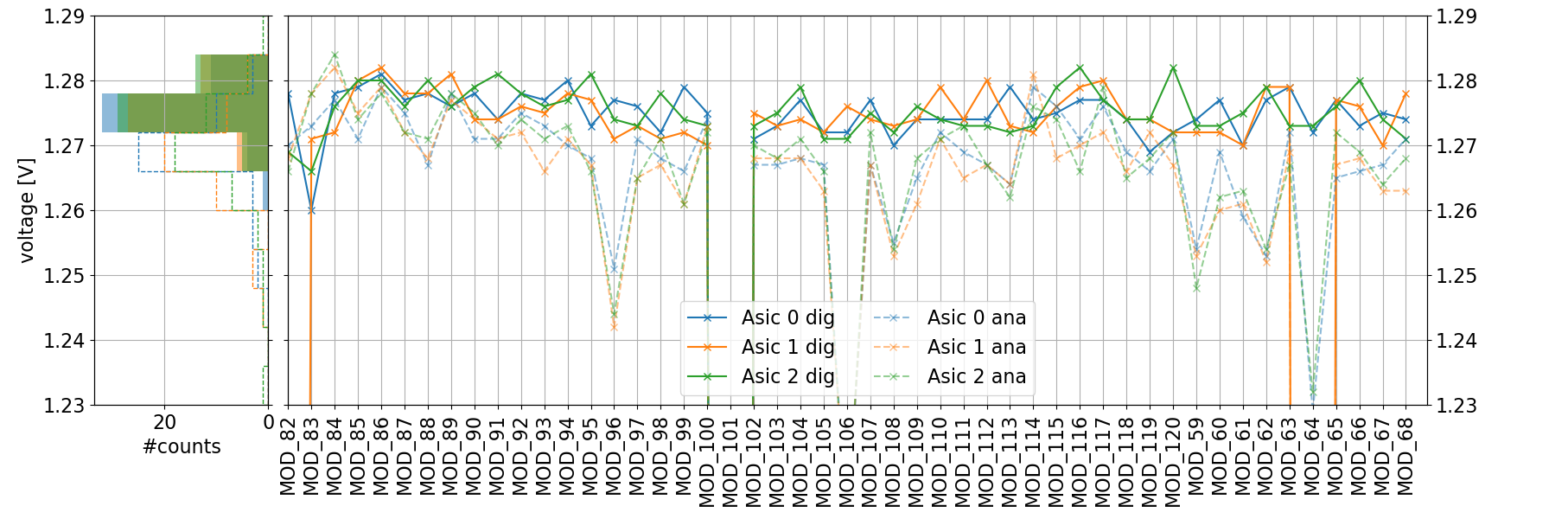}
    \caption{The analogue and digital currents, measured by the VeloPix chips, as a function of module number across the production. The plots are for the four tiles on each module, arranged from top to bottom as CLI, CSO, NLO, NSI.  In each plot the analogue and digital voltages of each of the three ASICs is shown.}
    \label{fig:summary_communication}
\end{figure}

 The purpose of measuring the analogue and digital current in each VeloPix chip is discussed in Sect.~\ref{sec:communication_test}. These measurements are performed three times, first in air at room temperature, and then in vacuum at -30$^{\circ}$~C before and after thermal cycling. The results for the measurements taken before thermal cycling are shown in Fig.~\ref{fig:summary_communication}. For a small number of tiles the link reading the ASIC voltage was not functional, although the voltage was correctly delivered to the ASIC. It was noted that the input voltage to the VeloPix chips increases at lower temperatures, as expected. No significant effect due to the thermal cycling was observed.

\subsubsection{Summary of PRBS tests}

\begin{figure}[htbp]
    \centering
    \includegraphics[width=0.95\linewidth]{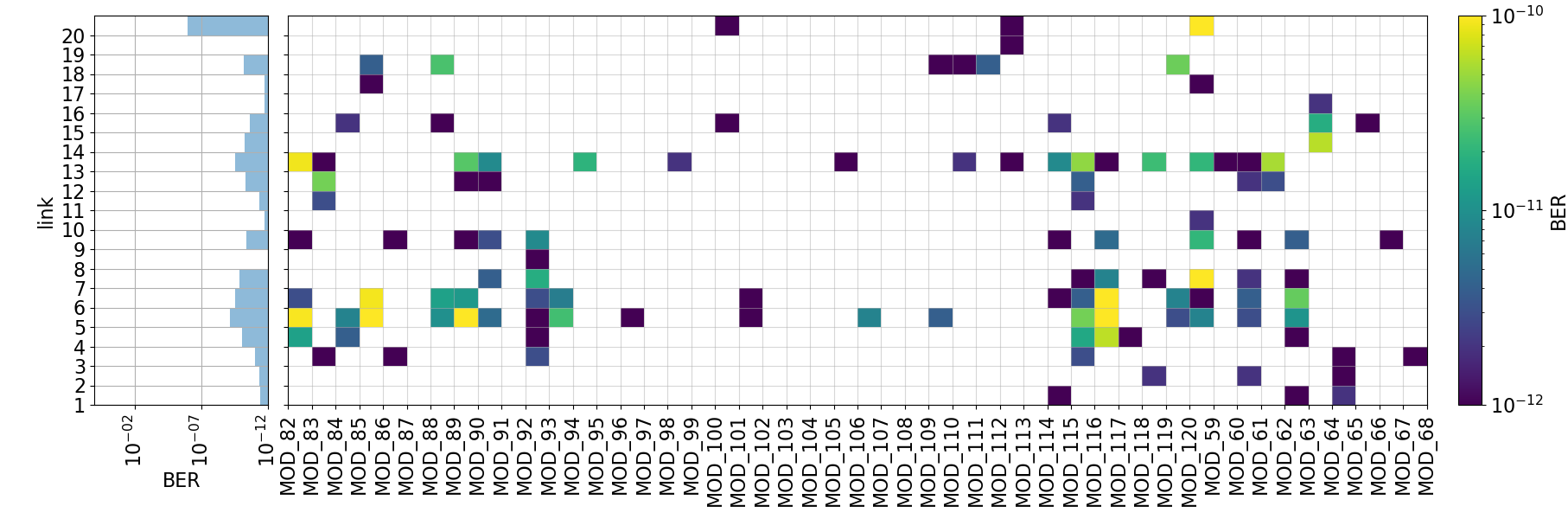}
    \includegraphics[width=0.95\linewidth]{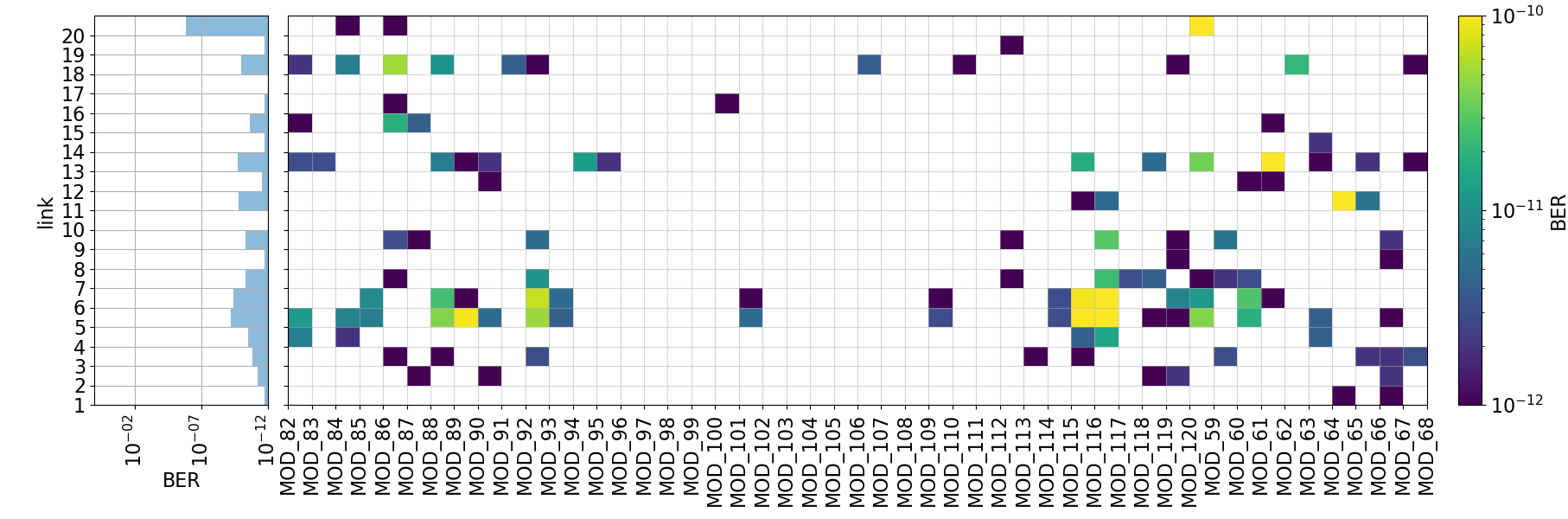}
    \caption{The bit error rates (BER) during PRBS tests performed before (top) and after (bottom) thermal cycling, as a function of module number across the production. The empty bins represent links that had no errors in the 10$^{12}$ bits tested. BER values above 10$^{-7}$ are not shown in order to enhance the granularity.}
    \label{fig:summary_prbs}
\end{figure}

The PRBS test is carried out before and after thermal cycling, as all the other electrical tests. The results for all the modules are shown in Fig.~\ref{fig:summary_prbs}. The majority of the links do not show any error in the 10$^{12}$ bits tested, while some specific links are noisier irrespective of the module under test. The issue seems to be due to the specific electrical test setup in which the measurement was performed.

\subsubsection{Summary of noise tests}
\begin{figure}[htbp]
    \centering
    \includegraphics[width=0.95\linewidth]{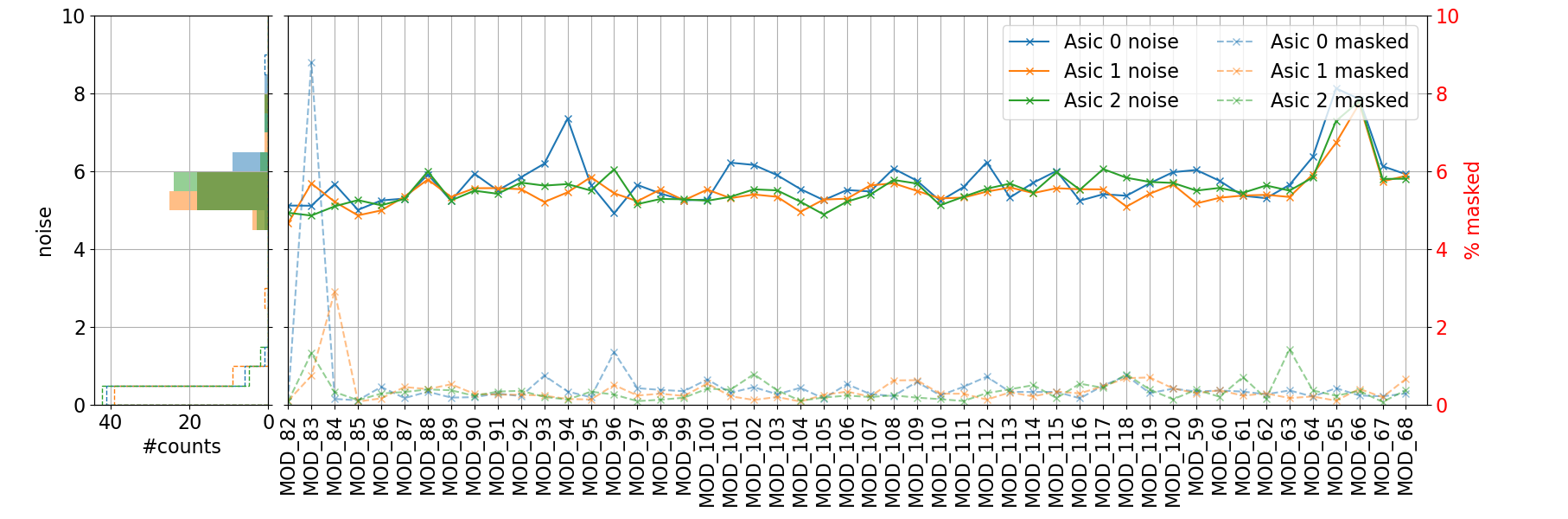}
    \includegraphics[width=0.95\linewidth]{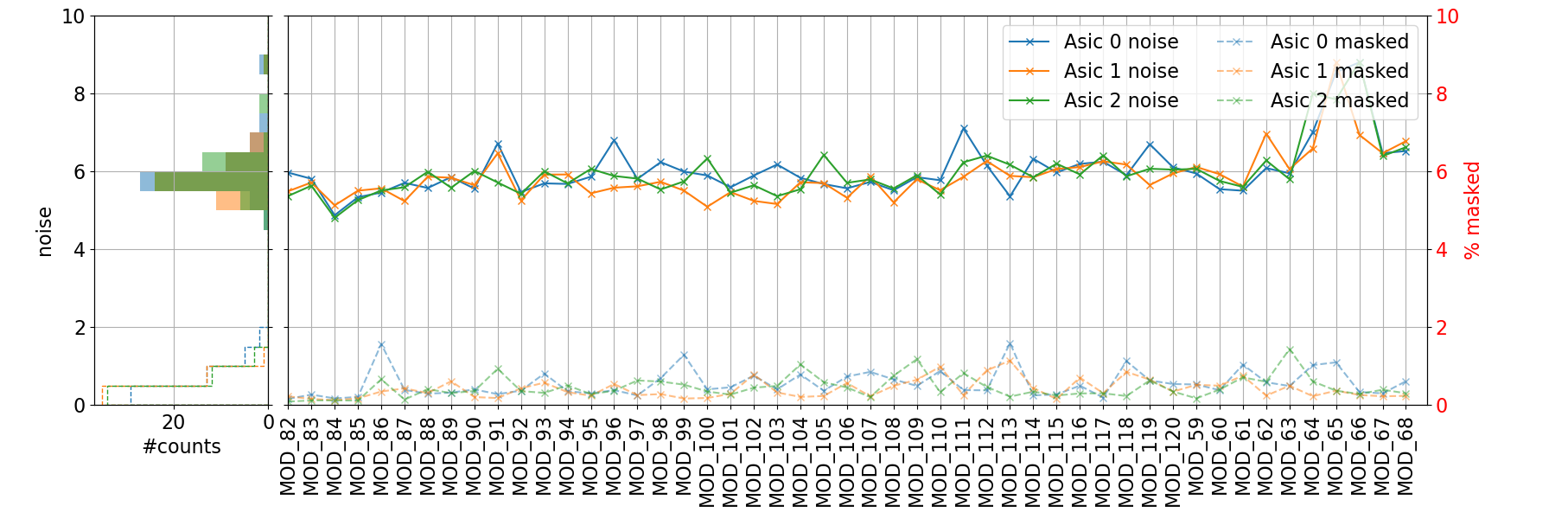}
    \includegraphics[width=0.95\linewidth]{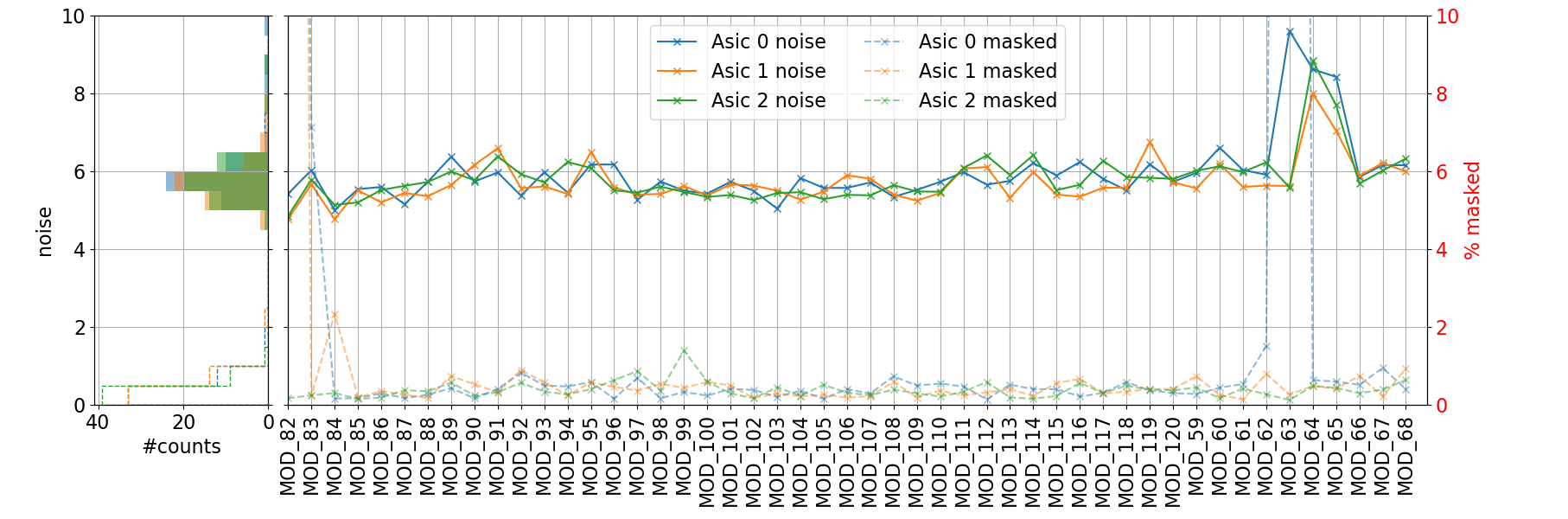}
    \includegraphics[width=0.95\linewidth]{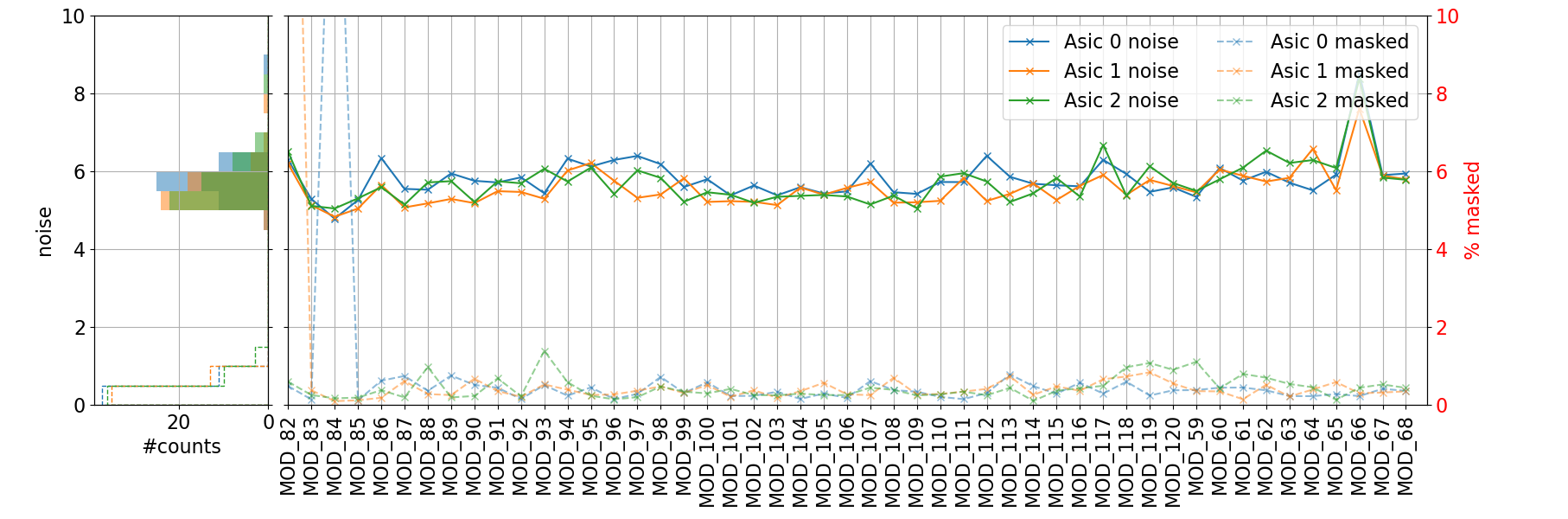}
    \caption{The average noise (in DAC counts, left axis) and the percentage of masked pixels (right axis), as a function of module number across the production. The plots are for the four tiles on each module, arranged from top to bottom as CLI, CSO, NLO, NSI.  In each plot the values for each of the three ASICs are shown.}
    \label{fig:summary_noise}
\end{figure}

The noise and equalisation scans are discussed in Sect.~\ref{sec:equalisation}.  
The average noise is an important parameter for the performance. For reference, a MIP in a 200$\mu$m thick silicon sensor generates about 16,000 electron-hole pairs. The average noise in each of the three ASICs in a tile are shown in Fig.~\ref{fig:summary_noise}, as function of the module number. As illustrated by the trends, the typical average noise in a VeloPix is just below 6\,DAC counts, which corresponds to about 90 e-h pairs. While this excellent performance gives a very high signal:noise ratio at the start of operations, the level of noise is expected to increase with radiation damage. The fraction of masked pixels is also shown in the plots.  The fraction of masked pixels, as measured during the module production, on the modules installed in the experiment is 0.5\%. No appreciable change in the results is observed after the thermal cycling.

\subsubsection{Summary of cooling performance tests}
\begin{figure}[htbp]
    \centering
    \includegraphics[width=0.6\linewidth]{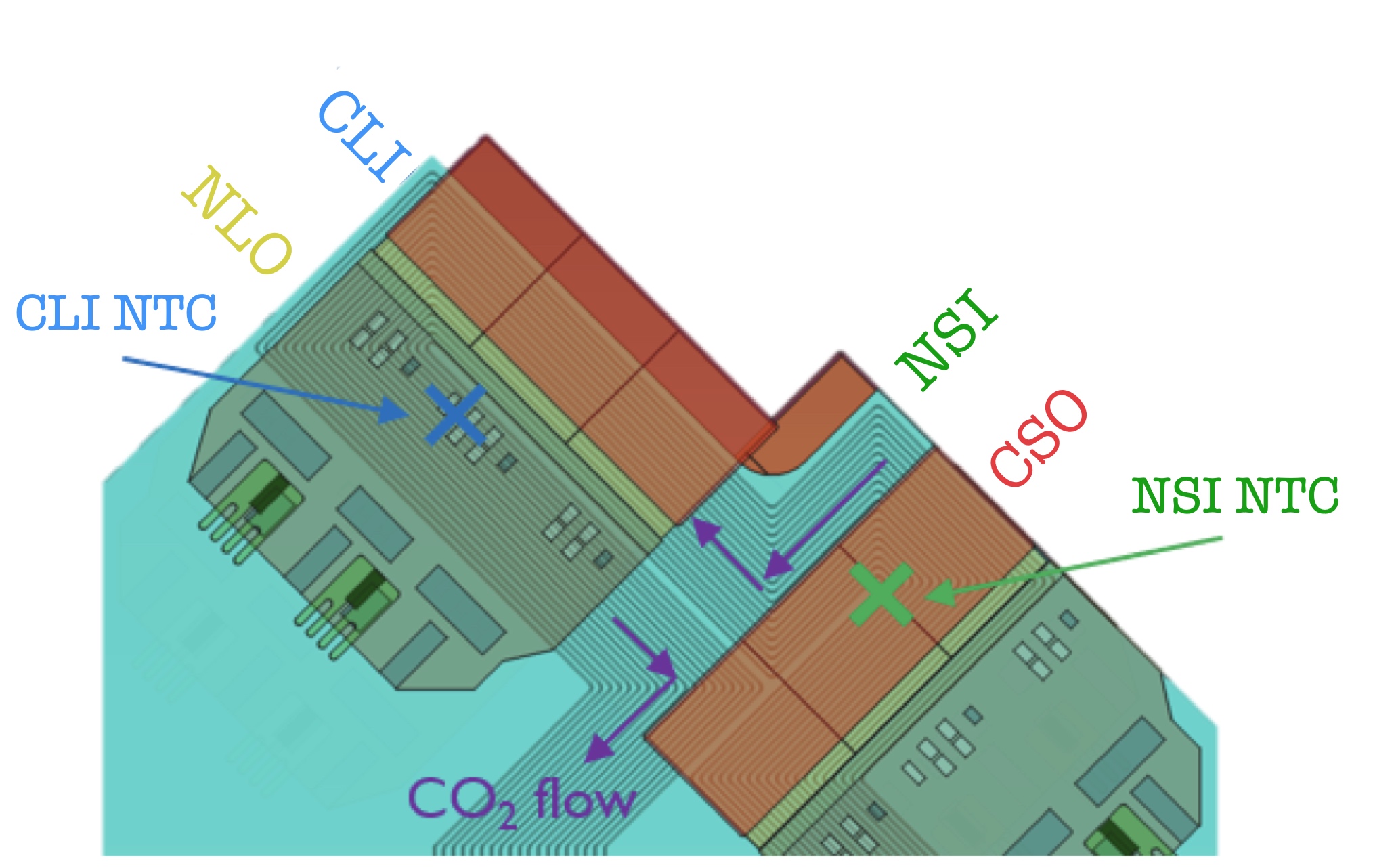}
    \caption{Location of the NTC sensors used to correct the temperature gradient measured in each tile.}
    \label{fig:NTC_location}
\end{figure}

The importance of measuring the cooling performance of each module is discussed in Sect.~\ref{sec:thermal_performance}. The main parameters affecting the thermal behaviour of a tile are the fluidic characteristics of the microchannel substrate and the pressure, flow and temperature of the coolant. Fig.~\ref{fig:cooling_performance} shows an example of the thermal performance of a module and the extrapolation at 26~W is used to evaluate the cooling performance of each VeloPix. In order to compare the behaviour of different modules at slightly different running conditions, a correction is applied to the measured temperatures. The change in temperature of the ASIC between 0\,W and that expected at 26\,W is taken. This is then corrected by subtracting the temperature value of the closest NTC sensor, which corresponds to NTC$_{\rm CLI}$ for tiles CLI and NLO, and to NTC$_{\rm NSI}$ for tiles CSO and NSI, to make a $\mathrm{\Delta T}$ result that is more independent of the operational conditions of the cooling system. The location of these NTC sensors on the front-end hybrids is shown in Fig.~\ref{fig:NTC_location}.  

\begin{figure}[htbp]
    \centering
    \includegraphics[width=0.95\linewidth]{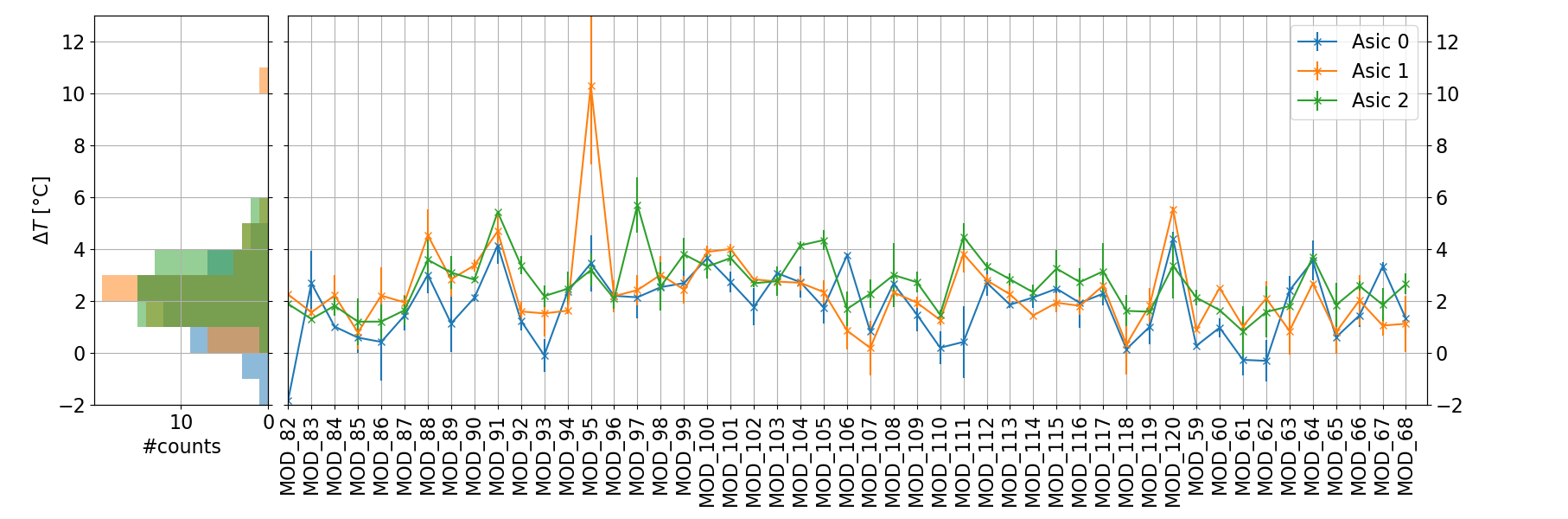}
    \includegraphics[width=0.95\linewidth]{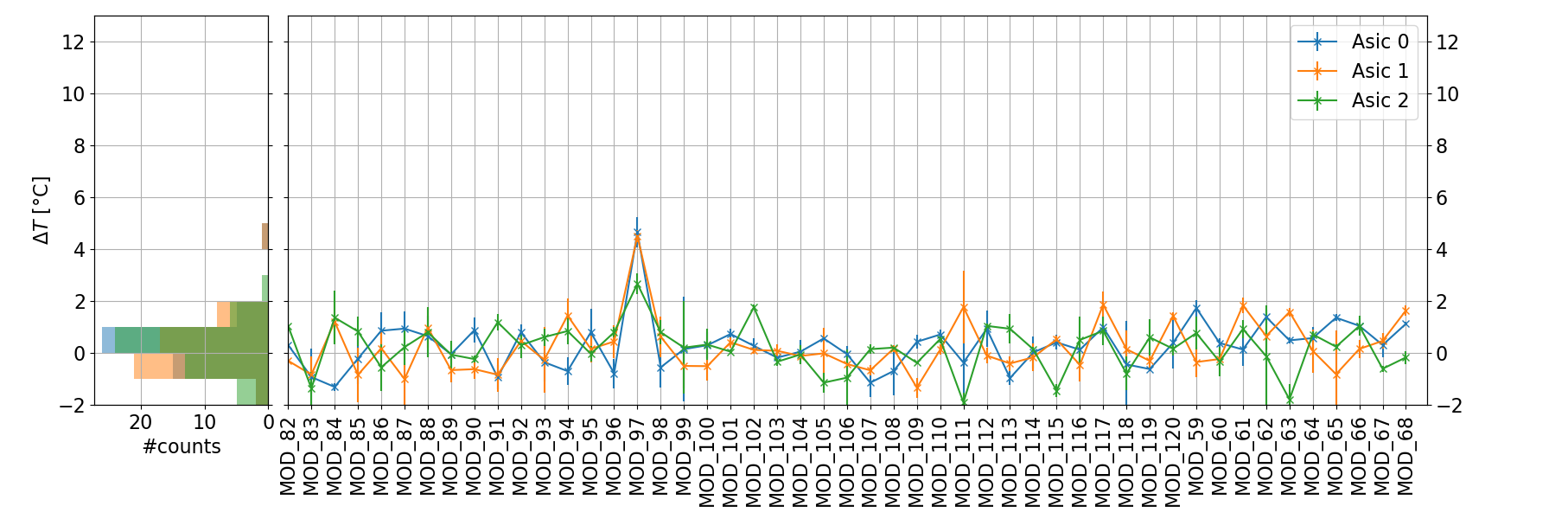}
    \includegraphics[width=0.95\linewidth]{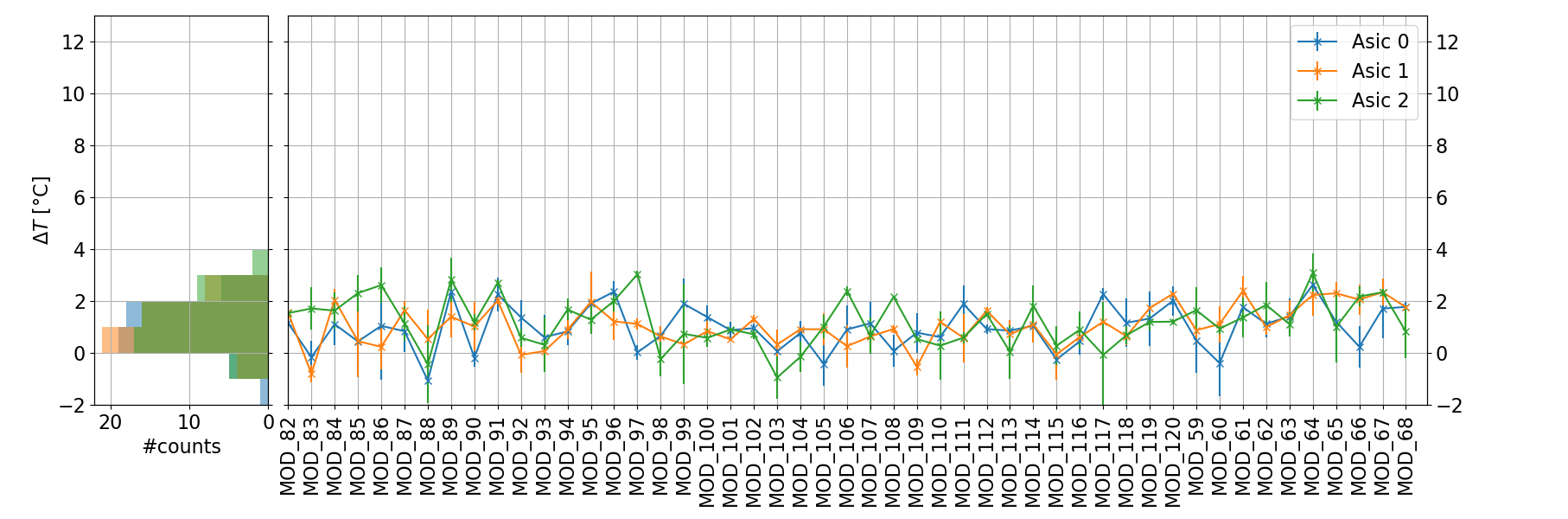}
    \includegraphics[width=0.95\linewidth]{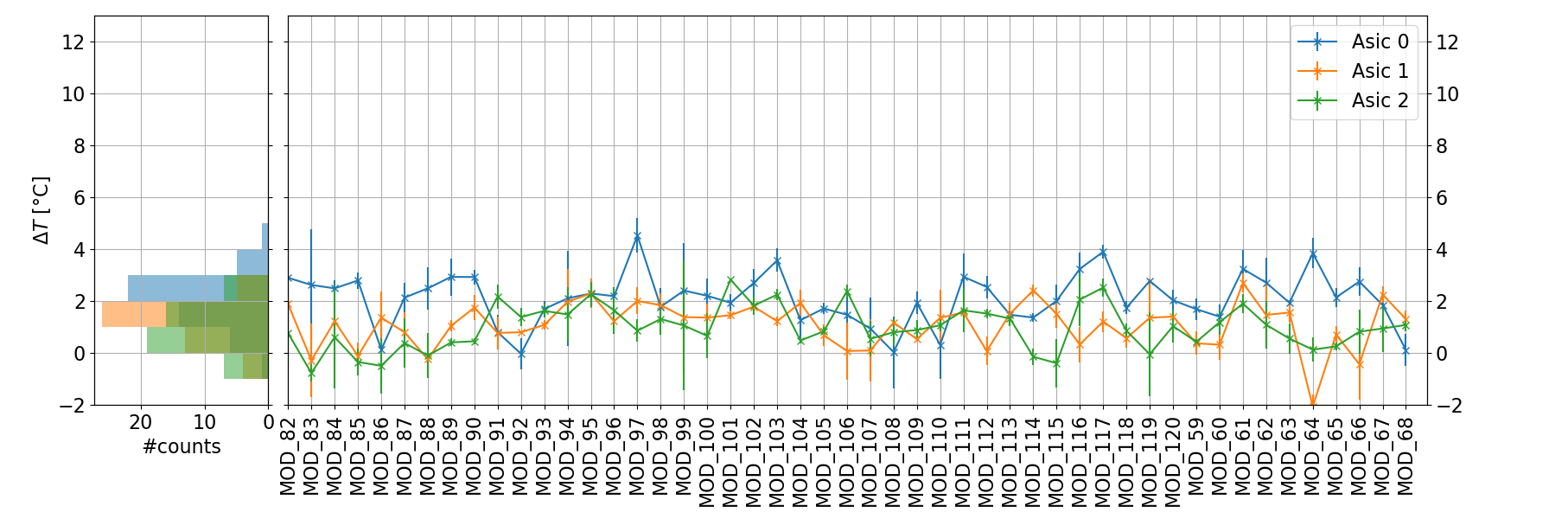}    
    \caption{The cooling performance of the modules (see text), as a function of the module number across the production.  The plots are for the four tiles on each module, arranged from top to bottom as CLI, CSO, NLO, NSI.  In each plot the values for each of the three ASICs are shown.}
    \label{fig:summary_bandgap_corrected}
\end{figure}

The corrected cooling performance results for the modules are shown in Fig.~\ref{fig:summary_bandgap_corrected}. The larger temperature gradient measured on ASIC~1 of tile CLI in module 95 is due to a malfunctioning ASIC which provides a wrong reading of its bandgap voltage. The worse cooling performance shown by module 97 is instead due to its larger glue layer thickness as discussed above in Section~\ref{sec:summary_glue}.

\begin{figure}[htbp]
    \centering
     \includegraphics[width=0.95\linewidth]{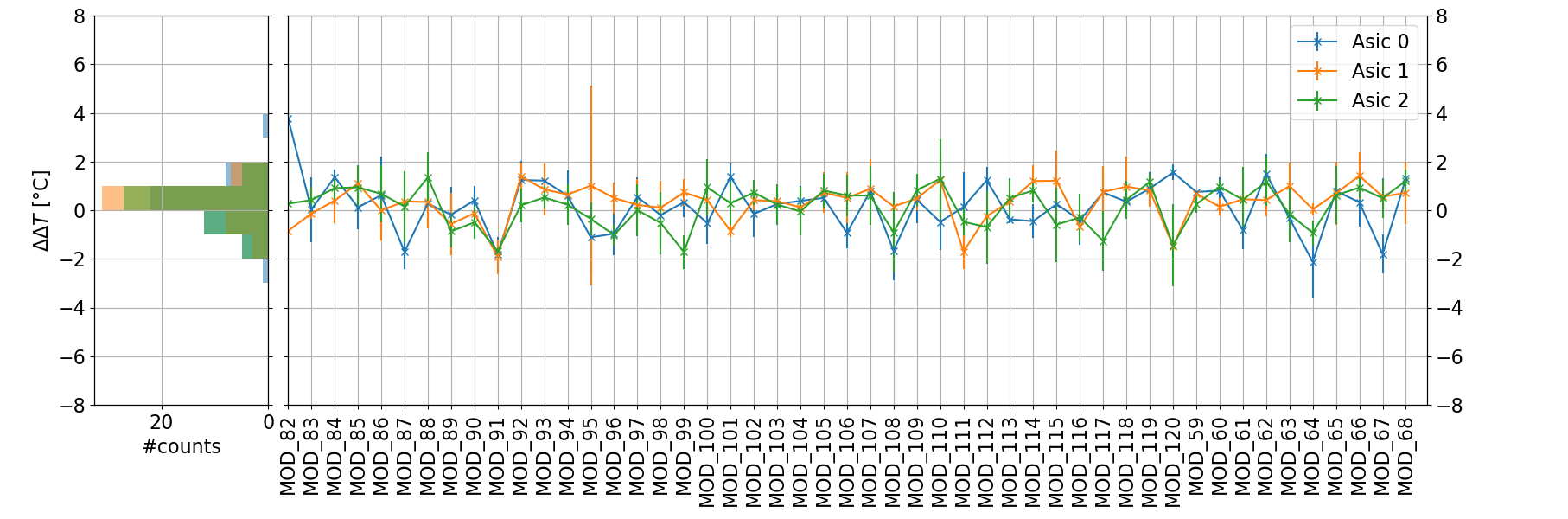}
     \includegraphics[width=0.95\linewidth]{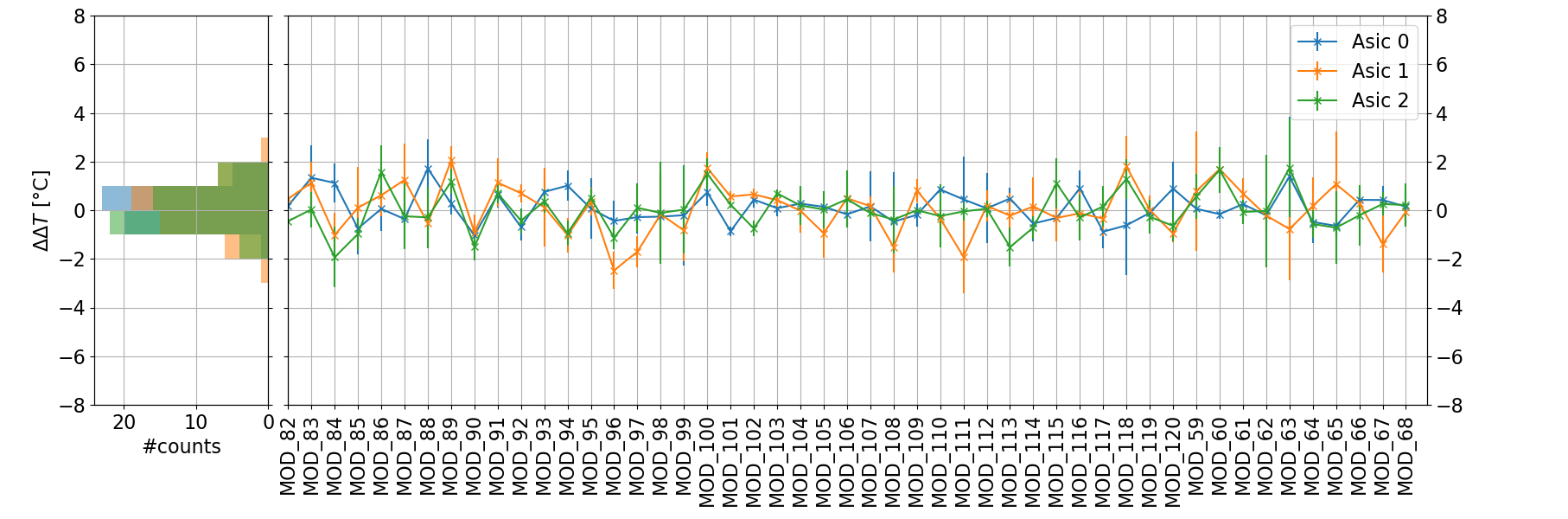}
     \includegraphics[width=0.95\linewidth]{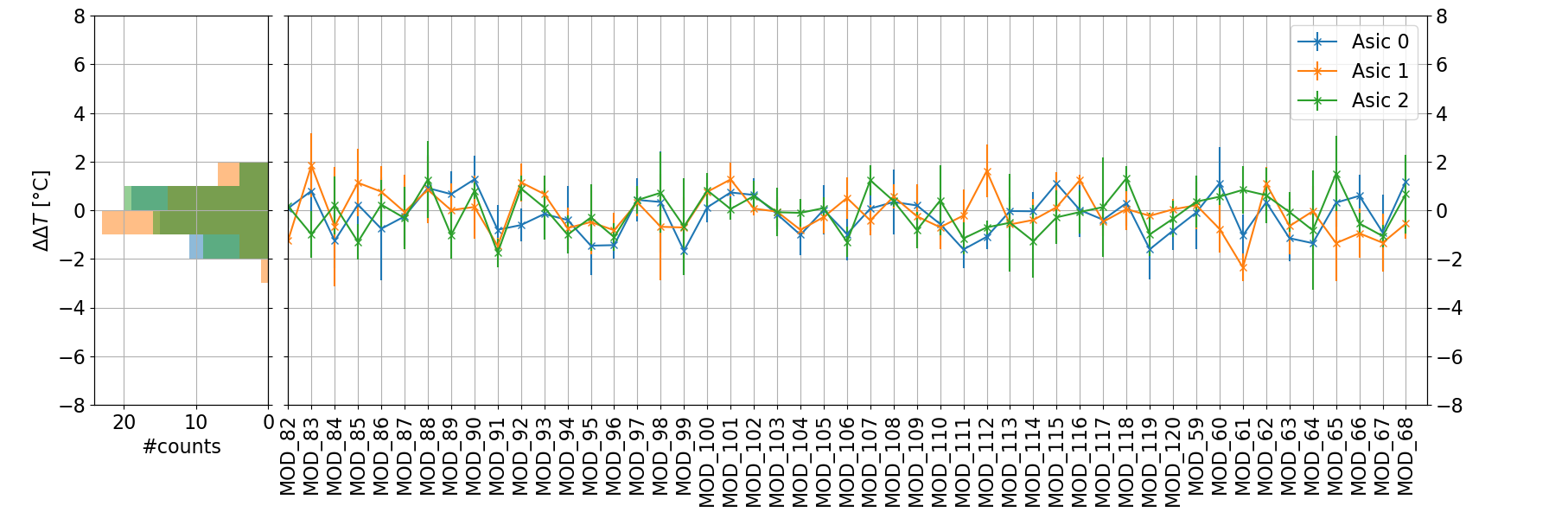}
     \includegraphics[width=0.95\linewidth]{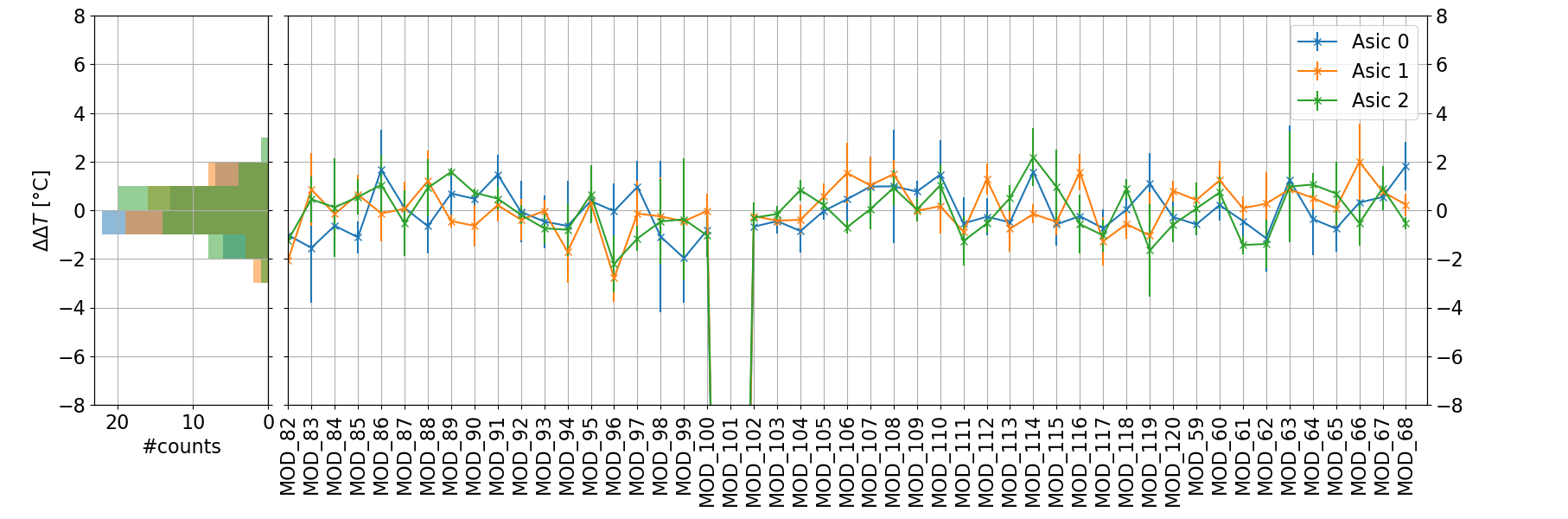}
    \caption{The change in cooling performance of the modules (see text) between the tests before and after thermal cycling, as a function of the module number across the production.  The plots are for the four tiles on each module, arranged from top to bottom as CLI, CSO, NLO, NSI.  In each plot the values that test for a potential degradation in cooling performance each of the three ASICs are shown.}
    \label{fig:summary_bandgap_degradation}
\end{figure}

As for all the other qualification tests, the thermal performance of a module is also evaluated twice, before and after the thermal cycling. The comparison between the two measurements provides an important check for any potential degradation of the glue layer due to thermally-induced mechanical stress, which may impact the cooling performance. The change of the thermal performance ($\mathrm{\Delta \Delta T}$) between the two tests for the modules is shown in Fig.~\ref{fig:summary_bandgap_degradation}. No significant change in the performance is observed.

\begin{figure}[htbp]
    \centering
     \includegraphics[width=0.95\linewidth]{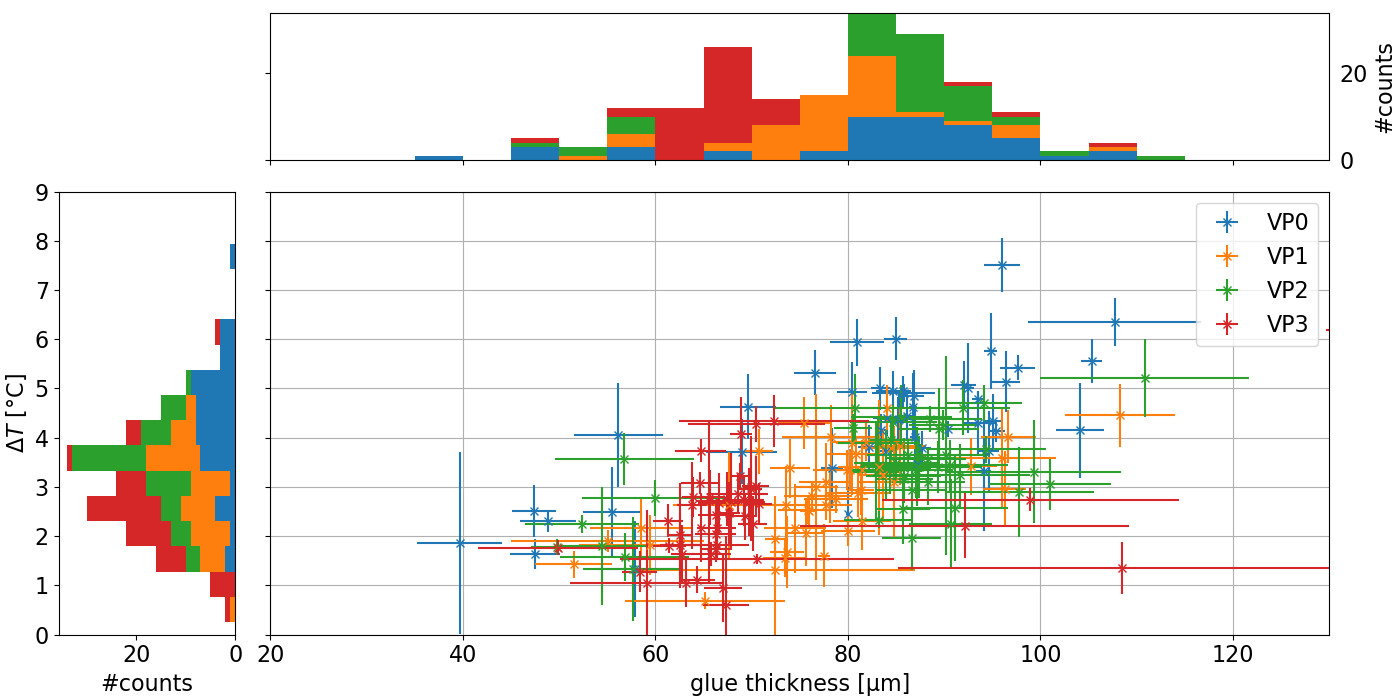}
    \caption{Cooling performance as function of the glue layer thickness for installed modules.}
    \label{fig:summary_bandgap_glue}
\end{figure}

\begin{figure}[htbp]
    \centering
     \includegraphics[width=0.95\linewidth]{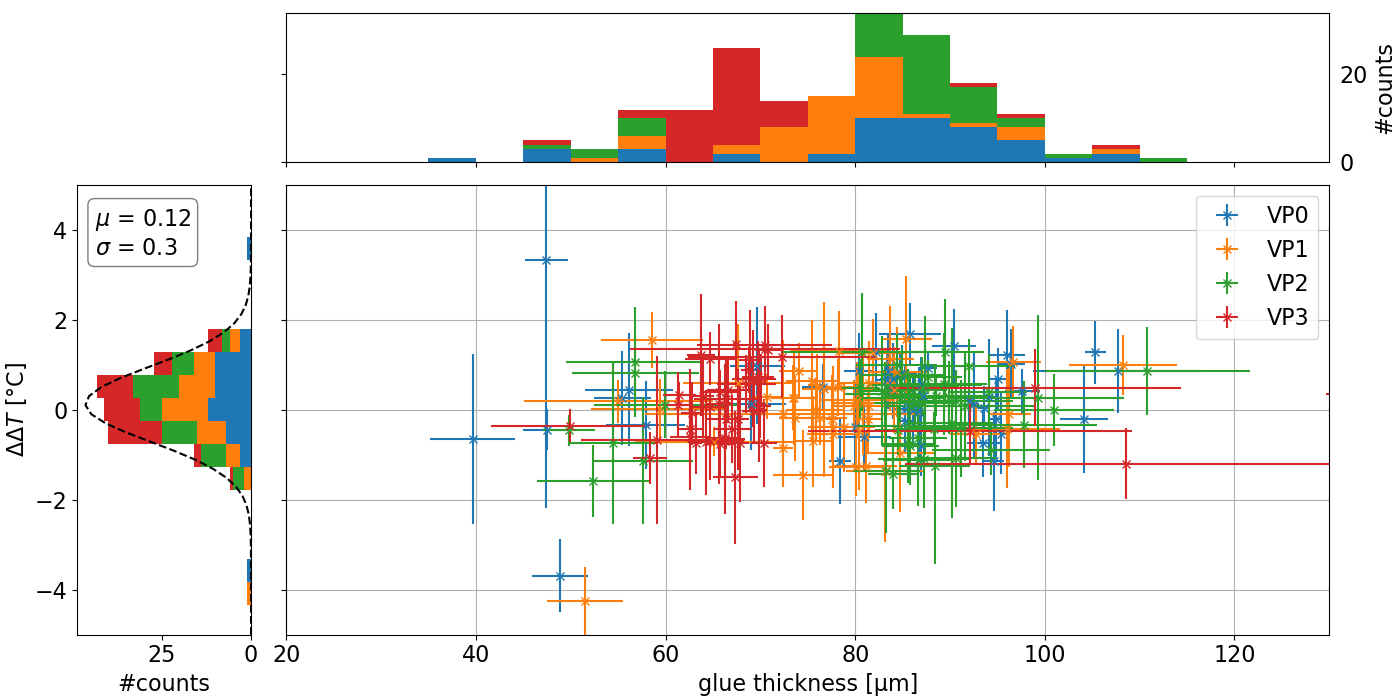}
    \caption{Cooling performance degradation as function of the glue layer thickness for installed modules.}
    \label{fig:summary_bandgap_glue_degradation}
\end{figure}

\begin{figure}[htbp]
    \centering
     \includegraphics[width=0.95\linewidth]{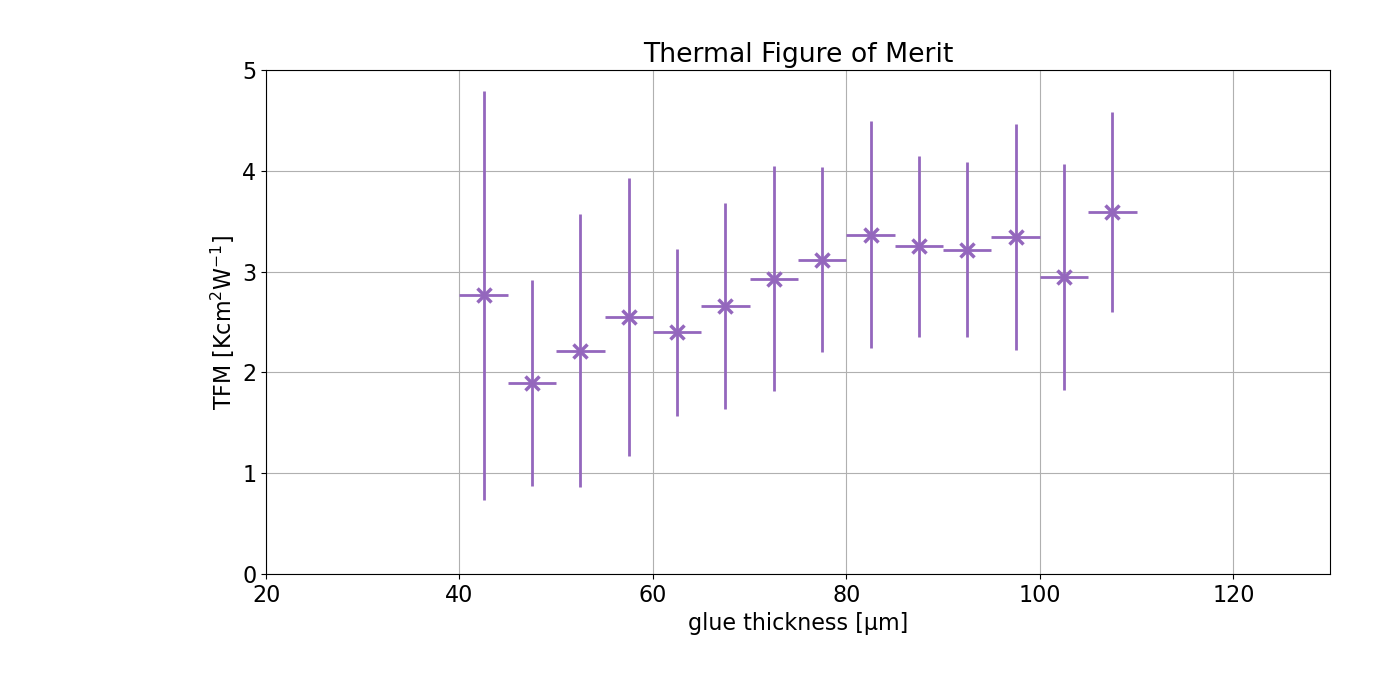}
    \caption{Thermal Figure of Merit as function of the glue layer thickness for installed modules.}
    \label{fig:summary_figure_of_merit}
\end{figure}

The cooling performance is also affected by the glue layer thickness underneath the tile. Fig.~\ref{fig:summary_bandgap_glue} shows this correlation for all the modules. As expected, a thinner glue layer typically leads to a better cooling performance. The adjustment of the assembly setup was such that tile CSO typically had a thinner glue layer than the other tiles. In addition, the change of the cooling performance before and after thermal cycling is shown in Fig.~\ref{fig:summary_bandgap_glue_degradation}, showing no dependence on the glue layer thickness. The missing data in the CSO trend (module 96) and NSI trend (module 101) are due to a damaged trace in the monitoring circuit of the middle ASIC, which prevents all the internal readings to reach the corresponding line in the front-end hybrid. 
The effectiveness of a cooling design can be quantified by the thermal figure of merit (TFM), which is defined as the temperature gradient between the hottest point on the surface of the silicon sensor and the coolant, divided by the thermal density power dissipated by the ASICs and sensors:
\begin{equation}
\nonumber
TFM = \frac{\Delta T_{\rm heater-output}}{\rm power~per~unit~area}.
\end{equation}
For reference, conventional designs that are based on cooling blocks can achieve a TFM of 20~Kcm$^2$W$^{-1}$, which can be improved to values around 12~Kcm$^2$W$^{-1}$ in designs with integrated pipe-structures~\cite{petagna}. For silicon microchannel solutions, the TFM is expected to reach values in the range 1.5-4.2~Kcm$^2$W$^{-1}$~\cite{nomerotski}.  The TFM for the VELO modules is shown in Fig.~\ref{fig:summary_figure_of_merit} as function of the glue layer thickness. The TFM varies between 1.5 and 3.5~Kcm$^2$W$^{-1}$, as expected; for the average glue thickness of 80~$\um$, the TFM is about 3~Kcm$^2$W$^{-1}$, which is the best achieved value in high energy physics experiments.

\subsubsection{Summary of I-V measurements}
\label{sec:summary_iv}
\begin{figure}[htbp]
    \centering
     \includegraphics[width=0.95\linewidth]{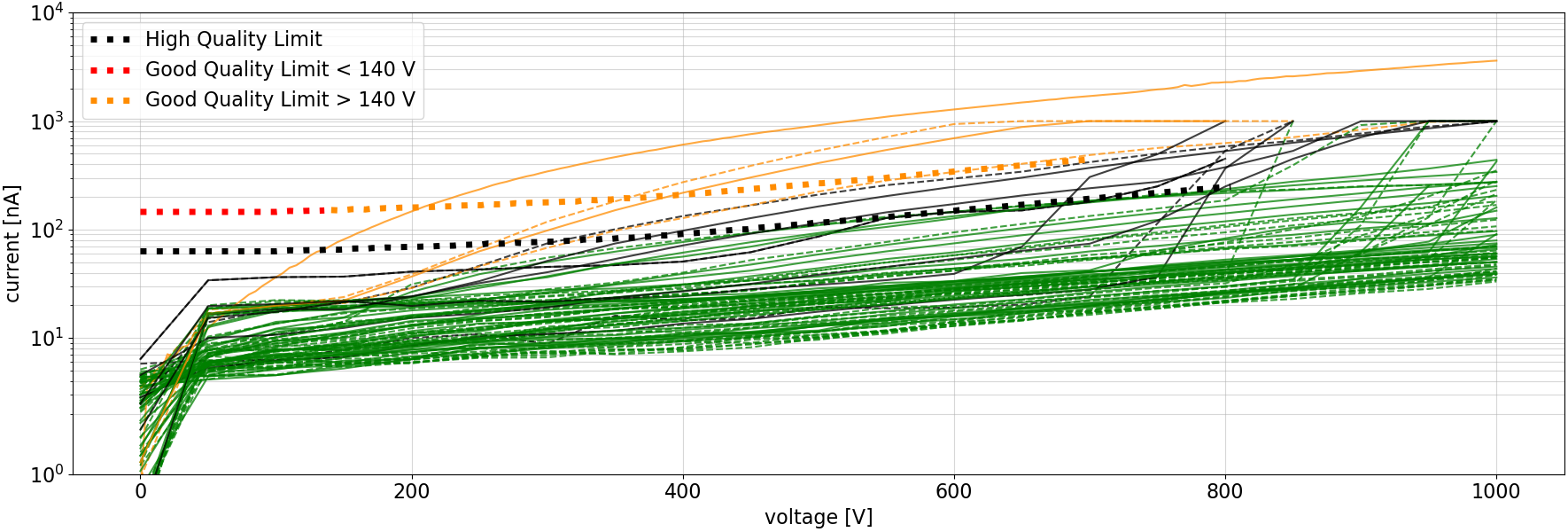}
    \caption{All IV curves for tile NLO. Solid (dotted) lines correspond to the measurements performed before (after) thermal cycling. The black, orange and red dotted lines define the quality grade assigned to the tile.}
    \label{fig:summary_iv_curves}
\end{figure}

The summary of all I-V measurements performed on tile NLO, as an example, is shown in Fig.~\ref{fig:summary_iv_curves}. The I-V characterisation of a tile on the assembled module is performed before (solid lines) and after (dotted lines) thermal cycling. The black, orange and red dotted lines represent the performance limits that define the grading of the tiles, which were defined based on measurements carried out in a probe station when the  tiles were received from the manufacturer. In this plot, the IV curves are coloured depending on their position relative to the performance limits (green in case they remain below them). As shown in the plot, the vast majority of the curves lie well within the high quality limit and therefore no issues are expected from increasing the bias voltage to compensate radiation damage.

\subsection{Module quality}
\label{sec:module_quality}
During module construction, each individual step described in Sect.~\ref{sec:metrology} and Sect.~\ref{sec:validation} is assigned a grade based on the criteria discussed. These grades are then combined together in two ways to assess the overall quality of each module, a final grade and a quality metric.  The final grade (A,B,C,D,F) is used to determine whether the module is installed in the experiment, with the quality metric (0-100) giving further information.

The final grade determines the module quality as good (grades A, B or C), acceptable (grade D) or bad (grade F). Good quality modules are assigned these three different grades (A, B or C) in order to have a finer assessment of their quality. Only modules with the good quality grades are used in the final detector. The position of the module within the VELO base is assigned using both the grade and the quality metric, such that the very highest quality modules are located in the most important regions for tracking.  Modules are assigned a grade D when one functionality is missing, e.g. a data link is not functioning or a VeloPix is dead; these modules are considered to be still available for installation in the detector if needed. Modules with more than just one missing functionality are assigned the grade F and they are not considered for installation in the VELO detector.

\begin{table}[h]
    \caption{Number of production modules assigned each final grading.}
    \label{tab:database_grade}
    \centering
    \begin{tabular}{c | c c c c c}
        Final grade & \textbf{A} & \textbf{B} & \textbf{C} & \textbf{D} & \textbf{F} \\
        \hline
        Number of modules & 24 & 28 & 6 & 3 & 2\\
    \end{tabular}
\end{table}

This approach provides a simple way to assess the overall quality of a module, in terms of its construction quality and performance. 
However, the rules listed above do not allow to distinguish between modules that have obtained the same final grade but have different results in the various tests performed during the assembly, which have different impact on the final performance of the detector.
For this reason, a quality metric was also developed. The quality metric is also based on the grades of the assembly and qualification tests, but it attributes points to the grades of all the individual measurements. In addition, a weighting mechanism allows the performance results to affect the final quality of a module more than the assembly steps. This goal is achieved by splitting the production into three groups: (1) the processes, containing all the assembly steps, (2) the metrology, thermal cycling and thermal degradation tests, and (3) the electrical tests, before and after thermal cycling. The quality number is then defined in a range from 0 to 100, based on the results obtained in these three groups, weighted according to the ratio of 2:3:5 respectively, thus giving more weight to the performance measurements.

\begin{figure}[htbp]
    \centering
    \includegraphics[width=0.99\linewidth]{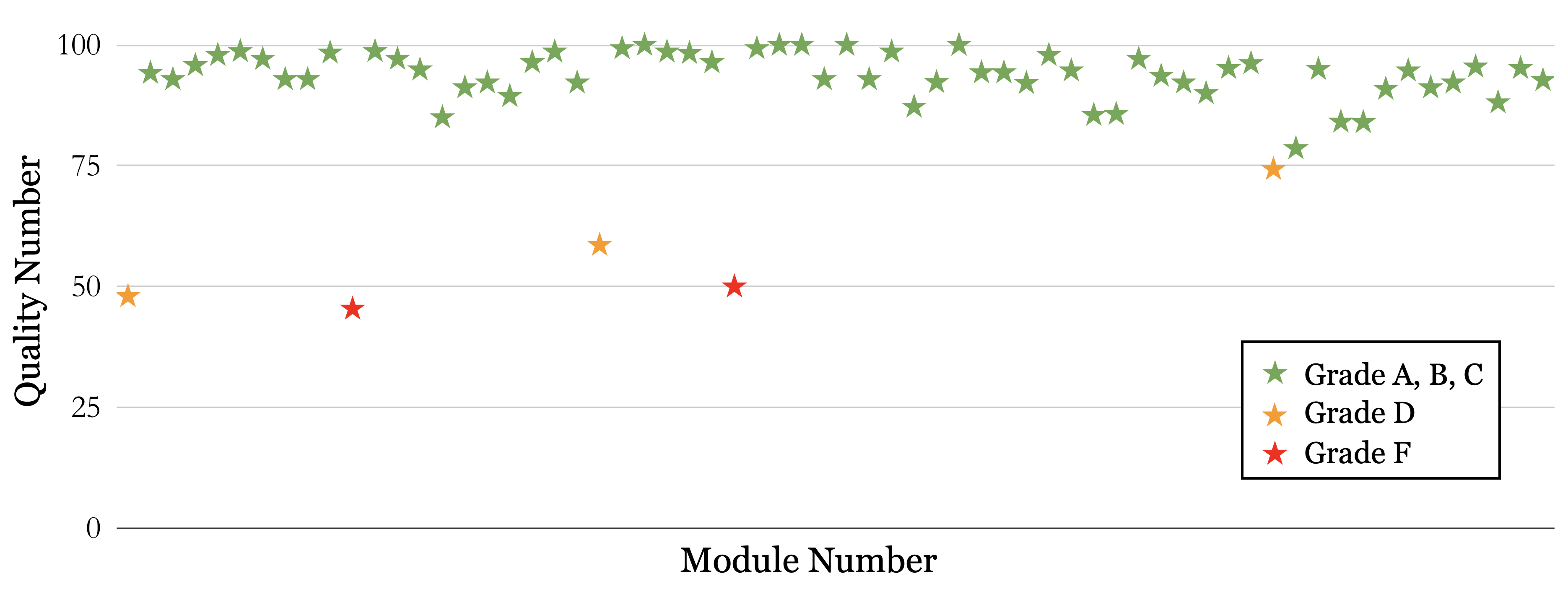}
    \caption{Final grade and quality number of all the 64 production modules.}
    \label{fig:summary_quality_grade}
\end{figure}

With the mechanisms described above, each module is assigned a grade and a quality number. The summary trend for all modules is shown in Fig.~\ref{fig:summary_quality_grade}. Modules graded A, B or C are shown in green, while D and F grade modules are shown in orange and red, respectively. 

%% file: VELO/conclusions.tex
The LHCb VELO Upgrade is one of the most advanced pixel detectors constructed in high-energy physics and the modules have been designed to meet a number of exacting specifications. The modules operate in vacuum and in close proximity to the LHC beams. The closest elements are expected to accumulate an integrated flux of $8 \times 10^{15}$~$n_{\mathrm{eq}}/\mathrm{cm}^{2}$. The VELO will reach a total data rate of 2.85\,Tbit/s with 25\,ns beam collision spacing. Up to 30\,W per module cooling is required and is achieved by evaporative CO$_2$ circulated within microchannels that are etched into the silicon substrate that provides mechanical support to the sensors. The detection element of the modules are 200\mum thick planar n$^+$-on-p sensors bump-bonded onto a 130\,nm technology readout ASIC.

The double-sided module is composed of a significant number of components: bump-bonded ASIC and sensor tiles; microchannel cooling substrate; front-end hybrids;  HV tapes; control (GBTx) hybrid; interconnect tapes; hurdle; bridge; carbon fibre legs; clamp; cooling pipes; LV cables; LV foot connector; aluminium module foot. Each component was designed, optimised and tested for its performance in the expected environment and to allow it to be incorporated during the module assembly.
The module assembly procedure included a number of steps: a bare-module was formed from a foot, rods and microchannel plate; bump-bonded ASIC and sensor tiles were attached with high precision; readout and control hybrids were placed; wire-bonding was performed; and cable attachments made. 

An extensive set of metrology and validation procedures were carried out, many in a realistic environment in a vacuum tank with CO$_2$ cooling and a near-final electrical readout chain. The measurements included: module flatness; tile positions and flatness; glue layer thickness; hybrid positions; wire-bond pull-tests; module displacements with cooling; thermal cycling; IV curves; ASIC communication tests; timing and control tests; data transmission error rate tests; module threshold equalisation and noise level determination. All assembly and characterisation steps were recorded in a custom database and an overall module grading was obtained. 

\begin{figure}[htbp]
    \centering
    \includegraphics[width=0.7\linewidth]{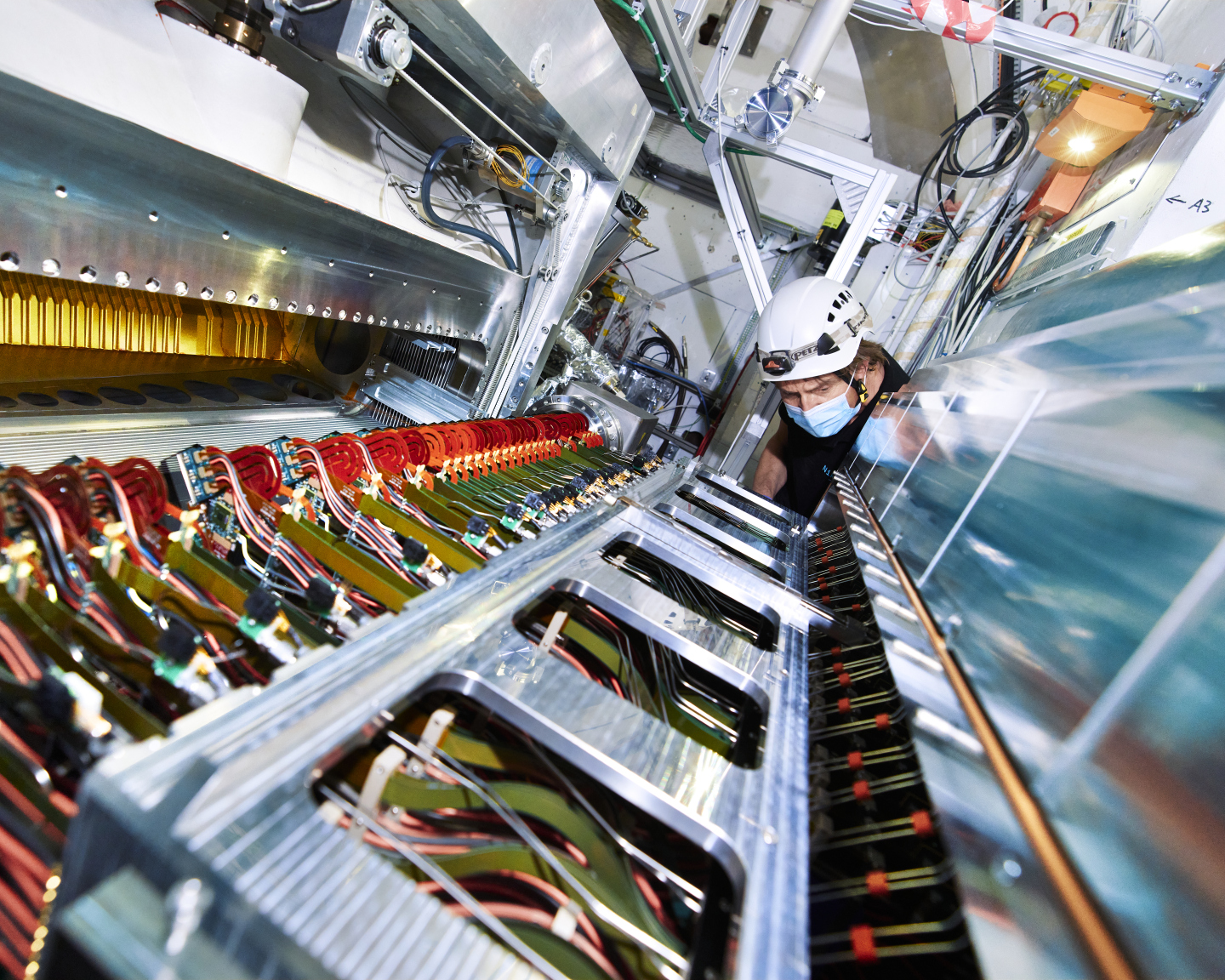}
    \caption{Installation of an LHCb VELO detector half into the vacuum tank in the LHCb cavern.}
    \label{fig:velo-insertion}
\end{figure}

The assembly and quality assurance of sufficient high-quality modules for the full detector was completed in December 2021 at the University on Manchester and Nikhef. The sensor attachment was achieved with an average precision of 21\,\um, more than 99.5\% of all pixels are fully functional, and a thermal figure of merit of $3\,\mathrm{Kcm^{2}W^{-1}}$ was obtained. The modules were integrated into the two detector-halves at the University of Liverpool, with system tests of the halves performed, and then transported to CERN. Following final tests and adjustments, each half was installed in LHCb in time for the start of operation of LHC Run~3. The insertion of one of the detector halves into the VELO tank at LHCb is shown in Fig.~\ref{fig:velo-insertion}. The detector is planned to operate for the next decade during Run~3 and Run~4 of the LHC.  

%% file: VELO/acknowledgements.tex
We thank the technical and administrative staff at the LHCb institutes. We acknowledge support from CERN and from the national agencies: NWO (Netherlands); MNiSW and NCN (Poland); MSHE (Russia); MICINN (Spain); SNSF and SER (Switzerland); STFC and The Royal Society (United Kingdom).